\pdfoutput=1

\documentclass[b5paper,10pt]{book}

\usepackage{phdthesis}
\usepackage[
  paperwidth=17cm
  ,paperheight=24cm
  ,inner=2.6cm
  ,outer=2.3cm
  ,top=2.3cm
  ,bottom=2.3cm
]{geometry}

\ifdefined\isprint
\usepackage[usenames,dvipsnames,table,gray]{xcolor}
\newcommand{\myincludegraphics}[2][]{\includegraphics[#1]{#2-grayscale.pdf}}

\else
\usepackage[usenames,dvipsnames,table]{xcolor}
\newcommand{\myincludegraphics}[2][]{\includegraphics[#1]{#2}}

\fi

\usepackage{bibentry}
\nobibliography*
\usepackage{environ}
\usepackage{dsfont}
\usepackage{times}
\usepackage[hyphens]{url}
\usepackage{tabularx,amsmath}
\usepackage{multirow}
\usepackage[np, autolanguage]{numprint}
\usepackage{booktabs}
\usepackage{amssymb}
\usepackage[square,comma,numbers,sort&compress,sectionbib]{natbib}
\usepackage{wrapfig}
\usepackage{enumerate}
\usepackage{supertabular}
\usepackage{multicol}
\usepackage{setspace}
\usepackage{import}
\usepackage{booktabs} 
\usepackage{graphicx,grffile}
\usepackage{url}
\usepackage{times}
\usepackage{microtype}
\usepackage{amssymb}
\usepackage{multirow}
\usepackage{color}
\usepackage[normalem]{ulem}
\usepackage[english]{babel}
\usepackage[noend]{algorithmic}
\usepackage{paralist}
\usepackage{flushend}
\usepackage{enumerate}
\usepackage{enumitem}
\usepackage{setspace}
\usepackage{textcomp}
\usepackage{listings}
\usepackage{keyval}
\usepackage{rotating}
\usepackage{acronym}
\usepackage{xintfrac}
\usepackage{subcaption}
\usepackage[ruled,ruled,noend,noline,slide]{algorithm2e}
\usepackage{xifthen}
\usepackage[flushleft]{threeparttable}
\usepackage{mathtools}
\usepackage[frozencache]{minted}
\usepackage{lscape}
\usepackage{adjustbox}
\usepackage{xstring}
\usepackage{tikz}
\usepackage{afterpage}
\usepackage{float}
\usetikzlibrary{arrows, arrows.meta, calc, decorations.markings, decorations.pathreplacing, decorations.pathmorphing, fadings, patterns, positioning, positioning, shapes, shapes.misc}
\tikzset{>=latex}
\tikzset{
	cross/.style = {cross out, draw,
					minimum size=2*(#1-\pgflinewidth),
         			inner sep=0pt, outer sep=0pt},
    cross/.default={1pt}
}
\usepackage{tikz-3dplot}
\usepackage{adjustbox}
\usepackage{hyperref} 
\hypersetup{%
pdfborder = {0 0 0},
pdftitle = {Remedies against the Vocabulary Gap in Information Retrieval},
pdfsubject = {PhD Thesis},
pdfkeywords = {semantic matching, query formulation, latent vector spaces, neural networks},
pdfauthor = {Christophe Van Gysel}
}
\usepackage{bookmark}
\usepackage{pdfpages}

\renewcommand*{\backref}[1]{} 
\renewcommand*{\backrefalt}[4]{%
\ifcase #1
\or (Cited on page~#2.)  %
\else 
(Cited on pages~#2.)  
\fi
}

\renewenvironment{table*}[1][]{\begin{sidewaystable}[htbp]}{\end{sidewaystable}\ignorespacesafterend}
\renewenvironment{figure*}[1][]{\begin{sidewaysfigure}[htbp]}{\end{sidewaysfigure}\ignorespacesafterend}

\let\oldincludegraphics\includegraphics
\let\oldfigure\figure
\let\endoldfigure\endfigure
\renewenvironment{figure}[1][ht]{
  \renewcommand{\includegraphics}[2][]{\centering\resizebox{0.80\columnwidth}{!}{\oldincludegraphics[##1]{##2}}}%
  \oldfigure[t!]
  \centering}%
  {\endoldfigure}

\let\oldinput\input
\let\oldtable\table
\let\endoldtable\endtable
\renewenvironment{table}[1][ht]{
  \renewcommand{\input}[1]{\centering\resizebox{0.80\columnwidth}{!}{\oldinput{##1}}}%
  \oldtable[t!]
  \centering}%
  {\endoldtable}

\newcommand{\ChapterRQ}[2][0]{
\newcommand{\RQ}[2]{%
    \newcount\rqid
    \rqid=\numexpr#1+##1\relax
    \begin{description}[topsep=2pt,leftmargin=0.8cm]%
    \item[RQ#2.\the\rqid] ##2%
    \end{description}}

\newcommand{\RQRef}[1]{%
    \newcount\rqid
    \rqid=\numexpr#1+##1\relax
    \textbf{RQ#2.\the\rqid}}

\newcommand{\RQAnswer}[3]{%
    \newcount\rqid
    \rqid=\numexpr#1+##1\relax
    \begin{description}%
    \item[RQ#2.\the\rqid] ##2%
    \end{description} ##3}
}

\newcommand{\currentscope}{}

\newcommand{\ScopeLabels}[1]{
\renewcommand{\currentscope}{#1}
\let\globallabel\label
\newcommand{\scopedlabel}[1]{\globallabel{#1:##1}}
\let\globalref\ref
\newcommand{\scopedref}[1]{\globalref{#1:##1}}
\renewcommand{\label}{\scopedlabel}
\renewcommand{\ref}{\scopedref}
}

\makeatletter
\def\label@in@display{%
    \ifx\df@label\@empty\else
        \@amsmath@err{Multiple \string\label's:
            label '\df@label' will be lost}\@eha
    \fi
    \scopedlabel
}
\makeatother

\makeatletter
\let\oldparagraph\paragraph%
\renewcommand{\paragraph}[1]{\oldparagraph{#1.}}
\makeatother

\renewcommand{\S}{Section~}

\acrodef{IR}{information retrieval}

\acrodef{QCM}{Query Change Model}
\acrodef{DPL}{Direct Policy Learning}

\acrodef{CNN}{Convolutional Neural Network}
\acrodef{SVM}{Support Vector Machine}

\acrodef{TF}{Term Frequency}
\acrodef{IDF}{Inverse Document Frequency}
\acrodef{RE}{Relative Entropy}
\acrodef{QLM}{Query-likelihood Language Model}
\acrodef{d}{Dirichlet}
\acrodef{jm}{Jelinek-Mercer}

\acrodef{PCA}{Principal Component Analysis}
\acrodef{LSI}{Latent Semantic Indexing}
\acrodef{LDA}{Latent Dirichlet Allocation}
\acrodef{SERT}{Semantic Entity Retrieval Toolkit}
\acrodef{LSE}{Latent Semantic Entities}
\acrodef{NVSM}{Neural Vector Space Model}
\acrodef{d2v}{doc2vec}
\acrodef{w2v}{word2vec}
\acrodef{si}{self-information}

\acrodef{MAP}{Mean Average Precision}
\acrodef{NDCG}{Normalized Discounted Cumulative Gain}
\acrodef{MRR}{Mean Reciprocal Rank}
\acrodef{P}{Precision}

\acrodef{LSM}{Latent Semantic Model}
\acrodef{NNLM}{Neural Network Language Model}

\acrodef{NLP}{Natural Language Processing}
\acrodef{ASR}{Automatic Speech Recognition}

\newcommand{\Length}[1]{{\left|#1\right|}}

\newcommand{\Apply}[2]{{#1\left(#2\right)}}
\newcommand{\Prob}[2][P]{\Apply{#1}{#2}}
\newcommand{\CondProb}[3][\Prob]{#1{#2 \mid #3}}

\newcommand{\ApplySquare}[2]{{#1\left[#2\right]}}
\newcommand{\Transpose}[1]{{{#1}^\intercal}}

\newcommand{\LossFn}{L}
\newcommand{\Parameters}{\theta}

\DeclareMathOperator*{\argmin}{arg\,min}
\DeclareMathOperator*{\argmax}{arg\,max}
\DeclarePairedDelimiter{\ceil}{\lceil}{\rceil}

\renewcommand{\log}[1]{\text{log}{\left(#1\right)}}
\renewcommand{\exp}[1]{e^{#1}}

\newcommand{\norm}[1]{\left\lVert{#1}\right\rVert}
\newcommand{\EuclideanNorm}{\norm}
\newcommand{\abs}[1]{|{#1}|}

\newcommand{\FullDocToVec}{doc2vec}
\newcommand{\FullWordToVec}{word2vec}
\newcommand{\Random}{random}
\newcommand{\GraphPCA}{Graph \acs{PCA}}
\newcommand{\LSI}{\acs{LSI}}
\newcommand{\LDA}{\acs{LDA}}
\newcommand{\SERT}{\acs{SERT}}
\newcommand{\LSE}{\acs{LSE}}
\newcommand{\QLM}{\acs{QLM}}

\newcommand{\splitatcommas}[1]{%
  \begingroup
  \begingroup\lccode`~=`, \lowercase{\endgroup
    \edef~{\mathchar\the\mathcode`, \penalty0 \noexpand\hspace{0pt plus 1em}}%
  }\mathcode`,="8000 #1%
  \endgroup
}
\newcommand{\MainFirstRQ}{How to formulate a query from complex textual structures---such as search sessions or email threads---in order to better answer an information need?}
\newcommand{\MainSecondRQ}{Can we learn a latent vector space of retrievable entities that performs well for retrieval?}
\newcommand{\MainThirdRQ}{Can we scale up latent vector spaces to larger entity domains that have less textual content per entity compared to the expert finding setting?}
\newcommand{\MainFourthRQ}{Can we further increase the retrieval effectiveness and scale latent vector spaces up to hundreds of thousands of documents?}

\newif\ifthesis
\thesistrue

\begin{document}

\ifdefined\isprint
\else
\includepdf[fitpaper=false, noautoscale, offset=-9.178cm -0.293cm]{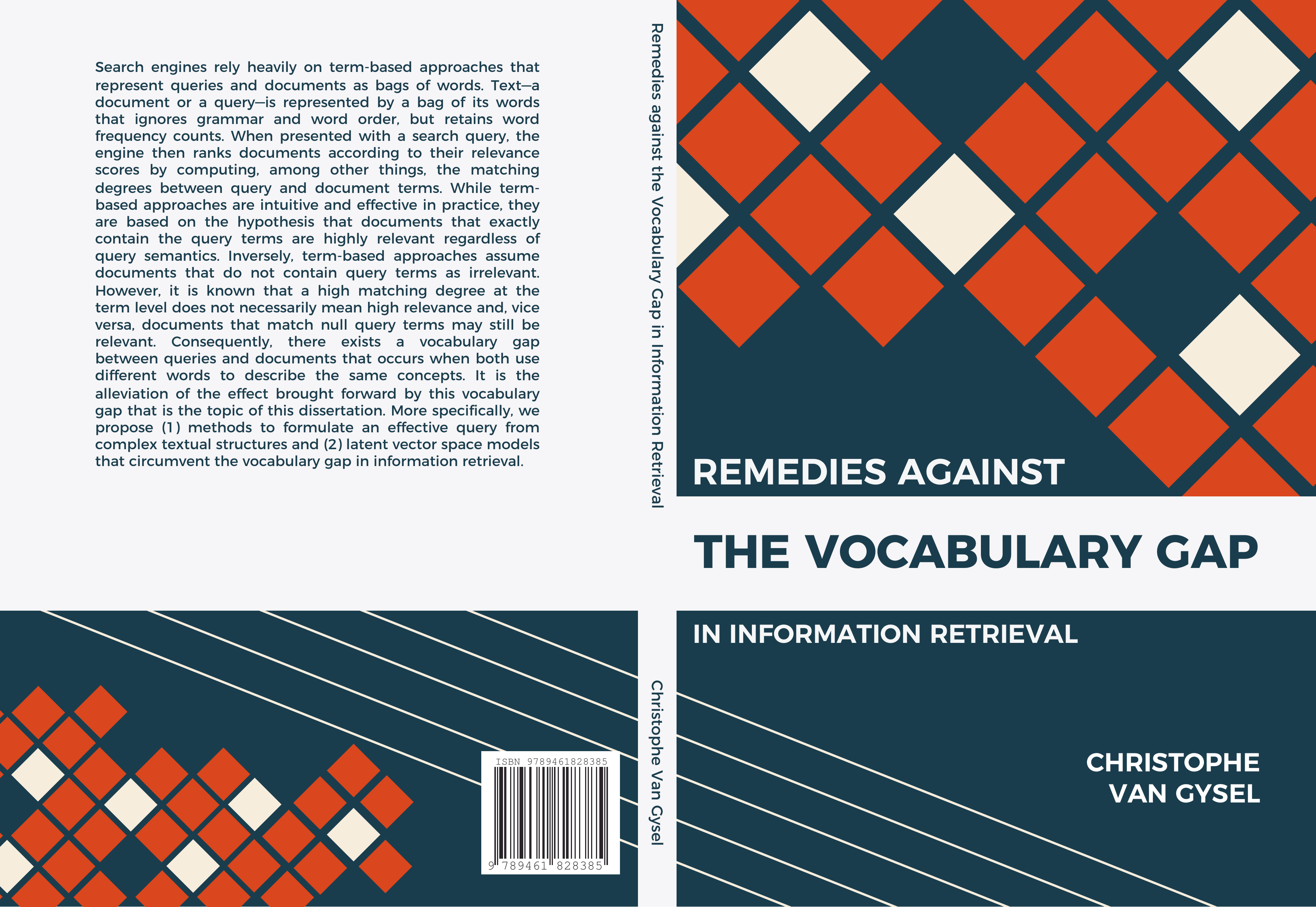}
\fi

\frontmatter

{\pagestyle{empty}
\newcommand{\printtitle}[1][\huge]{%
{#1\bf Remedies against the Vocabulary Gap\\in Information Retrieval \\[0.8cm]
}}

\begin{titlepage}
\par\vskip 2cm
\begin{center}
\printtitle
\vfill
{\LARGE\bf Christophe Van Gysel}
\vskip 2cm
\end{center}
\end{titlepage}

\mbox{}

\begin{center}
This version differs from the official version submitted to the university\\through minor changes in Chapters~\ref{chapter:research-02}~and~\ref{chapter:research-06}.
\end{center}

\newpage\setcounter{page}{1}

\par\vskip 2cm
\begin{center}
\printtitle{}
\par\vspace {4cm}
{\large \sc Academisch Proefschrift}
\par\vspace {1cm}
{\large ter verkrijging van de graad van doctor aan de \\
Universiteit van Amsterdam\\
op gezag van de Rector Magnificus\\
prof.\ dr.\ ir.\ K.I.J. Maex\\
ten overstaan van een door het College voor Promoties ingestelde \\
commissie, in het openbaar te verdedigen in \\
de Agnietenkapel\\
op vrijdag 17 november 2017, te 12:00 uur \\ }
\par\vspace {1cm} {\large door}
\par \vspace {1cm}
{\Large Christophe Jacky Henri Van Gysel}
\par\vspace {1cm}
{\large geboren te Lier, Belgi\"{e}}
\end{center}

\clearpage
\noindent%
\textbf{Promotiecommissie} \\\\
\begin{tabular}{@{}l l l}
Promotor: \\
& Prof.\ dr.\ M.\ de Rijke & Universiteit van Amsterdam \\
Co-promotor: \\
& Dr. E. Kanoulas & Universiteit van Amsterdam \\
Overige leden: \\
& Prof.\ dr.\ B.\ Goethals & Universiteit Antwerpen \\
& Dr.\ K.\ Hofmann & Microsoft Research Cambridge \\
& Dr.\ C.\ Monz & Universiteit van Amsterdam \\
& Prof.\ dr.\ M.\ Welling & Universiteit van Amsterdam \\
& Prof.\ dr.\ M.\ Worring & Universiteit van Amsterdam \\
\end{tabular}

\bigskip\noindent%
Faculteit der Natuurwetenschappen, Wiskunde en Informatica\\

\vfill

\noindent
The research was supported by
the European Community's Seventh Framework Programme (FP7/2007-2013) under
grant agreement nr 312827 (VOX-Pol),
the Google Faculty Research Award scheme
and
the Bloomberg Research Grant program.
\\\\
All content represents the opinion of the author, which is not necessarily shared or endorsed by his respective employers and/or sponsors.
\\\\
Computing resources were provided by the Netherlands Organisation for Scientific Research (NWO) through allocation SH-322-15 of the Cartesius system, the Advanced School for Computing and Imaging (ASCII) by allocation of the Distributed ASCII Supercomputer 4 (DAS-4) system and the Information and Language Processing Systems group.
\bigskip

\noindent
Copyright \copyright~2017 Christophe Van Gysel, Amsterdam, The Netherlands\\
Cover by Samira Abnar and Mostafa Dehghani\\
Printed by Off Page, Amsterdam\\
\\
ISBN: 978-94-6182-838-5\\

\clearpage
}

{\pagestyle{empty}

{
\begin{center}

\noindent
\textbf{Acknowledgements} \\ \vskip-4ex
\end{center}
}

\noindent I never intended to pursue a doctorate degree.
\vskip1ex

\noindent Senior researchers---academics and industry folks alike---kept repeating that a masters degree and an engineering position at a high-end technology giant would be the better career path. After all, postgraduate studies are often described as one of the few remaining forms of modern slavery in the Western world.\footnote{\url{http://www.independent.co.uk/news/education/education-news/postgraduate-students-are-being-used-as-slave-labour-7791509.html}}
Given that this dissertation completes my postgraduate degree, I can only conclude that I'm susceptible to reverse psychology.
\vskip1ex

\noindent At the end of August 2014, I was spending a few months in New York. It was there that I was told that the postgraduate student experience has just one significant factor: the student's advisor. This was only a few days after I had first met with my prospective advisor---Maarten. The postgraduate experience boiled down to the following question: \emph{``Is the student's advisor a lion or a wolf?''}\footnote{\url{https://www.math.ku.edu/~jmartin/fun/grad.html}}
Will your advisor fight for you---as a lion---or misguide you and stab you in the back---as a wolf? After these three years, I can say with certainty that Maarten resembles something closer to a griffin: a mythical lion with wings.
Consequently, it goes without saying that I express my deepest gratitude towards Maarten for his guidance and ideas.
\vskip1ex

\noindent I also thank my co-advisor, Evangelos, for his insightful feedback and the inspiring conversations we shared. Before Evangelos became my co-advisor, I was working closely with Marcel. Marcel, thank you for the guidance during those first few months.
\vskip1ex

\noindent I am honoured to have a graduation committee consisting of very talented researchers from a wide variety of backgrounds. Bart, Christof, Katja, Marcel, Max, thank you for taking the time to read my dissertation and your valuable feedback. Likewise, I thank my paranymphs, Hosein and Rolf, for standing by me during the defence of this dissertation.
\vskip1ex

\noindent I thank the vocabulary gap for being an interesting problem that I could write this dissertation on.
\vskip1ex

\noindent Science cannot exist without collaborations and I thank my co-authors for all the hard work.
Alexey, %
Bart, %
Bhaskar, %
Evangelos, %
Fran{\c{c}}oise, %
Grzegorz, %
Ian, %
Leonid, %
Maarten, %
Marcel, %
Matteo, %
Mostafa, %
Nicola, %
Piotr, %
Roy and %
Tom, %
thank you for discussions, the modelling, the experiments, the writing, the polishing and the publishing.
\vskip1ex

\noindent We all have to start somewhere and I would express my gratitude to the people who helped me during the early days of my post-graduate studies. Thank you, Daan, Manos, Tom and Zhaochun.
\vskip1ex

\noindent Science, and consequently life, would be boring without the social aspect of it. Thanks to my colleagues at ILPS for the insightful discussions, the support and the fun evenings: %
Abdo, %
Adith, %
Aldo, %
Aleksandr, %
Alexey, %
Ana, %
Anna, %
Anne, %
Arianna, %
Artem, %
Bob, %
Boris, %
Chang, %
Chuan, %
Cristina, %
Daan, %
Damien, %
Dan, %
Dat, %
David, %
David, %
Dilek, %
Eva, %
Evgeny, %
Fei, %
Hamid, %
Harrie, %
Hendrik, %
Hendrike, %
Hosein, %
Ilya, %
Isaac, %
Iva, %
Ivan, %
Julia, %
Kaspar, %
Katya, %
Ke, %
Lars, %
Maarten, %
Marlies, %
Marzieh, %
Masrour, %
Mostafa, %
Nikos, %
Praveen, %
Richard, %
Ridho, %
Rolf, %
Shangsong, %
Svitlana, %
Tobias, %
Tom, %
Xiaojuan, %
Xinyi, %
Yaser, %
Zhaochun and %
Ziming.
\vskip1ex

\noindent Petra, thank you for everything. You are invaluable within ILPS.
\vskip1ex

\noindent During the course of my masters studies at the University of Antwerp and my doctoral studies at the University of Amsterdam I gained priceless experience as part of my 7 internships: %
(1 \& 2) Gus, Sam and Sam, with whom I worked at the ads ranking quality team of Google, thank you. It was during my time there that I first started wondering about pursuing a doctoral degree.
(3) Alec, Chad and Christopher, who supervised me at the infrastructure security team at Facebook. Thanks for the hacking.
(4) Thanks to Leonid, Ian and Fran{\c{c}}oise who provided me with my first industrial research experience at the speech team of Google.
(5) Yi, Xiaochuan and Ilya, who motivated me to push the limits of speech recognition, at the language modelling team of Apple.
(6) Bhaskar, Matteo and Nicola, whom I worked with at the query formulation team of Microsoft Bing. Thanks for including me into the team and showing me that research and product development go hand in hand.
(7) Finally, thanks to Sushobhan, Xinran and Jie, at the search team of Snap Inc., who had a \emph{vocabulary gap} that needed to be bridged.
\vskip1ex

\noindent Many friends supported my throughout my doctoral studies. I thank my friends that I met during my journey at the University of Antwerp and the Association of Mathematics, Informatics and Physics. It has been an amazing few years and I know that I should visit more often. Beyond Antwerp, I thank my friends that I met while travelling and participating in foreign exchanges.
\vskip1ex

\noindent I thank my parents, Catherine and Marc, for their eternal support. Without their guidance, I would not be where I am today. My brother, Cedric, thank you for your support and the discussions. My grandparents, Harry, Henri, Jacky, Lisette, thank you for always looking after me. My aunt and uncle, Ingrid and Patrick, thank you for conversations and gatherings. My cousins, Alexandar, Amelie, Laurens, Sarah, thanks for all the good times.
\vskip1ex

\noindent Last, but not least, I thank my girlfriend, Katya, for all her love, support and understanding.

\clearpage
}
\tableofcontents

\mainmatter

\chapter{Introduction}
\label{chapter:introduction}

Search engines heavily rely on term-based approaches according to which queries and documents are represented as bags of words. Text---a document or a query---is represented by a bag of its words that ignores grammar and word order, but retains word frequency counts. When presented with a search query, the engine then ranks documents according to their relevance scores by computing, among other things, the matching degrees between query and document terms. While term-based approaches are intuitive and effective in practice, they are based on the hypothesis that documents that exactly contain the query terms are highly relevant regardless of query semantics. Inversely, term-based approaches assume documents that do not contain query terms to be irrelevant. \citet{Li2014} note that a high matching degree at the term level does not necessarily mean high relevance and, vice versa, documents that match null query terms may still be relevant. Consequently, there exists a \emph{vocabulary gap} between queries and documents that occurs when both use different words to describe the same concepts.

In addition to the crude heuristic that mandates document relevance to be a function of query/document term overlap, the ubiquity of term-based approaches can be explained by efficiency constraints imposed on retrieval engines. The typical concise nature of queries, together with term-based matching, is used to filter out documents without query terms using a specialized data structure called the \emph{inverted index} \citep[p. 129: \S5.3]{Croft2015book}. An inverted index operates in a similar way as a subject index in a reference book. For example, in an encyclopedia, the subject index contains references to pages that discuss a particular subject. Consequently, the reader is redeemed of the cumbersome task of determining the relevance of every page individually w.r.t. her information need. Search engines employ a filter-and-refine strategy where an initial method, typically the lookup in an inverted index, is used to generate a pool of candidate documents \citep[p. 135: \S10.3]{Liu2011}. The pool of candidate documents is subsequently re-ranked using more expensive methods \citep{Matveeva2006nestedranker}.

There are, however, two major drawbacks of the inverted index that are relevant to this dissertation:
\begin{inparaenum}[(1)]
	\item Following up on our earlier discussion, term-based retrieval may incorrectly classify relevant documents that do not contain query terms as irrelevant. This negatively affects the recall of the retrieval results. Given that term-based retrieval is used for an initial filtering of documents, semantic matching re-rankers applied during the refining step of the retrieval pipeline are powerless when it comes to relevant documents whose terms overlap little with the textual query issued by the user.
	\item To alleviate the filtering of false negatives, search engines often automatically reformulate queries by adding terms (i.e., expansion) \citep[p. 199: \S6.2.3]{Croft2015book}. However, query expansion is a double-edged sword as \citet{Grossman1994invertedindex} show that the number of CPU operations and the disk I/O grow superlinearly---while the number of novel discovered relevant documents diminishes rapidly---with the query size. In fact, incautious and overzealous query expansion causes query drift \citep{Mitra1998drift} and generates a candidate document set that is unrelated to the original query \citep[p. 187: \S9.1.6]{Manning2008book}.
\end{inparaenum}

This dissertation directly targets these two pitfalls of the inverted index and approaches the shortcomings of the inverted index from two opposite angles.
\begin{inparaenum}[(1)]
\item In Part~\ref{part:formulation}, we formulate queries from complex and heterogeneous textual structures (search sessions and email threads) so as to fulfill the user's information need. Search sessions consist of a sequence of user interactions (i.e., query reformulations, clicks), search engine responses (i.e., ranked documents in a result page, also referred to as SERP) and are indicative of a complex information need. The task is then to formulate a textual query to satisfy the overall information need that characterizes the session. In the case of email threads, the explicit information need is less straightforward. We focus on a particular case where an incoming email message is a request for content. The task is then to formulate a query that correctly retrieves the appropriate item from a document repository. Overall, the common goal of the research performed in Part~\ref{part:formulation} is to formulate queries that are
\begin{inparaenum}[(a)]
	\item shorter and
	\item exhibit more effective retrieval than taking the full textual structure as a query (i.e., queries/SERPs and full email messages for the session and email domains, respectively).
\end{inparaenum}
\item Query expansion is a slippery slope as adding more terms increases latency and can cause query drift. However, nearest neighbour algorithms can be more efficient than classical term-based retrieval \citep{Boytsov2016knn} and, consequently, may be used as an alternative \citep{Li2014twostage} or in addition \citep{Petrovic2010streaming} to the inverted index. Therefore, Part~\ref{part:latent} is dedicated to the modelling of latent vector spaces for information retrieval (IR). In particular, we focus on retrieval domains where semantic matching is known to be important: entities \citep{Balog2006experts,DeVries2007inex} and news articles \citep{Harman1992tipster}.
\end{inparaenum}


\section{Research outline and questions}
\label{section:introduction:rqs}

\newcommand{\MainRQ}[2]{\begin{description}\item[RQ#1] #2\end{description}}
\newcommand{\MainRQRef}[1]{{\textbf{RQ#1}}}

The broad theme of this dissertation involves fulfilling information needs that are
\begin{inparaenum}[(1)]
	\item embedded within complex textual structures, such as sessions or email threads through \textbf{query formulation}, and
	\item expressed as textual queries that require \textbf{semantic matching} (e.g., informational queries \citep{Broder2002taxonomy}).
\end{inparaenum}
Below, we introduce the high-level research questions answered in the respective chapters. In each chapter, we pose multiple sub-questions whose answers are combined to answer the higher-level questions below.

\subsection{Query formulation}

Information needs are often posed in a form that is more complex than a single short user-formulated textual query. What if we know the user is trying to accomplish a complex task by issuing multiple short queries? At first, the user issues an initial query. Subsequently, at each step the user observes feedback from the search engine and reformulates her request. How can we use the previously-issued queries and the documents observed by the user to improve retrieval effectiveness? In the case of email threads, many emails contain implicit or explicit requests for content. Can we formulate a query from email threads that retrieves the relevant content to be attached? This bring us to our first research question:

\MainRQ{1}{\MainFirstRQ{}}

We perform an analysis on session search logs from TREC and provide empirical insight in the potential of lexical language models (Chapter~\ref{chapter:research-01}). Building upon this insight, we propose a \ac{CNN} that formulates a query from email requests for the task of proactive attachment recommendation (Chapter~\ref{chapter:research-02}).

\subsection{Latent vector spaces}

Entities (e.g., people, products) are often characterized by large bodies of text \citep{Balog2012survey,McAuley2015-1}. Bag of words approaches may degrade for long documents due to their verbosity and scope \citep{Cummins2016longdocuments}. In addition, entity domains require more semantic matching than domains where navigational queries are prevalent (e.g., Web search) as entity-oriented queries often describe the entity rather than searching for a known-item. For example, in expert finding, users describe the expertise of the expert they are searching for instead of the name of the expert herself \citep{TREC2010}. When searching for products on an e-commerce website, users often formulate queries by listing the characteristics of the product they are interested in  \citep{Rowley2000}.

\MainRQ{2}{\MainSecondRQ{}}

We introduce the log-linear model for expert finding and show its effectiveness on three expert finding benchmarks (Chapter~\ref{chapter:research-03}). In expert finding, the user issues a topical query and is presented with a ranking of people. We perform an analysis of the semantic matching performed by the model and give insight in the regularities contained within latent expert representations (Chapter~\ref{chapter:research-04}). For example, we show that experts who operate in similar domains have similar representations.

\bigskip

In \MainRQRef{2} we consider a particular entity domain (i.e., expert finding) that consists of a small number of entities that each have a large body of associated text. In addition, the queries in the expert domain require semantic matching, as users describe the domain of the expert and known-item queries are inherently not part of the expert finding task.
However, the training procedure proposed in response to \MainRQRef{2} is linear in the number of entities. This is impractical as it severely limits the training speed. In the next research question, we address this impracticality and investigate sampling methods to scale up training to large entity spaces.

\MainRQ{3}{\MainThirdRQ{}}

To answer \MainRQRef{3} we scale up the model training to large entity spaces by sampling negative examples (Chapter~\ref{chapter:research-05}). However, its retrieval effectiveness diminishes as the number of retrievable entities increases. Furthermore, how do our latent vector spaces perform on non-entity domains? We address these concerns in \MainRQRef{4}.

\MainRQ{4}{\MainFourthRQ{}}

Our final research question, \MainRQRef{4}, is answered in Chapter~\ref{chapter:research-06} by the introduction of the \acl{NVSM} (\acs{NVSM}). We improve the loss function of our latent vector spaces by incorporating \acs{IR}-specific regularities and evaluate \acs{NVSM} on article retrieval benchmarks from TREC.


\section{Main contributions}
\label{section:introduction:contributions}

The main contributions of this dissertation are listed in this section. Our contributions
come in the form of algorithmic, theoretical, empirical and open-source software contributions. For each contribution, we list the chapter where the contribution is made or, in the case of software packages, where the package was used to generate experimental results.

\newcommand{\Contribution}[2]{\item #1 [\textbf{#2}]}
\newcommand{\SoftwareContribution}[3]{\item #1 ({\footnotesize\url{https://github.com/cvangysel/#1}}) --- #2 [\textbf{#3}]}

\subsection{Algorithmic contributions}

\begin{enumerate}

\Contribution{A frequency-based approach to represent a user's information need in session search.}{Ch.~\ref{chapter:research-01}}

\Contribution{A Convolutional Neural Network (\acs{CNN}) that learns to formulate queries that retrieve attachments in accordance to incoming request messages.}{Ch.~\ref{chapter:research-02}}

\Contribution{Three latent vector space models: the log-linear model for expert finding, Latent Semantic Entities (\acs{LSE}) and the Neural Vector Space Model (\acs{NVSM}).}{Ch.~\ref{chapter:research-03}, \ref{chapter:research-05} and \ref{chapter:research-06}}
\end{enumerate}

\subsection{Theoretical contributions}

\begin{enumerate}[resume]

\Contribution{A proactive email recommendation task, including a methodology for creating pseudo collections for model training and testing.}{Ch.~\ref{chapter:research-02}}
\Contribution{A formal framework for ranking attachments given a ranking over email messages that contain them.}{Ch.~\ref{chapter:research-02}}

\Contribution{Analyses of the time/space complexity of the log-linear model for expert finding and the \acs{NVSM}.}{Ch.~\ref{chapter:research-03} and \ref{chapter:research-06}}

\Contribution{A framework for analysing the quality of different entity representations, independent of the textual matching component.}{Ch.~\ref{chapter:research-05}}
\end{enumerate}

\subsection{Empirical contributions}

\begin{enumerate}[resume]
\Contribution{%
\begin{inparaenum}[(a)]
\item Analysis of the TREC Session Track that shows the prominence of short search sessions within the benchmarks.
\item Investigation of the effectiveness of specialized session search methods compared to our naive frequency-based approach.
\item Investigation of the viability of lexical query matching in session search.
\end{inparaenum}}{Ch.~\ref{chapter:research-01}}

\Contribution{Comparison of different query term formulation methods, their effectiveness and an analysis of the formulated queries and errors.}{Ch.~\ref{chapter:research-02}}

\Contribution{%
\begin{inparaenum}[(a)]
\item Comparison of the log-linear model for expert finding with state-of-the-art retrieval methods, including traditional vector space models and language models.
\item Insight in how the uncertainty of the predictions of the log-linear model can be used to determine the effectiveness of the model.
\item Comparative error analysis between the semantic log-linear model and traditional generative language models that perform exact matching.
\item Insight in the relative strengths of semantic matching and exact matching for the expert retrieval task through an ensemble of the log-linear model and lexical language models.
\end{inparaenum}}{Ch.~\ref{chapter:research-03}}

\Contribution{Insight in the domain regularities (i.e., clusterings, similarity, importance) contained within latent entity representations (with an application to expert finding).}{Ch.~\ref{chapter:research-04}}

\Contribution{%
\begin{inparaenum}[(a)]
\item A parameter sensitivity analysis of \acs{LSE} models, with a focus on representation dimensionality and the amount of word context used to train the model.
\item Comparison of \acs{LSE} models with state-of-the-art latent vector spaces in terms of retrieval effectiveness and according to the quality of the entity representations.
\item Insight in how \acs{LSE} can benefit the retrieval performance in entity-oriented search engines that combine query-independent, lexical and semantic signals in a learning to rank model.
\end{inparaenum}}{Ch.~\ref{chapter:research-05}}

\Contribution{%
\begin{inparaenum}[(a)]
\item Comparison of \ac{NVSM} with lexical language models and state-of-the-art latent vector space models on article retrieval collections.
\item Analysis of the internals of \ac{NVSM} and how it encodes word importance in the word representations.
\item Insight in the judgement bias inherent in TREC test collections.
\item Advice on how to configure the hyperparameters of \ac{NVSM}.
\end{inparaenum}}{Ch.~\ref{chapter:research-06}}
\end{enumerate}

\subsection{Open-source software contributions}

\begin{enumerate}[resume]
\SoftwareContribution{sesh}{a testbed for evaluating session search.}{Ch.~\ref{chapter:research-01}}

\SoftwareContribution{SERT}{the Semantic Entity Retrieval Toolkit that contains implementations of the log-linear model for expert finding and \ac{LSE}.}{Ch.~\ref{chapter:research-03}, \ref{chapter:research-04}, \ref{chapter:research-05} and App.~\ref{chapter:research-08}}

\SoftwareContribution{cuNVSM}{a highly-optimized CUDA implementation of \ac{LSE} and \ac{NVSM} that results in fast training and efficient memory usage.}{Ch.~\ref{chapter:research-06}}

\SoftwareContribution{pyndri}{a Python interface to the Indri search engine.}{Ch.~~\ref{chapter:research-01}, \ref{chapter:research-02}, \ref{chapter:research-06} and App.~\ref{chapter:research-07}}
\end{enumerate}

\noindent Only the software used for the experiments of Chapter~\ref{chapter:research-02} has not been released open-source as it is intellectual property of Microsoft Corporation.


\section{Thesis overview}
\label{section:introduction:overview}

In this section we give an overview of the dissertation and provide recommendations for reading directions.

The chapter you are currently enjoying (Chapter~\ref{chapter:introduction}) gives an introduction to the subject of this dissertation. In addition, the chapter provides an overview of the research questions and contributions. Chapter~\ref{chapter:background} discusses related work for the Chapters~\ref{chapter:research-01}~to~\ref{chapter:research-06} that follow.

Part~\ref{part:formulation} of this dissertation contains research chapters related to query formulation from complex textual structures. In particular, Chapter~\ref{chapter:research-01} investigates the potential of lexical query formulation methods in session search. Chapter~\ref{chapter:research-02} introduces the task of proactive email attachment recommendation. In addition, it proposes a method that formulates a lexical query from an email thread

Part~\ref{part:latent} of this dissertation introduces novel latent vector spaces for Information Retrieval. Chapter~\ref{chapter:research-03} targets a particular instance of the entity ranking task: expert finding. Chapter~\ref{chapter:research-04} performs additional analysis on the learned latent expert representations. The expert finding model is then adapted to larger entity domains (i.e., product search) in Chapter~\ref{chapter:research-05}. Chapter~\ref{chapter:research-06} introduces \acs{NVSM}, an extension to \acs{LSE} that brings qualitative improvements. We evaluate \acs{NVSM} on article retrieval and perform an in-depth analysis of its matching signal. In addition, we investigate the pool bias in off-line test collections and give practical advice on how to configure the hyperparameters of \acs{NVSM}.

We conclude this dissertation in Chapter~\ref{chapter:conclusions} and give directions for future work. Appendices~\ref{chapter:research-07}~and~\ref{chapter:research-08} provide a description of some of the software that was developed as part of this dissertation.

Readers familiar with retrieval models and latent vector spaces may skip the appropriate parts of Chapter~\ref{chapter:background}. Part~\ref{part:formulation} and Part~\ref{part:latent} can be read independently of each other. If time is of the essence, only read Chapters~\ref{chapter:research-02},~\ref{chapter:research-06}~and~\ref{chapter:conclusions}.

\section{Origins}
\label{section:introduction:origins}

We list for each chapter the publications on which it is based. The dissertation is based on, in total, 9 publications \citep{VanGysel2015ESAIR,VanGysel2016experts,VanGysel2016sessions,VanGysel2016products,VanGysel2017pyndri,VanGysel2017expertregularities,VanGysel2017proactive,VanGysel2017nvsm,VanGysel2017sert}.

\newcommand{\PublishedOriginEntry}[7][Chapter]{\item[#1~\ref{#2}] is based on the #7 paper \emph{#3} published at #4 by \citeauthor*{#5} \citep{#6}.\smallskip\\}
\newcommand{\UnpublishedOriginEntry}[7][Chapter]{\item[#1~\ref{#2}] is based on the #7 paper \emph{#3} under review at #4 by \citeauthor*{#5} \citep{#6}.\smallskip\\}

\begin{description}
	\PublishedOriginEntry{chapter:research-01}{Lexical Query Modeling in Session Search}{ICTIR'16}{VanGysel2016sessions}{VanGysel2016sessions}{conference}%
	The naive baseline method was designed by Van Gysel, experiments and analyses were performed by Van Gysel. All authors contributed to the text, Van Gysel did most of the writing.

	\PublishedOriginEntry{chapter:research-02}{Reply With: Proactive Recommendation of Email Attachments}{CIKM'17}{VanGysel2017proactive}{VanGysel2017proactive}{conference}%
	The research was performed during a research internship at Microsoft Bing in London. The task was proposed by Mitra, Venanzi and Cancedda. The model was designed by Van Gysel and was inspired by ideas of Mitra. Some parts of the model were inspired through suggestions by Venanzi, Cancedda and Rosemarin. Kukla and Rosemarin helped by providing data. Grudzien performed additional analysis. Van Gysel did most of the writing, with the help of Mitra. Venanzi, Cancedda and Rosemarin also contributed to the text.

	\PublishedOriginEntry{chapter:research-03}{Unsupervised, Efficient and Semantic Expertise Retrieval}{WWW'16}{VanGysel2016experts}{VanGysel2016experts,VanGysel2015ESAIR}{conference}%
	The model was designed by Van Gysel, experiments and analyses were performed by Van Gysel. All authors contributed to the text, Van Gysel did most of the writing.

	\PublishedOriginEntry{chapter:research-04}{Structural Regularities in Expert Vector Spaces}{ICTIR'17}{VanGysel2017expertregularities}{VanGysel2017expertregularities}{conference}%
	Experiments and analyses were performed by Van Gysel. All authors contributed to the text, Van Gysel did most of the writing.

	\PublishedOriginEntry{chapter:research-05}{Learning Latent Vector Spaces for Product Search}{CIKM'16}{VanGysel2016products}{VanGysel2016products}{conference}%
	The model was designed by Van Gysel, experiments and analyses were performed by Van Gysel. All authors contributed to the text, Van Gysel did most of the writing.

	\UnpublishedOriginEntry{chapter:research-06}{Neural Vector Spaces for Unsupervised Information Retrieval}{TOIS}{VanGysel2017nvsm}{VanGysel2017nvsm}{journal}%
	The model was designed by Van Gysel, experiments and analyses were performed by Van Gysel. All authors contributed to the text, Van Gysel did most of the writing.

	\PublishedOriginEntry[Appendix]{chapter:research-07}{Pyndri: a Python Interface to the Indri Search Engine}{ECIR'17}{VanGysel2017pyndri}{VanGysel2017pyndri}{conference}%
	The software was implemented by Van Gysel. The software upon which Pyndri builds, Indri \citep{Strohman2005indri}, was contributed by the Lemur project. All authors contributed to the text, Van Gysel did most of the writing.

	\PublishedOriginEntry[Appendix]{chapter:research-08}{\acl{SERT}}{NeuIR'17}{VanGysel2017sert}{VanGysel2017sert}{workshop}%
	The software was implemented by Van Gysel. All authors contributed to the text, Van Gysel did most of the writing.

\end{description}

Work performed as part of this dissertation also contributed to and benefited from insights gained through research that led to the following publications:%
\newcommand{\ContributedEntry}[1]{\item \bibentry{#1}}%
\begin{itemize}
	\ContributedEntry{VanGysel2014thesis}
	\ContributedEntry{VanGysel2015politicalpresence}
	\ContributedEntry{VanGysel2015garbagemodeling}
	\ContributedEntry{VanGysel2017rankreduced}
	\ContributedEntry{Kenter2017nn4ir}
\end{itemize}


\chapter{Background}
\label{chapter:background}

In this chapter, we discuss the background for the research presented in this dissertation. The related work concerning Part~\ref{part:formulation} and Part~\ref{part:latent} is covered in Section~\ref{chapter:background:formulation} and Section~\ref{chapter:background:latent}, respectively. We assume that the reader is familiar with the basic principles underlying modern information retrieval, as can be found in, e.g., \citep{Croft2015book,Manning2008book}.

{
\section{Query formulation}
\label{chapter:background:formulation}

We first cover work related to application domains: session search (\S\ref{chapter:background:related:sessions}), proactive information retrieval (\S\ref{chapter:background:related:proactive}) and email (\S\ref{chapter:background:related:email}).

{
\newcommand{\entity}{item}
\newcommand{\entities}{items}

\subsection{Session search}
\label{chapter:background:related:sessions}

Many complex information seeking tasks, such as planning a trip or buying a car, cannot sufficiently be expressed in a single query \citep{Hassan2014}. These multi-faceted tasks are exploratory, comprehensive, survey-like or comparative in nature \citep{Raman2013} and require multiple search iterations to be adequately answered \citep{Kotov2011}. \citet{Donato2010} note that 10\% of the user sessions of a web search engine (more than 25\% of query volume) consists of such complex information needs.

The TREC Session Track \citep{TREC2011-2014} created an environment for researchers ``to test whether systems can improve their performance for a given query by using previous queries and user interactions with the retrieval system.'' The track's existence led to an increasing number of methods aimed at improving session search. \citet{Yang2015} introduce the \acl{QCM} (\acs{QCM}), which uses lexical editing changes between consecutive queries in addition to query terms occurring in previously retrieved documents, to improve session search. They heuristically construct a lexicon-based query model for every query in a session. Query models are then linearly combined for every document, based on query recency \citep{Yang2015} or document satisfaction \citep{Luo2014-TREC2014,TREC2014}, into a session-wide lexical query model. However, there has been a clear trend towards the use of supervised learning \citep{TREC2014,Yang2015,Luo2015} and external data sources \citep{Guan2012,Luo2014}. \citet{Guan2012} perform lexical query expansion by adding higher-order n-grams to queries by mining document snippets. In addition, they expand query representations by including anchor texts to previously top-ranked documents in the session. \citet{TREC2014} expand document representations by including incoming anchor texts. \citet{Luo2015} introduce a linear point-wise learning-to-rank model that predicts relevance given a document and query change features. They incorporate document-independent session features in their ranker. The use of machine-learned ranking and the expansion of query and document representations is meant to address a specific instance of a wider problem in information retrieval, namely the query document mismatch \citep{Li2014}. In Chapter~\ref{chapter:research-01} of this dissertation, we analyse the session query logs made available by TREC and compare the performance of different lexical query modelling approaches for session search.

\subsection{Proactive information retrieval}
\label{chapter:background:related:proactive}

\emph{Zero-query} search---or proactive IR---scenarios have received increasing attention recently \cite{Allan2012frontiers}. However, similar approaches have also been studied in the past under other names, such as \emph{just-in-time} \cite{Rhodes2000just, Rhodes1996remembrance, Rhodes2000margin, Rhodes1997wearable}, query-free \cite{Hart1997query} or anticipatory \cite{Liebling2012anticipatory, Budzik1999watson} IR. According to \citet{Hart1997query}, the goal of the proactive retrieval system is to surface information that helps the user in a broader task. While some of these works focus on displaying contextually relevant information next to Web pages \cite{Rhodes1996remembrance, Rhodes2000margin, Crabtree1998adaptive, Budzik1999watson, Maglio2000suitor} or multimedia \citep{Odijk2015dynamic}, others use audio cues \cite{Mynatt1998designing, Sawhney2000nomadic} or signals from other sensors \cite{Ryan1999enhanced, Rhodes1997wearable} to trigger the retrieval.
In more recent years, proactive IR systems have re-emerged in the form of intelligent assistant applications on mobile devices, such as Siri, Google Now and Cortana. The retrieval in these systems may involve modelling repetitive usage patterns to proactively show concise information cards \cite{Shokouhi2015queries, Song2016query} or surface them in response to change in user context such as location \cite{Benetka2017anticipating}.
\citet{Hart1997query}, \citet{Budzik1999watson} and \citet{Liebling2012anticipatory} propose to proactively formulate a query based on the user's predicted information need.

In contrast to previous work on proactive contextual recommendation, we formulate a query to retrieve attachable items to assist users with composing emails instead of supplying information to support content or triggering information cards in mobile assistants. We propose a novel proactive retrieval task for email attachment recommendation in Chapter~\ref{chapter:research-02} of this dissertation.

\subsection{Predictive models for email}
\label{chapter:background:related:email}

Email overload is the inability to effectively manage communication due to the large quantity of incoming messages \citep{Whittaker1996overload}. \citet{Grevet2014overload} find that work email tends to be overloaded due to outstanding tasks or reference emails saved for future use. \citet{Ai2017emailsearch} find that \numprint{85}\% of email searches are targeted at retrieving known items (e.g., reference emails, attachments) in mailboxes. \citet{Horvitz1999principles} argues that a combination of two approaches---%
\begin{inparaenum}[(1)]
	\item providing the user with powerful tools, and
	\item predicting the user's next activity and taking automated actions on her behalf
\end{inparaenum}%
---is effective in many scenarios. Modern email clients may better alleviate email overload and improve user experience by incorporating predictive models that try to anticipate the user's need and act on their behalf.

Missing attachments in email generates a wave of responses notifying the sender of her error. \citet{Dredze2006forgotattachment} present a method that notifies the user when a file should be attached before the email is sent. \citet{DiCastro2016youvegotmail} proposed a learning framework to predict the action that will be performed on an email by the user, with the aim of prioritizing actionable emails. \citet{Carvalho2005collective} classify emails according to their speech acts. \citet{Graus2014recipientrecommendation,Qadir2016activity} recommend recipients to send an email message to. \citet{Kannan2016smart} propose an end-to-end method for automatically generating email responses that can be sent by the user with a single click. %
In Chapter~\ref{chapter:research-02} of this dissertation, we introduce a neural network architecture that learns to formulate a query to predict items to attach to an email reply.

\subsection{Query formulation and reformulation}
\label{chapter:background:related:formulation}

The task we study in Chapter~\ref{chapter:research-02}, i.e., the task of contextual recommendation of attachable \entities{} by means of query formulation, has not received much attention. However, there is work on query extraction from verbose queries and query construction for related patent search. Similar to our work in Part~\ref{part:formulation}, the methods below consider the search engine as a black box.

\subsubsection{Prior art search}
\label{chapter:background:related:patents}

Establishing novelty is an important part of the patenting process. Patent practitioners (e.g., lawyers, patent office examiners) employ a search strategy where they construct a query based on a new patent application in order to find prior art. However, patents are different from typical documents due to their length and lack of mid-frequency terms \citep{Lupu2013patentsurvey}.

Automated query generation methods have been designed to help practitioners search for prior art. \citet{Xue2009transformingpatents} use TF-IDF to generate a ranking of candidate query terms, considering different patent fields, to rank similar patents. In later work \citep{Xue2009querygenerationpatents}, they incorporate a feature combination approach to further improve prior art retrieval performance. Alternative term ranking features, such as relative entropy \citep{Mahdabi2011buildingqueriespriorart}, term frequency and log TF-IDF \citep{Cetintas2012querygenerationpriorart}, have also been explored. \citet{Kim2011automaticbooleanquerysuggestion} suggest boolean queries by extracting terms from a pseudo-relevant document set. \citet{Golestan2015patent} find that an interactive relevance feedback approach outperforms state-of-the-art automated methods in prior art search.

\subsubsection{Improving verbose queries}
\label{chapter:background:related:verbosequeries}

\citet{Bendersky2008discoveringkeyconcepts} point out that search engines do not perform well with verbose queries \citep{Balasubramanian2010webqueryreductions}. \citet{Kumaran2009reducinglongqueries} propose a sub-query extraction method that obtains ground truth by considering every sub-query of a verbose query and cast it as a learning to rank problem. \citet{Xue2010verbosequeries} use a Conditional Random Field (CRF) to predict whether a term should be included. However, inference using their method becomes intractable in the case of long queries. \citet{Lee2009querytermsranking} learn to rank query terms instead of sub-queries with a focus on term dependency. \citet{Huston2010evaluatingverbosequeryprocessing} find that removing the stop structure in collaborative question answering queries increases retrieval performance. \citet{Maxwell2013compactquerytermselection} propose a method that selects query terms based on a pseudo-relevance feedback document set. \citet{Meij2009learningsemanticquery} identify semantic concepts within queries to suggest query alternatives. Related to the task of improving verbose queries is the identification of important terms \citep{Zhao2010termnecessity}. \citet{He2004scs} note that the use of relevance scores for query performance prediction is expensive to compute and focus on a set of pre-retrieval features that are strong predictors of the query's ability to retrieve relevant documents. \citet{Arguello2016improvespokenqueries} apply query subset selection on spoken queries. \citet{Nogueira2017reformulationrl} apply reinforcement learning to query reformulation. See \cite{Gupta2015verboseyqueriessurvey} for an overview on information retrieval with verbose queries.

In Chapter~\ref{chapter:research-01}, we investigate how terms appearing in web search engines user sessions (e.g., query terms, terms in result page snippets) can be used to reformulate user queries in order to improve retrieval effectiveness. Chapter~\ref{chapter:research-02} introduces a neural network that formulates a query from email requests for the task of proactive attachment recommendation.
}
\section{Latent semantic models}
\label{chapter:background:latent}

We first cover related work on Latent Semantic Models (\acs{LSM})---the subject of Part~\ref{part:latent}---in Section~\ref{chapter:background:related:latent}, followed by the entity retrieval task---the application domain of Chapters~\ref{chapter:research-03} to \ref{chapter:research-05}---in Section~\ref{chapter:background:related:entities}. Finally, we review work from neural language modelling---which inspired the latent vector spaces introduced in this dissertation---in Section~\ref{chapter:background:related:nlm}, and neural information retrieval.

\subsection{Latent semantic models for information retrieval}
\label{chapter:background:related:latent}

The mismatch between queries and documents is a critical challenge in search \citep{Li2014}. Latent Semantic Models (LSMs) enable retrieval based on conceptual content, instead of exact term matches. Especially relevant to this dissertation is the class of unsupervised latent semantic models. We distinguish between count-based approaches (\S\ref{chapter:background:related:latent:count}) and approaches where representations are learned using neural networks. The latter class of methods is split between methods that combine pre-trained word embeddings (\S\ref{chapter:background:related:latent:pretrained}) from neural language models (\S\ref{chapter:background:related:nlm}) and representations that are learned from scratch (\S\ref{chapter:background:related:latent:scratch}) specifically for the task at hand.

\subsubsection{Count-based approaches}
\label{chapter:background:related:latent:count}

Latent Semantic Indexing (LSI) \citep{Deerwester1990lsi} and probabilistic LSI (pLSI) \citep{Hofmann1999} were introduced in order to mitigate the mismatch between documents and queries \citep{Li2014}. \citet{Blei2003} proposed Latent Dirichlet Allocation (LDA), a topic model that generalizes to unseen documents.

\subsubsection{Combining pre-trained embeddings}
\label{chapter:background:related:latent:pretrained}

\citet{Vulic2015monolingual} are the first to aggregate word embeddings learned with a context-predicting distributional semantic model (DSM); query and document are represented as a sum of word embeddings learned from a pseudo-bilingual document collection with a Skip-gram model.
\citet{Kenter2015shorttext} extract features from embeddings for the task of determining short text similarity. \citet{Zuccon2015nntm} use embeddings to estimate probabilities in a translation model that is combined with traditional retrieval models (similar to \citep{Ganguly2015generalizedlm,Tu2016semantic}). \citet{Zamani2016queryexpansion,Zamani2016embeddinglm} investigate the use of pre-trained word embeddings for query expansion and as a relevance model to improve retrieval. \citet{Guo2016wordtransport} introduce the Bag-of-Word-Embeddings (BoWE) representation where every document is represented as a matrix of the embeddings occurring in the document; their non-linear word transportation model compares all combinations of query/document term representations at retrieval time. They incorporate lexical matching into their model by exactly comparing embedding vector components for specific terms (i.e., specific terms occurring in both document and query are matched based on the equality of their vector components, contrary to their lexical identity).

\subsubsection{Learning from scratch}
\label{chapter:background:related:latent:scratch}

The methods discussed above incorporate features from neural language models. The recent deep learning revival, however, was due to the end-to-end optimization of objectives and representation learning \citep{LeCun1998gradient,Krizhevsky2012net,Sutskever2014seq2seq} in contrast to feature engineering or the stacking of independently-estimated models. The following neural methods learn representations of words and documents from scratch. \citet{Salakhutdinov2009} introduce semantic hashing for the document similarity task. \citet{Le2014} propose \FullDocToVec{}, a method that learns representations of words and documents. \citet{Ai2016doc2veclm} evaluate the effectiveness of \FullDocToVec{} representations for ad-hoc retrieval, but obtain dissappointing results that are further analysed in \citep{Ai2016doc2vecanalysis}. In Part~\ref{part:latent} of this dissertation, we introduce three novel models that learn document representations from scratch.

\subsection{Entity retrieval}
\label{chapter:background:related:entities}

Around 40\% of web queries \cite{Pound2010objectretrieval} concern entities. Entity-oriented queries express an information need that is better answered by returning specific entities as opposed to documents \cite{Balog2011entitytrack}. The entity retrieval task is characterized by a combination of (noisy) textual data and semi-structured knowledge graphs that encode relations between entities \citep{Dietz2016tutorial}. Entity and document retrieval \citep[p.~224]{Balog2012survey} are closely related as performance of the latter can greatly impact that of the former \citep{Macdonald2008}. 

\subsubsection{Expert retrieval}
\label{chapter:background:related:experts}

Early expert retrieval systems were often referred to as expert locator and expertise management systems~\citep{Maybury2006}. These database systems typically relied on people to self-assess their expertise against a predefined set of topics~\citep{McDonald2000}, which is known to generate unreliable results~\citep{BecerraFernandez2000}.

With the introduction of the P@NOPTIC system~\citep{Craswell2001}, and later the TREC Enterprise track~\citep{TREC2010}, there has been an active research interest in automated expertise profiling methods. It is useful to distinguish between \emph{profile-based} methods, which create a textual representation of a candidate's knowledge, and \emph{document-based} methods, which represent candidates as a weighted combination of documents. The latter generally perform better at ranking, while the former is more efficient as it avoids retrieving all documents relevant to a query~\citep[p.~221]{Balog2012survey}.

There has been much research on generative probabilistic models for expert retrieval \citep{Petkova2006,Fang2007}. Such models have been categorized in candidate generation models \citep{Cao2005}, topic generation models \citep{Balog2006experts,Balog2009lmframework} and proximity-based variants \citep{Serdyukov2008,Balog2009lmframework}. Of special relevance to us are the unsupervised profile-centric (Model~1) and document-centric (Model~2) models of \citet{Balog2006experts}, which focus on raw textual evidence without incorporating collection-specific information (e.g., query modelling, document importance or document structure). Supervised discriminative models \citep{Fang2010,Moreira2011,Sorg2011} are preferred when query-candidate relevance pairs are available for training. Unlike their generative counterparts these models have no issue combining complex and heterogeneous features (e.g., link-based features, document importance features, etc.); they resemble Learning to Rank (L2R) methods for document retrieval \citep{Balog2012survey,Liu2011}. However, a lack of training data may greatly hinder their applicability \citep[p.~179]{Balog2012survey}. Beyond unsupervised generative and supervised discriminative approaches, there are graph-based approaches based on random walks \citep{Serdyukov2008-2} and voting-based approaches based on data fusion~\citep{MacDonald2006}. \citet{Demartini2009expertspaces} propose a vector space-based method for the entity ranking task; their framework extends vector spaces operating on documents to entities. Closely related to expert finding is the task of expert profiling, of which the goal is to describe an expert by her areas of expertise \cite{Balog2007similarexperts}, and similar expert finding \citep{Balog2007similarexperts}; see \cite{Balog2012survey} for an overview. In Chapter~\ref{chapter:research-03} of this dissertation we introduce a log-linear model that learns the relations between experts and words in an unsupervised manner from scratch. Compared to generative language models that perform term-based matching, our model contributes a complementary semantic matching signal. We show that an unsupervised example of the term-based lexical methods and our methods performs best.

\subsubsection{Product retrieval}
\label{chapter:background:related:products}

Product search engines are an important source of traffic in the e-commerce market \citep{Jansen2006}. Specialized solutions are needed to maximize the utilization of these platforms. \citet{Nurmi2008} note a discrepancy between buyers' shopping lists and how retail stores maintain information. They introduce a grocery retrieval system that retrieves products using shopping lists written in natural language. Product resolution \citep{Balog2011-2} is an important task for e-commerce aggregation platforms, such as verticals of major web search engines and price comparison websites. \citet{Duan2013} propose a probabilistic mixture model for the attribute-level analysis of product search logs. They focus on structured aspects of product entities, while in this work we learn representations from unstructured documents. \citet{Duan2013-2} extend the language modelling approach to product databases by incorporating the ability to condition on specification (e.g., lightweight products only). They note that while languages such as SQL can be used effectively to query these databases, their use is difficult for non-experienced end users. \citet{Duan2015} study the problem of learning query intent representation for structured product entities. In Chapter~\ref{chapter:research-05} of this dissertation we introduce a latent vector space model that learns representations of entities, words and the connection in between. We show how our latent vector space contributes a complementary matching signal that can be incorporated in product search engines.

\subsubsection{Representation learning}
\label{chapter:background:related:entityreprs}

Part~\ref{part:latent} of this dissertation covers the learning of entity representations. There already exists some work in this area. \citet{Bordes2011} leverage structured relations captured in Knowledge Bases (KB) for entity representation learning and evaluate their representations on the link prediction task. Our approach has a strong focus on modelling the language of all entities collaboratively, without the need for explicit entity relations during training. \citet{Zhao2015} employ matrix factorization methods to construct low-dimensional continuous representations of entities, categories and words for determining similarity of Wikipedia entities. They employ a word pair similarity evaluation set and only evaluate on pairs referring to Wikipedia entities; they learn a single semantic space for widely-differing concepts (entities, categories and words) of different cardinalities and make extensive use of an underlying Knowledge Graph (KG) to initialize their parameters.

In Part~\ref{part:latent} of this dissertation, we learn low-dimensional representations of words, entities and documents. Entity representations are the subject of Chapter~\ref{chapter:research-03} (people) and Chapter~\ref{chapter:research-05} (products). The representations learned for people in Chapter~\ref{chapter:research-03} are analysed in Chapter~\ref{chapter:research-04}. In Chapter~\ref{chapter:research-06}, we learn representations of news article documents.

\subsection{Neural language modelling}
\label{chapter:background:related:nlm}

Large-vocabulary neural probabilistic language models for modelling word sequence distributions have become very popular recently \citep{Bengio2003,Mnih2007,Mnih2009}. These models learn continuous-valued distributed representations for words, also known as embeddings \citep{Mnih2013,Mikolov2013word2vec,Pennington2014}, in order to fight the curse of dimensionality and increase generalization by introducing the expectation that similar word vectors signify semantically or syntactically similar words. \acl{NNLM}s (\acs{NNLM}) \citep{Mikolov2010rnn,Bengio2003} have shown promising results in \acl{NLP} (\acs{NLP}) \citep{Sutskever2014seq2seq,Jozefowicz2015rnn,Tran2016rmn} and \acl{ASR} (\acs{ASR}) \citep{Graves2014speech,Sak2015acousticlstm} compared to Markovian models. \citet{Collobert2011scratch} apply \acs{NNLM}s to arbitrary \acs{NLP} tasks by learning one set of word representations in a multi-task setting. Even more recently, there has been a surge in multi-modal neural language models \citep{Kiros2014}, which lend themselves to the task of automated image captioning.

\subsubsection{Representations and regularities}
\label{chapter:background:related:representations}

The idea that representations may capture linguistic or semantic regularities has received considerable attention. More generally, the idea of learning a representation of the elements of a discrete set of objects (e.g., words) is not new \cite{Rumelhart1985,Hinton1986,Bengio2013representations}. However, it has only been since the turn of the last century that \acs{NNLM}s, which learn word embeddings as a side-effect of dealing with high-dimensionality, were shown to be better at modelling language than Markovian models \cite{Bengio2003,Mnih2007,Mnih2009}. 

\citet{Turian2010} compare word representations learned by neural networks, distributional semantics and cluster-based methods as features in Named Entity Recognition (NER) and chunking. They find that both cluster-based methods and distributed word representations learned by \acs{NNLM}s improve performance, although cluster-based methods yield better representations for infrequent words. \citet{Baroni2014} confirm the superiority of context-predicting (word embeddings) over context-counting (distributional semantics) representations. 

Later algorithms are specifically designed for learning word embeddings \cite{Mnih2013,Mikolov2013word2vec,Pennington2014}, such that, somewhat ironically, \acs{NNLM}s became a side-product. These embeddings contain linguistic regularities~\cite{Mikolov2013regularities,Levy2014}, as evidenced in syntactic analogy and semantic similarity tasks. Multiple \emph{word} representations can be combined to form \emph{phrase} representations \cite{Mikolov2013compositionality}. Clusterings of word embeddings can be used to discover word classes \cite{Mikolov2013compositionality}. And insights gathered from word embedding algorithms can be used to improve distributional semantics \citep{Levy2015}.

\subsection{Neural information retrieval}
\label{chapter:background:related:nir}

The recent revival of neural networks due to advances in computer vision \citep{Krizhevsky2012net}, \acs{NLP} \citep{Collobert2011scratch,Mikolov2013word2vec} and \acs{ASR} \citep{Graves2014speech} has led to an increasing interest in these technologies from the information retrieval community.

Beyond representation learning that we extensively discussed in Section~\ref{chapter:background:related:latent}, there are more applications of neural models in IR \citep{Craswell2016neuir,Onal2017neural}. In machine-learned ranking \citep{Liu2011}, we have RankNet \citep{Burges2005ranknet}. In the class of supervised learning-to-match approaches, where clicks are available, there are DSSM~\citep{Huang2013,Shen2014} and DSN~\citep{Deng2013}. \citet{Guo2016relevance} learn a relevance model by extracting features from BoWE representations in addition to corpus statistics such as inverse document frequency. Recently, \citet{Mitra2017distributed} have introduced a supervised document ranking model that matches using both local and distributed representations. Next to retrieval models there has been work on modelling user interactions with neural methods. \citet{Borisov2016click} introduce a neural click model that represents user interactions as a vector representation; in \citep{Borisov2016contextaware}, they extend their work by taking into account click dwell time.
}

{

\renewcommand{\\}{}
\renewcommand{\pagebreak}{}
\renewcommand{\break}{}

{
\newcommand{\paper}{chapter}

\part{Query Formulation}
\label{part:formulation}

\chapter{Lexical Query Modelling in~Session~Search}
\label{chapter:research-01}

{
\ScopeLabels{research-01}
\ChapterRQ{1}

\def \paperImplementationUrl {\url{https://github.com/cvangysel/sesh}}

\def \QCM {\acs{QCM}}
\def \Nugget {Nugget}
\def \NuggetLexical {Nugget (RL2)}
\def \NuggetAnchors {Nugget (RL3)}
\def \NuggetAnchorClicks {Nugget (RL4)}
\def \TF {\acs{TF}}
\def \DPL {\acs{DPL}}

\def \NDCG {\acs{NDCG}@10}
\def \MeanReciprocalRank {\acs{MRR}}

\renewcommand{\log}{\text{log}}

\newcommand{\lengthplotinner}[2][]{%
	\IfFileExists{#2}{
		\includegraphics[height=0.235\textheight#1]{#2}}{
		\resizebox{0.185\paperheight}{!}{\missingfigure{#2}}}%
}%
\newcommand{\lengthplot}[1]{%
	\def \PlotPath {resources/per_session_length/#1--exclude_method_names.pdf}%
	\begin{subfigure}[t]{0.175\paperheight}%
	\lengthplotinner{\PlotPath}%
	\caption{#1\label{fig:session_length:#1}}%
	\end{subfigure}%
}

\renewcommand{\newcommand}{\providecommand}


\section{Introduction}

In Section~\globalref{chapter:background:related:sessions}, we discussed the expansion of query and document representations in the session search domain in order to bridge the vocabulary gap \citep{Li2014}.
In this \paper{}, we analyse the session query logs made available by TREC and compare the performance of different lexical query modelling approaches for session search, taking into account session length.\footnote{An open-source implementation of our testbed for evaluating session search is available at \paperImplementationUrl{}.} In addition, we investigate the viability of lexical query models in a session search setting.

\medskip
\noindent
The main purpose of this \paper{} is to investigate the potential of lexical methods in session search and provide foundations for future research. We ask the following questions towards answering \MainRQRef{1}:
\RQ{1}{Increasingly complex methods for session search are being developed, but how do naive methods perform?}
\RQ{2}{How well can lexical methods perform?}
\RQ{3}{Can we solve the session search task using lexical matching only?}

\section{Lexical matching for sessions}
\label{section:methodology}

\newcommand{\Prob}[1]{P(#1)}
\newcommand{\CondProb}[2]{\Prob{#1 \mid #2}}

\newcommand{\Vocabulary}{V}

\newcommand{\Sessions}{S}

\newcommand{\Corpus}{D}
\newcommand{\Session}{\MakeLowercase{\Sessions{}}}
\newcommand{\Queries}{Q}
\newcommand{\Terms}{T}
\newcommand{\SERPs}{R}

\newcommand{\Time}{i}

\newcommand{\Query}{\MakeLowercase{\Queries{}}}
\newcommand{\SERP}{\MakeLowercase{\SERPs{}}}
\newcommand{\Document}{\MakeLowercase{\Corpus{}}}

\newcommand{\Term}{\MakeLowercase{\Terms{}}}
\newcommand{\Word}{w}

\newcommand{\Length}[1]{|#1|}
\newcommand{\SERPLength}[1]{\Length{\SERP{}#1}}
\newcommand{\QueryLength}[1]{\Length{\Query{}#1}}

\newcommand{\NumInteractions}{n}
\newcommand{\TopK}{k}

\newcommand{\SessionModel}[1][\Session{}]{\theta^{#1}}
\newcommand{\SessionModelParameter}[2][\Session{}]{\theta^{#1}_{#2}}

\newcommand{\DocumentModel}[1][]{\theta^{\Document{}{#1}}}
\newcommand{\DocumentModelParameter}[2][]{\theta^{\Document{}{#1}}_{#2}}

We define a search session $\Session{}$ as a sequence of $\NumInteractions{}$ interactions $(\Query{}_i, \SERP{}_i)$ between user and search engine, where $\Query{}_i$ denotes a user-issued query consisting of $\QueryLength{_i}$ terms $\Term{}_{i,1}$, \ldots, $\Term{}_{i,{\QueryLength{_i}}}$ and $\SERP{}_i$ denotes a result page consisting of $\SERPLength{_i}$ documents $\SERP{}_{i, 1}$, \ldots, $\SERP{}_{i, \SERPLength{_i}}$ returned by the search engine (also referred to as SERP). The goal, then, is to return a SERP $\SERP{}_{\NumInteractions{} + 1}$ given a query $\Query{}_{\NumInteractions{} + 1}$ and the session history that maximizes the user's utility function.

In this \paper{}, we formalize session search by modelling an observed session $\Session{}$ as a query model parametrized by $\SessionModel{} = \{ \SessionModelParameter{1}$, \ldots, $\SessionModelParameter{\Length{\Vocabulary{}}} \}$, where $\SessionModelParameter{i}$ denotes the weight associated with term $\Term{}_i \in \Vocabulary{}$ (specified below). Documents $\Document{}_j$ are then ranked in decreasing order of
\begin{equation*}
\log\CondProb{\Document{}_j}{\Session{}} = \sum^{\Length{\Vocabulary{}}}_{k=1} \SessionModelParameter{k} \log\DocumentModelParameter[_j]{k},
\end{equation*}
where $\DocumentModel[_j]$ is a lexical model of document $\Document{}_j$, which can be a language model (LM), a vector space model or a specialized model using hand-engineered features. Query model $\SessionModel{}$ is a function of the query models of the interactions $\Time{}$ in the session, $\SessionModel[\Session{}_\Time{}]{}$ (e.g., for a uniform aggregation scheme, $\SessionModel{} = \sum_\Time{} \SessionModel[\Session{}_\Time{}]{}$). Existing session search methods \citep{Yang2015,Guan2012} can be expressed in this formalism as follows:
\newcommand{\BestDocument}[1][1]{\SERP{}_{\Time{} - 1, #1}}
\newcommand{\idf}[1]{\text{idf}({#1})}
\begin{description}[topsep=0pt,itemsep=0pt,parsep=0pt]
	\item[Term frequency (\TF{})] Terms in a query are weighted according to their frequency in the query (i.e., $\SessionModelParameter[\Session{}_\Time{}]{k}$ becomes the frequency of term $\Term{}_k$ in $\Query{}_\Time{}$). Queries $\Query{}_\Time{}$ that are part of the same session $\Session{}$ are then aggregated uniformly for a subset of queries. In this \paper{}, we consider the following subsets: the \emph{first query}, the \emph{last query} and the \emph{concatenation of all queries} in a session. Using the last query corresponds to the official baseline of the TREC Session track \citep{TREC2014}.

	\item[\Nugget{}]
	\Nugget{} \citep{Guan2012} is a method for effective structured query formulation for session search. Queries $\Query{}_i$, part of session $\Session{}$, are expanded using higher order n-grams occurring in both $\Query{}_i$ and snippets of the top-$k$ documents in the previous interaction, $\BestDocument[1]{}$, \ldots, $\BestDocument[k]{}$. This effectively expands the vocabulary by additionally considering n-grams next to unigram terms. The query models of individual queries in the session are then aggregated using one of the aggregation schemes.
	\Nugget{} is primarily targeted at resolving the query-document mismatch by incorporating structure and external data and does not model query transitions. The method can be extended to include external evidence by expanding $\SessionModel{}$ to include anchor texts pointing to (clicked) documents in previous SERPs.

	\item[Query Change Model (\QCM{})]
	QCM \citep{Yang2015} uses syntactic editing changes between consecutive queries in addition to query changes and previous SERPs to enhance session search.
	In QCM \citep[Section~6.3]{Yang2015}, document model $\DocumentModel{}$ is provided by a language model with Dirichlet smoothing and the query model at interaction $\Time{}$, $\SessionModel[\Session{}_\Time{}]$, in session $\Session{}$ is given by%
	\begin{equation*}
	\small
	{\SessionModelParameter[\Session{}_\Time{}]{k}} = \begin{cases}
	1 + \alpha (1 - \CondProb{\Term{}_k}{\BestDocument{}}), & \!\Term{}_k \in \Query{}_{\text{theme}} \\

	1 - \beta \CondProb{\Term{}_k}{\BestDocument{}}, & \!\Term{}_k \in + \Delta \Query{} \wedge \Term{}_k \in \BestDocument{} \\
	1 + \epsilon \, \idf{\Term{}_k}, & \!\Term{}_k \in + \Delta \Query{} \wedge \Term{}_k \notin \BestDocument{} \\

	- \delta \CondProb{\Term{}_k}{\BestDocument{}}, & \!\Term{}_k \in - \Delta \Query{},
	\end{cases}
	\end{equation*}
	where $\Query{}_\text{theme}$ are the session's theme terms, $+ \Delta \Query{}$ ($- \Delta \Query{}$, resp.) are the added (removed) terms, $\CondProb{\Term{}_k}{\BestDocument{}}$ denotes the probability of $\Term{}_k$ occurring in SAT clicks, $\idf{\Term{}_k}$ is the inverse document frequency of term $\Term{}_k$ and $\alpha$, $\beta$, $\epsilon$, $\delta$ are parameters. The $\SessionModel[\Session{}_\Time{}]$ are then aggregated into $\SessionModel{}$ using one of the aggregation schemes, such as the uniform aggregation scheme (i.e., the sum of the $\SessionModel[\Session{}_\Time{}]$).
\end{description}
In \S\ref{section:discussion}, we analyse the methods listed above in terms of their ability to handle sessions of different lengths and contextual history.


\section{Experiments}
\label{section:experiments}

\begin{table*}[t]
	\centering

	\caption{Overview of 2011, 2012, 2013 and 2014 TREC session tracks. For the 2014 track, we report the total number of sessions in addition to those sessions with judgements. We report the mean and standard deviation where appropriate; M denotes the median.\label{tbl:statistics}}

	\resizebox{\textwidth}{!}{\begin{tabular}{lcccc}

\toprule
& 2011 & 2012 & 2013 & 2014 \\

\midrule

\textbf{Sessions} \\

Sessions & \numprint{76}  & \numprint{98}  & \numprint{87}  & \numprint{100} (\numprint{1021} total) \\
Queries per session & \nprounddigits{2} \npdecimalsign{.} \numprint{3.6842} $\pm$ \numprint{1.7933}; M=\numprint{3.0000}  & \nprounddigits{2} \npdecimalsign{.} \numprint{3.0306} $\pm$ \numprint{1.5744}; M=\numprint{2.0000}  & \nprounddigits{2} \npdecimalsign{.} \numprint{5.0805} $\pm$ \numprint{3.5983}; M=\numprint{4.0000}  & \nprounddigits{2} \npdecimalsign{.} \numprint{4.3389} $\pm$ \numprint{2.2228}; M=\numprint{4.0000} \\
Unique terms per session & \nprounddigits{2} \npdecimalsign{.} \numprint{7.0132} $\pm$ \numprint{3.2827}; M=\numprint{6.5000}  & \nprounddigits{2} \npdecimalsign{.} \numprint{5.7551} $\pm$ \numprint{2.9522}; M=\numprint{5.0000}  & \nprounddigits{2} \npdecimalsign{.} \numprint{8.8621} $\pm$ \numprint{4.3791}; M=\numprint{8.0000}  & \nprounddigits{2} \npdecimalsign{.}\numprint{7.7855} $\pm$ \numprint{4.0786}; M=\numprint{7.0000} \\

\midrule

\textbf{Topics} \\

Session per topic & \nprounddigits{2} \npdecimalsign{.} \phantom{00}\numprint{1.2258} $\pm$ \phantom{00}\numprint{0.4551}; M=\phantom{00}\numprint{1.0000}  & \nprounddigits{2} \npdecimalsign{.} \phantom{00}\numprint{2.0417} $\pm$ \phantom{00}\numprint{0.9781}; M=\phantom{00}\numprint{2.0000}  & \nprounddigits{2} \npdecimalsign{.} \phantom{00}\numprint{2.1803} $\pm$ \phantom{00}\numprint{0.9323}; M=\phantom{00}\numprint{2.0000}  & \nprounddigits{2} \npdecimalsign{.} \phantom{0}\numprint{20.9500} $\pm$ \phantom{00}\numprint{4.8146}; M=\phantom{0}\numprint{21.0000} \\
Document judgments per topic & \nprounddigits{2} \npdecimalsign{.} \numprint{313.1129} $\pm$ \numprint{114.6277}; M=\numprint{292.0000}  & \nprounddigits{2} \npdecimalsign{.} \numprint{372.1042} $\pm$ \numprint{162.6306}; M=\numprint{336.5000}  & \nprounddigits{2} \npdecimalsign{.} \numprint{268.0000} $\pm$ \numprint{116.8587}; M=\numprint{247.0000}  & \nprounddigits{2} \npdecimalsign{.} \numprint{332.3333} $\pm$ \numprint{149.0269}; M=\numprint{322.0000} \\

\midrule

\textbf{Collection} \\

Documents & \multicolumn{2}{c}{\numprint{21258800}} & \multicolumn{2}{c}{\numprint{15702181}} \\
Document length & \multicolumn{2}{c}{\nprounddigits{2} \npdecimalsign{.} \numprint{1096.18328043} $\pm$ \numprint{1502.45255801}} & \multicolumn{2}{c}{\nprounddigits{2} \npdecimalsign{.} \numprint{649.074948569} $\pm$ \numprint{1635.29462878}} \\
Terms & \multicolumn{2}{c}{\numprint{\xintFloat [3]{34015925}} (\numprint{\xintFloat [3]{23303541122}} total)} & \multicolumn{2}{c}{\numprint{\xintFloat [3]{23575957}} (\numprint{\xintFloat [3]{10191892325}} total)} \\
Spam scores & \multicolumn{2}{c}{GroupX} & \multicolumn{2}{c}{Fusion} \\

\bottomrule

\end{tabular}}

\end{table*}

\subsection{Benchmarks}

We evaluate the lexical query modelling methods listed in \S\ref{section:methodology} on the session search task (G1) of the TREC Session track from 2011 to 2014 \citep{TREC2011-2014}. We report performance on each track edition independently and on the track aggregate. Given a query, the task is to improve retrieval performance by using previous queries and user interactions with the retrieval system. To accomplish this, we first retrieve the \numprint{2000} most relevant documents for the given query and then re-rank these documents using the methods described in \S\ref{section:methodology}. We use the ``Category B'' subsets of ClueWeb09 (2011/2012) and ClueWeb12 (2013/2014) as document collections. Both collections consist of approximately 50 million documents. Spam documents are removed before indexing by filtering out documents with scores (GroupX and Fusion, respectively) below 70 \citep{Cormack2012}. Table~\ref{tbl:statistics} shows an overview of the benchmarks and document collections.

\subsection{Evaluation measures}

To measure retrieval effectiveness, we report Normalized Discounted Cumulative Gain at rank 10 (\NDCG{}) in addition to Mean Reciprocal Rank (\MeanReciprocalRank{}). The relevance judgements of the tracks were converted from topic-centric to session-centric according to the mappings provided by the track organizers.\footnote{We take into account the mapping between judgements and actual relevance grades for the 2012 edition.} Evaluation measures are then computed using TREC's official evaluation tool, {\tt trec\_eval}.\footnote{\url{https://github.com/usnistgov/trec_eval}}

\subsection{Systems under comparison}
\label{section:experiments:methods}

We compare the lexical query model methods outlined in \S\ref{section:methodology}. All methods compute weights for lexical entities (e.g., unigram terms) on a per-session basis, construct a structured Indri query \citep{Metzler2004} and query the document collection using {\tt pyndri}.\footnote{\url{https://github.com/cvangysel/pyndri}} For fair comparison, we use Indri's default smoothing configuration (i.e., Dirichlet smoothing with $\mu = 2500$) and uniform query aggregation for all methods (different from the smoothing used for \QCM{} in \citep{Yang2015}). This allows us to separate query aggregation techniques from query modelling approaches in the case of session search.

For \Nugget{}, we use the default parameter configuration (\splitatcommas{$k_\text{snippet}=10,\, \theta=0.97,\, k_\text{anchor}=5$} and $\beta=0.1$), using the strict expansion method. We report the performance of \Nugget{} without the use of external resources (RL2), with anchor texts (RL3) and with click data (RL4). For \QCM{}, we use the parameter configuration as described in \citep{Yang2015,Luo2015}: $\alpha=2.2,\, \beta=1.8,\, \epsilon=0.07$ and $\delta=0.4$.

In addition to the methods above, we report the performance of an oracle that always ranks in decreasing order of ground-truth relevance. This oracle will give us an upper-bound on the achievable ranking performance.

\subsection{Ideal lexical term weighting}
\label{section:experiments:ideal}

We investigate the maximally achievable performance by weighting query terms. Inspired by \citet{Bendersky2012}, we optimize \NDCG{} for every session using a grid search over the term weight space. We sweep the weight of every term between $-1.0$ and $1.0$ (inclusive) with increments of $0.1$, resulting in a total of $21$ weight assignments per term. Due to the exponential time complexity of the grid search, we limit our analysis to the \numprint{230} sessions with $7$ unique query terms or less (see Table~\ref{tbl:statistics}). This experiment will tell us the maximally achievable retrieval performance in session search by the re-weighting of lexical terms only.


\section{Results \& discussion}
\label{section:discussion}

\begin{table*}[t]
	\centering

	\caption{Overview of experimental results on 2011--2014 TREC Session tracks of the \TF{}, \Nugget{} and \QCM{} methods (see \S\ref{section:methodology}). The ground-truth oracle shows the ideal performance (\S\ref{section:experiments:methods}).\label{tbl:results}}

	{\begin{tabular}{c@{ }c@{ }c@{ }c@{ }c@{ }c@{ }c@{ }c@{ }c}
\toprule
\multirow{2}{*}{} & \multicolumn{2}{c}{2011} & \multicolumn{2}{c}{2012} & \multicolumn{2}{c}{2013} & \multicolumn{2}{c}{2014} \\ 
& NDCG@10 & MRR & NDCG@10 & MRR & NDCG@10 & MRR & NDCG@10 & MRR \\ 
\cmidrule(lr){2-3}
\cmidrule(lr){4-5}
\cmidrule(lr){6-7}
\cmidrule(lr){8-9}
\multicolumn{1}{l}{Ground-truth oracle} & $0.777$ & $0.868$ & $0.695$ & $0.865$ & $0.517$ & $0.920$ & $0.410$ & $0.800$ \\ 
\multicolumn{1}{l}{TF (first query)} & $0.371$ & $0.568$ & $0.302$ & $0.523$ & $0.121$ & $0.379$ & $0.120$ & $0.336$ \\ 
\multicolumn{1}{l}{TF (last query)} & $0.358$ & $0.598$ & $0.316$ & $0.586$ & $0.133$ & $0.358$ & $0.156$ & $0.458$ \\ 
\multicolumn{1}{l}{TF (all queries)} & $\phantom{}\textbf{0.448}$ & $\phantom{}\textbf{0.685}$ & $0.348$ & $0.604$ & $0.162$ & $0.477$ & $\phantom{}\textbf{0.174}$ & $\phantom{}\textbf{0.478}$ \\ 
\multicolumn{1}{l}{Nugget (RL2)} & $0.437$ & $0.677$ & $0.352$ & $0.609$ & $\phantom{}\textbf{0.163}$ & $0.488$ & $0.173$ & $0.476$ \\ 
\multicolumn{1}{l}{Nugget (RL3)} & $0.442$ & $0.678$ & $\phantom{}\textbf{0.360}$ & $\phantom{}\textbf{0.619}$ & $0.162$ & $\phantom{}\textbf{0.488}$ & $0.172$ & $0.477$ \\ 
\multicolumn{1}{l}{Nugget (RL4)} & $0.437$ & $0.677$ & $0.352$ & $0.609$ & $0.163$ & $0.488$ & $0.173$ & $0.476$ \\ 
\multicolumn{1}{l}{QCM} & $0.440$ & $0.661$ & $0.342$ & $0.575$ & $0.160$ & $0.484$ & $0.162$ & $0.450$ \\ 
\bottomrule
\end{tabular}
}

\vspace*{-0.75\baselineskip}
\end{table*}

\begin{figure}[t]
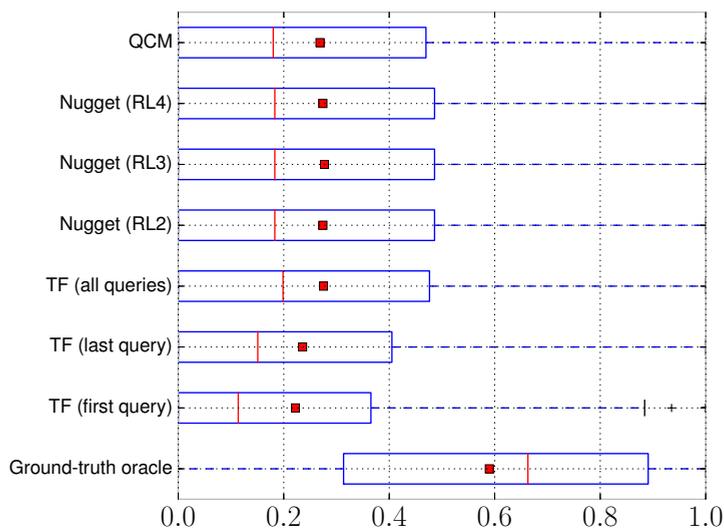


\newcommand{\boxplotinner}[2][]{%
	\IfFileExists{#2}{
		\myincludegraphics[width=\columnwidth#1]{#2}}{
		\resizebox{\columnwidth}{!}{\missingfigure{#2}}}%
}%

\boxplotinner{resources/boxplot/all.pdf}

\caption{Box plot of \NDCG{} on all sessions of the TREC Session track (2011--2014). The box depicts the first, second (median) and third quartiles. The whiskers are located at 1.5 times the interquartile range on both sides of the box. The square and crosses depict the average and outliers respectively.\label{fig:boxplot}}

\vspace*{-0.75\baselineskip}
\end{figure}

In this section, we report and discuss our experimental results. Of special interest to us are the methods that perform lexical matching based on a user's queries in a single session: \QCM{}, \NuggetLexical{} and the three variants of \TF{}.
Table~\ref{tbl:results} shows the methods' performance on the TREC Session track editions from 2011 to 2014. No single method consistently outperforms the other methods. Interestingly enough, the methods based on term frequency (\TF{}) perform quite competitively compared to the specialized session search methods (\Nugget{} and \QCM{}). In addition, the \TF{} variant using all queries in a session even outperforms \NuggetLexical{} on the 2011 and 2014 editions and \QCM{} on nearly all editions. Using the concatenation of all queries in a session, while being an obvious baseline, has not received much attention in recent literature or by TREC \citep{TREC2011-2014}. In addition, note that the best-performing (unsupervised) \TF{} method achieves better results than the supervised method of \citet{Luo2015} on the 2012 and 2013 tracks. Fig.~\ref{fig:boxplot} depicts the boxplot of the \NDCG{} distribution over all track editions (2011--2014). The term frequency approach using all queries achieves the highest mean/median overall. Given this peculiar finding, where a generic retrieval model performs better than specialized session search models, we continue with an analysis of the TREC Session search logs.

\begin{figure*}[t]
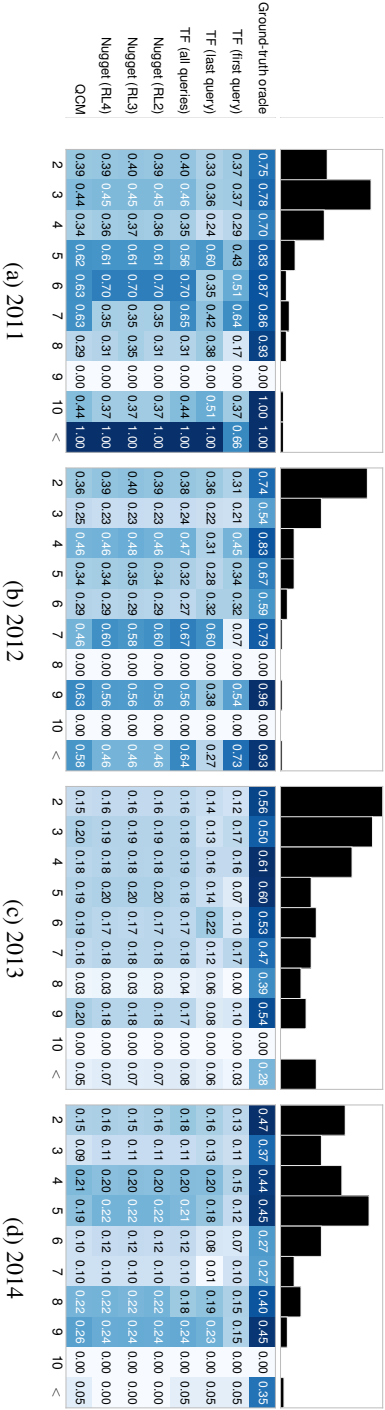


\newcommand{\lengthplotinner}[2][]{%
	\IfFileExists{#2}{
		\includegraphics[height=0.185\textheight#1]{#2}}{
		\resizebox{0.245\textwidth}{!}{\missingfigure{#2}}}%
}%
\newcommand{\lengthplot}[1]{%
	\def \PlotPath {resources/per_session_length/#1--exclude_method_names.pdf}%
	\begin{subfigure}[t]{0.225\textwidth}%
	\lengthplotinner{\PlotPath}%
	\caption{#1\label{fig:session_length:#1}}%
	\end{subfigure}%
}
\mbox{
\begin{subfigure}[t]{0.10\textwidth}%
\lengthplotinner[,trim=0 0 9.525cm 0, clip]{resources/per_session_length/2011.pdf}%
\end{subfigure}%
\hfill%
\lengthplot{2011}%
\hfill%
\lengthplot{2012}%
\hfill%
\lengthplot{2013}%
\hfill%
\lengthplot{2014}%
}

\caption{The top row depicts the distribution of session lengths for the 2011--2014 TREC Session tracks, while the bottom row shows the performance of the \TF{}, \Nugget{} and \QCM{} models for different session lengths.\label{fig:session_length}}

\vspace*{-0.75\baselineskip}
\end{figure*}

In Fig.~\ref{fig:session_length} we investigate the effect of varying session lengths in the session logs. The distribution of session lengths is shown in the top row of Fig.~\ref{fig:session_length}. For the 2011--2013 track editions, most sessions consisted of only two queries. The mode of the 2014 edition lies at 5 queries per session. If we examine the performance of the methods on a per-session length basis, we observe that the \TF{} methods perform well for short sessions. This does not come as a surprise, as for these sessions there is only a limited history that specialized methods can use. However, the \TF{} method using the concatenation of all queries still performs competitively for longer sessions. This can be explained by the fact that as queries are aggregated over time, a better representation of the user's information need is created. This aggregated representation naturally emphasizes important \emph{theme terms} of the session, which is a key component in the \QCM{} \citep{Yang2015}.

\begin{figure}[th!]

\begin{subfigure}[b]{\columnwidth}
	\includegraphics[width=\columnwidth]{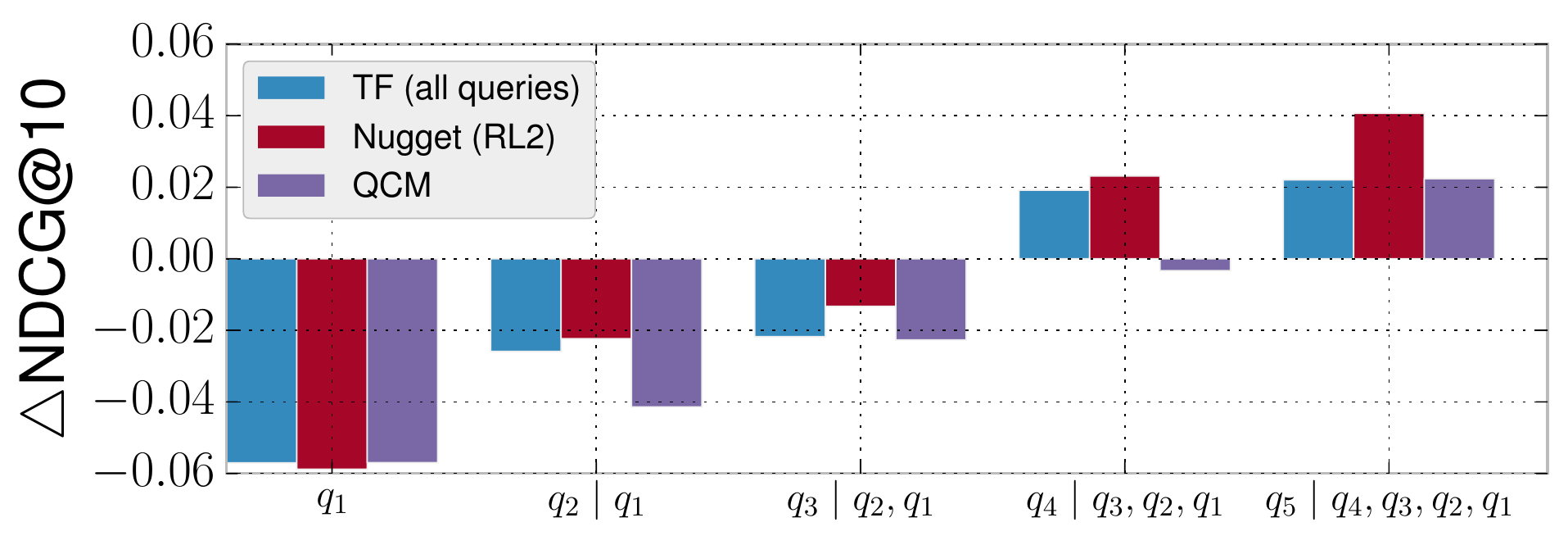}

	\caption{Full history of session\label{fig:progressing_session:all}}
\end{subfigure}

\begin{subfigure}[b]{\columnwidth}
	\includegraphics[width=\columnwidth]{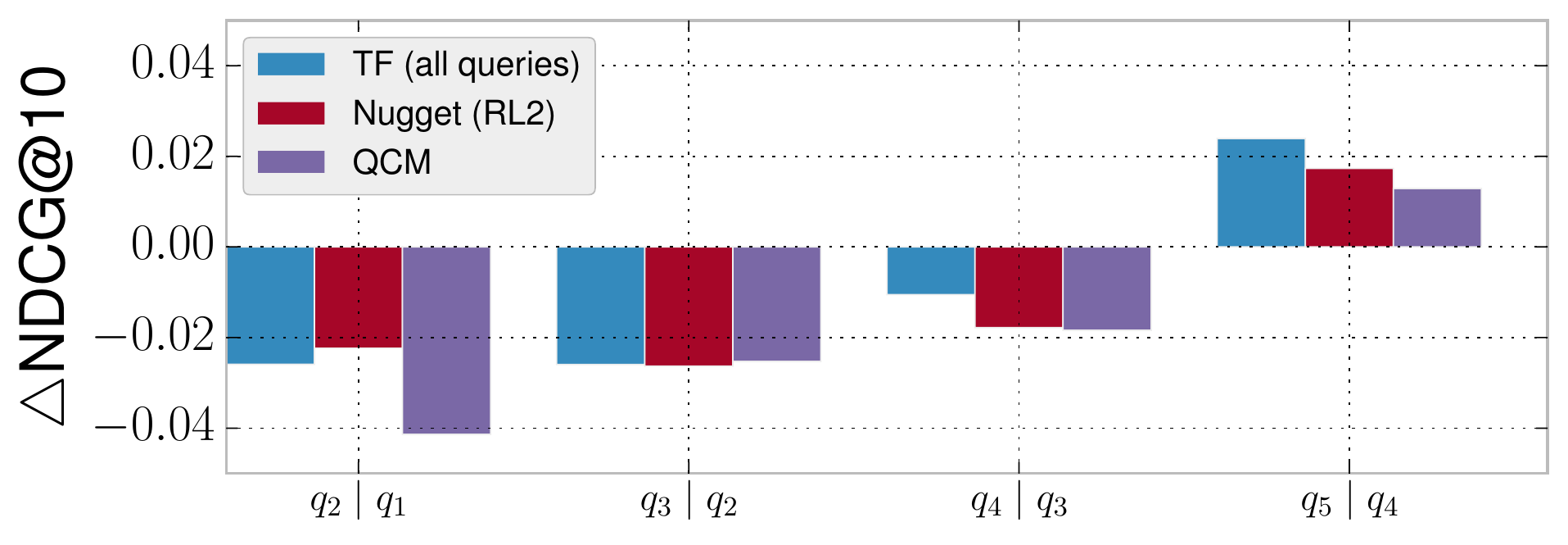}

	\caption{Previous query in session only\label{fig:progressing_session:previous}}
\end{subfigure}

\medskip

\caption{Difference in \NDCG{} with the official TREC baseline (\TF{} using the last query only) of $5$-query sessions (\numprint{45} instances) with different history configurations for the 2011--2014 TREC Session tracks.\label{fig:progressing_session}}

\vspace*{-0.75\baselineskip}

\end{figure}

\begin{table}[th!]
	\centering

	\caption{\NDCG{} for \TF{} weighting (\S\ref{section:methodology}), ideal term weighting (\S\ref{section:experiments:ideal}) and the ground-truth oracle (\S\ref{section:experiments:methods}).\label{tbl:lexical_results}}

	\resizebox{0.55\columnwidth}{!}
	{\begin{tabular}{c@{ }c@{ }c@{ }c@{ }c}
\toprule
\multirow{2}{*}{} & \multicolumn{1}{c}{2011} & \multicolumn{1}{c}{2012} & \multicolumn{1}{c}{2013} & \multicolumn{1}{c}{2014} \\ 
\cmidrule(lr){2-2}
\cmidrule(lr){3-3}
\cmidrule(lr){4-4}
\cmidrule(lr){5-5}
\multicolumn{1}{l}{TF (all queries)} & $0.391$ & $0.333$ & $0.179$ & $0.183$ \\ 
\multicolumn{1}{l}{Ideal term weighing} & $0.589$ & $0.528$ & $0.361$ & $0.296$ \\ 
\multicolumn{1}{l}{Ground-truth oracle} & $0.716$ & $0.682$ & $0.593$ & $0.453$ \\ 
\bottomrule
\end{tabular}
}

\end{table}

\break

How do these methods perform as the search session progresses? Fig.~\ref{fig:progressing_session} shows the performance of sessions of length five after every user interaction, when using all queries in a session (Fig.~\ref{fig:progressing_session:all}) and when using only the previous query (Fig.~\ref{fig:progressing_session:previous}). We can see that \NDCG{} increases as the session progresses for all methods. Beyond half of the session, the session search methods outperform retrieving according to the last query in the session. We see that, for longer sessions, specialized methods (\Nugget{}, \QCM{}) outperform generic term frequency models. This comes as no surprise. \citet{Bennett2012} note that users tend to reformulate and adapt their information needs based on observed results and this is essentially the observation upon which \QCM{} builds.

Fig.~\ref{fig:boxplot} and Table~\ref{tbl:results} reveal a large \NDCG{} gap between the compared methods and the ground-truth oracle. How can we bridge this gap? Table~\ref{tbl:lexical_results} shows a comparison between frequency-based term weighting, the ideal term weighting (\S\ref{section:experiments:ideal}) and the ground-truth oracle (\S\ref{section:experiments:methods}) for all sessions consisting of 7 unique terms or less (\S\ref{section:experiments:ideal}). Two important observations. There is still plenty of room for improvement using lexical query modelling only. Relatively speaking, around half of the gap between weighting according to term frequency and the ground-truth can be bridged by predicting better term weights. However, the other half of the performance gap cannot be bridged using lexical matching only, but instead requires a notion of semantic matching \cite{Li2014}.


\section{Summary}
\label{section:conclusions}

We have shown that naive frequency-based term weighting methods perform on par with specialized session search methods on the TREC Session track (2011--2014).\footnote{An open-source implementation of our testbed for evaluating session search is available at \paperImplementationUrl{}.} This is due to the fact that shorter sessions are more prominent in the session query logs. On longer sessions, specialized models are able to exploit session history more effectively. Future work should focus on creating benchmarks consisting of longer sessions with complex information needs.

In the next chapter (Chapter~\globalref{chapter:research-02}), we introduce a neural network model that learns to formulate a query from complex textual structures (i.e., email threads). In \S\ref{section:discussion}, we observed that the query/document mismatch is prevalent in session search and methods restricted to lexical query modelling face a very strict performance ceiling. Therefore, Part~\globalref{part:latent} of this dissertation is dedicated to the modelling of latent vector spaces that bridge the vocabulary gap between query and document.
}

\chapter{Reply With: Proactive Recommendation of Email Attachments}
\label{chapter:research-02}

{
\ScopeLabels{research-02}
\ChapterRQ[3]{1}

\SetKw{Continue}{continue}

\renewcommand{\DeclareMathOperator}[2]{}

\newcommand{\RelativeFrequencyAttachmentAlreadyPresent}{35}
\newcommand{\RelativeFrequencyAttachmentInReply}{14}

\newcommand{\RetrievalSystem}{R}

\newcommand{\Time}{t}

\newcommand{\Threads}{C}
\newcommand{\Thread}{\MakeLowercase{\Threads{}}}

\newcommand{\Users}{U}
\newcommand{\User}{\MakeLowercase{\Users{}}}

\newcommand{\AttachableEntities}{E}
\newcommand{\AttachableEntity}{\MakeLowercase{\AttachableEntities{}}}

\newcommand{\UserTimeConstraintedAttachableEntities}[1][\CurrentTime{}]{{\AttachableEntities{}_{\User{}, {#1}}}}

\newcommand{\Messages}{M}
\newcommand{\Mailbox}{\Messages}
\newcommand{\Message}{\MakeLowercase{\Messages{}}}
\newcommand{\MessageTime}{{\Time{}}_{\Message{}}}

\newcommand{\RequestMessage}{\Message{}_{\text{req}}}
\newcommand{\ReplyMessage}{\Message{}_{\text{res}}}

\newcommand{\RequestReplyPair}{\langle\RequestMessage{}, \ReplyMessage{}\rangle}

\newcommand{\TargetAttachableEntity}{\AttachableEntity{}_{actual}}

\newcommand{\RequestEntityPair}{\langle\RequestMessage{}, \TargetAttachableEntity{}\rangle}

\newcommand{\RequestMessageCandidateQueries}{\CandidateQueries{}_{\RequestMessage{}}}

\newcommand{\Queries}{Q}
\newcommand{\Query}{\MakeLowercase{\Queries{}}}

\newcommand{\Documents}{D}
\newcommand{\Document}{\MakeLowercase{\Documents{}}}

\newcommand{\Vocabulary}{V}

\newcommand{\FrequencyFn}{\#}

\newcommand{\CurrentTime}{\Time{}^\prime{}}

\newcommand{\CandidateQueries}{\tilde{\Queries{}}}
\newcommand{\CandidateQuery}{\tilde{\Query{}}}
\newcommand{\CandidateQueryTermBudget}{k}

\newcommand{\entity}{item}
\newcommand{\entities}{items}

\newcommand{\RankModel}{\NeuralRankCutoff{}}

\newcommand{\Neural}{\acs{CNN}}
\newcommand{\NeuralLogistic}{\Neural{}-p}
\newcommand{\NeuralRankCutoff}{\Neural{}}
\newcommand{\LeaveOutRankSVM}{Rank\acs{SVM}}

\newcommand{\AllTerms}{Full}
\newcommand{\TF}{\acs{TF}}
\newcommand{\TFIDF}{\acs{TF}-\acs{IDF}}
\newcommand{\logTFIDF}{log\acs{TF}-\acs{IDF}}
\newcommand{\RelativeEntropy}{\acs{RE}}
\newcommand{\Multinomial}{Random $k$}
\newcommand{\BernouilliProcess}{Random \%}

\newcommand{\SubjectAllTerms}{\AllTerms{}}
\newcommand{\BodyAllTerms}{\AllTerms{}}
\newcommand{\SubjectBodyAllTerms}{\AllTerms{}}

\newcommand{\SubjectTF}{\TF{}}
\newcommand{\BodyTF}{\TF{}}
\newcommand{\SubjectBodyTF}{\TF{}}

\newcommand{\SubjectTFIDF}{\TFIDF{}}
\newcommand{\BodyTFIDF}{\TFIDF{}}
\newcommand{\SubjectBodyTFIDF}{\TFIDF{}}

\newcommand{\SubjectlogTFIDF}{\logTFIDF{}}
\newcommand{\BodylogTFIDF}{\logTFIDF{}}
\newcommand{\SubjectBodylogTFIDF}{\logTFIDF{}}

\newcommand{\SubjectMultinomial}{\Multinomial{}}
\newcommand{\BodyMultinomial}{\Multinomial{}}
\newcommand{\SubjectBodyMultinomial}{\Multinomial{}}

\newcommand{\SubjectBernouilliProcess}{\BernouilliProcess{}}
\newcommand{\BodyBernouilliProcess}{\BernouilliProcess{}}
\newcommand{\SubjectBodyBernouilliProcess}{\BernouilliProcess{}}

\newcommand{\SubjectBernoulliProcess}{\SubjectBernouilliProcess{}}
\newcommand{\BodyBernoulliProcess}{\BodyBernouilliProcess{}}
\newcommand{\SubjectBodyBernoulliProcess}{\SubjectBodyBernouilliProcess{}}

\newcommand{\SubjectRelativeEntropy}{\RelativeEntropy{}}
\newcommand{\BodyRelativeEntropy}{\RelativeEntropy{}}
\newcommand{\SubjectBodyRelativeEntropy}{\RelativeEntropy{}}

\newcommand{\SubjectRelativEntropy}{\SubjectRelativeEntropy{}}
\newcommand{\BodyRelativEntropy}{\BodyRelativeEntropy{}}
\newcommand{\SubjectBodyRelativEntropy}{\SubjectBodyRelativeEntropy{}}

\newcommand{\AvocadoCollection}{Avocado}
\newcommand{\InternalCollection}{PIE}

\newcommand{\ResearchQuestionOne}{Do convolutional neural networks (\Neural{}) improve ranking efficacy over state-of-the-art query formulation methods?}
\newcommand{\ResearchQuestionTwo}{When do \Neural{}s work better than non-neural methods on the attachable item recommendation task?}
\newcommand{\ResearchQuestionThree}{What features are most important when training \Neural{}s?}

\newcommand{\RecipRank}{\acs{MRR}}
\newcommand{\NDCG}{\acs{NDCG}}
\newcommand{\PrecisionCut}{\acs{P}@5}

\newcommand{\circled}[2][yscale=0.6]{%
\begin{tikzpicture}[#1, xscale=1.0, baseline=(char.base)]%
\node[shape=ellipse, draw, inner sep=2.5pt, text width=0.35cm] (char) {#2};
\end{tikzpicture}%
}

\newcommand{\EndOfRanking}[1]{\scalebox{1}[0.8]{\circled{\footnotesize{}\normalfont{}\hspace*{-1.40pt}EoR}}}

\renewcommand{\newcommand}{\providecommand}


\section{Introduction}
\label{sec:intro}

In the previous chapter, we looked at formulating a query from user sessions in web search engines---a particular complex textual structure. In this chapter, we move our focus to formulating queries from a different complex textual structure: email threads. Email is still pervasive in the enterprise space \cite{PewResearch2014dominant}, in spite of the growing popularity of social networks and other modern online communication tools.
Users typically respond to incoming emails with textual responses. However, an analysis of the publicly available Avocado dataset \cite{Oard2015avocado} reveals that \RelativeFrequencyAttachmentInReply{}\% of those messages also contain items, such as a file or a hyperlink to an external document. Popular email clients already detect when users forget to attach files by analysing the text of the response message \cite{Dredze2006forgotattachment, Dredze2008intelligent}. On Avocado, we find that in \RelativeFrequencyAttachmentAlreadyPresent{}\% of the cases where the response contains attachments, the item being attached is also present in the sender's mailbox at the time of composing the response. This implies that modern email clients could help users compose their responses faster by proactively retrieving and recommending relevant items that the user may want to include with their message.

In \emph{proactive} information retrieval (IR) systems \cite{Liebling2012anticipatory, Song2016query, Benetka2017anticipating, Shokouhi2015queries}, the user does not initiate the search. Instead, retrieval is triggered automatically based on a user's current context. The context may include the time of day \cite{Song2016query}, the user's geographic location \cite{Benetka2017anticipating}, recent online activities \cite{Shokouhi2015queries} or some other criteria. In our scenario, retrieval is based on the context of the current conversation, and in particular, the message the user is responding to. In a typical IR scenario, items are ranked based on query-dependent feature representations. In the absence of an explicit search query from the user, proactive IR models may formulate a keyword-based search query using the available context information and retrieve results for the query using a standard IR model \cite{Liebling2012anticipatory}. Search functionalities are available from most commercial email providers and email search has been studied in the literature \cite{Ai2017emailsearch}. Therefore, we cast the email attachment recommendation problem as a query formulation task and use an existing IR system to retrieve emails. Attachable items are extracted from the retrieved emails and a ranking is presented to the user.

Fig.~\ref{fig:example} shows an example of an email containing an explicit request for a file. In general, there may or may not be an explicit request, but it may be appropriate to attach a relevant file with the response. Our task is to recommend the correct ``transition document'' as an attachment when Beth or Changjiu is responding to this email. In order to recommend an attachment, the model should formulate a query, such as ``\emph{Initech transition}'', based on the context of the request message, that retrieves the correct document from Beth's or Changjiu's mailbox. To formulate an effective query, the model must identify the discriminative terms in the message from Anand that are relevant to the actual file request.

\begin{figure}[t]
\center
\center
\includegraphics[width=1.0\linewidth]{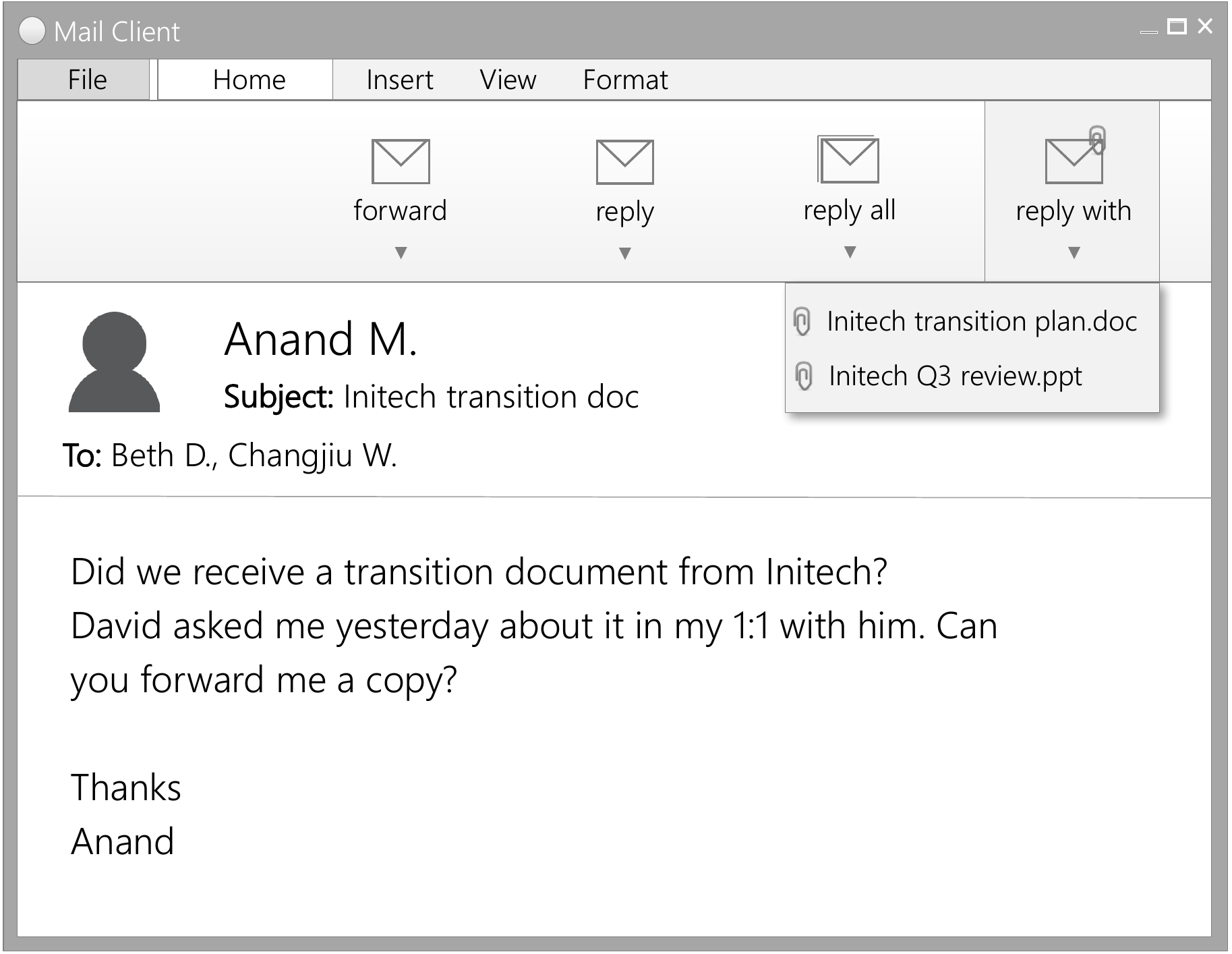}
\caption{Anand asks Beth and Changjiu to forward him a copy of the Initech\protect\footnotemark{} transition document. Beth's email client recommends two files, part of earlier emails in Beth's mailbox, for her to attach to her reply.}
\label{fig:example}
\end{figure}
\footnotetext{Initech is a fictional company from a popular 1990s comedy film. Any resemblance to real organizations is purely coincidental.}

Machine learning models that aim to solve the query formulation task need reliable feedback on what constitutes a good query. One option for generating labeled data for training and evaluation involves collecting manual assessments of proposed queries or individual query terms. However, it is difficult for human annotators to determine the ability of a proposed query to retrieve relevant items given only the request message. In fact, the efficacy of the query depends on the target message that should be retrieved, as well as how the IR system being employed functions. The relevance of the target message, in turn, is determined by whether they include the correct item that should be attached to the response message. Therefore, instead we propose an evaluation framework that requires an email corpus but no manual assessments. Request/response message pairs are extracted from the corpus and the model, that takes the request message as input, is evaluated based on its ability to retrieve the items attached to the response message. An IR system is employed for the message retrieval step, but is treated as a \emph{black box} in the context of evaluation. Our framework provides a concise specification for the email attachment recommendation task (\S\ref{sec:method}).

Our proposed approach for training a deep convolutional neural network (CNN) for the query formulation step is covered in \S\ref{sec:model}. The model predicts a distribution over all the terms in the request message and terms with high predicted probability are selected to form a query. Model training involves generating a dataset of request/attachment pairs similar to the case of evaluation. Candidate queries are algorithmically synthesized for each request/attachment pair such that a message from the user's mailbox with the correct item attached is ranked highly. We refer to synthetic queries as the \emph{silver-standard queries} (or silver queries for brevity) to emphasize that they achieve reasonable performance on the task, but are potentially sub-optimal. The neural model is trained to minimize the prediction loss w.r.t. the silver queries given the request message as input.

\medskip
\noindent
The research questions we ask in this \paper{} towards answering \MainRQRef{1} are as follows:
\RQ{1}{\ResearchQuestionOne{}}
\RQ{2}{\ResearchQuestionTwo{}}
\RQ{3}{\ResearchQuestionThree{}}


\section{Related work}
\label{sec:related_work}

We refer to Section~\globalref{chapter:background:formulation} of our background chapter (Chapter~\globalref{chapter:background}) where the sections on proactive information retrieval (\S\globalref{chapter:background:related:proactive}) and predictive models in email (\S\globalref{chapter:background:related:email}) provide background on the problem domain of this \paper{}. Section~\globalref{chapter:background:related:formulation} provides background on query formulation methods, with an emphasis on prior art search and sub-query selection in ad-hoc document retrieval. We elaborate briefly on the similarities and differences between the email domain and the domains of prior art and ad-hoc document retrieval.

Query extraction methods used for prior art search (\S\globalref{chapter:background:related:patents}) can also be applied to the task of attachable \entity{} recommendation considered in this \paper{}. Consequently, we consider the methods mentioned above as our baselines (\S\ref{sec:baselines}). However, there are a few notable differences between the patent and email domains:
\begin{inparaenum}[(1)]
	\item Email messages are much shorter in length than patents.
	\item Patents are more structured (e.g., US patents contain more than 50 fields) than email messages.
	\item Patents are linked together by a static citation graph that grows slowly, whereas email messages are linked by means of a dynamic conversation that is fast-paced and transient in nature.
	\item In the case of email, there is a social graph between email users that can act as an additional source of information.
\end{inparaenum}

The task we consider in this \paper{} differs from search sub-query selection (\S\globalref{chapter:background:related:verbosequeries}) as follows.
\begin{inparaenum}[(1)]
	\item Search queries are formulated by users as a way to interface with a search engine. Requests in emails may be more complex as they are formulated to retrieve information from a human recipient, rather than an automated search engine. In other words, email requests are more likely to contain natural language and figurative speech than search engine queries. This is because the sender of the request does not expect their message to be parsed by an automated system.
	\item Search sub-query extraction aims to improve retrieval effectiveness while the query intent remains fixed. This is not necessarily the case in our task, as a request message like has the intent to retrieve information from the recipient (rather than a retrieval system operating on top of the recipient's mailbox).
	\item Work on search sub-query selection \citep{Kumaran2009reducinglongqueries,Xue2010verbosequeries} takes advantage of the fact that 99.9\% of search queries consist of 12 terms or less \citep{Bendersky2009analysis} by relying on computations that are intractable otherwise. As emails are longer (Table~\ref{tbl:statistics}), many of the methods designed for search sub-query selection are not applicable in our setting.
\end{inparaenum}


\section{Proactive attachable item recommendation}
\label{sec:method}

\begin{figure}[t!]
\centering
\includegraphics[width=\linewidth]{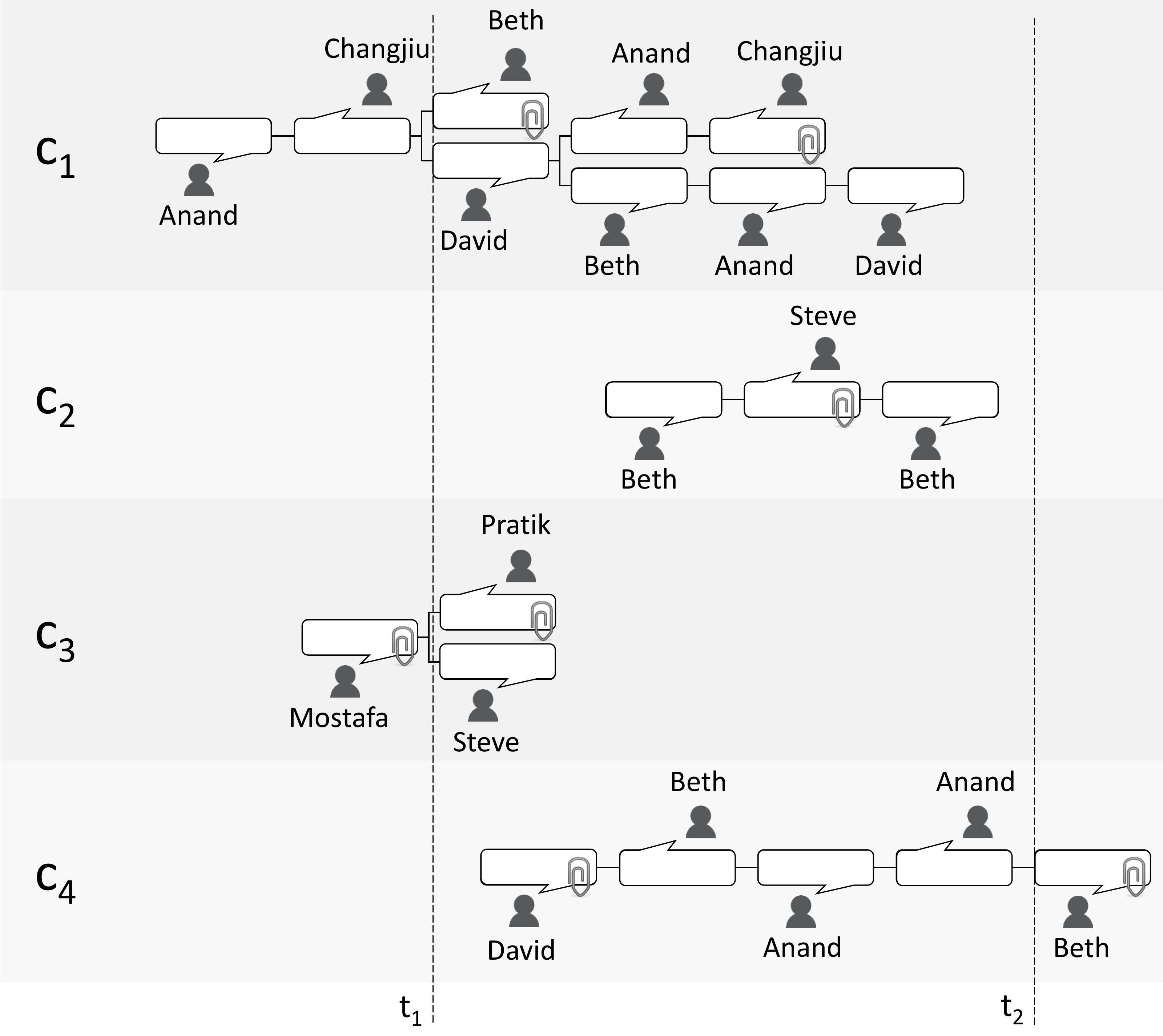}
\caption{Beth's mailbox has four on-going conversations. When Beth replies with an attachment in conversation $\Thread{}_{1}$, at time $\Time{}_{1}$, only the attachment she received from Mostafa (conversation $\Thread{}_{3}$) is present in her mailbox.}
\label{fig:framework}
\end{figure}

\newcommand{\FirstUser}{\User{}_{a}}
\newcommand{\SecondUser}{\User{}_{b}}

\newcommand{\ExampleRequest}{\Message{}_{a\rightarrow b}}
\newcommand{\ExampleResponse}{\Message{}_{b\rightarrow a}}

\newcommand{\ExampleRequestReplyPair}{\left\langle\Message_{a\rightarrow b}, \Message_{b\rightarrow a}\right\rangle}

Given message $\ExampleRequest{}$ from user $\FirstUser{}$ to user $\SecondUser{}$, we want to recommend an item $\AttachableEntity{}$ that the receiver $\User{}_{b}$ may want to attach (or include) in the response $\ExampleResponse$. Email corpora, such as Avocado \cite{Oard2015avocado}, contain many conversation threads where each conversation $\Thread{}$ contains messages $\Message_{i} \in \Thread{}$ exchanged between several participants. From these conversations, we can identify pairs of request-response messages $\ExampleRequestReplyPair{}$ where $\ExampleResponse{}$ contains an attachment $\TargetAttachableEntity{}$. We assume that user $\SecondUser{}$ included $\TargetAttachableEntity{}$ in $\ExampleResponse{}$ in response to an explicit or an implicit request in the message $\ExampleRequest{}$. Such pairs of request message and attachment $\ExampleRequestReplyPair{}$ form the ground-truth in our evaluation framework.

\newcommand{\BethUser}{\User{}_{\text{Beth}}}
\newcommand{\DavidUser}{\User{}_{\text{David}}}
\newcommand{\ChangjiuUser}{\User{}_{\text{Changjiu}}}
\newcommand{\MostafaUser}{\User{}_{\text{Mostafa}}}
\newcommand{\SteveUser}{\User{}_{\text{Steve}}}

Fig.~\ref{fig:framework} shows a sample mailbox $\Mailbox{}_{\text{Beth}}$ of user $\BethUser{}$ containing four conversations $\{\Thread{}_{1}, \Thread{}_{2}, \Thread{}_{3}, \Thread{}_{4}\}$. During these conversations, $\BethUser{}$ responds with an attachment twice---at time $\Time{}_{1}$ and $\Time{}_{2}$. At time $\Time{}_{1}$, in this toy example, the set of candidate items that are available in the user's mailbox for recommendation contains only the attachment from $\MostafaUser{}$ received during the conversation $\Thread{}_{3}$. At $\Time{}_{2}$, however, the set of candidates includes attachments received on all four conversation threads---from $\ChangjiuUser{}$ ($\Thread{}_{1}$), $\SteveUser{}$ ($\Thread{}_{2}$), $\MostafaUser{}$ ($\Thread{}_{3}$), $\User{}_{\text{Pratik}}$ ($\Thread{}_{3}$), and $\DavidUser{}$ ($\Thread{}_{4}$)---as well as the item sent by $\BethUser{}$ previously on the conversation thread $\Thread{}_{1}$.

It is important to emphasize that our problem setting has two important constraints when recommending items, that any model should adhere to
\begin{inparaenum}[(1)]
	\item a \textbf{privacy} constraint: the model can only recommend items from a user's own mailbox, and
	\item a \textbf{temporal} constraint: the model can only recommend items that are already present in the user's mailbox at the time of recommendation.
\end{inparaenum}

\subsection{Attachment retrieval}

In addition to the above domain-specific constraints, we limit our setup to using a standard IR system $\RetrievalSystem{}$ for retrieval, and cast the problem that the model needs to solve as a query formulation task. Using an existing IR system has the practical benefit that one only needs to maintain a single system in contrast to the alternative where a separate attachment recommendation engine needs to be maintained. The model is presented with a message $\RequestMessage{}$ containing an explicit or an implicit content request. The model is tasked with generating a query that can be submitted to the retrieval system $\RetrievalSystem{}$ that retrieves a set of ranked messages $\Messages{}_{\RetrievalSystem{}}$ from the user's mailbox. Under this assumption, the retrieval system $\RetrievalSystem{}$ is treated as a black box, and we are only interested in optimizing the query formulation model. Note that a query is only formulated when it is clear that an item needs to be attached to a reply message (\S\ref{sec:rq}), such as is the topic of \citep{Dredze2006forgotattachment,Dredze2008intelligent}.

To extract a ranked list of attachable items from search engine $\RetrievalSystem{}$, we adopt an approach popular in entity retrieval frameworks \citep{Balog2006experts} where an entity model is the mixture of document models that the entity is associated with. For a given query $\Query{}$ issued at time $\CurrentTime{}$ in the mailbox of user $\User{}$, attachable items $\AttachableEntity{} \in \AttachableEntities{}$ are then ranked in decreasing order of
\newcommand{\RelevanceProb}{\CondProb{\AttachableEntity{}}{\Query{}, \User{}, \CurrentTime{}}}%
\newcommand{\RetrievalSystemProb}{\Prob[S_\RetrievalSystem{}]}%
\newcommand{\AssociationStrength}{\Apply{f}{\AttachableEntity{} \mid \Message{}}}
\begin{equation}
\label{eq:ranking}
\RelevanceProb{} \propto \frac{1}{\Apply{Z_1}{\AttachableEntity{}, \User{}, \CurrentTime{}}} \sum_{\substack{\Message{} \in \Mailbox{} \\ \MessageTime{} < \CurrentTime{}}} \CondProb[\RetrievalSystemProb]{\Message{}}{\Query{}} \AssociationStrength{}
\end{equation}
where $\CondProb[\RetrievalSystemProb]{\Message{}}{\Query{}}$ is the relevance score for message $\Message{}$ given query $\Query{}$ according to retrieval system $\RetrievalSystem{}$, $\MessageTime{}$ is the timestamp when the message $\Message{}$ appeared first in the mailbox $\Mailbox{}$, $\CurrentTime{}$ is time when the model needs to make the recommendation, $\AssociationStrength{}$ denotes the association strength between message $\Message{}$ and \entity{} $\AttachableEntity{}$, and
\newcommand{\NormalizationConstant}{\Apply{Z_1}{\AttachableEntity{}, \User{}, \CurrentTime{}}}%
\[
\NormalizationConstant{} = \sum_{\substack{\Message{} \in \Mailbox{} \\ \MessageTime{} < \CurrentTime{}}} \AssociationStrength{}
\]
is a normalization constant. The normalization constant $\NormalizationConstant$ avoids a bias towards attachable items that are associated with many messages (e.g., electronic business cards).

We associate messages with an attachable item according to the presence of the item within a message and its surrounding messages within the conversation:
$
\AssociationStrength{} = \Apply{\mathds{1}_{\Apply{\text{context}}{\Message{}}}}{\AttachableEntity{}}.
$
In this \paper{}, we take $\Apply{\text{context}}{\Message{}}$ to be all messages $\Message{}^\prime$ in the same conversation $\Thread{}_\Message{}$ as message $\Message{}$ that occurred before the time of recommendation, i.e., $\Time{}_{\Message{}^\prime} < \CurrentTime{}$. Note that the exact definition of an attachable \entity{} depends on the email domain and can include individual files, file bundles and hyperlinks to external documents amongst others (see \S\ref{sec:collections}).

\subsection{Evaluating query formulations}
\label{sec:framework-eval}

Once we have extracted $\RequestEntityPair{}$ pairs from an email corpus, each request message $\RequestMessage{}$ is presented to the query formulation model that we want to evaluate. The model generates a query $\Query{}$ conditioned on the message $\RequestMessage{}$. The query $\Query{}$ is submitted to retrieval system $\RetrievalSystem{}$ and attachable items extracted from the retrieved messages are determined according to Eq.~\ref{eq:ranking}. Given the ranked list of attachable \entities{} $\AttachableEntities{}_{\text{retrieved}}$ and the expected item $\TargetAttachableEntity{}$ we can compute standard rank-based IR metrics such as MRR and NDCG (\S\ref{sec:metrics}). We report the mean metric over all $\RequestEntityPair{}$ pairs extracted from the corpus. 

Our approach of using the historical information from an email corpus for evaluation is comparable to the application of \emph{click-through} data for similar purposes in Web search.
In the document retrieval scenario, a user's click on a document $\Document{}$ on the search result page is considered an implicit vote of confidence on the relevance of $\Document{}$ to the query. Learning to rank models can be trained on this \emph{click-through} data \cite{Joachims2002svm, xu2010improving, macdonald2009usefulness} if human relevance judgements are not available in adequate quantity. By explicitly attaching a file $\TargetAttachableEntity{}$, similarly, the user of an email system provides a strong indication that recommending $\TargetAttachableEntity{}$ at the time of composing $\ReplyMessage{}$ would have been useful. We can use this information to train and evaluate supervised models for ranking attachments at the time of email composition.


\section{Query formulation model}
\label{sec:model}

We first introduce a method for generating pseudo training data \citep{Asadi2011pseudo,Berendsen2013pseudo,azzopardi-building-2007,Beitzel2003automatic,Huurnink2010simulating,Huurnink2010validating,Kim2009,Tague1981simulation,Tague1980problems} without the need for manual assessments. Silver-standard queries are algorithmically synthesized for each request/attachment pair and consequently scored by measuring the query's ability to retrieve the relevant attachment (\S\ref{sec:model:subquery-generation}). The request/query pairs part of the pseudo training collection are then used to train a convolutional neural network (\S\ref{sec:model:neural}) that learns to extract query terms from a request message. In this \paper{}, we use a convolutional architecture rather than a recurrent one, as we intend to model term importance by term context without relying on the exact ordering of terms.

\subsection{Model training}
\label{sec:model:subquery-generation}

The candidate silver queries are extracted for request-response pairs $\RequestReplyPair{}$ in a training set. Given a request message $\RequestMessage{}$ and its associated target attachable \entity{} $\TargetAttachableEntity{} \in \AttachableEntities{}_{\ReplyMessage{}}$ that is attached to reply $\ReplyMessage{}$, where $\AttachableEntities{}_{\ReplyMessage{}}$ is the set of \entities{} attached to $\ReplyMessage{}$, the objective is to select the $\CandidateQueryTermBudget{}$ terms that are most likely to retrieve \entity{} $\TargetAttachableEntity{}$ according to Eq.~\ref{eq:ranking}.

In the ideal case, one considers the powerset of all terms within request message $\RequestMessage{}$ as candidate silver queries \citep{Kumaran2009reducinglongqueries,Xue2010verbosequeries}. However, considering all terms is computationally intractable in our case as email messages tend to average between \numprint{70} to \numprint{110} tokens (Table~\ref{tbl:statistics}).

In order to circumvent the intractability accompanied with computing the powerset of all terms in a message, we use the following stochastic strategy to select a fixed number of candidate query terms that we compute the powerset of. We consider two sources of query terms.
\newcommand{\RecallableMessagesFn}{\Messages{}_{\text{recallable}}}
\newcommand{\RecallableMessages}{\Apply{\RecallableMessagesFn{}}{\TargetAttachableEntity{}, \CurrentTime{}}}
The first source of candidate query terms consists of \textbf{subject} terms: topic terms in the subject of the request message. Email subjects convey relevance and context \citep{Weil2004overcomingoverload} and can be seen as a topical summary of the message. For the second source of query terms, we consider \textbf{recallable} terms: infrequent terms that occur frequently in messages $\RecallableMessages{} = \{\Message{} \in \Mailbox{} \mid \MessageTime{} < \CurrentTime{}, \TargetAttachableEntity{} \in \AttachableEntities{}_\Message{} \}$ that contained \entity{} $\TargetAttachableEntity{}$ and occur at least once in the request message.. That is, we gather all terms that have the potential to retrieve $\TargetAttachableEntity{}$ (according to Eq.~\ref{eq:ranking}) and select those terms that occur in at least 30\% of messages $\RecallableMessages{}$ and occur in less than 1\% of all messages.

\begin{algorithm}[t!]
\newcommand{\CandidateTerms}{T}
\newcommand{\CandidateTerm}{t}

\SetAlgoLined

\SetKwFunction{isUnwanted}{isUnwanted}

\KwData{request message $\RequestMessage{}$, query term budget $\CandidateQueryTermBudget{}$}
\KwResult{candidate silver queries $\RequestMessageCandidateQueries{}$}
set of candidate terms $\CandidateTerms{} \gets$ \{\}\;
\While{$($term candidates left $\wedge$ $\Length{\CandidateTerms{}} < \CandidateQueryTermBudget{}$$)$}{
	$S \gets$ uniformly random choose \textbf{subject} or \textbf{recallable terms}\;

	\If{$S = \varnothing$}{
		\Continue
	}

	$\CandidateTerm{} \gets \argmin_{\CandidateTerm{} \in S} \Apply{\text{df}}{\CandidateTerm{}}$\;
	\If{$\neg$\isUnwanted{$\CandidateTerm{}$}}{
		$\CandidateTerms{} = \CandidateTerms{} \, \cup \, \{t\}$\;
	}
	$S \gets S \setminus \{\CandidateTerm{}\}$
}
$\RequestMessageCandidateQueries \gets 2^{\CandidateTerms{}} - \{\varnothing\}$
\caption{Candidate query terms are selected by choosing a random query term source and selecting the query term with the lowest document frequency while ignoring unwanted query terms \citep{Salton1983boolean}. The \texttt{isUnwanted} predicate is true when the term is a stopword, contains a digit, contains punctuation or equals the names of the email sender/recipients. After selecting $\CandidateQueryTermBudget{}$ terms, we consider the powerset of selected terms as silver queries.\label{alg:subquery-generation}}
\end{algorithm}

To construct candidate silver queries for a request message $\RequestMessage{}$, we follow the strategy as outlined in Algorithm~\ref{alg:subquery-generation} that mimics the boolean query formulation process of \citet{Salton1983boolean}. Candidate terms are selected from either the \textbf{subject} or \textbf{recallable} source in increasing order of document frequency (i.e., infrequent terms first). Unwanted terms, such as stopwords, digits, punctuation and the names of the sender and recipients that occur in the email headers, are removed. Afterwards, we take the candidate queries $\RequestMessageCandidateQueries{}$ to be all possible subsets of candidate terms (excluding the empty set).

Once we have obtained the set of candidate queries $\RequestMessageCandidateQueries{}$ for request message $\RequestMessage{}$ we score the candidate queries as follows. For every $\CandidateQuery{} \in \RequestMessageCandidateQueries{}$ we rank email messages using retrieval system $\RetrievalSystem{}$ according to $\CandidateQuery{}$. We then apply Eq.~\ref{eq:ranking} to obtain a ranking over \entities{} $\UserTimeConstraintedAttachableEntities{}$ in the mailbox of user $\User{}$ at time $\CurrentTime{}$. As we know the target \entity{} $\TargetAttachableEntity{}$ to be retrieved for request message $\RequestMessage{}$, we quantify the performance of candidate query $\CandidateQuery{}$ by its reciprocal rank,
\newcommand{\CandidateQueryScore}[1]{\Apply{\text{score}}{#1}}
$
\CandidateQueryScore{\CandidateQuery{}} = \frac{1}{\Apply{\text{rank}}{\TargetAttachableEntity{}}} \in (0, 1]
$
where $\Apply{\text{rank}}{\TargetAttachableEntity{}}$ denotes the position of \entity{} $\TargetAttachableEntity{}$ (Eq.~\ref{eq:ranking}) in the item ranking.

After computing the score for every candidate silver query, we group queries that perform at the same level (i.e., that have the same score) for a particular request message $\RequestMessage{}$. We then apply two post-processing steps that improve silver-standard query quality based on the trade-off between query broadness and specificness. Following \citet{Salton1983boolean} on boolean query formulation, specific queries are preferred over broad queries to avoid loss in precision. Queries can be made more specific by adding terms. Consequently, within every group of equally-performing queries, we remove subset queries whose union results in another query that performs at the same level as the subsets. For example, if the queries \emph{``barack obama''}, \emph{``obama family''} and \emph{``barack obama family''} all achieve the same reciprocal rank, then we only consider the latter three-term query and discard the two shorter, broader queries. An additional argument for the strategy above follows from the observation that any term not part of the query is considered as undesirable during learning. Therefore, including all queries listed above as training material would introduce a negative bias against the terms \emph{``barack''} and \emph{``family''}. However, queries that are too specific can reduce the result set \citep{Salton1983boolean} or cause query drift \citep{Mitra1998drift}. Therefore, the second post-processing step constitutes the removal of supersets of queries that perform equal or worse. The intuition behind this is that the inclusion of the additional terms in the superset query did not improve retrieval performance. For example, if queries \emph{``barack obama''} and \emph{``barack obama president''} perform equally well, then the addition of the term \emph{``president''} had no positive impact on retrieval. Consequently, including the superset query (i.e., \emph{``barack obama president''}) in the training set is likely to motivate the inclusion of superfluous terms that negatively impact retrieval effectiveness.

\subsection{A convolutional neural network for ranking query terms}
\label{sec:model:neural}

\newcommand{\circled}[2][yscale=0.6]{%
\begin{tikzpicture}[#1, xscale=1.0, baseline=(char.base)]%
\node[shape=ellipse, draw, inner sep=1.5pt, text width=0.35cm] (char) {#2};
\end{tikzpicture}%
}

\newcommand{\EndOfRanking}[1]{\scalebox{1}[0.8]{\circled{\footnotesize{}\normalfont{}\hspace*{-0.5pt}EoR}}}
\newcommand{\CaptionEndOfRanking}{\EndOfRanking}

\begin{figure}[t]
\small
\begin{tikzpicture}[yscale=-0.40]
\node[anchor=west, align=center, text width=1cm] at (0.0, 0) {\textbf{Rank}};
\node[anchor=west, align=center, text width=4cm] at (1.0, 0) {\textbf{Term}};
\node[anchor=west, align=center, text width=1cm] at (5.2, 0) {\textbf{Score}};
\newcommand{\RankedItem}[3]{
	\node[anchor=west, align=center, text width=1cm] at (0.0, #1) {\texttt{#1.}};
	\node[anchor=west, align=center, text width=4cm] at (1.0, #1) {\texttt{#2}};
	\node[anchor=west, align=center, text width=1cm] at (5.2, #1) {\texttt{#3}};
}
\RankedItem{1}{initech}{0.20}
\RankedItem{2}{initech}{0.18}
\RankedItem{3}{transition}{0.15}
\RankedItem{4}{- - - - - -\EndOfRanking{}- - - - - -}{0.10}
\RankedItem{5}{david}{0.07}
\RankedItem{6}{...}{}
\end{tikzpicture}
\vspace*{-\baselineskip}
\caption{Terms in the request message of Fig.~\ref{fig:example} are ranked by our model. In addition to the terms, the model also ranks a \protect\CaptionEndOfRanking{} token that specifies the query end. The final query becomes \emph{``initech transition''} as duplicate terms are ignored.\label{fig:example_ranking}}

\end{figure}

After obtaining a set of candidate queries $\RequestMessageCandidateQueries{}$ for every request/\entity{} pair $\RequestEntityPair{}$ in the training set, we learn to select query terms from email threads using a convolutional neural network model that convolves over the terms contained in the email thread. Every term is characterized by its context and term importance features that have been used to formulate queries in previous work \citep{Bendersky2008discoveringkeyconcepts,Cetintas2012querygenerationpriorart,Xue2009querygenerationpatents,Mahdabi2011buildingqueriespriorart,Zhao2008sqs,Kumaran2009reducinglongqueries,He2004scs}. Our model jointly learns to
\begin{inparaenum}[(1)]
	\item generate a ranking of message terms, and
	\item determine how many terms of the message term ranking should be included in the query.
\end{inparaenum}
In order to determine the number of terms included in the query, the model learns to rank an end-of-ranking token \EndOfRanking{} in addition to the message terms. Fig.~\ref{fig:example_ranking} shows a term ranking for the example in Fig.~\ref{fig:example}. Terms in the request message are ranked in decreasing order of the score predicted by our model. Terms appearing at a lower rank than the \EndOfRanking{} are not included in the query.

\newcommand{\WeightMatrix}[1]{W_{#1}}
\newcommand{\TermEmbeddingMatrix}{\WeightMatrix{\text{repr}}}

\newcommand{\ReLU}{
	\begin{tikzpicture}
		\draw (0, 0) -- (0.5, 0);
		\draw (0.3, 0) -- (0.60, 0.50);
	\end{tikzpicture}
}

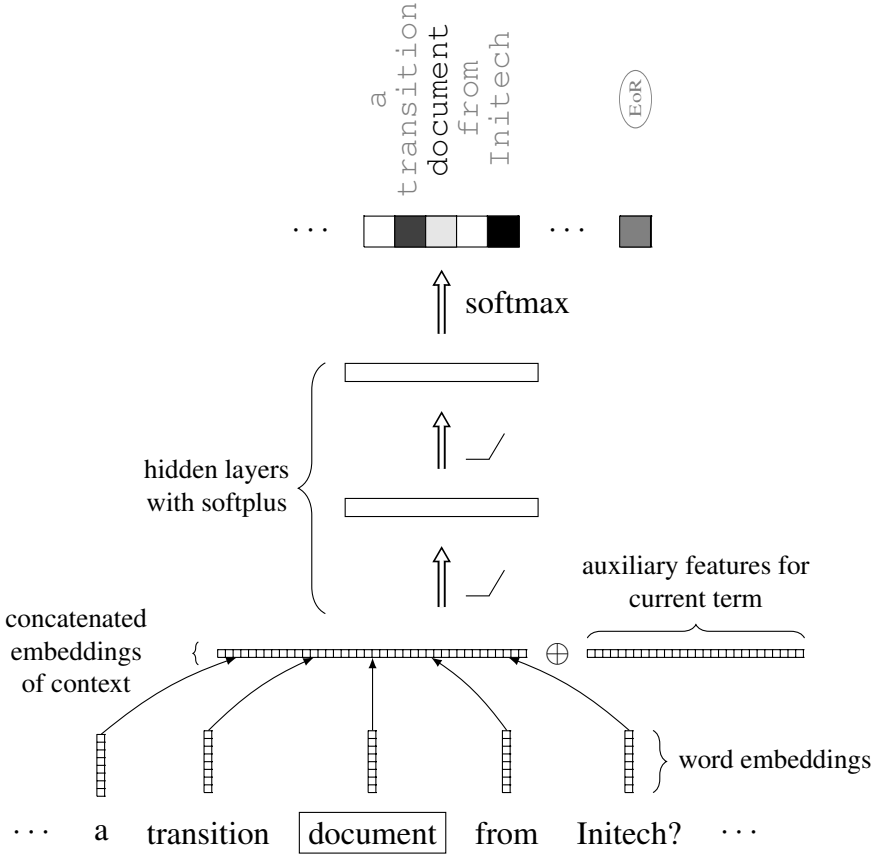
\begin{figure}[t]

\resizebox{\columnwidth}{!}{%
\begin{tikzpicture}[font=\large]

\tikzstyle{vecArrow} = [thick, decoration={markings,mark=at position
   1 with {\arrow[semithick]{open triangle 60}}},
   double distance=1.4pt, shorten >= 5.5pt,
   preaction = {decorate},
   postaction = {draw,line width=1.4pt, white,shorten >= 4.5pt}]
]

\node [rectangle, minimum height = 0.09cm, minimum width = 4cm, inner sep = 0] (concat_context) {};
\draw [anchor = bottom left, step=0.10cm, black, shift={(concat_context.south west)}] (concat_context.south west) grid (concat_context.north east);

\newcommand{\Input}[3]{
\node [anchor=base, align=center, #3] (#1) {#2};
}
\newcommand{\Word}[4]{
\Input{#1}{#2}{#3}
\node [rectangle, above = 0.25cm of #1, minimum height = 0.79cm, minimum width = 0.10cm, inner sep = 0] (#1_repr) {};
\draw [anchor = bottom left, step=0.10cm, black, shift={(#1_repr.south west)}] (#1_repr.south west) grid (#1_repr.north east);

\draw [->] (#1_repr.north) to #4;
}

\Word{document}{document}{draw, below = 2cm of concat_context}{(concat_context.south)}

\Word{transition}{transition}{left = 0.25 of document}{[bend left=10] ($ (concat_context.south west) + (1.25, 0) $)}
\Word{a}{a}{left = 0.25 of transition}{[bend left=10] ($ (concat_context.south west) + (0.25, 0) $)}
\Input{ellipsis_begin}{$\cdots$}{left = 0.25cm of a}

\Word{from}{from}{right = 0.25 of document}{[bend right=10] ($ (concat_context.south east) - (1.25, 0) $)}
\Word{initech}{Initech?}{right = 0.25 of from}{[bend right=10] ($ (concat_context.south east) - (0.25, 0) $)}
\Input{ellipsis_end}{$\cdots$}{right = 0.25 of initech}

\node [right = 0.1cm of concat_context] (concat_ellipsis) {$\oplus$};

\node [rectangle, right = 0.1cm of concat_ellipsis, minimum height = 0.09cm, minimum width = 2.8cm, inner sep = 0] (auxiliary) {};
\draw [anchor = bottom left, step=0.10cm, black, shift={(auxiliary.south west)}] (auxiliary.south west) grid (auxiliary.north east);

\node [rectangle, draw, above = 1.5cm of concat_ellipsis, minimum width=2.5cm, minimum height=0.2cm, xshift=-1.5cm] (hidden_first) {};
\draw[vecArrow] ($ (concat_ellipsis) + (-1.5, 0.60) $) to node [right, xshift=5pt] {\ReLU{}} ($ (hidden_first) - (0, 0.5) $);

\node [rectangle, draw, above = 1.5cm of hidden_first, minimum width=2.5cm, minimum height=0.2cm] (hidden_second) {};
\draw[vecArrow] ($ (hidden_first) + (0, 0.5) $) to node [right, xshift=5pt] {\ReLU{}} ($ (hidden_second) - (0, 0.5) $);

\node [rectangle, above = 1.5cm of hidden_second, minimum height = 0.40cm, minimum width = 2.0cm, inner sep = 0] (softmax) {};
\draw [anchor = bottom left, step=0.40cm, black, shift={(softmax.south west)}] (softmax.south west) grid (softmax.north east);

\node [anchor=base, align=center, left = 0.25cm of softmax] {$\cdots{}$};
\node [anchor=base, align=center, right = 0.25cm of softmax] (softmax_right_ellipsis) {$\cdots{}$};

\draw[vecArrow] ($ (hidden_second) + (0, 0.5) $) to node [right, xshift=5pt] {softmax} ($ (softmax) - (0, 0.50) $);

\node [rectangle, right = 0.25cm of softmax_right_ellipsis, minimum height = 0.40cm, minimum width = 0.40cm, inner sep = 0] (softmax_eor) {};
\draw [anchor = bottom left, step=0.40cm, black, shift={(softmax_eor.south west)}] (softmax_eor.south west) grid (softmax_eor.north east);

\newcommand{\SoftmaxLabel}[2]{
\node [anchor=base, align=center, above = 1.5cm of softmax, rotate = 90, #2] {\texttt{#1}};
}

\SoftmaxLabel{a}{yshift=0.60cm, text=gray}
\SoftmaxLabel{transition}{yshift=0.20cm, text=gray}
\SoftmaxLabel{document}{yshift=-0.20cm}
\SoftmaxLabel{from}{yshift=-0.60cm, text=gray}
\SoftmaxLabel{Initech}{yshift=-1.0cm, text=gray}

\SoftmaxLabel{\EndOfRanking{}}{yshift=-2.85cm, text=gray}

\draw [fill=black] ($ (softmax.south east) + (0, 0) $) rectangle ($ (softmax.south east) + (-0.40cm, 0.40cm) $);
\draw [fill=black!75] ($ (softmax.south west) + (0.40cm, 0) $) rectangle ($ (softmax.south west) + (0.80cm, 0.40cm) $);
\draw [fill=black!50] ($ (softmax.south east) + (1.30cm, 0) $) rectangle ($ (softmax.south east) + (1.70cm, 0.40cm) $);
\draw [fill=black!10] ($ (softmax.south west) + (0.80cm, 0) $) rectangle ($ (softmax.south west) + (1.20cm, 0.40cm) $);

\draw [decorate, decoration={brace,amplitude=10pt}, xshift=-4pt, yshift=0pt] ($ (hidden_first.south west) - (0.25, 1.25) $) -- ($ (hidden_second.north west) - (0.25, 0) $) node [black, midway, xshift=-40pt, rotate=0] {\normalsize \begin{tabular}{c}hidden layers\\with softplus\end{tabular}};

\draw [decorate, decoration={brace,amplitude=5pt,mirror}, xshift=40pt, yshift=0pt] ($ (initech_repr.south east) + (0.25, 0) $) -- ($ (initech_repr.north east) + (0.25, 0) $) node [black, midway, xshift=45pt, rotate=0] {\normalsize \begin{tabular}{c}word embeddings\end{tabular}};

\draw [decorate, decoration={brace,amplitude=2pt}, xshift=-4pt, yshift=0pt] ($ (concat_context.north west) + (-0.25, -0.20) $) -- ($ (concat_context.south west) + (-0.25, 0.20) $) node [black, midway, xshift=-45pt, rotate=0] {\normalsize \begin{tabular}{c}concatenated\\embeddings\\of context\end{tabular}};

\draw [decorate, decoration={brace,amplitude=5pt,mirror}, xshift=-4pt, yshift=0pt] ($ (auxiliary.south east) + (0, 0.25) $) -- ($ (auxiliary.south west) + (0, 0.25) $) node [black, midway, yshift=20pt, rotate=0] {\normalsize \begin{tabular}{c}auxiliary features for\\current term\end{tabular}};

\end{tikzpicture}}
\caption{The model convolves over the terms in the message. For every term, we represent it using the word representations of its context. These representations are learned as part of the model. After creating a representation of the term's context, we concatenate auxiliary features (Table~\ref{tbl:features}). At the output layer, a score is returned as output for every term in the message. The softmax function converts the raw scores to a distribution over the message terms and the \protect\CaptionEndOfRanking{} token. Grayscale intensity in the distribution depicts probability mass.\label{fig:model}}

\end{figure}

\newcommand{\Term}{w}
\newcommand{\ContextWindow}{L}

\begin{table}[t]

\caption{Overview of term representation (learned as part of the model) and auxiliary features.\label{tbl:features}}
\small
\resizebox{\columnwidth}{!}{%
\begin{tabularx}{\linewidth}{lX}
\toprule

\multicolumn{2}{l}{\textbf{Context features (learned representations)}} \\

\texttt{term} & Representation of the term. \\
\texttt{context} & Representations of the context surrounding the term. \\

\midrule

\multicolumn{2}{l}{\textbf{Part-of-Speech features}} \\

\texttt{is\_noun} & POS tagged as a noun \citep{Bendersky2008discoveringkeyconcepts} \\
\texttt{is\_verb} & POS tagged as a verb \\
\texttt{is\_other} & POS tagged as neither a noun or a verb \\

\midrule

\multicolumn{2}{l}{\textbf{Message features}} \\

\texttt{is\_subject} & Term occurrence is part of the subject \citep{Cetintas2012querygenerationpriorart} \\
\texttt{is\_body} & Term occurrence is part of the body \citep{Cetintas2012querygenerationpriorart} \\
\texttt{Abs. TF} & Abs. term freq. within the message \citep{Xue2009querygenerationpatents} \\
\texttt{Rel. TF} & Rel. term freq. within the message \citep{Xue2009querygenerationpatents} \\
\texttt{Rel. pos.} & Rel. position of the term within the message \\
\texttt{is\_oov\_repr} & Term does not have a learned representation \\

\midrule

\multicolumn{2}{l}{\textbf{Collection statistics features}} \\

\texttt{IDF} & Inverse document frequency of the term \citep{Xue2009querygenerationpatents} \\
\texttt{TF-IDF} & \texttt{TF} $\times$ \texttt{IDF} \citep{Xue2009querygenerationpatents} \\

\texttt{Abs. CF} & Abs. collection freq. within the collection \\
\texttt{Rel. CF} & Rel. collection freq. within the collection \\

\texttt{Rel. Entropy} & KL divergence from the unsmoothed collection term distribution to the smoothed ($\lambda = 0.5$) document term distribution \citep{Mahdabi2011buildingqueriespriorart} \\

\texttt{SCQ} & Similarity Collection/Query \citep{Zhao2008sqs} \\
\texttt{ICTF} & Inverse Collection Term Frequency \citep{Kumaran2009reducinglongqueries} \\
\texttt{Pointwise SCS} & Pointwise Simplified Clarity Score \citep{He2004scs} \\

\bottomrule
\end{tabularx}%
}

\end{table}

Our convolutional neural network (\Neural{}) term ranking model is organized as follows; see Fig.~\ref{fig:model} for an overview. Given request message $\RequestMessage{}$, we perform a convolution over the $n$ message terms $\Term{}_1, \ldots, \Term{}_n$. Every term $\Term{}_k$ is characterized by
\begin{inparaenum}[(1)]
	\item the term $\Term{}_k$ itself,
	\item the $2 \cdot \ContextWindow{}$ terms, $\Term{}_{k - \ContextWindow{}}, \ldots, \Term{}_{k - 1}, \Term{}_{k + 1}, \ldots, \Term{}_{k + \ContextWindow{}}$, surrounding term $\Term{}_k$ where $\ContextWindow$ is a context width hyperparameter, and
	\item auxiliary query term quality features (see Table~\ref{tbl:features}).
\end{inparaenum}
\newcommand{\TermRankingFn}{g}
\newcommand{\TermRankingScore}[2]{\Apply{\TermRankingFn}{#1, #2}}
\newcommand{\EndOfRankFn}{h}
\newcommand{\EndOfRankScore}[1]{\Apply{\EndOfRankFn}{#1}}

For every term in the message, the local context features (1st part of Table~\ref{tbl:features}) are looked up in term embedding matrix $\TermEmbeddingMatrix{}$ (learned as part of the model) and the auxiliary features (part 2-4 of Table~\ref{tbl:features}) are computed. For the auxiliary features, we apply min-max feature scaling on the message-level such that they fall between $0$ and $1$. The flattened embeddings, concatenated with the auxiliary feature vector, are fed to the neural network. At the output layer, the network predicts a term ranking score, $\TermRankingScore{\Term{}_k}{\RequestMessage{}}$, for every term. In addition, a score for the \EndOfRanking{} token, $\EndOfRankScore{\RequestMessage{}}$, is predicted as well. The \EndOfRanking{} score function $\EndOfRankFn{}$ takes the same form as the term score function $\TermRankingFn{}$, but has a separate set of parameters and takes as input an aggregated vector that represents the whole message. More specifically, the input to the \EndOfRanking{} score function is the average of the term representations and their auxiliary features.

The ranking scores are then transformed into a distribution over message terms and the \EndOfRanking{} token as follows:
\newcommand{\PartitionFn}{\exp{\EndOfRankScore{\RequestMessage{}}} + \sum^{\Length{\RequestMessage{}}}_{l = 1} \exp{\TermRankingScore{\Term{}_l}{\RequestMessage{}}}}
\begin{eqnarray*}
\CondProb{\Term{}_k}{\RequestMessage{}} & = & \frac{1}{Z_2} {\exp{\TermRankingScore{\Term{}_k}{\RequestMessage{}}}} \\
\CondProb{\EndOfRanking{}}{\RequestMessage{}} & = & \frac{1}{Z_2} {\exp{\EndOfRankScore{\RequestMessage{}}}} \nonumber
\end{eqnarray*}
with
$
\Apply{Z_2}{\RequestMessage} = \PartitionFn{}
$ as a normalization constant.
\newcommand{\NormalizedGroundTruthProb}[1]{\Apply{Q}{#1}}
\newcommand{\UnnormalizedGroundTruthProb}[1]{\Apply{\hat{Q}}{#1}}
For every query $\CandidateQuery{} \in \CandidateQueries{}$, the ground-truth distribution equals:
\begin{eqnarray}
\label{eq:rank_groundtruth}
\CondProb[\NormalizedGroundTruthProb]{\Term{}_k}{\CandidateQuery{}} & = & \alpha \cdot \frac{\Apply{\mathds{1}_{\CandidateQuery{}}}{\Term{}_k}}{\Apply{\FrequencyFn{}}{\Term{}_k, \RequestMessage{}} \cdot \Length{\CandidateQuery{}}}  \\
\CondProb[\NormalizedGroundTruthProb]{\EndOfRanking{}}{\CandidateQuery{}} & = & (1 - \alpha) \nonumber
\end{eqnarray}
where $\alpha = 0.95$ is a hyperparameter that determines the probability mass assigned to the \EndOfRanking{} token and $\Apply{\mathds{1}_{\CandidateQuery{}}}{\Term{}_k}$ is the indicator function that evaluates to $1$ when term $\Term{}_k$ is part of silver query $\CandidateQuery{}$. The frequency count $\Apply{\FrequencyFn{}}{\Term{}_k, \RequestMessage{}}$ denotes the number of times term $\Term{}_k$ occurs in message $\RequestMessage{}$ and is included such that frequent and infrequent message terms are equally important.

Eq.~\ref{eq:rank_groundtruth} assigns an equal probability to every unique term in message $\RequestMessage{}$ that occurs in silver query $\CandidateQuery{}$. Our cost function consists of two objectives. The first objective aims to make $\CondProb{\;\cdot}{\RequestMessage{}}$ close to $\CondProb[\NormalizedGroundTruthProb]{\;\cdot}{\CandidateQuery{}}$ by minimizing the cross entropy:
\newcommand{\CELossFn}{\LossFn{}_{\text{xent}}}
\newcommand{\CutOffLossFn}{\LossFn{}_{\text{cutoff}}}
\newcommand{\CELoss}[1][\RequestMessage{}]{\Apply{\CELossFn{}}{\Parameters{} \mid #1, \CandidateQuery{}}}
\begin{equation}
\label{eq:ce_loss_fn}
\CELoss{} = - \sum_{\omega \in \Omega} \CondProb[\NormalizedGroundTruthProb]{\omega}{\CandidateQuery{}} \log{\CondProb{\omega}{\RequestMessage{}}}
\end{equation}
where $\Omega = \left(\Term{}_1, \ldots, \Term{}_n, \EndOfRanking{}\right)$ is the sequence of all terms in the message $\RequestMessage{}$ concatenated with the end-of-ranking token. Eq.~\ref{eq:ce_loss_fn} promotes term ranking precision as it causes terms in the silver query to be ranked highly, immediately followed by the end-of-ranking token. The second objective encourages term ranking recall by dictating that the \EndOfRanking{} token should occur at the same rank as the lowest-ranked silver query term:
\newcommand{\CutOffLoss}[1][\RequestMessage{}]{\Apply{\CutOffLossFn{}}{\Parameters{} \mid #1, \CandidateQuery{}}}
\begin{equation}
\label{eq:cutoff_loss_fn}
\CutOffLoss{} = \left(\Apply{\text{min}_{\Term{} \in \CandidateQuery{}}}{\TermRankingScore{\Term{}}{\RequestMessage{}}} - \EndOfRankScore{\RequestMessage{}}\right)^2
\end{equation}
The two objectives (Eq.~\ref{eq:ce_loss_fn}-\ref{eq:cutoff_loss_fn}) are then combined in a batch objective:
\newcommand{\Batch}{B}
\begin{eqnarray}
\label{eq:loss}
\Apply{\LossFn{}}{\Parameters{} \mid \Batch{}} & = & \frac{1}{\Length{\Batch{}}} \sum_{\left(\Message{}, \CandidateQuery{}\right) \in \Batch{}} \CandidateQueryScore{\CandidateQuery{}} \big(\CELoss[\Message{}]{} + \CutOffLoss[\Message{}]{} \big) \nonumber \\
& &  + \frac{1}{2 \lambda} \sum_{W \in \Parameters{}_W} \sum_{ij} W^2_{ij}
\end{eqnarray}%
where $\Batch{}$ is a uniformly random sampled batch of message/query pairs, $\Parameters{}_W$ is the set of parameter matrices and $\lambda$ is a weight regularization parameter. Objective~\ref{eq:ce_loss_fn} resembles a list-wise learning to rank method \citep{Cao2007listwise} where a softmax over the top-ranked items is used. Eq.~\ref{eq:loss} is then optimized using gradient descent.


\section{Experimental set-up}
\label{sec:experiment}

\subsection{Research questions}
\label{sec:rq}

\newcommand{\RQ}[2]{%
    \begin{description}[topsep=2pt,leftmargin=0.8cm]%
    \phantomsection\label{section:setup:rq#1}%
    \item[RQ#1] #2%
    \end{description}%
}

\newcommand{\RQRef}[1]{\textbf{\hyperref[section:setup:rq#1]{RQ#1}}}

As indicated in the introduction of this \paper{}, we seek to answer the following research questions:

\RQ{1}{\ResearchQuestionOne{}}
What if we consider the different fields (subject and body) in the email message when selecting query terms? To what extent do methods based on selecting the top ranked terms according to term scoring methods (e.g., \TFIDF{}, \RelativeEntropy{}) perform? Can \Neural{}s outperform state-of-the-art learning to rank methods? What can we say about the length of the queries extracted by the different methods?

\RQ{2}{\ResearchQuestionTwo{}}
In the case that \Neural{}s improve retrieval effectiveness over query extraction methods: what can we say about the errors made by \Neural{}s? In particular, in what cases do our deep convolutional neural networks perform better or worse compared to the query term ranking methods under comparison?

\RQ{3}{\ResearchQuestionThree{}}
Are all types of features useful? Can we make any inferences about the email domain or the attachable \entity{} recommendation task?

\subsection{Experimental design}
\label{sec:design}

We operate under the assumption that an incoming message has been identified as a request for content. A query is then formulated from the message using one of the query formulation methods (\S\ref{sec:baselines}). To answer the research questions posed in \S\ref{sec:rq}, we compare \Neural{}s with existing state-of-the-art query term selection methods on enterprise email collections (\RQRef{1}). In addition, we look at the query lengths generated by the formulation methods that perform best. \RQRef{2} is answered by examining the per-instance difference in Reciprocal Rank (RR) (\S\ref{sec:metrics}). After that, we perform a qualitative analysis where we examine the outlier examples. For \RQRef{3} we perform a feature ablation study where we systematically leave out a feature category (Table~\ref{tbl:features}).

\subsection{Data collections and pre-processing}
\label{sec:collections}

\begin{table}[t!]
\caption{Overview of the enterprise email collections used in this \paper{}: \AvocadoCollection{} (public) and \InternalCollection{} (proprietary).\label{tbl:statistics}}
\resizebox{0.80\columnwidth}{!}{%
}
\end{table}

We answer our research questions (\S\ref{sec:rq}) using two enterprise email collections that each constitute a single tenant (i.e., an organization):
\begin{inparaenum}[(1)]
\item the \AvocadoCollection{} collection \citep{Oard2015avocado} is a public data set that consists of emails taken from \numprint{279} custodians of a defunct information technology company, and
\item the Proprietary Internal Emails (\InternalCollection{}) collection is a proprietary dataset of Microsoft internal emails obtained through an employee participation program.
\end{inparaenum}
We perform cross-validation on the collection level. That is, when testing on one collection, models are trained and hyperparameters are selected on the other collection (i.e., train/validate on \AvocadoCollection{}, test on \InternalCollection{} and vice versa). Models should generalize over multiple tenants (i.e., organizations) as maintaining specialized models is cumbersome. In addition, model effectiveness should remain constant over time to avoid frequent model retraining. Consequently, topical regularities contained within a tenant should not influence our comparison. Furthermore, privacy concerns may dictate that training and test tenants are different. On the training set, we create a temporal 95/5 split for training and validation/model selection. On the test set, all instances are used for testing only. Attachable \entities{} consist of file attachments and URLs; see Table~\ref{tbl:statistics}.

The training and test instances are extracted, for every collection independently, in the following unsupervised manner. File attachments and normalized URLs are extracted from all messages. We remove outlier \entities{} by trimming the bottom and top 5\% of the attachable \entity{} frequency distribution. Infrequent \entities{} are non-retrievable and are removed in accordance to our experimental design (\S\ref{sec:design}). However, in this \paper{} we are interested in measuring the performance on retrieving attachable \entities{} that are in the ``torso'' of the distribution and, consequently, frequent \entities{} (e.g., electronic business cards) are removed as well. Any message that links to an attachable \entity{} (i.e., URL or attachment) and the message preceding it is considered a request/reply instance. In addition, we filter request/reply instances containing attachable \entities{} that
\begin{inparaenum}[(a)]
	\item occurred previously in the same thread, or
	\item contain attachable \entities{} that do not occur in the user's mailbox before the time of the request message (see \S\ref{sec:design}).
\end{inparaenum}
Mailboxes are indexed and searched using Indri \cite{Strohman2005indri,VanGysel2017pyndri}. For message retrieval, we use the Query-Likelihood Model (QLM) with Dirichlet smoothing \citep{Zhai2004smoothing} where the smoothing parameter ($\mu$) is set to the average message length \citep{Balog2006experts,Weerkamp2009contextualinformationemailsearch}. At test time, query formulation methods extract query terms from the request message, queries are executed using the email search engine of the user (i.e., Indri) and attachable \entities{} are ranked according to Eq.~\ref{eq:ranking}. Rankings are truncated such that they only contain the top-1000 messages and top-100 attachable \entities{}. The ground truth consists of binary relevance labels where \entities{} linked in the reply message are relevant.

\newcommand{\RQAnswer}[2]{
\begin{description}[topsep=0pt,parsep=0pt,leftmargin=0.8cm]
	\item[\RQRef{#1}] #2
\end{description}}

\newcommand{\Significant}{$^{*\phantom{*}}$}
\newcommand{\MoreSignificant}{$^{**}$}
\newcommand{\HighlySignificant}{$^{**}$}

\subsection{Methods under comparison}
\label{sec:baselines}

As the attachable \entity{} recommendation task is first introduced in this \paper{}, there exist no methods directly aimed at solving this task. However, as mentioned in the related work section (\S\ref{sec:related_work}), there are two areas (prior art search and verbose query reduction) that focus on extracting queries from texts. Consequently, we use computationally tractable methods (\S\ref{sec:related_work}) from these areas for comparison:
\newcommand{\Subject}{subject}
\newcommand{\Body}{body}
\begin{inparaenum}[(1)]
	\item single features, i.e., term frequency (\TF{}), \TFIDF{}, \logTFIDF{}, relative entropy (\RelativeEntropy{}), used for prior art retrieval \citep{Xue2009transformingpatents,Xue2009querygenerationpatents,Mahdabi2011buildingqueriespriorart,Cetintas2012querygenerationpriorart} where the top-$k$ unique terms are selected from either the \Subject{}, the \Body{} or both. Hyperparameters $1 \leq k \leq 15$ and, in the case of \RelativeEntropy{}, $\lambda = 0.1, \ldots, 0.9$ are optimized on the validation set,
	\item the learning to rank method for query term ranking proposed by \citet{Lee2009querytermsranking} for the verbose query reduction task. To adapt this method for our purposes, we use the domain-specific features listed in Table~\ref{tbl:features} (where the representations are obtained by training a Skip-Gram word2vec model with default parameters on the email collection), only consider single-term groups (as higher order term groups are computationally impractical during inference) and use a more-powerful pairwise \LeaveOutRankSVM{} \citep{Joachims2002svm} (with default parameters \citep{Sculley2009ranksvm}) instead of a pointwise approach. Feature value min-max normalization is performed on the instance-level. The context window width $\ContextWindow{} = 3, 5, \ldots, 15$ is optimized on the validation set.
	In addition, we consider the following baselines:
	\item all terms (\AllTerms{}) are selected from either the \Subject{}, the \Body{} or both,
	\item random terms, selected from the \Subject{}, the \Body{} or both, where we either select $k$ unique terms randomly (\Multinomial{}) or a random percentage $p$ of terms (\BernouilliProcess{}). Hyperparameters $1 \leq k \leq 15$ and $p = 10\%, 20\%, \ldots, 50\%$ are optimized on the validation set.
	Finally, we consider a pointwise alternative to the \NeuralRankCutoff{} model:
	\item \NeuralLogistic{} with the logistic function at the output layer (instead of the softmax) and terms are selected if their score exceeds a threshold optimized on the validation set F1 score (instead of the \EndOfRanking{} token).
\end{inparaenum}

The \Neural{} models are trained for \numprint{30} iterations using Adam \citep{Kingma2014adam} with $\alpha = 10^{-5}$, $\beta_1 = 0.9$, $\beta_2 = 0.999$ and $\epsilon = 10^{-8}$. The iteration with the lowest data loss on the validation set is selected. Word embeddings are \numprint{128}-dimensional, the two hidden layers have \numprint{512} hidden units each, with dropout ($p = 0.50$) and the softplus activation function. Weights are initialized according to \citep{Glorot2010}. We set the batch size $\Length{\Batch{}} = 128$ and regularization lambda $\lambda = 0.1$. The context window width $\ContextWindow{} = 3, 5, \ldots, 15$ is optimized on the validation set. For word embeddings (both as part of the \Neural{} and \LeaveOutRankSVM{}), we consider the top-$60\text{k}$ terms. Infrequent terms are represented by a shared representation for the unknown token.

\subsection{Evaluation measures and significance}
\label{sec:metrics}

To answer \RQRef{1}, we report the Mean Reciprocal Rank (\RecipRank{}), Normalized Discounted Cumulative Gain (\NDCG{}) and Precision at rank 5 (\PrecisionCut{}) evaluation measures computed using \texttt{trec\_eval}.\footnote{\url{https://github.com/usnistgov/trec_eval}} For \RQRef{2}, we examine the pairwise differences in terms of Reciprocal Rank (RR). In the case of \RQRef{3}, we measure the relative difference in \RecipRank{} when removing a feature category. Wherever reported, significance is determined using a two-tailed paired Student t-test.


\section{Results \& discussion}
\label{sec:discussion}

We start by presenting a comparison between methods (\RQRef{1}) on attachable \entity{} recommendation, provide an error analysis (\RQRef{2}) and perform a feature importance study (\RQRef{3}) (see \S\ref{sec:design} for an overview of the experimental design).

\subsection{Overview of experimental results}

\begin{table}[t!]
\setlength{\tabcolsep}{0pt} %
\renewcommand{\arraystretch}{1.0} %
\caption{Comparison of \RankModel{} with state-of-the-art query formulation methods (\S\ref{sec:baselines}) on the \AvocadoCollection{} and \InternalCollection{} collections. The numbers reported on \AvocadoCollection{} were obtained using models trained/validated on \InternalCollection{} and vice versa (\S\ref{sec:collections}). Significance is determined using a paired two-tailed Student t-test ($^{*}$ $p < 0.10$; \MoreSignificant{} $p < 0.05$) \citep{Smucker2007significance} between \NeuralRankCutoff{} and the second best performing method (in italic).\label{tbl:results}}
\centering
\resizebox{0.90\columnwidth}{!}{}
\end{table}
\begin{figure}[t!]
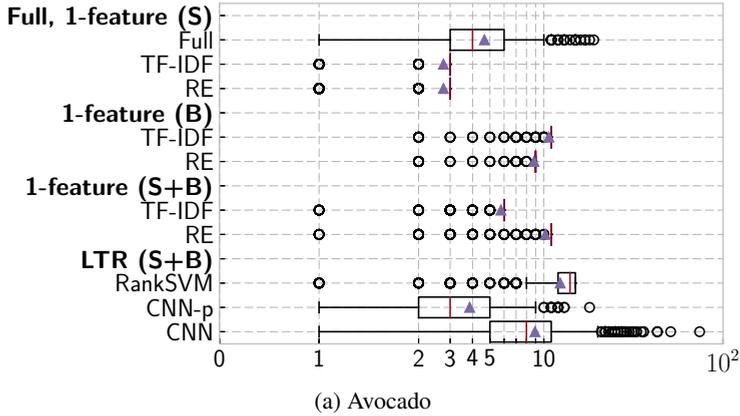
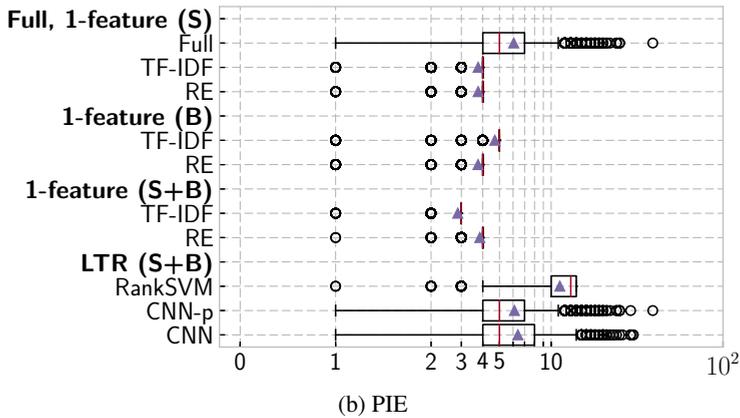

\centering
\begin{subfigure}[b]{\columnwidth}
    \centering
    \myincludegraphics[width=0.80\textwidth]{resources/generated_queries-AvocadoCollection.pdf}
    \caption{\AvocadoCollection{}}
    \label{fig:query_boxplot:avocado}
\end{subfigure}
~
\begin{subfigure}[b]{\columnwidth}
    \centering
    \myincludegraphics[width=0.80\textwidth]{resources/generated_queries-InternalCollection.pdf}
    \caption{\InternalCollection{}}
    \label{fig:query_boxplot:internal}
\end{subfigure}
\caption{Query length distribution according to the most prominent methods (\textbf{S} and \textbf{B} denote subject and body fields, resp.). \TFIDF{}, \RelativeEntropy{} and \LeaveOutRankSVM{} select a fixed number of terms for all queries, whereas \Neural{}s select a variable number of terms.\label{fig:query_boxplot}}
\end{figure}

\RQAnswer{1}{Table~\ref{tbl:results} shows the recommendation results of attachable \entities{} in enterprise email collections (\S\ref{sec:collections}).}

We see that \RankModel{} outperforms all other query formulation methods on both enterprise email collections. Significance is achieved (\RecipRank{}) between \RankModel{} and the second-best performing methods: \NeuralLogistic{} and \SubjectRelativeEntropy{} (subject), respectively, on the \AvocadoCollection{} and \InternalCollection{}. The methods that select terms only from the email subject perform strongly on both collections. Within the set of subject methods (1st part of Table~\ref{tbl:results}), we also observe that there is little difference between the methods. In fact, for \AvocadoCollection{}, simply taking the subject as query performs better than any of the remaining subject-based methods. Subjects convey relevance and context \citep{Weil2004overcomingoverload} and can compactly describe the topic of a content request. However, in order to generate better queries, we need to extract additional terms from the email body as email subjects tend to be short.

The same methods that we used to extract terms from the subject perform poorly when only presented with the body of the email (2nd part of Table~\ref{tbl:results}). When allowing the methods to select terms from the full email (subject and body), we see a small increase in retrieval performance (3rd part of Table~\ref{tbl:results}) compared to body-only terms. However, none of the methods operating on the full email message manage to outperform the subject by itself.

Our learning to rank (LTR) query term methods (last part of Table~\ref{tbl:results}) outperform the subject-based methods (ignoring \LeaveOutRankSVM{}). This comes as little surprise, as the presence of a term in the subject is incorporated as a feature in our models (Table~\ref{tbl:features}). The reason why \LeaveOutRankSVM{}, originally introduced for reducing long search queries, performs poorly is due to the fact that its training procedure fails to deal with the long length of emails. That is, \LeaveOutRankSVM{} is trained by measuring the decrease in retrieval effectiveness that occurs from leaving one term out of the full email message (i.e., top-down). This is an error-prone way to score query terms as emails tend to be relatively long (Table~\ref{tbl:statistics}).
Conversely, our approach to generate silver query training data (\S\ref{sec:model:subquery-generation}) considers groups of query terms and the reference query is constructed in a bottom-up fashion.

Fig.~\ref{fig:query_boxplot} shows the distribution of generated query lengths for the most prominent methods (\TFIDF{}, \RelativeEntropy{}, \LeaveOutRankSVM{} and the neural networks). On both collections, subject-oriented methods (\TFIDF{}, \RelativeEntropy{}) seem to extract queries of nearly all the same length. When considering the full subject field, we see that its length varies greatly with outliers of up to 70 query terms. Methods that extract terms from the email body or the full email generate slightly longer queries than the \TFIDF{} and \RelativeEntropy{} subject methods. The methods that learn to rank query terms generate the longest queries. While \LeaveOutRankSVM{} selects query terms according to a global rank cut-off, the \Neural{}s select a variable number of terms for every request message. Consequently, it comes as no surprise that we observe high variance within the \Neural{}-generated query lengths. In addition, we observe that \Neural{}s have a similar query length distribution as the subject queries. This is due to the fact that the \Neural{}s actually expand the subject terms, as we will see in the next section.

\subsection{Analysis of differences}

\begin{figure}[t]
\centering
\begin{subfigure}[b]{\columnwidth}
    \centering
    \includegraphics[width=0.85\textwidth]{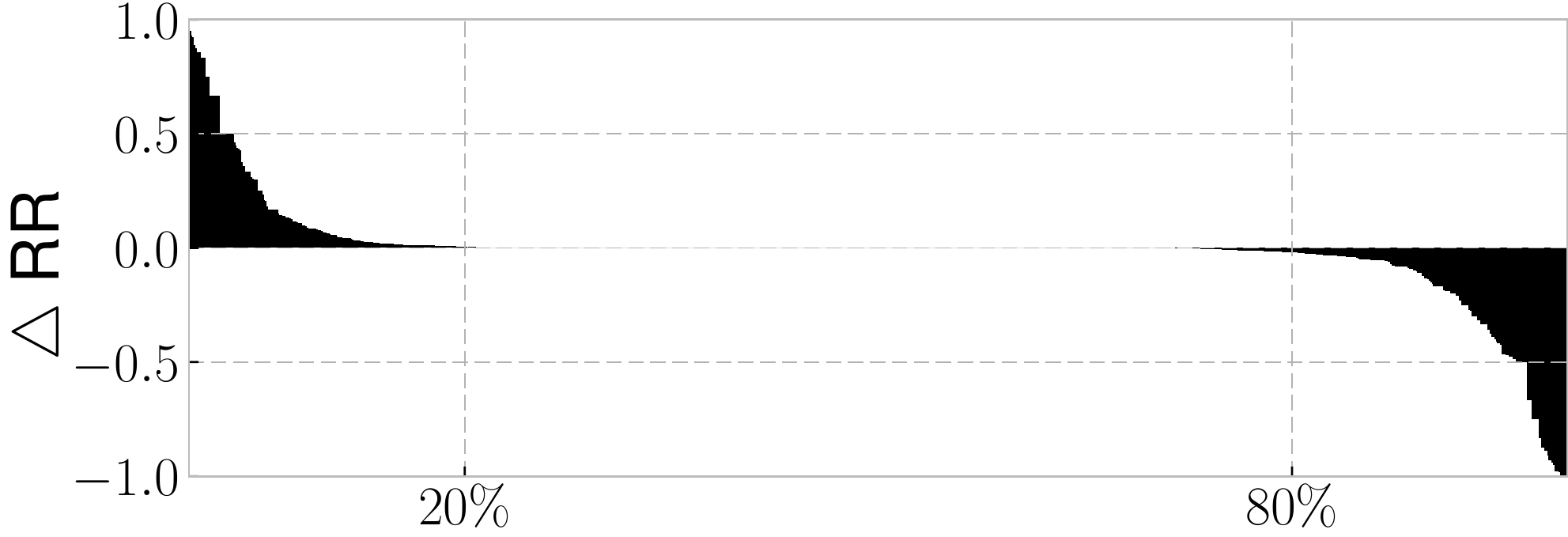}
\end{subfigure}%
\vspace*{-\baselineskip}
\caption{Per-instance differences in Reciprocal Rank (RR) between the email subject query and the \NeuralRankCutoff{} query on \AvocadoCollection{}. The plot for \InternalCollection{} is qualitatively similar to the one shown. Positive bars (left) indicate instances where the subject performs better than \NeuralRankCutoff{} and vice versa for negative (right).\label{fig:delta}}

\end{figure}

\RQAnswer{2}{Fig.~\ref{fig:delta} shows the per-instance differences between \NeuralRankCutoff{} and the full email subject as query.}

For about \numprint{60}\% of request messages both query generation methods (full \Subject{} and \Neural{}) generate queries that perform at the same level. In fact, \RecipRank{} on this subset of instances is \numprint{9}\% (\AvocadoCollection{}) and \numprint{17}\% (\InternalCollection{}) better than the best performance over the full test set (Table~\ref{tbl:results}). Do the two methods generate identical queries on this subset? The average Jaccard similarity between the queries extracted by both methods is $0.55$ (\AvocadoCollection{}) and $0.62$ (\InternalCollection{}). This indicates that, while query terms extracted by either method overlap, there is a difference in query terms that does not impact retrieval. Upon examining the differences we find that, for the subject, these terms are stopwords and email subject abbreviations (e.g., RE indicating a reply). In the case of \Neural{}, the difference comes from email body terms that further clarify the request. We find that the \Neural{} builds upon the subject query (excluding stopwords) using terms of the body. However, for the \numprint{60}\% of request instances where no difference in performance is observed (Fig.~\ref{fig:delta}), the subject by itself suffices to describe the request.

What can we say about the remaining \numprint{40}\% of queries where there is an observable difference? A closer look at Fig.~\ref{fig:delta} shows that there are extreme peaks at both sides of the graph and that neither method fully dominates the other. Upon examining the outliers, we find the following trends:
\begin{inparaenum}[(1)]
	\item When the subject is indescriptive of the email content (e.g., the subject is \emph{``important issue''}) then the \Neural{} can extract better terms from the email body.
	\item Topic drift within conversations negatively impacts the retrieval effectiveness of the subject as a query. However, long threads do not necessarily exhibit topic drift as in some cases the subject remains a topical representation of the conversation.
	\item Mentions of named entities constitute effective query terms. For example, the name of a person who is responsible for an attachable item tends to improve retrieval effectiveness over using the subject as a query,
	\item Long queries generated by the \Neural{} can often disambiguate a request and perform much better than the subject query.
	\item In most cases where the subject query outperforms the \Neural{}, this is due to the fact that the \Neural{} model extracts too many noisy terms and creates query drift.
\end{inparaenum}

\subsection{Feature importance}

\begin{figure}[t]
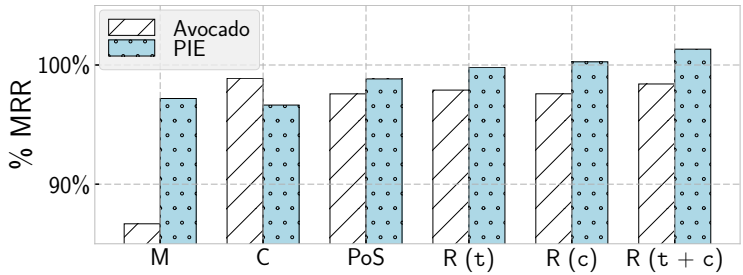


\myincludegraphics[width=0.85\columnwidth]{resources/feature_ablation.pdf}
\caption{Feature ablation study for the \NeuralRankCutoff{} model on the \AvocadoCollection{} (stripes) and \InternalCollection{} (dots) benchmarks. One of the following feature categories (Table~\ref{tbl:features}) is systematically left out: Message (M), Collection statistics (C), Part-of-Speech (PoS), \texttt{term} (\texttt{t}), \texttt{context} (\texttt{c}) or all representations (\texttt{t} + \texttt{c}). \label{fig:feature_ablation}}

\end{figure}

\RQAnswer{3}{Fig.~\ref{fig:feature_ablation} depicts a feature ablation study where we systematically leave out one of the feature categories.}

We observe that both local (message and part-of-speech) and global (collection) features are of importance. When comparing the behavior of the enterprise email collections (\S\ref{sec:collections}), we see that the message (M) features have a significant ($p < 0.10$) impact on both collections. The collection statistics (C) yield a significant ($p<0.10$) difference in the case of \InternalCollection{}; no significant differences were observed in the remaining cases. In addition, while \AvocadoCollection{} benefits from the learned representations, the inclusion of the representations of the context slightly decreases performance on \InternalCollection{}. This can be explained by our evaluation setup (\S\ref{sec:design}) where models evaluated on \InternalCollection{} are trained using \AvocadoCollection{} and vice versa. Therefore, it is likely that the model learns certain patterns present from the data-scarce \AvocadoCollection{} collection (Table~\ref{tbl:statistics}) that causes false positive terms to be selected for \InternalCollection{}.


\section{Summary}
\label{sec:conclusion}

We introduced a novel proactive retrieval task for recommending email attachments that involves formulating a query from an email request message. An evaluation framework was proposed that extracts labeled request/attachment instances from an email corpus containing request/reply pairs automatically. Candidate queries, which we refer to as \emph{silver queries}, are algorithmically synthesized for request/attachment instances and a deep convolutional neural network (\Neural{}) is trained using the silver queries that learns to extract query terms from request messages.

We find that our framework extracts instances that are usable for training and testing. Our \Neural{}, which we train using silver queries, significantly outperforms existing methods for extracting query terms from verbose texts. Terms occurring in the subject of the email are representative of the request and formulating a query using the subject is a strong baseline. A study of the per-instance \RecipRank{} differences show that the \Neural{} and subject query perform quite differently for $40\%$ of instances. A qualitative analysis suggests that our \Neural{} outperforms the subject query in cases where the subject is indescriptive. In addition, mentions of named entities constitute good query terms and lengthy queries disambiguate the request. In cases when the subject query outperforms the \Neural{}, it is due to noisy terms being selected from the email body. A feature ablation study shows that both local (i.e., message) and global (i.e., collection) features are important.

Our work has the following limitations.
\begin{inparaenum}[(1)]
\item In this \paper{} we only consider terms occurring in the request message as candidates. While this prevents the term candidate set from becoming too large, it does limit the ability for methods to formulate expressive queries in the case where request messages are concise.
\item The retrieval model used in this \paper{}, a language model with Dirichlet smoothing, is ubiquitous in retrieval systems. However, smoothing allows the search engine to deal with verbose queries \citep{Zhai2004smoothing} that contain terms absent from the messages. Subjects often contain superfluous terms (e.g., email clients prepend \emph{FW} to the subjects of forwarded messages). Consequently, our findings may change when considering other retrieval model classes, such as boolean models or semantic matching models.
\end{inparaenum}
}

\part{Latent Vector Spaces}
\label{part:latent}

\chapter{Unsupervised, Efficient and Semantic Expertise Retrieval}
\label{chapter:research-03}

{
\ScopeLabels{research-03}
\ChapterRQ{2}

\def \paperImplementationUrl {\url{https://github.com/cvangysel/SERT}}

\def \ResearchQuestionOne {How does our discriminative log-linear model compare to vector space-based methods and generative language models for the expert retrieval task in terms of retrieval performance?}
\def \ResearchQuestionTwo {What can we learn regarding the different types of errors made by generative and discriminative language models?}
\def \ResearchQuestionThree {How does the complexity of inference in our log-linear model compare to vector-space based and generative models?}
\def \ResearchQuestionFour {How does the log-linear model handle incremental indexing and what are its limitations?}

\renewcommand{\boldsymbol}[1]{\mathbf{#1}}

\newcommand{\Document}{d}
\newcommand{\EmbeddingSize}{e}

\newcommand{\VocabularyMatrix}[1][]{
\ifthenelse{\isempty{#1}}%
{W_{p}}%
{W_{p_{#1}}}%
}

\newcommand{\CandidateMatrix}[1][]{
\ifthenelse{\isempty{#1}}%
{W_{c}}%
{W_{c_{#1}}}%
}

\newcommand{\CandidateBias}[1][]{
\ifthenelse{\isempty{#1}}%
{\boldsymbol{b_c}}%
{b_{c_#1}}%
}

\newcommand{\CandidateVector}[1][]{
\ifthenelse{\isempty{#1}}%
{\boldsymbol{c}}%
{c_#1}%
}

\newcommand{\assoc}[1][]{C_{\Document{}#1}}

\def \OldLossFn {\LossFn}
\renewcommand{\LossFn}[1][]{\Apply{L}{\VocabularyMatrix{}#1, \CandidateMatrix{}#1, \boldsymbol{\CandidateBias{}}#1}}

\renewcommand{\log}{\text{log}}
\renewcommand{\exp}{\text{exp}}

\newcommand{\ergkrap}{}

\newcommand{\halftoprow}{& \multicolumn{2}{@{}l@{}}{MAP} & \multicolumn{2}{@{}l@{~~}}{NDCG@100} & \multicolumn{2}{@{}l@{}}{MRR} & \multicolumn{2}{@{}l@{}}{P@5} & \multicolumn{2}{@{}l@{}}{P@10} }
\newcommand{\toprow}{\halftoprow\halftoprow\\\cmidrule(r){2-11}\cmidrule(r){12-21}}

\newcommand{\subfloat}[2][]{%
\begin{subfigure}[b]{0.25\paperheight}%
\centering%
{\setlength{\textwidth}{0.75\paperheight}#2}%
\caption{#1}%
\end{subfigure}%
}

\renewcommand{\newcommand}{\providecommand}


\section{Introduction}
\label{section:introduction}

The transition to the knowledge and information economy~\citep{OECD1996} introduces a great reliance on cognitive capabilities~\citep{Powell2004}. It is crucial for employers to facilitate information exchange and to stimulate collaboration~\citep{Davenport1998}. In the past, organizations would set-up special-purpose database systems for their members to maintain a profile \citep{BecerraFernandez2000}. However, these systems required employees to be proactive. In addition, self-assessments are known to diverge from reality \citep{Kruger1991,Berendsen2013profileassessment} and document collections can quickly become practically infeasible to manage manually. Therefore, there has been an active interest in automated approaches for constructing expertise profiles~\citep{BecerraFernandez2000,TREC2010} and retrieving experts from an organization's heterogeneous document repository~\citep{Craswell2001}. \emph{Expert finding} (also known as expertise retrieval or expert search) addresses the task of finding the right person with the appropriate skills and knowledge~\citep{Balog2012survey}. It attempts to provide an answer to the question:
\begin{quote}
\noindent Given a topic X, who are the candidates with the most expertise w.r.t.\ X?
\end{quote}

\noindent%
The expertise retrieval task gained popularity in the research community during the TREC Enterprise Track \citep{TREC2010} and has remained relevant ever since, while broadening to social media and to tracking the dynamics of expertise~\citep{Petkova2006,Fang2007,Balog2009lmframework,Demartini2009expertspaces,Fang2010,Moreira2011,Balog2012survey,Berendsen2013profileassessment,Fang2014,vanDijk2015}. Existing methods fail to address key challenges:
\begin{inparaenum}[(1)]
	\item Queries and expert documents use different representations to describe the same concepts \citep{Hinton1986,Li2014}. Term mismatches between queries and experts \citep{Li2014} occur due to the inability of widely used maximum-likelihood language models to make use of \emph{semantic} similarities between words \citep{Salakhutdinov2009}.
	\item As the amount of available data increases, the need for more powerful approaches with greater learning capabilities than smoothed maximum-likeli\-hood language models is obvious~\citep{Vapnik1998}.
	\item Supervised methods for expertise retrieval \citep{Fang2010,Moreira2011} were introduced at the turn of the last decade. However, the acceleration of data availability has the major disadvantage that, in the case of supervised methods, manual annotation efforts need to sustain a similar order of growth. This calls for the further development of \emph{unsupervised} methods.
	\item In some expertise retrieval methods, a language model is constructed for every document in the collection. These methods lack \emph{efficient} query capabilities for large document collections, as each query term needs to be matched against every document~\citep{Balog2012survey}.
\end{inparaenum}
Our proposed solution has a strong emphasis on \emph{unsupervised model construction}, \emph{efficient query capabilities} and \emph{semantic matching} between query terms and candidate experts.

Specifically, we propose an unsupervised log-linear model with efficient inference capabilities for the expertise retrieval task. We show that our approach improves retrieval performance compared to vector space-based and generative language models, mainly due to its ability to perform semantic matching \citep{Li2014}. Our method does not require supervised relevance judgements and is able to learn from raw textual evidence and document-candidate associations alone. The purpose of this \paper{} is to provide insight in how discriminative language models can improve performance of core retrieval tasks compared to maximum-likelihood language models. Therefore, we avoid explicit feature engineering and the incorporation of external evidence in this \paper{}.
In terms of performance, the current best-performing formal language model \citep{Balog2006experts} exhibits a worst-case time complexity linear in the size of the document collection. In contrast, the inference time complexity of our approach is asymptotically bounded by the number of candidate experts.

\medskip
\noindent
The research questions we ask in this \paper{} towards answering \MainRQRef{2} are as follows:
\RQ{1}{\ResearchQuestionOne{}}
\RQ{2}{\ResearchQuestionTwo{}}
\RQ{3}{\ResearchQuestionThree{}}
\RQ{4}{\ResearchQuestionFour{}}


\section{Related work}
\label{section:related}

We refer to Section~\globalref{chapter:background:latent} of our background chapter (Chapter~\globalref{chapter:background}). Particularly relevant to this \paper{} is the subsection that covers prior work on expert retrieval (\S\globalref{chapter:background:related:experts}) and its relation to document retrieval, followed by semantic matching methods (\S\globalref{chapter:background:related:latent}) and neural language models (\S\globalref{chapter:background:related:nlm}).

\medskip\noindent%
What we add on top of the related work described above is the following. In this \paper{} we model the conditional probability of the expertise of a candidate given a single query term (contrary to binary relevance given a character-based n-gram \citep{Huang2013}). In the process we learn a distributed vector representation (similar to LSI, pLSI and semantic hashing) for both words and candidates such that nearby representations indicate semantically similar concepts.

We propose a log-linear model that is similar to neural language models. The important difference is that we predict a candidate expert instead of the next word. To the best of our knowledge, we are the first to propose such a solution. We employ an embedding layer in our shallow model for the same reasons as mentioned above: we learn continuous word representations that incorporate semantic and syntactic similarity tailored to an expert's domain.

\break
\section{A log-linear model for expert search}
\label{section:methodology}

In the setting of this chapter we have a document collection $D$ and a predefined set of candidate experts $C$ (entities to be retrieved). Documents $\Document{} \in D$ are represented as a sequence of words $w_1$, \ldots, $w_{|\Document{}|}$ originating from a vocabulary $V$, where $w_i \in V$ and the operator $|\cdot|$ denotes the document length in tokens. For every document $\Document{} \in D$ we write $\assoc$ to denote the set of candidates $c \in C$ associated with it (i.e., $C = \bigcup_{\Document{} \in D} \assoc{}$). These document-candidate associations can be obtained explicitly from document meta-data (e.g., the author of an e-mail) or implicitly by mining references to candidates from the document text. Notice that some documents might not be associated with any candidate. When presented with a query $q$ of constituent terms $t_1$, \ldots, $t_{|q|}$, the expert retrieval task is to return a list of candidates $\rho(C)$ ordered according to topical expertise. We generate this ranking using a relatively shallow neural network which directly models $P(\CandidateVector[] \mid q)$.

We employ vector-based distributed representations \citep{Hinton1986}, for both words (i.e., word embeddings) and candidate experts, in a way that motivates the unsupervised construction of features that express regularities of the expertise finding domain. These representations can capture the similarity between concepts (e.g., words and candidate experts) by the closeness of their representations in vector space. That is, concepts with similar feature activations are interpreted by the model as being similar, or even interchangeable.

\subsection{The model}
To address the expert search task, we model $P(\CandidateVector[j] \mid q)$ and rank candidates $c_j$ accordingly for a given $q$. We propose an unsupervised, discriminative approach to obtain these probabilities. We construct our model solely from textual evidence: we do not require query-candidate relevance assessments for training and do not consider external evidence about the corpus (e.g., different weightings for different sub-collections), the document (e.g., considering certain parts of the document more useful) nor link-based features.

Let $\EmbeddingSize{}$ denote the size of the vector-based distributed representations of both words in $V$ and candidate experts in $C$. These representations will be learned by the model using gradient descent \cite{Mikolov2013word2vec} (Section~\ref{section:learning}). For notational convenience, we write $P(\CandidateVector[] \mid \cdot)$ for the (conditional) probability distribution over candidates, which is the result of vector arithmetic. We define the probability of a candidate $c_j$ given a single word $w_i \in V$ as the log-linear model
\newcommand{\OneHotVector}[1]{\boldsymbol{v_#1}}
\newcommand{\eqUnigram}{\exp\left(\CandidateMatrix[] \cdot (\VocabularyMatrix[] \cdot \OneHotVector{i}) + \CandidateBias[]\right)}
\begin{equation}
\label{eq:unigram}
P(\CandidateVector[] \mid w_i) = \frac{1}{Z_1} \eqUnigram{},
\end{equation}
where $\VocabularyMatrix[]$ is the $\EmbeddingSize{} \times |V|$ projection matrix that maps the one-hot representation (i.e., 1-of-$|V|$) of word $w_i$, $\OneHotVector{i}$, to its $\EmbeddingSize{}$-dimensional distributed representation, $\CandidateBias[]$ is a $|C|$-dimensional bias vector and $\CandidateMatrix[]$ is the $|C| \times \EmbeddingSize{}$ matrix that maps the word embedding to an unnormalized distribution over candidates $C$, which is then normalized by $Z_1 = \sum^{|C|}_{j=1} \left[\eqUnigram\right]_j$. If we consider Bayes' theorem, the transformation matrix $\CandidateMatrix[]$ and bias vector $\CandidateBias[]$ can be interpreted as the term log-likelihood $\log P(w_i \mid \CandidateVector[])$ and candidate log-prior $\log P(\CandidateVector[])$, respectively. The projection matrix $\VocabularyMatrix[]$ attempts to soften the curse of dimensionality introduced by large vocabularies $V$ and maps words to word feature vectors \citep{Bengio2003}. Support for large vocabularies is crucial for retrieval tasks \citep{Salakhutdinov2009,Huang2013}.

We then assume conditional independence of a candidate's expertise given an observation of data (i.e., a word). Given a sequence of words $w_1$, \ldots, $w_k$ we have:
\newcommand{\eqBag}[1][]{\exp\left(\sum_{i = 1}^{k} \log\left(P(\CandidateVector[#1] \mid w_i)\right)\right)}
\newcommand{\ergkrap}{\mbox{}\hspace*{-3mm}}
\begin{eqnarray}
P(\CandidateVector[] \mid w_1, \ldots, w_k) & \ergkrap= & \ergkrap\frac{1}{Z_2} \tilde{P}(\CandidateVector[] \mid w_1, \ldots, w_k) = \frac{1}{Z_2} \prod_{i = 1}^{k} P(\CandidateVector[] \mid w_i)
\nonumber\\
& \ergkrap= &\ergkrap\frac{1}{Z_2} \eqBag[]
\label{eq:bag}
\end{eqnarray}
where $\tilde{P}(\CandidateVector[] \mid w_1, \ldots, w_k)$ denotes the unnormalized score and $$Z_2 = \sum^{|C|}_{j = 1} \eqBag[j]$$ is a normalizing term. The transformation to log-space in \eqref{eq:bag} is a well-known trick to prevent floating point underflow \citep[p. 445]{Montavon2012}. Given \eqref{eq:bag}, inference is straight-forward. That is, given query $q = t_1$, \ldots, $t_k$, we compute $P(\CandidateVector[] \mid t_1$, \ldots, $t_k)$ and rank the candidate experts in descending order of probability.

Eq.~\ref{eq:unigram} defines a neural network with a single hidden layer. We can add additional layers. Preliminary experiments, however, show that the shallow log-linear model \eqref{eq:unigram} performs well-enough in most cases. Only for larger data sets did we notice a marginal gain from adding an additional layer between projection matrix $\VocabularyMatrix[]$ and the softmax layer over $C$ ($\CandidateMatrix[]$ and bias $\CandidateBias[]$), at the expense of longer training times and loss of transparency.

\subsection{Parameter estimation}
\label{section:learning}
\newcommand{\CrossEntropyTarget}[1][]{
\ifthenelse{\isempty{#1}}%
{\boldsymbol{p}}%
{\boldsymbol{p}^{(#1)}}%
}
\newcommand{\CrossEntropyApprox}[1][]{
\ifthenelse{\isempty{#1}}%
{\boldsymbol{\tilde{p}}}%
{\boldsymbol{\tilde{p}}^{(#1)}}%
}
The matrices $\VocabularyMatrix[]$, $\CandidateMatrix[]$ and the vector $\CandidateBias[]$ in \eqref{eq:unigram} constitute the parameters of our model. We estimate them using error back propagation \citep{Rumelhart1986} as follows. For every document $\Document{}_j \in D$ we construct an ideal distribution over candidates $\CrossEntropyTarget[] = P(\CandidateVector[] \mid \Document{}_j)$ based on the document-candidate associations $\assoc[_j]$ such that
\[
P(\CandidateVector[] \mid d_j) =
\begin{cases}
\frac{1}{|\assoc[_j]|}, & c \in \assoc[_j] \\
0, & c \not\in \assoc[_j]
\end{cases}
\]
We continue by extracting n-grams where $n$ remains fixed during training. For every n-gram $w_1$, \ldots, $w_n$ generated from document $\Document{}$ we compute $\CrossEntropyApprox[] = P(\CandidateVector[] \mid w_1$, \ldots, $w_n)$ using \eqref{eq:bag}. During model constructing we then optimize the cross-entropy $H(\CrossEntropyTarget[], \CrossEntropyApprox[])$ (i.e., the joint probability of the training data if $|\assoc[_j]| = 1$ for all $j$) using batched gradient descent. The loss function for a single batch of $m$ instances with associated targets $(\CrossEntropyTarget[i], \CrossEntropyApprox[i])$ is as follows:
\newcommand{\Context}[1][]{
\ifthenelse{\isempty{#1}}%
{w_{1}, \ldots, w_{n}}%
{w^{(#1)}_{1}, \ldots, w^{(#1)}_{n}}%
}
\newcommand{\SourceDocument}[1]{
\Document{}^{(#1)}
}
\begin{eqnarray}
\label{eq:loss}
\lefteqn{\LossFn{}}
\nonumber \\
& = & \ergkrap\frac{1}{m} \sum_{i = 1}^m \frac{|\Document{}_{\max}|}{|\SourceDocument{i}|}  H(\CrossEntropyTarget[i], \CrossEntropyApprox[i])
\nonumber \\
&   & \ergkrap+ \frac{\lambda}{2 m} \left( \sum_{i,j} \VocabularyMatrix[i,j]^2 + \sum_{i,j} \CandidateMatrix[i,j]^2 \right) \\
& = & \ergkrap - \frac{1}{m} \sum_{i = 1}^m  \frac{|\Document{}_{\max}|}{|\SourceDocument{i}|} \sum_{j = 1}^{|C|} P(\CandidateVector[j] \mid \SourceDocument{i}) \log\left(P(\CandidateVector[j] \mid \Context[i])\right)
\nonumber \\
&   & \ergkrap+ \frac{\lambda}{2 m} \left( \sum_{i,j} \VocabularyMatrix[i,j]^2 + \sum_{i,j} \CandidateMatrix[i,j]^2 \right),
\nonumber
\end{eqnarray}
where $\SourceDocument{i}$ refers to the document from which n-gram $\Context[i]$ was extracted, $\Document{}_{\max} = \argmax_{\Document{} \in D}{|\Document{}|}$ indicates the longest document in the collection, and $\lambda$ is a weight regularization parameter. The update rule for a particular parameter $\theta$ ($\VocabularyMatrix[]$, $\CandidateMatrix[]$ or $\CandidateBias[]$) given a single batch of size $m$ is:
\newcommand{\LearningRate}{\boldsymbol{\alpha}}
\begin{equation}
\label{eq:update}
\theta^{(t+1)} = \theta^{(t)} - \LearningRate{}^{(t)} \odot \frac{\partial \LossFn[^{(t)}]}{\partial \theta},
\end{equation}
where $\LearningRate{}^{(t)}$ and $\theta^{(t)}$ denote the per-parameter learning rate and parameter $\theta$ at time $t$, respectively. The learning rate $\LearningRate{}$ consists of the same number of elements as there are parameters; in the case of a global learning rate, all elements of $\LearningRate{}$ are equal to each other.
The derivatives of the loss function \eqref{eq:loss} are given in the Appendix.

In the next section we will discuss our experimental setup, followed by an overview of our experimental results and further analysis in Section~\ref{section:discussion}.


\section{Experimental setup}
\label{section:setup}

\subsection{Research questions}
As indicated in the introduction of this \paper{}, we seek to answer the following research questions:

\RQ{1}{\ResearchQuestionOne{}}
\noindent%
In particular, how does the model perform when compared against vector space-based (LSI and TF-IDF) and generative approaches (profile-centric Model~1 and document-centric Model~2)?

\RQ{2}{\ResearchQuestionTwo{}}
\noindent%
Does the best-performing generative model simply perform slightly better on the topics for which the other models perform decent as well, or do they make very different errors? If the latter holds, an ensemble of the rankings produced by both model types might exceed performance of the individual rankings.

\RQ{3}{\ResearchQuestionThree{}}
\noindent%
The worst-case inference cost of document-centric models makes them unattractive in online settings where the set of topics is not defined beforehand and the document collection is large. Profile-centric methods are preferred in such settings as they infer from one language model per candidate expert for every topic (i.e., a pseudo-document consisting of a concatenation of all documents associated with an expert) \citep{Balog2012survey}. Vector space-based methods \citep{Demartini2009expertspaces} have similar problems due to the curse of dimensionality \citep{Indyk1998} and consequently their inferential time complexity is likewise asymptotically bounded by the number of experts.

\RQ{4}{\ResearchQuestionFour{}}

\subsection{Benchmarks}
\label{sec:benchmarks}

\newcommand{\specialcell}[3][c]{\begin{tabular}[#1]{@{}#2@{}}#3\end{tabular}}

\begin{table*}
\centering
\begin{threeparttable}
	\caption{An overview of the three datasets (W3C, CERC and TU) used for evaluation and analysis.\label{tbl:benchmarks}}
	\begin{tabular}{lc@{$\,$}lc@{$\,$}lc@{$\,$}l}
	\toprule
	\multicolumn{1}{c}{} & \multicolumn{2}{c}{W3C} & \multicolumn{2}{c}{CERC} & \multicolumn{2}{c}{TU} \\
	\midrule
	Number of documents & \numprint{331037}\phantom{.00} & & \numprint{370715}\phantom{.00} & & \phantom{0}\numprint{31209}\phantom{.00} &  \\
	Average document length$^a$ & \phantom{00}\numprint{1237.23} & & \phantom{000,}\numprint{460.48} & & \phantom{00}\numprint{2454.93} & \\

	\\

	Number of candidates$^b$ & \phantom{000,}\numprint{715}\phantom{.00} & & \phantom{00}\numprint{3479}\phantom{.00} & & \phantom{000,}\numprint{977}\phantom{.00} & \\

	\\

	\specialcell{l}{Number of document-candidate associations} & \numprint{200939}\phantom{.00} & & \numprint{236958}\phantom{.00} & & \phantom{0}\numprint{36566}\phantom{.00} & \\
	\specialcell{l}{Number of documents (with $\assoc > 0$)} & \phantom{0}\numprint{93826}\phantom{.00} & & \numprint{123934}\phantom{.00} & & \phantom{0}\numprint{27834}\phantom{.00} &  \\
	\specialcell{l}{Number of associations per document$^c$} & \phantom{000,00}\numprint{2.14} & $\pm\, 3.29$ & \phantom{000,00}\numprint{1.91} & $\pm\, 3.70$ & \phantom{000,00}\numprint{1.13} & $\pm\, 0.39$ \\
	\specialcell{l}{Number of associations per candidate} & \phantom{000,}\numprint{281.03} & $\pm\, 666.63$ & \phantom{000,0}\numprint{68.11} & $\pm\,$\numprint{1120.74} & \phantom{000,0}\numprint{37.43} & $\pm\, 61.00$ \\

	\\

	Queries & \phantom{000,0}\numprint{49}\phantom{.00} & (2005) & \phantom{000,0}\numprint{50}\phantom{.00} & (2007) & \phantom{00}\numprint{1662}\phantom{.00} & (GT1) \\
	& \phantom{000,0}\numprint{50}\phantom{.00} & (2006) & \phantom{000,0}\numprint{77}\phantom{.00} & (2008) & \phantom{00}\numprint{1266}\phantom{.00} & (GT5) \\
	\bottomrule
	\end{tabular}
	\begin{tablenotes}
	\footnotesize
	\item[$a$] Measured in number of tokens.
	\item[$b$] Only candidates with at least a single document association are considered.
	\item[$c$] Only documents with at least one association are considered.
	\end{tablenotes}
\end{threeparttable}
\vspace*{-.5\baselineskip}
\end{table*}%

The proposed method is applicable in the setting of the Expert Search task of the TREC Enterprise track from 2005 to 2008 \citep{TREC2010}. We therefore evaluate on the W3C and CERC benchmarks released by the track. The W3C dataset \citep{Craswell2005} is a crawl of the W3C's sites in June 2004 (mailing lists, web pages, etc.). The CSIRO Enterprise Research Collection (CERC) \citep{Bailey2007} is a dump of the intranet of Australia's national science agency. Additionally, we evaluate our method on a smaller, more recent benchmark based on the employee database of Tilburg University (TU) \citep{Berendsen2013profileassessment}, which consists of bi-lingual, heterogeneous documents. See Table~\ref{tbl:benchmarks}.

Mining document-candidate associations and how they influence performance has been extensively covered in previous work \citep{Balog2006experts,Balog2012survey} and is beyond the scope of this \paper{}. For TU, the associations are part of the benchmark. For W3C, a list of possible candidates is given and we extract the associations ourselves by performing a case-insensitive match of full name or e-mail address \citep{Balog2006experts}. For CERC, we make use of publicly released associations~\citep{Balog2008thesis}.

As evaluation measures we use Mean Average Precision (MAP), Mean Reciprocal Rank (MRR), Normalized Discounted Cumulative Gain at rank 100 (NDCG@100) and Precision at rank 5 (P@5) and rank 10 (P@10).

\subsection{Baselines}
\label{sec:baseline}

We compare our approach to existing unsupervised methods for expert retrieval that solely rely on textual evidence and static doc\-ument-candidate associations.
\begin{inparaenum}[(1)]%
\item \citet{Demartini2009expertspaces} propose a generic framework to adapt vector spaces operating on documents to entities. We compare our method to TF-IDF (raw frequency and inverse document frequency) and LSI (300 latent topics) variants of their vector space model for entity ranking (using cosine similarity).
\item In terms of language modelling, \citet{Balog2006experts} propose two models for expert finding based on generative language models. The first takes a profile-centric approach comparing the language model of every expert to the query, while the second is document-centric. We consider both models with different smoothing configurations: Jelinek-Mercer (jm) smoothing with $\lambda=0.5$ \citep{Balog2006experts} and Dirichlet (d) smoothing with $\beta$ equal to the average document length \citep{Balog2009lmframework} (see Table~\ref{tbl:benchmarks}).
\end{inparaenum}
Significance of results produced by the baselines (compared to our method) is determined using a two-tailed paired randomization test \citep{Smucker2007significance}.

\subsection{Implementation details}
\label{sec:implementation}

\newcommand{\m}{\sqrt{\frac{6.0}{m + n}}}

The vocabulary $V$ is constructed from each corpus by ignoring punctuation, stop words and case; numbers are replaced by a numerical placeholder token. During our experiments we prune $V$ by only retaining the $2^{16}$ most-frequent words so that each word can be encoded by a 16-bit unsigned integer. Incomplete n-gram instances are padded by a special-purpose token.

In terms of parameter initialization, we sample the initial matrices $\CandidateMatrix{}$ and $\VocabularyMatrix{}$ \eqref{eq:unigram} uniformly in the range
\begin{equation*}
\left[ -\m, \m \; \right]
\end{equation*}
for an $m \times n$ matrix, as this initialization scheme improves model training convergence \citep{Glorot2010}, and take the bias vector $\CandidateBias{}$ to be null. The projection layer $\VocabularyMatrix{}$ is initialized with pre-trained word representations trained on Google News data \citep{Mikolov2013compositionality}; the number of word features is set to $\EmbeddingSize{}=300$, similar to pre-trained representations.

We used adadelta ($\rho=0.95$, $\epsilon=10^{-6}$) \citep{Zeiler2012} with batched gradient descent ($m=1024$) and weight decay $\lambda=0.01$ during training on NVidia GTX480 and NVidia Tesla K20 GPUs. We only iterate once over the entire training set for each experiment.


\section{Results \& discussion}
\label{section:discussion}

We start by giving a high-level overview of our experimental results and then address issues of scalability, provide an error analysis and discuss the issue of incremental indexing.

\subsection{Overview of experimental results}

\newcommand{\Significant}{^{*}}
\newcommand{\MoreSignificant}{^{**}}
\newcommand{\HighlySignificant}{^{***}}

\begin{table*}[t]
	\centering
	\caption{Evaluation results for models trained on the W3C, CERC and TU benchmarks. Suffixes (d) and (jm) denote Dirichlet and Jelinek-Mercer smoothing,  respectively (Section~\ref{sec:baseline}). Significance of results is determined using a two-tailed paired randomization test \citep{Smucker2007significance} ($\HighlySignificant{} \, p < 0.01$;  $\MoreSignificant{} \, p < 0.05$; $\Significant{} \, p < 0.1$) with respect to the log-linear model (adjusted using the Benjamini-Hochberg procedure for multiple testing \citep{Benjamini1995}).\label{tbl:all_results}}
	\resizebox{0.80\columnwidth}{!}{\begin{tabular}{c@{ }c@{ }c@{ }c@{ }c@{ }c@{ }c@{ }c@{ }c@{ }c@{ }c}
\toprule
\multirow{2}{*}{W3C} & \multicolumn{5}{c}{2005} & \multicolumn{5}{c}{2006} \\ 
& MAP & NDCG@100 & MRR & P@5 & P@10 & MAP & NDCG@100 & MRR & P@5 & P@10 \\ 
\cmidrule(lr){2-6}
\cmidrule(lr){7-11}
\multicolumn{1}{l}{LSI} & $\phantom{}0.135$ & $\phantom{}0.266$ & $\phantom{}0.306$ & $\phantom{}0.192$ & $\phantom{}0.196$ & $\phantom{}0.245$ & $\phantom{}0.371$ & $\phantom{}0.482$ & $\phantom{}0.287$ & $\phantom{}0.338$ \\ 
\multicolumn{1}{l}{TF-IDF} & $\phantom{}0.243$ & $\phantom{}0.426$ & $\phantom{}0.541$ & $\phantom{}0.384$ & $\phantom{}0.350$ & $\phantom{}0.343$ & $\phantom{}0.531$ & $\phantom{}0.650$ & $\phantom{}0.492$ & $\phantom{}0.498$ \\ 
\multicolumn{1}{l}{Model 1 (d)} & $\phantom{}0.192$ & $\phantom{}0.358$ & $\phantom{}0.433$ & $\phantom{}0.276$ & $\phantom{}0.266$ & $\phantom{}0.321$ & $\phantom{}0.491$ & $\phantom{}0.635$ & $\phantom{}0.478$ & $\phantom{}0.449$ \\ 
\multicolumn{1}{l}{Model 1 (jm)} & $\phantom{}0.190$ & $\phantom{}0.352$ & $\phantom{}0.390$ & $\phantom{}0.272$ & $\phantom{}0.276$ & $\phantom{}0.311$ & $\phantom{}0.483$ & $\phantom{}0.596$ & $\phantom{}0.502$ & $\phantom{}0.437$ \\ 
\multicolumn{1}{l}{Model 2 (d)} & $\phantom{}0.198$ & $\phantom{}0.369$ & $\phantom{}0.429$ & $\phantom{}0.288$ & $\phantom{}0.272$ & $\phantom{}0.261$ & $\phantom{}0.419$ & $\phantom{}0.551$ & $\phantom{}0.441$ & $\phantom{}0.404$ \\ 
\multicolumn{1}{l}{Model 2 (jm)} & $\phantom{}0.211$ & $\phantom{}0.380$ & $\phantom{}0.451$ & $\phantom{}0.332$ & $\phantom{}0.296$ & $\phantom{}0.260$ & $\phantom{}0.423$ & $\phantom{}0.599$ & $\phantom{}0.449$ & $\phantom{}0.429$ \\ 
\multicolumn{1}{l}{Log-linear (ours)} & $\phantom{}\textbf{0.248}$ & $\phantom{}\textbf{0.444}$ & $\phantom{\Significant{}}\textbf{0.618}\Significant{}$ & $\phantom{}\textbf{0.412}$ & $\phantom{}\textbf{0.361}$ & $\phantom{\HighlySignificant{}}\textbf{0.484}\HighlySignificant{}$ & $\phantom{\HighlySignificant{}}\textbf{0.667}\HighlySignificant{}$ & $\phantom{\HighlySignificant{}}\textbf{0.833}\HighlySignificant{}$ & $\phantom{\HighlySignificant{}}\textbf{0.713}\HighlySignificant{}$ & $\phantom{\HighlySignificant{}}\textbf{0.644}\HighlySignificant{}$ \\ 
\cmidrule{1-11}
\multirow{2}{*}{CERC} & \multicolumn{5}{c}{2007} & \multicolumn{5}{c}{2008} \\ 
& MAP & NDCG@100 & MRR & P@5 & P@10 & MAP & NDCG@100 & MRR & P@5 & P@10 \\ 
\cmidrule(lr){2-6}
\cmidrule(lr){7-11}
\multicolumn{1}{l}{LSI} & $\phantom{}0.031$ & $\phantom{}0.107$ & $\phantom{}0.060$ & $\phantom{}0.016$ & $\phantom{}0.014$ & $\phantom{}0.038$ & $\phantom{}0.099$ & $\phantom{}0.106$ & $\phantom{}0.042$ & $\phantom{}0.055$ \\ 
\multicolumn{1}{l}{TF-IDF} & $\phantom{}0.332$ & $\phantom{}0.486$ & $\phantom{}0.463$ & $\phantom{}0.196$ & $\phantom{}0.141$ & $\phantom{}0.269$ & $\phantom{}0.465$ & $\phantom{}0.525$ & $\phantom{}0.332$ & $\phantom{}0.277$ \\ 
\multicolumn{1}{l}{Model 1 (d)} & $\phantom{}0.287$ & $\phantom{}0.427$ & $\phantom{}0.384$ & $\phantom{}0.156$ & $\phantom{}0.096$ & $\phantom{}0.181$ & $\phantom{}0.355$ & $\phantom{}0.388$ & $\phantom{}0.200$ & $\phantom{}0.172$ \\ 
\multicolumn{1}{l}{Model 1 (jm)} & $\phantom{}0.278$ & $\phantom{}0.420$ & $\phantom{}0.384$ & $\phantom{}0.156$ & $\phantom{}0.084$ & $\phantom{}0.170$ & $\phantom{}0.347$ & $\phantom{}0.339$ & $\phantom{}0.181$ & $\phantom{}0.159$ \\ 
\multicolumn{1}{l}{Model 2 (d)} & $\phantom{}0.352$ & $\phantom{}0.495$ & $\phantom{}0.454$ & $\phantom{}0.180$ & $\phantom{}0.138$ & $\phantom{}0.264$ & $\phantom{}0.461$ & $\phantom{}0.510$ & $\phantom{}0.281$ & $\phantom{}0.244$ \\ 
\multicolumn{1}{l}{Model 2 (jm)} & $\phantom{}\textbf{0.361}$ & $\phantom{}\textbf{0.500}$ & $\phantom{}0.467$ & $\phantom{}0.192$ & $\phantom{}0.138$ & $\phantom{}0.274$ & $\phantom{}0.463$ & $\phantom{}0.517$ & $\phantom{}0.278$ & $\phantom{}0.239$ \\ 
\multicolumn{1}{l}{Log-linear (ours)} & $\phantom{}0.344$ & $\phantom{}0.493$ & $\phantom{}\textbf{0.513}$ & $\phantom{}\textbf{0.215}$ & $\phantom{}\textbf{0.150}$ & $\phantom{\HighlySignificant{}}\textbf{0.342}\HighlySignificant{}$ & $\phantom{\MoreSignificant{}}\textbf{0.519}\MoreSignificant{}$ & $\phantom{\MoreSignificant{}}\textbf{0.656}\MoreSignificant{}$ & $\phantom{\Significant{}}\textbf{0.381}\Significant{}$ & $\phantom{}\textbf{0.299}$ \\ 
\cmidrule{1-11}
\multirow{2}{*}{TU} & \multicolumn{5}{c}{GT1} & \multicolumn{5}{c}{GT5} \\ 
& MAP & NDCG@100 & MRR & P@5 & P@10 & MAP & NDCG@100 & MRR & P@5 & P@10 \\ 
\cmidrule(lr){2-6}
\cmidrule(lr){7-11}
\multicolumn{1}{l}{LSI} & $\phantom{}0.095$ & $\phantom{}0.205$ & $\phantom{}0.153$ & $\phantom{}0.060$ & $\phantom{}0.051$ & $\phantom{}0.097$ & $\phantom{}0.208$ & $\phantom{}0.129$ & $\phantom{}0.043$ & $\phantom{}0.036$ \\ 
\multicolumn{1}{l}{TF-IDF} & $\phantom{}0.216$ & $\phantom{}0.356$ & $\phantom{}0.324$ & $\phantom{}0.131$ & $\phantom{}0.097$ & $\phantom{}0.233$ & $\phantom{}0.378$ & $\phantom{}0.288$ & $\phantom{}0.108$ & $\phantom{}0.079$ \\ 
\multicolumn{1}{l}{Model 1 (d)} & $\phantom{}0.171$ & $\phantom{}0.308$ & $\phantom{}0.258$ & $\phantom{}0.103$ & $\phantom{}0.082$ & $\phantom{}0.241$ & $\phantom{}0.385$ & $\phantom{}0.292$ & $\phantom{}0.109$ & $\phantom{}0.081$ \\ 
\multicolumn{1}{l}{Model 1 (jm)} & $\phantom{}0.189$ & $\phantom{}0.325$ & $\phantom{}0.277$ & $\phantom{}0.112$ & $\phantom{}0.085$ & $\phantom{}0.231$ & $\phantom{}0.373$ & $\phantom{}0.271$ & $\phantom{}0.100$ & $\phantom{}0.075$ \\ 
\multicolumn{1}{l}{Model 2 (d)} & $\phantom{}0.154$ & $\phantom{}0.284$ & $\phantom{}0.228$ & $\phantom{}0.087$ & $\phantom{}0.070$ & $\phantom{}0.191$ & $\phantom{}0.334$ & $\phantom{}0.233$ & $\phantom{}0.084$ & $\phantom{}0.065$ \\ 
\multicolumn{1}{l}{Model 2 (jm)} & $\phantom{}\textbf{0.234}$ & $\phantom{}\textbf{0.370}$ & $\phantom{}0.342$ & $\phantom{}0.136$ & $\phantom{}0.101$ & $\phantom{}0.253$ & $\phantom{}0.394$ & $\phantom{}0.302$ & $\phantom{}0.108$ & $\phantom{}0.081$ \\ 
\multicolumn{1}{l}{Log-linear (ours)} & $\phantom{}0.219$ & $\phantom{}0.356$ & $\phantom{}\textbf{0.351}$ & $\phantom{\Significant{}}\textbf{0.145}\Significant{}$ & $\phantom{}\textbf{0.105}$ & $\phantom{\HighlySignificant{}}\textbf{0.287}\HighlySignificant{}$ & $\phantom{\HighlySignificant{}}\textbf{0.425}\HighlySignificant{}$ & $\phantom{\HighlySignificant{}}\textbf{0.363}\HighlySignificant{}$ & $\phantom{\HighlySignificant{}}\textbf{0.134}\HighlySignificant{}$ & $\phantom{\HighlySignificant{}}\textbf{0.092}\HighlySignificant{}$ \\ 
\bottomrule
\end{tabular}
}
\end{table*}

\begin{figure*}[t]
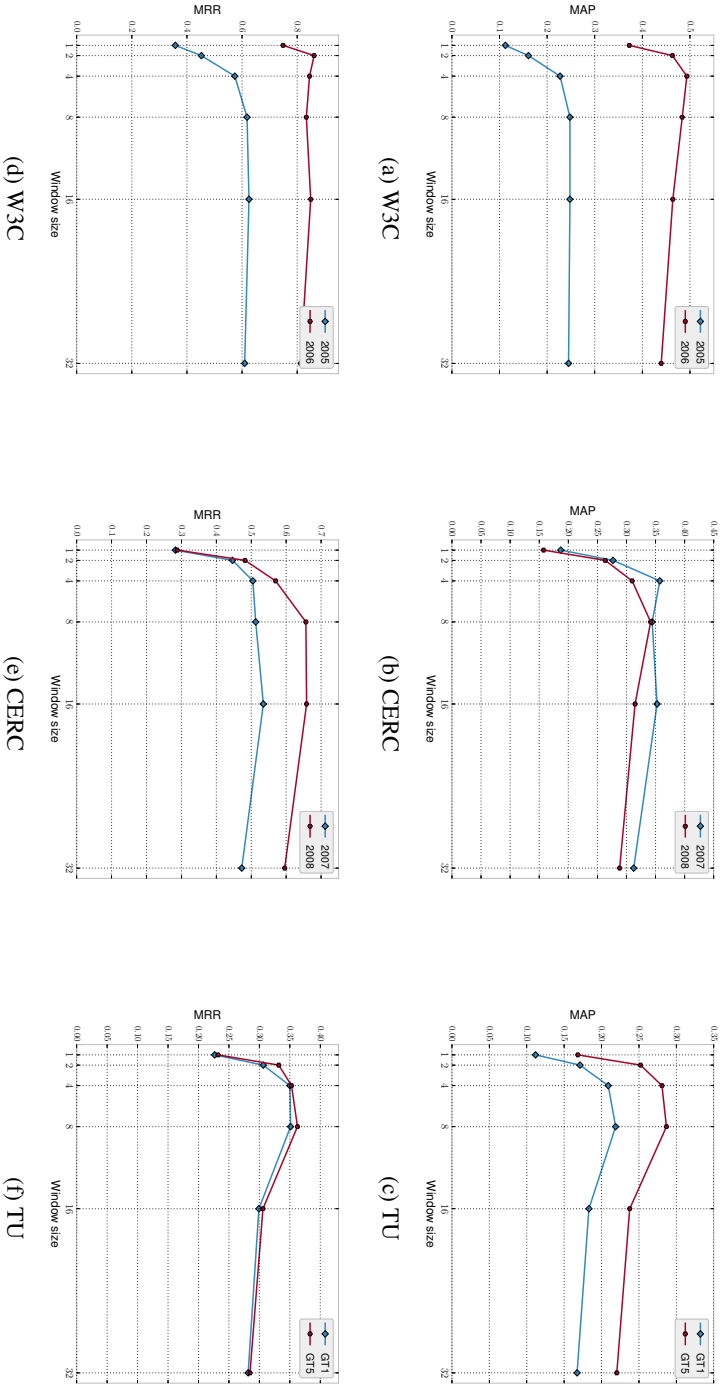

	\centering

	\newcommand{\windowsweepplot}[4]{
		\subfloat[#3\label{fig:windowsweep:#1:#2}]{%
			\myincludegraphics[width=0.32\textwidth]{resources/window_sweep_results/#4_#2.pdf}
		}
	}

	\windowsweepplot{w3c}{map}{W3C}{W3C}
	\hfill
	\windowsweepplot{cerc}{map}{CERC}{CERC}
	\hfill
	\windowsweepplot{tu}{map}{TU}{TU}

	\windowsweepplot{w3c}{mrr}{W3C}{W3C}
	\hfill
	\windowsweepplot{cerc}{mrr}{CERC}{CERC}
	\hfill
	\windowsweepplot{tu}{mrr}{TU}{TU}

	\caption{Sensitivity analysis for window size (n-gram) during parameter estimation \eqref{eq:loss} for W3C, CERC and TU benchmarks.\label{fig:windowsweep}}
\vspace*{-.5\baselineskip}
\end{figure*}

We evaluate the log-linear model on the W3C, CERC and TU benchmarks (Section~\ref{sec:benchmarks}). During training we extract non-overlap\-ping n-grams for the W3C and CERC benchmarks and overlapping n-grams for the TU benchmark. As the TU benchmark is considerably smaller, we opted to use overlapping n-grams to counter data sparsity. The architecture of each benchmark model (e.g., number of candidate experts) is inherently specified by the benchmarks themselves (see Table~\ref{tbl:benchmarks}). However, the choice of n-gram size during training remains open. Errors for input $w_1$, \ldots, $w_n$ are propagated back through $\CandidateMatrix{}$ until the projection matrix $\VocabularyMatrix{}$ is reached; if a single word $w_i$ causes a large prediction error, then this will influence its neighbouring words $w_1$, \ldots, $w_{i-1}$, $w_{i+1}$, \ldots, $w_{n}$ as well. This allows the model to learn continuous word representations tailored to the expert retrieval task and the benchmark domain.

A larger window size has a negative impact on batch throughput during training. We are thus presented with the classic trade-off between model performance and construction efficiency. Notice, however, that the number of n-grams decreases as the window size increases if we extract non-overlapping instances. Therefore, larger values of $n$ lead to faster wall-clock time model construction for the W3C and CERC benchmarks in our experiments.

We sweep over the window width $n = 2^i$ ($0 \leq i < 6$) for all three benchmarks and their corresponding relevance assessments. We report MAP and MRR for every configuration (see Fig.~\ref{fig:windowsweep}). We observe a significant performance increase between $n = 1$ and $n = 2$ on all benchmarks, which underlines the importance of the window size parameter. The increase in MAP implies that the performance achieved is not solely due to initialization with pre-trained representations (Section~\ref{sec:implementation}), but that the model efficiently learns word representations tailored to the problem domain. The highest MAP scores are attained for relatively low $n$. As $n$ increases beyond $n = 8$ a gradual decrease in MAP is observed on all benchmarks. In our remaining experiments we choose $n = 8$ regardless of the benchmark.

\begin{figure*}[t]
	\centering

	\newcommand{\entropyprecisionplot}[5]{
		\subfloat[#2 ($R=#4\HighlySignificant{}$)\label{fig:entropyprecision:#1}]{%
			\myincludegraphics[width=0.32\textwidth]{resources/entropy_precision_results/#3_entropy_ap.pdf}
		}
	}

	\entropyprecisionplot{w3c}{W3C}{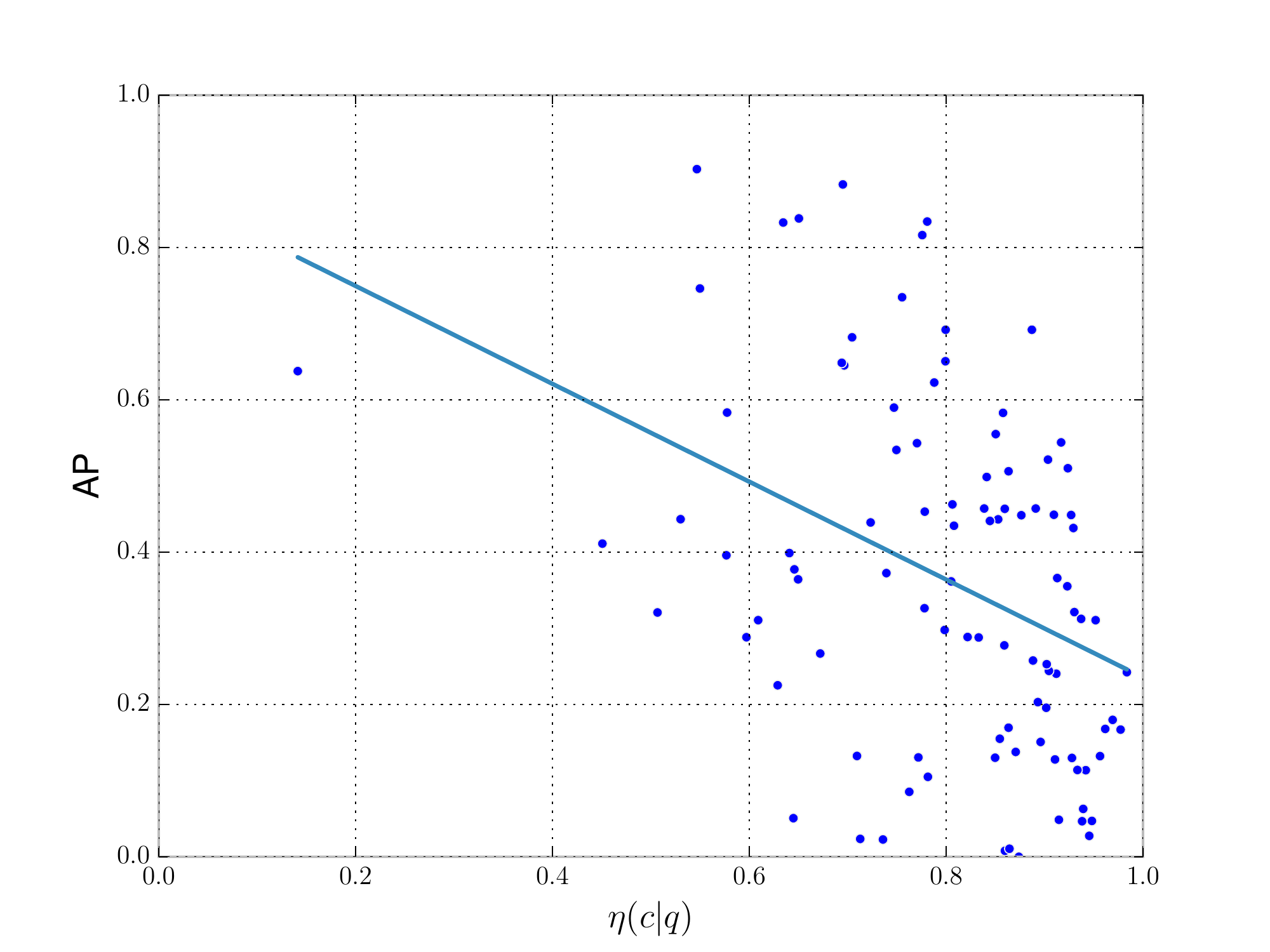}{-0.39}{0.000065}
	\hfill
	\entropyprecisionplot{cerc}{CERC}{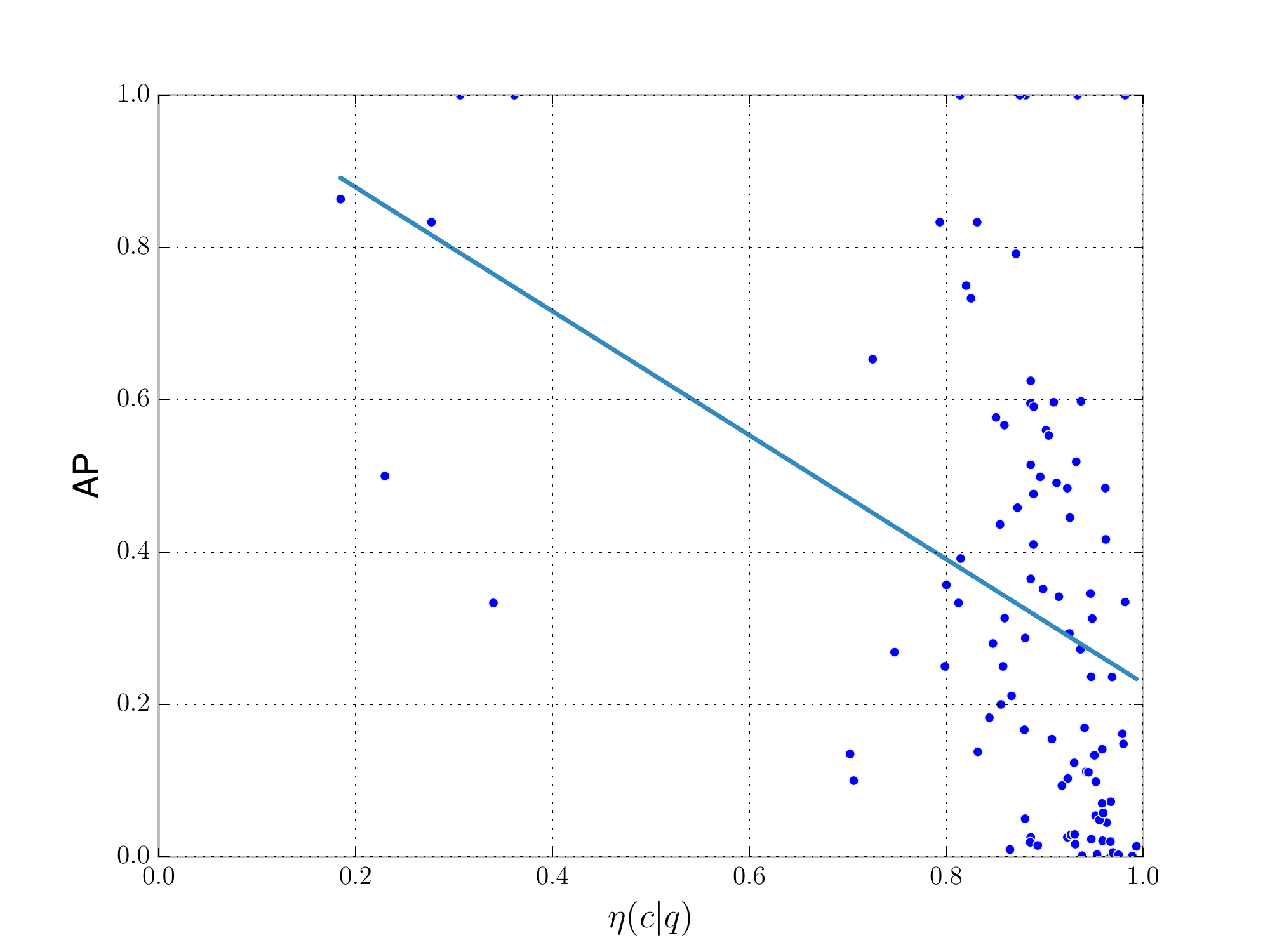}{-0.44}{0.000020}
	\hfill
	\entropyprecisionplot{tu}{TU}{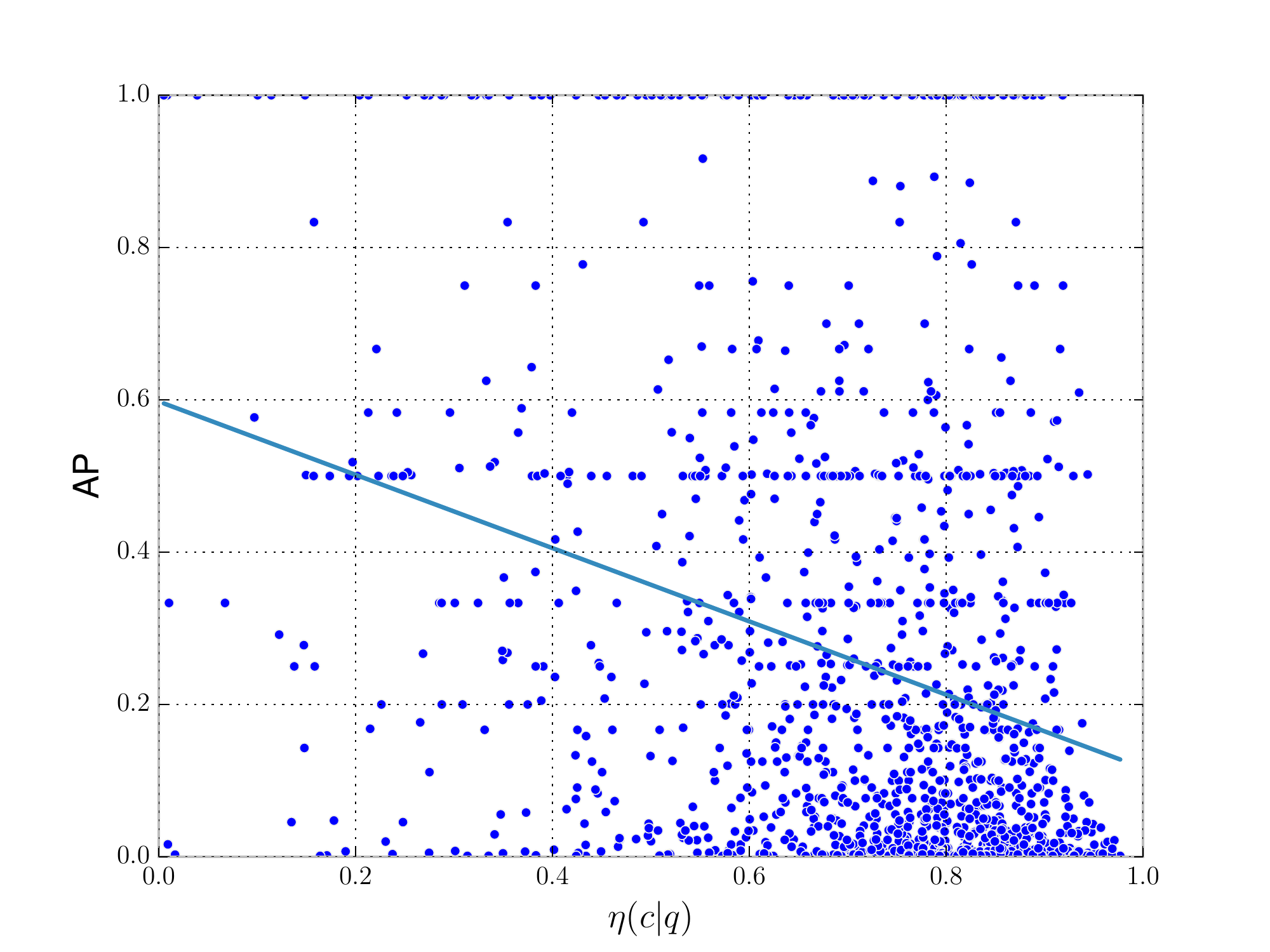}{-0.30}{0.000000}

\caption{Scatter plot of the normalized entropy of distribution $P(\CandidateVector[] \mid q)$ \eqref{eq:bag} returned by the log-linear model and per-query average precision for W3C, CERC and TU benchmarks. Pearson's $R$ and associated $p$-value (two-tailed paired permutation test: $\HighlySignificant{} \, p < 0.01; \, \MoreSignificant{} \, p < 0.05; \, \Significant{} \, p < 0.1$) are between parentheses. The depicted linear fit was obtained using an ordinary least squares regression. \label{fig:entropyprecision}}
\end{figure*}

Words $w_i$ that mainly occur in documents associated with a particular expert are learned to produce distributions $P(\CandidateVector[] \mid w_i)$ with less uncertainty than words associated with many experts in \eqref{eq:unigram}. The product of $P(\CandidateVector[] \mid w_i)$ in \eqref{eq:bag} aggregates this expertise evidence generated by query terms. Hence, queries with strong evidence for a particular expert should be more predictable than very generic queries. To quantify uncertainty we measure the normalized entropy \citep{Shannon1948} of $P(\CandidateVector[] \mid  q)$:
\begin{equation}
\label{eq:normentropy}
\eta(c \mid  q) = - \frac{1}{\log(|C|)} \sum_{j=1}^{|C|} p(\CandidateVector[j] \mid  q) \log (p(\CandidateVector[j] \mid  q)).
\end{equation}
Equation~\ref{eq:normentropy} can be interpreted as a similarity measure between the given distribution and the uniform distribution. Importantly, Fig.~\ref{fig:entropyprecision} shows that there is a statistically significant negative correlation between query-wise normalized entropy and average precision for all benchmarks.

Table~\ref{tbl:all_results} presents a comparison between the log-linear model and the various baselines (Section~\ref{sec:baseline}). Our unsupervised method significantly ($p < 0.01$) outperforms the LSI-based method consistently. In the case of the TF-IDF method and the profile-centric generative language models (Model~1), we always perform better and statistical significance is achieved in the majority of cases. The document-centric language models (Model~2) perform slightly better than our method on two (out of six) benchmarks in terms of MAP and NDCG@$100$:
\begin{inparaenum}[(a)]
\item For the CERC 2007 assessment we match performance of the document-centric generative model with Jelinek-Mercer smoothing.
\item For TU GT1 the generative counterpart seems to outperform our method.
\end{inparaenum}

Notice that over all assessments, the log-linear model consistently outperforms all profile-centric approaches and is only challenged by the smoothed document-centric approach. In addition, for the precision-based measures (P@$k$ and MRR), the log-linear model consistently outperforms all other methods we compare to.

Next, we turn to a topic-wise comparative analysis of discriminative and generative models. After that, we analyse the scalability and efficiency of the log-linear model and compare it to that of the generative counterparts, and address incremental indexing.

\subsection{Error analysis}
\label{sec:erroranalysis}

\begin{figure*}[t]
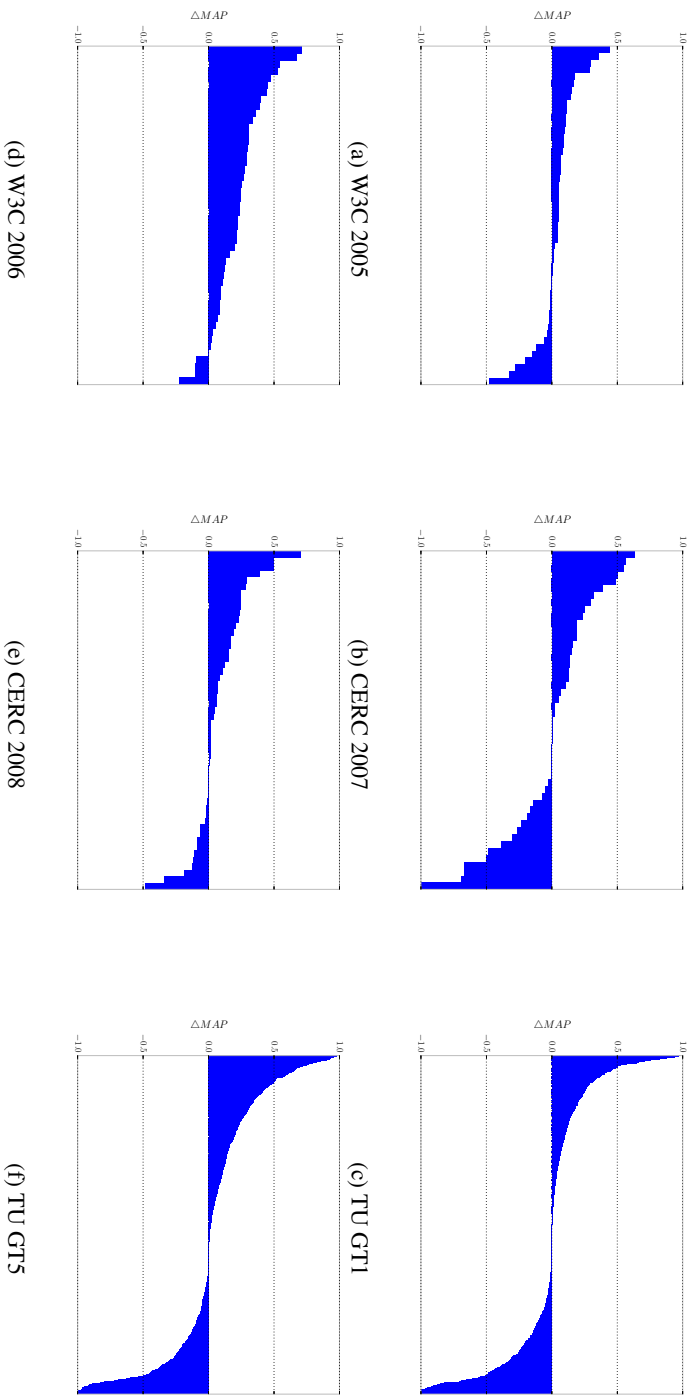

	\centering

	\newcommand{\errorplot}[4]{
		\subfloat[#3\label{fig:error:#1}]{%
			\myincludegraphics[width=0.32\textwidth]{resources/error_analysis_results/#4.pdf}
		}
	}

	\errorplot{w3c-2005}{map}{W3C 2005}{W3C_window_sweep/full_8window_allexperts_uint16_doclength_padding_ent05.expert.qrels_model2.jm.run}
	\hfill
	\errorplot{cerc-2007}{map}{CERC 2007}{CERC_window_sweep/full_8window_allexperts_uint16_doclength_padding_ent07.expert.qrels_model2.jm.run}
	\hfill
	\errorplot{tu-gt1}{map}{TU GT1}{TU_window_sweep_overlapping/GT1}

\vspace*{-\baselineskip}
	\errorplot{w3c-2006}{map}{W3C 2006}{W3C_window_sweep/full_8window_allexperts_uint16_doclength_padding_ent06.expert.qrels_model2.jm.run}
	\hfill
	\errorplot{cerc-2008}{map}{CERC 2008}{CERC_window_sweep/full_8window_allexperts_uint16_doclength_padding_ent08.expert.qrels_model2.jm.run}
	\hfill
	\errorplot{tu-gt5}{map}{TU GT5}{TU_window_sweep_overlapping/GT5}

	\caption{Difference of average precision between log-linear model and Model 2 \citep{Balog2006experts} with Jelinek-Mercer smoothing per topic for W3C, CERC and TU benchmarks.\label{fig:error}}
	 \vspace*{-.5\baselineskip}
\end{figure*}

How does our log-linear model achieve its superior performance over established generative models? Fig.~\ref{fig:error} depicts the per-topic differences in average precision between the log-linear model and Model~2 (with Jelinek-Mercer smoothing) on all benchmarks. For each plot, the vertical bars with a positive AP difference correspond to test topics for which the log-linear model outperforms Model~2 and vice versa for bars with a negative AP difference.

The benefit gained from the projection matrix $\VocabularyMatrix{}$ is two-fold. First, it avoids the curse of dimensionality introduced by large vocabularies. Second, term similarity with respect to the expertise domain is encoded in latent word features. When examining words nearby query terms in the embedding space, we found words to be related to the query term. For example, word vector representations of \emph{xml} and \emph{nonterminal} are very similar for the W3C benchmark ($l_2$ norm). This can be further observed in Fig.~\ref{fig:windowsweep}: log-linear models trained on single words perform significantly worse compared to those that are able to learn from neighbouring words.

We now take a closer look at the topics for which the log-linear model outperforms Model~2 and vice versa. More specifically, we investigate textual evidence related to a topic and whether it is considered relevant by the benchmark. For the log-linear model, we examine terms nearby topic terms in $\VocabularyMatrix[]$ ($l_2$-norm), as these terms are considered semantically similar by the model and provide a means for semantic matching. For every benchmark, we first consider topics where exact matches (Model~2) perform best, followed by examples which benefit from semantic matching (log-linear model). Topic identifiers are between parentheses.

\begin{description}[itemsep=0pt,topsep=0pt]
\item[W3C] Topics \emph{P3P specification} and \emph{CSS3} (EX8 and EX69, respectively) should return candidates associated with the definition of these standards. The log-linear model, however, considers these close to related technologies such as CSS2 for \emph{CSS3} and UTF-8 for P3P. Semantic matching works for topics \emph{Semantic Web Coordination} and \emph{Annotea server protocol} (EX1 and EX103), where the former is associated with RDF libraries, RDF-related jargon and the names of researchers in the field, while the latter is associated with implementations of the protocol and the maintainer of the project.
\item[CERC] For CSIRO, topic \emph{nanohouse} (CE-035) is mentioned in many irrelevant contexts (i.e., spam) and therefore semantic matching fails. The term \emph{fish oil} (CE-126) is quickly associated with different kinds of fish, oils and organizations related to marines and fisheries. On the other hand, we observe \emph{sensor networks} (CE-018) to be associated with sensor/networking jargon and sensor platforms. Topic \emph{forensic science workshop} (CE-103) expands to syntactically-similar terms (e.g., plural), the names of science laboratories and references to support/law-protection organizations.
\break
\item[TU] The TU benchmark contains both English and Dutch textual evidence. Topics \emph{sustainable tourism} and \emph{interpolation} (1411 and 4882) do not benefit from semantic matching due to a semantic gap: \emph{interpolation} is associated with the polynomial kind while the relevance assessments focus on stochastic methods. Interestingly, for the topic \emph{law and informatization/computerization} (1719) we see that the Dutch translation of \emph{law} is very closely related. Similar terms to \emph{informatization} are, according to the log-linear model, Dutch words related to cryptography. Similar dynamics are at work for \emph{legal-political space} (12603), where translated terms and semantic-syntactic relations aid performance.
\end{description}

\begin{figure*}[t]
	\centering

	\newcommand{\queryexpansionplot}[4]{
		\subfloat[#3\label{fig:queryexpansion:#1}]{%
			\myincludegraphics[width=0.32\textwidth]{resources/semantic_expansion/#4_#2.pdf}
		}
	}

	\queryexpansionplot{w3c}{map}{W3C}{W3C}
	\hfill
	\queryexpansionplot{cerc}{map}{CERC}{CERC}
	\hfill
	\queryexpansionplot{tu}{map}{TU}{TU}

	\caption{Effect of query expansion by adding nearby terms in $\VocabularyMatrix{}$ \eqref{eq:unigram} in traditional language models (Model 1 \citep{Balog2006experts} with Jelinek-Mercer smoothing) for W3C, CERC and TU benchmarks.\label{fig:queryexpansion}}

\end{figure*}

\noindent%
In order to further quantify the effect of the embedding matrix $\VocabularyMatrix{}$, we artificially expand benchmark topic terms by $k$ nearby terms. We then examine how the performance of a profile-centric generative language model \citep[Model~1]{Balog2006experts} evolves for different values of $k$ (Fig.~\ref{fig:queryexpansion}). The purpose of this analysis is to provide further insight in the differences between maximum-likelihood language models and the log-linear model. Fig.~\ref{fig:queryexpansion} shows that, for most benchmarks, MAP increases as $k$ goes up. Interestingly enough, the two benchmarks that exhibit a decrease in MAP for larger $k$ (CERC 2007 and TU GT1) are likewise those for which generative language models outperform the log-linear model in Table~\ref{tbl:all_results}. This suggests that the CERC 2007 and TU GT1 benchmarks require exact term matching, while the remaining four benchmarks benefit greatly from the semantic matching provided by our model.

\begin{table*}[t]
	\centering
	\caption{Comparison of Model~2, the log-linear model and an ensemble of the former on W3C, CERC and TU benchmarks. Significance of results is determined using a two-tailed paired randomization test \citep{Smucker2007significance} ($\HighlySignificant{} \, p < 0.01$; $\MoreSignificant{} \, p < 0.05$; $\Significant{} \, p < 0.1$) with respect to the ensemble ranking  (adjusted using the Benjamini-Hochberg procedure for multiple testing \citep{Benjamini1995}).\label{tbl:ensemble}}
	\resizebox{\textwidth}{!}{\begin{tabular}{c@{ }c@{ }c@{ }c@{ }c@{ }c@{ }c@{ }c@{ }c@{ }c@{ }c}
\toprule
\multirow{2}{*}{W3C} & \multicolumn{5}{c}{2005} & \multicolumn{5}{c}{2006} \\ 
& MAP & NDCG@100 & MRR & P@5 & P@10 & MAP & NDCG@100 & MRR & P@5 & P@10 \\ 
\cmidrule(lr){2-6}
\cmidrule(lr){7-11}
\multicolumn{1}{l}{Model 2 (jm)} & $\phantom{}0.211$ & $\phantom{}0.380$ & $\phantom{}0.451$ & $\phantom{}0.332$ & $\phantom{}0.296$ & $\phantom{}0.260$ & $\phantom{}0.423$ & $\phantom{}0.599$ & $\phantom{}0.449$ & $\phantom{}0.429$ \\ 
\multicolumn{1}{l}{Log-linear (ours)} & $\phantom{}0.248$ & $\phantom{}0.444$ & $\phantom{}0.618$ & $\phantom{}0.412$ & $\phantom{}0.361$ & $\phantom{\HighlySignificant{}}\textbf{0.484}\HighlySignificant{}$ & $\phantom{\MoreSignificant{}}\textbf{0.667}\MoreSignificant{}$ & $\phantom{}\textbf{0.833}$ & $\phantom{\MoreSignificant{}}\textbf{0.713}\MoreSignificant{}$ & $\phantom{\MoreSignificant{}}\textbf{0.644}\MoreSignificant{}$ \\ 
\multicolumn{1}{l}{Ensemble} & $\phantom{\HighlySignificant{}}\textbf{0.291}\HighlySignificant{}$ & $\phantom{\MoreSignificant{}}\textbf{0.479}\MoreSignificant{}$ & $\phantom{}\textbf{0.668}$ & $\phantom{}\textbf{0.440}$ & $\phantom{}\textbf{0.378}$ & $\phantom{}0.433$ & $\phantom{}0.634$ & $\phantom{}0.825$ & $\phantom{}0.657$ & $\phantom{}0.586$ \\ 
\cmidrule{1-11}
\multirow{2}{*}{CERC} & \multicolumn{5}{c}{2007} & \multicolumn{5}{c}{2008} \\ 
& MAP & NDCG@100 & MRR & P@5 & P@10 & MAP & NDCG@100 & MRR & P@5 & P@10 \\ 
\cmidrule(lr){2-6}
\cmidrule(lr){7-11}
\multicolumn{1}{l}{Model 2 (jm)} & $\phantom{}0.361$ & $\phantom{}0.500$ & $\phantom{}0.467$ & $\phantom{}0.192$ & $\phantom{}0.138$ & $\phantom{}0.274$ & $\phantom{}0.463$ & $\phantom{}0.517$ & $\phantom{}0.278$ & $\phantom{}0.239$ \\ 
\multicolumn{1}{l}{Log-linear (ours)} & $\phantom{}0.344$ & $\phantom{}0.493$ & $\phantom{}0.513$ & $\phantom{}0.215$ & $\phantom{}0.150$ & $\phantom{}0.342$ & $\phantom{}0.519$ & $\phantom{}0.656$ & $\phantom{}0.381$ & $\phantom{}0.299$ \\ 
\multicolumn{1}{l}{Ensemble} & $\phantom{\MoreSignificant{}}\textbf{0.452}\MoreSignificant{}$ & $\phantom{\HighlySignificant{}}\textbf{0.589}\HighlySignificant{}$ & $\phantom{\HighlySignificant{}}\textbf{0.627}\HighlySignificant{}$ & $\phantom{\Significant{}}\textbf{0.248}\Significant{}$ & $\phantom{}\textbf{0.160}$ & $\phantom{\HighlySignificant{}}\textbf{0.395}\HighlySignificant{}$ & $\phantom{\HighlySignificant{}}\textbf{0.593}\HighlySignificant{}$ & $\phantom{}\textbf{0.716}$ & $\phantom{\MoreSignificant{}}\textbf{0.459}\MoreSignificant{}$ & $\phantom{\HighlySignificant{}}\textbf{0.357}\HighlySignificant{}$ \\ 
\cmidrule{1-11}
\multirow{2}{*}{TU} & \multicolumn{5}{c}{GT1} & \multicolumn{5}{c}{GT5} \\ 
& MAP & NDCG@100 & MRR & P@5 & P@10 & MAP & NDCG@100 & MRR & P@5 & P@10 \\ 
\cmidrule(lr){2-6}
\cmidrule(lr){7-11}
\multicolumn{1}{l}{Model 2 (jm)} & $\phantom{}0.234$ & $\phantom{}0.370$ & $\phantom{}0.342$ & $\phantom{}0.136$ & $\phantom{}0.101$ & $\phantom{}0.253$ & $\phantom{}0.394$ & $\phantom{}0.302$ & $\phantom{}0.108$ & $\phantom{}0.081$ \\ 
\multicolumn{1}{l}{Log-linear (ours)} & $\phantom{}0.219$ & $\phantom{}0.356$ & $\phantom{}0.351$ & $\phantom{}0.145$ & $\phantom{}0.105$ & $\phantom{}0.287$ & $\phantom{}0.425$ & $\phantom{}0.363$ & $\phantom{}0.134$ & $\phantom{}0.092$ \\ 
\multicolumn{1}{l}{Ensemble} & $\phantom{\HighlySignificant{}}\textbf{0.271}\HighlySignificant{}$ & $\phantom{\HighlySignificant{}}\textbf{0.417}\HighlySignificant{}$ & $\phantom{\HighlySignificant{}}\textbf{0.403}\HighlySignificant{}$ & $\phantom{\HighlySignificant{}}\textbf{0.165}\HighlySignificant{}$ & $\phantom{\HighlySignificant{}}\textbf{0.121}\HighlySignificant{}$ & $\phantom{\HighlySignificant{}}\textbf{0.331}\HighlySignificant{}$ & $\phantom{\HighlySignificant{}}\textbf{0.477}\HighlySignificant{}$ & $\phantom{\HighlySignificant{}}\textbf{0.402}\HighlySignificant{}$ & $\phantom{\HighlySignificant{}}\textbf{0.156}\HighlySignificant{}$ & $\phantom{\HighlySignificant{}}\textbf{0.105}\HighlySignificant{}$ \\ 
\bottomrule
\end{tabular}
}
	\vspace*{-.5\baselineskip}
\end{table*}

The per-topic differences suggest that Model~2 and the log-linear model make very different errors: Model~2 excels at retrieving exact query matches, while the log-linear model performs semantic matching. Based on these observations we hypothesize that a combination of the two approaches will raise retrieval performance even further. To test this hypothesis, we propose a simple ensemble of rankings generated by Model~2 and the log-linear model by re-ranking candidates according to the multiplicatively-combined reciprocal rank:
\begin{equation}
\label{eq:ensemble}
\text{rank}_\text{ensemble}(c_j, q_i) \propto
 \frac{1}{\text{rank}_\text{model~2}(c_j, q_i)} \cdot \frac{1}{\text{rank}_\text{log-linear}(c_j, q_i)},
\end{equation}
where $\text{rank}_{M}(c_j, q_i)$ denotes the position of candidate $c_j$ in a ranking generated by model $M$ for answering query $q_i$. Equation~\eqref{eq:ensemble} is equivalent to performing data fusion using CombSUM \citep{Shaw1994} where the scores are given by the logarithm of the reciprocal ranks of the experts. Table~\ref{tbl:ensemble} compares the result of this ensemble to that of its constituents. Compared to the supervised methods of \citet{Fang2010}, we conclude that our fully unsupervised ensemble matches the performance of their method on the CERC 2007 benchmark and outperforms their method on the W3C 2005 benchmark. The superior performance of the ensemble suggests the viability of hybrid methods that combine semantic and exact matching.

\subsection{Scalability and efficiency}

Inference in the log-linear model is expressed in linear algebra operations (Section~\ref{section:methodology}). These operations can be efficiently performed by highly optimized software libraries and special-purpose hardware (i.e., GPUs). But the baseline methods against which we compare do not benefit from these speed-ups. Furthermore, many implementation-specific details and choice of parameter values can influence runtime considerably (e.g. size of the latent representations). Therefore, we opt for a theoretical comparison of the inference complexity of the log-linear model and compare these to the baselines (Section~\ref{sec:baseline}).

The log-linear model generates a ranking of candidate experts by straight-forward matrix operations. The look-up operation in the projection matrix $\VocabularyMatrix{}$ occurs in constant time complexity, as the multiplication with the one-hot vector $v_i$ comes down to selecting the $i$-th column from $\VocabularyMatrix{}$. Multiplication of the $|C| \times \EmbeddingSize{}$ matrix $\CandidateMatrix{}$ with the $\EmbeddingSize{}$-dimensional word feature vector exhibits $O(|C| \cdot \EmbeddingSize{})$ runtime complexity. If we consider addition of the bias term and division by the normalizing function $Z_1$, the time complexity of \eqref{eq:unigram} becomes
\begin{equation*}
O(\underbrace{|C| \cdot (\EmbeddingSize{} + (\EmbeddingSize{} - 1))}_{\text{matrix-vector multiplication}} + \underbrace{|C|}_{\text{bias term}} + \underbrace{2 \cdot |C| - 1}_{Z_1}).
\end{equation*}
Notice, however, that the above analysis considers sequential execution. Modern computing hardware has the ability to parallelize common matrix operations \citep{Kruger2003,Fatahalian2004}. The number of candidate experts $|C|$ is the term that impacts performance most in the log-linear model (under the assumption that $|C| \gg \EmbeddingSize{}$).

If we consider $n$ terms, where $n$ is the query length during inference or the window size during training, then the complexity of~\eqref{eq:bag} becomes
\begin{eqnarray*}
\lefteqn{O(\underbrace{n \cdot |C| \cdot (2 \cdot \EmbeddingSize{} - 1) + n \cdot (3  \cdot |C| - 1)}_{\text{n forward-passes}}} \\
&&{} + \underbrace{(n - 1) \cdot |C|}_{\text{factor product}} + \underbrace{2 \cdot |C| - 1}_{Z_2})
\end{eqnarray*}
Notice that $Z_2$ does not need to be computed during inference as it does not affect the candidate expert ranking.

In terms of space complexity, parameters $\VocabularyMatrix{}$, $\CandidateMatrix{}$ and $\CandidateBias{}$, in addition to the intermediate results, all require memory space proportional to their size. Considering \eqref{eq:bag} for a sequence of $k$ words and batches of $m$ instances, we require $O(m \cdot k \cdot |C|)$ floating point numbers for every forward-pass to fit in-memory. While such an upper bound seems reasonable by modern computing standards, it is a severely limiting factor when considering large-scale communities and while utilizing limited-memory GPUs for fast computation.

The inferential complexity of the vector space-based models for entity retrieval \citep{Demartini2009expertspaces} depends mainly on the dimensionality of the vectors and the number of candidate experts. The dimensionality of the latent entity representations is too high for efficient nearest neighbour retrieval \citep{Indyk1998} due to the curse of dimensionality. Therefore, the time complexity for the LSI- and TF-IDF-based vector space models are respectively $O(\gamma \cdot |C|)$ and $O(|V| \cdot |C|)$, where $\gamma$ denotes the number of latent topics in the LSI-based method. As hyperparameters $e$ and $\gamma$ both indicate the dimensionality of latent entity representations, the time complexity of the LSI-based method is comparable to that of the log-linear model. We note that $|V| \gg |C|$ for all benchmarks ($|V|$ is between $18$ to $91$ times larger than $|C|$) we consider in this section and therefore conclude that the TF-IDF method loses to the log-linear model in terms of efficiency.

Compared to the unsupervised generative models of \citeauthor{Balog2006experts}, we have the profile-centric Model~1 and the document-centric Mo\-del~2 with inference time complexity $O(n \cdot |C|)$ and $O(n \cdot |D|)$, respectively, with $|D| \gg |C|$. In the previous section we showed that the log-linear model always performs better than Model~1 and nearly always outperforms Model~2. Hence, our log-linear model generally achieves the expertise retrieval performance of Model~2 (or higher) at the complexity cost of Model~1 during inference.

\subsection{Incremental indexing}

Existing unsupervised methods use well-understood maximum-likelihood language models that support incremental indexing. We now briefly discuss the incremental indexing capabilities of our proposed method. Extending the set of candidate experts $C$ requires the log-linear model to be re-trained from scratch as it changes the topology of the network. Moreover, every document associated with a candidate expert is considered as a negative example for all other candidates. While it is possible to reiterate over all past documents and only learn an additional row in matrix $\CandidateMatrix{}$, the final outcome is unpredictable.

If we consider a stream of documents instead of a predefined set $D$, the log-linear model can be learned in an online fashion. However, stochastic gradient descent requires that training examples are picked at random such that the batched update rule \eqref{eq:update} behaves like the empirical expectation over the full training set \citep{Bottou2010}. While we might be able to justify the assumption that documents arrive randomly, the $n$-grams extracted from those documents clearly violate this requirement.

Considering a stream of documents leads to the model forgetting expertise evidence as an (artificial) shift in the underlying distribution of the training data occurs. While such behaviour is undesirable for the task considered in this section, it might be well-suited for temporal expert finding \citep{Rybak2014,Fang2014}, where expertise drift over time is considered. However, temporal expertise finding is beyond the scope for this section and left for future work.


\section{Summary}
\label{section:conclusions}

We have introduced an unsupervised discriminative, log-linear model for the expert retrieval task. Our approach exclusively employs raw textual evidence. Future work can focus on improving performance by feature engineering and incorporation of external evidence. Furthermore, no relevance feedback is required during training. This renders the model suitable for a broad range of applications and domains.

We evaluated our model on the W3C, CERC and TU benchmarks and compared it to state-of-the-art vector space-based entity ranking (based on LSI and TF-IDF) and language modelling (profile-centric and document-centric) approaches. The log-linear model combines the ranking performance of the best maximum-likelihood language modelling approach (document-centric) with inference time complexity linear in the number of candidate experts. We observed a notable increase in precision over existing methods. Analysis of our model's output reveals a negative correlation between the per-query performance and ranking uncertainty: higher confidence (i.e., lower entropy) in the rankings produced by our approach often occurs together with higher rank quality.

An error analysis of the log-linear model and traditional language models shows that the two make very different errors. These errors are mainly due to the semantic gap between query intent and the raw textual evidence. Some benchmarks expect exact query matches, others are helped by our semantic matching. An ensemble of methods employing exact and semantic matching generally outperforms the individual methods. This observation calls for further research in the area of combining exact and semantic matching.

In the next chapter (Chapter~\globalref{chapter:research-04}), we further investigate the representations learned as part of the log-linear model. In particular, we investigate whether
\begin{inparaenum}[(a)]
	\item a clustering of experts corresponds to the working groups within organizations, and
	\item experts that work in the same fields have similar representations.
\end{inparaenum}%
Chapter~\globalref{chapter:research-05} and Chapter~\globalref{chapter:research-06} are dedicated to scaling up the latent vector space model introduced in this chapter to retrieval problems that consist of more retrievable objects and with different characteristics than expert finding.

{
\renewcommand{\newpage}{}
\renewcommand{\appendix}{\section{Appendix}}

\newpage
\appendix

{

\newcommand{\BigSkip}{\qquad\qquad\qquad\qquad\qquad\qquad}

\newcommand{\LossGradNoReg}[1]{\frac{\partial \LossFn{}}{\partial #1} & = & - \frac{1}{m}  \Bigg( \sum_{i = 1}^m \frac{|\Document{}_{\max}|}{|\SourceDocument{i}|} \sum_{j = 1}^{|C|} P(\CandidateVector[j] | \SourceDocument{i}) \frac{\partial \log\left(P(\CandidateVector[j] \mid \Context[i])\right)}{\partial #1} \Bigg)}

\newcommand{\RescaleEqn}[1]{
	\scalebox{0.85}{%
	\begin{minipage}{0.85\linewidth}%
	#1
	\end{minipage}}}

The derivative of \eqref{eq:loss} w.r.t. bias term $\CandidateBias[]$ equals
\begin{center}
\RescaleEqn{
\begin{eqnarray*}
\hspace*{-1cm}
\LossGradNoReg{\CandidateBias[]} \nonumber
\end{eqnarray*}
}
\end{center}
and w.r.t. an arbitrary matrix parameter $\theta$ ($\VocabularyMatrix[]$ or $\CandidateMatrix[]$):
\begin{center}
\RescaleEqn{
\begin{eqnarray*}
\hspace*{-1cm}
\LossGradNoReg{\theta} \nonumber \\
& & + \frac{\lambda}{m} \sum_{i,j} \theta_{i,j} \nonumber.
\end{eqnarray*}
}
\end{center}
\renewcommand{\RescaleEqn}[1]{#1}
Further differentiation for parameter $\theta$ ($\VocabularyMatrix[]$, $\CandidateMatrix[]$ or $\CandidateBias[]$):
\begin{center}
\RescaleEqn{
\begin{eqnarray*}
\frac{\partial \log\left(P(\CandidateVector[j] \mid w_{1}, \ldots, w_{n})\right)}{\partial \theta} \nonumber
& =& \frac{1}{P(\CandidateVector[j] \mid w_{1}, \ldots, w_{n})} \frac{\partial P(\CandidateVector[j] \mid w_{1}, \ldots, w_{n})}{\partial \theta} \\
\frac{\partial P(\CandidateVector[j] \mid w_{1}, \ldots, w_{n})}{\partial \theta} & = &\frac{\frac{\partial \tilde{P}(c_j \mid w_{1}, \ldots, w_{n})}{\partial \theta} Z_2 - \tilde{P}(c_j \mid w_{1}, \ldots, w_{n}) \frac{\partial Z_2}{\partial \theta}}{Z^2_2} \\
\frac{\partial Z_2}{\partial \theta} &=& \sum_k \frac{\partial \tilde{P}(c_k \mid w_{1}, \ldots, w_{n})}{\partial \theta} \\
\frac{\partial \tilde{P}(c_j \mid w_{1}, \ldots, w_{n})}{\partial \theta}
&=& \sum_k \frac{\partial P(\CandidateVector[j] \mid w_k)}{\partial \theta} \prod_{i \neq k} P(\CandidateVector[j] \mid w_i)
\end{eqnarray*}
}
\end{center}
For a given candidate $c_j$ and word $w_i$, following \eqref{eq:unigram} we have
\newcommand{\NormalizedUnigramProb}{P(\CandidateVector[j] \mid w_i)}
\newcommand{\UnigramProb}{\tilde{P}(\CandidateVector[j] \mid w_i)}
\begin{center}
\RescaleEqn{
\begin{eqnarray*}
\NormalizedUnigramProb &=& \frac{\UnigramProb{}}{Z_1}\\
 &=& \frac{\exp \left( \left( \sum^{\EmbeddingSize{}}_{k=1} \CandidateMatrix[j,k] \VocabularyMatrix[k,i] \right) + \CandidateBias[j] \right)}{\sum^{|C|}_{l=1} \exp \left( \left( \sum^{\EmbeddingSize{}}_{k=1} \CandidateMatrix[l,k] \VocabularyMatrix[k,i] \right) + \CandidateBias[l] \right)}
\end{eqnarray*}
}
\end{center}
and consequently, with $\boldsymbol{\VocabularyMatrix[i]^\top}$ denoting the $i$-th column of matrix $\VocabularyMatrix[]$,
\begin{center}
\RescaleEqn{
\begin{equation}
\label{eq:dist_repr_grad}
\begin{split}
\frac{\partial \NormalizedUnigramProb}{\partial \boldsymbol{\CandidateMatrix[j]}} & = \frac{\left( Z_1 - \UnigramProb{} \right) \UnigramProb{} \boldsymbol{\VocabularyMatrix[i]^\top} }{Z^2_1} \\
\frac{\partial \NormalizedUnigramProb}{\partial \CandidateBias[j]} & = \frac{\left( Z_1 - \UnigramProb{} \right) \UnigramProb{}}{Z^2_1} \\
\frac{\partial \NormalizedUnigramProb}{\partial \boldsymbol{\VocabularyMatrix[i]^\top}} & = \frac{\left( \boldsymbol{\CandidateMatrix[j]} - \sum^{|C|}_{l=1} \boldsymbol{\CandidateMatrix[l]} \right) \UnigramProb{}}{Z_1}
\end{split}
\end{equation}
}
\end{center}
As can be seen in \eqref{eq:dist_repr_grad}, the distributed representations of candidates $c_j$ at time $t + 1$ are updated using the representation of words $w_i$ at time $t$ and vice versa.

}
}
}


\chapter{Structural Regularities in Text-based Entity Vector Spaces}
\label{chapter:research-04}

{
\ScopeLabels{research-04}
\ChapterRQ[4]{2}

\newcommand{\BenchmarkTU}{TU}
\newcommand{\BenchmarkWThreeC}{W3C}

\newcommand{\NDCG}{\ac{NDCG}}
\newcommand{\RPrec}{R-Precision}

\newcommand{\ResearchQuestionOne}{Do clusterings of text-based entity representations reflect the structure of their domains?}
\newcommand{\ResearchQuestionTwo}{To what extent do different text-based entity representation methods encode relations between entities?}

\newcommand{\DocToVec}{\FullDocToVec}
\newcommand{\WordToVec}{\FullWordToVec}

\renewcommand{\newcommand}{\providecommand}


\section{Introduction}
\label{sec:introduction}

The construction of latent entity representations is a recurring problem \citep{Bordes2011,He2013learning,Zhao2015,Clark2016improving,Demartini2009expertspaces} in natural language processing and information retrieval. So far, entity representations are mostly learned from relations between entities \citep{Bordes2011,Zhao2015} for a particular task in a supervised setting \citep{He2013learning}. How can we learn latent entity representations if
\begin{inparaenum}[(i)]
	\item entities only have relations to documents in contrast to other entities (e.g., scholars are represented by the papers they authored), and
	\item there is a lack of labeled data?
\end{inparaenum}

As entities are characterized by documents that consist of words, can we use word embeddings to construct a latent entity representation? Distributed representations of words \cite{Hinton1986}, i.e., word embeddings, are learned as part of a neural language model and have been shown to capture semantic \cite{Collobert2008} and syntactic regularities \cite{Mikolov2013word2vec,Pennington2014}. In addition, word embeddings have proven to be useful as feature vectors for natural language processing tasks \cite{Turian2010}, where they have been shown to outperform representations based on count-based distributional semantics \cite{Baroni2014}. A down-side of word embeddings \cite{Bengio2003} is that they do not take into account the document a word sequence occurred in or the entity that generated it.

\citeauthor{Le2014} address this problem by extending \WordToVec{} models to \DocToVec{} by additionally modelling the document a phrase occurred in. That is, besides word embeddings they learn embeddings for documents as well. We can apply \DocToVec{} to the entity representation problem by representing an entity as a pseudo-document consisting of all documents the entity is associated with. The neural model we introduced in the previous \paper{}---which we refer to as \SERT{} in this \paper{}---incorporates real-world structural relations between represented entities even though the representations are learned from text only. In addition to word embeddings, we learned a representation for entities such that the words that are highly discriminative for an entity have a representation similar to that entity. 

In this \paper{}, we study the regularities contained within entity representations that are estimated, in an unsupervised manner, from texts and associations alone. Do they correspond to structural real-world relations between the represented entities? E.g., if the entities we represent are people, do these regularities correspond to collaborative and hierarchical structures in their domain (industrial, governmental or academic organizations in the case of experts)? Answers to these questions are valuable because if they allow us to better understand the inner workings of entity retrieval models and give important insights into the entity-oriented  tasks they are used for~\citep{Le2014}. In addition, future work can build upon these insights to extract structure within entity domains given only a document collection and entity-document relations so to complement or support structured information.

Our working hypothesis is that text-based entity representations encode regularities within their domain. To test this hypothesis
we compare latent text-based entity representations learned by neural networks (\WordToVec{}, \DocToVec{}, \SERT{}), count-based entity vector representations constructed using Latent Semantic Indexing (\LSI{}) and Latent Dirichlet Allocation (\LDA{}), dimensionality-reduced adjacency representations (\GraphPCA{}) and \Random{} representations sampled from a standard multivariate normal distribution. For evaluation purposes we focus on expert finding, a particular case of entity ranking. Expert finding is the task of finding the right person with the appropriate skills or knowledge~\cite{Balog2012survey}, based on a document collection and associations between people and documents. These associations can be extracted using entity linking methods or from document meta-data (e.g., authorship). Typical queries are descriptions of expertise areas, such as \emph{distributed computing}, and expert search engines answer the question ``Who are experts on \emph{distributed computing}?'' asked by people unfamiliar with the field.

Our main finding is that, indeed, semantic entity representations encode domain regularities. Entity representations can be used as feature vectors for clustering and that those partitions correspond to structural groups within the entity domain. We also find that similarity between entity representations correlates with relations between entities. In particular, we show how representations of experts in the academic domain encode the co-author graph. Lastly, we show that one of the semantic representation learning methods, \SERT{}, additionally encodes importance amongst entities and, more specifically, the hierarchy of scholars in academic institutions.

\medskip
\noindent
The research questions we ask in this \paper{} towards answering \MainRQRef{2} are as follows:
\RQ{1}{\ResearchQuestionOne{}}
\RQ{2}{\ResearchQuestionTwo{}}


\section{Related work}

We refer to Section~\globalref{chapter:background:latent} of our background chapter (Chapter~\globalref{chapter:background}). The subsection covering entity retrieval (\S\globalref{chapter:background:related:entities}) and regularities in language representations (\S\globalref{chapter:background:related:representations}) are of special interest to this \paper{}.

In the maximum-likelihood language modelling paradigm, experts are represented as a normalized bag-of-words vector with additional smoothing. These vectors are high-dimensional and sparse due to the large vocabularies used in expert domains. Therefore, bag-of-words vectors are unsuited for use as representations---the topic of study in this \paper{}---as lower-dimensional and continuous vector spaces are preferred in machine learning algorithms \cite{Weber1998}. 


\section{Text-based entity vector spaces}
\label{sec:methodology}

\newcommand{\Word}{w}
\newcommand{\Term}{t}

\newcommand{\Vocabulary}{V}

\newcommand{\Document}{d}
\newcommand{\Documents}{\MakeUppercase{\Document{}}}

\newcommand{\Query}{q}

\newcommand{\Candidates}{X}
\newcommand{\Candidate}{\MakeLowercase{\Candidates{}}}

\newcommand{\DocumentCandidates}[1][]{\Candidates{}_{\Document{}#1}}
\newcommand{\CandidateDocuments}[1][]{\Documents{}_{\Candidate{}#1}}

\newcommand{\Length}[1]{{|#1|}}

\newcommand{\WordRepresentation}{v}
\newcommand{\CandidateRepresentation}{e}

\newcommand{\CandidateBias}{b}

\newcommand{\VectorDim}{k}
\newcommand{\VectorMap}[1][]{g}

\newcommand{\ScoreFn}[2]{\text{score}(#1, #2)}

\newcommand{\Prob}[2][P]{#1(#2)}
\newcommand{\CondProb}[3][P]{\Prob[#1]{#2 \mid #3}}

\newcommand{\CosineSimilarity}[1]{S_C}

For text-based entity retrieval tasks we are given a document collection $\Documents{}$ and a set of entities $\Candidates{}$. Documents $\Document{} \in \Documents{}$ consist of a sequence of words $\Word{}_{1}, \ldots, \Word{}_\Length{\Document{}}$ originating from a vocabulary $\Vocabulary{}$, where $\Length{\cdot}$ denotes the document length in number of words. For every document $\Document{}$ we have a set $\DocumentCandidates{} \subset \Candidates{}$ of associated entities ($\DocumentCandidates{}$ can be empty for some documents) and conversely $\CandidateDocuments{} \subset \Documents{}$ consists of all documents associated with entity $\Candidate{}$. The associations between documents and experts can be obtained in multiple ways. E.g., named-entity recognition can be applied to the documents and mentions can subsequently be linked to entities. Or associations can be extracted from document meta-data (e.g., authorship).

Once determined, the associations between entities $\Candidates{}$ and documents $\Documents{}$ encode a bipartite graph. If two entities $\Candidate{}_i, \Candidate{}_j \in \Candidates{}$ are associated with the same document, we say that $\Candidate{}_i$ and $\Candidate{}_j$ are co-associated. However, the semantics of a co-association are equivocal as the semantics of an association are ambiguous by itself (e.g., author vs. editor). Therefore, instead of relying solely on document associations, we use the textual data of associated documents to construct an entity representation.

Vector space models for document retrieval, such as LSI \cite{Deerwester1990lsi} or LDA \cite{Blei2003}, can be adapted to entity retrieval. We substantiate this for a specific entity retrieval task: expert finding. As there are many more documents than experts, it is not ideal to estimate a vector space directly on the expert-level using bag-of-word vectors (e.g., by representing every expert as a concatenation of its documents) due to data sparsity. Therefore, it is preferable to first estimate a vector space on the document collection and then use the obtained document representations to construct an entity vector. \citet{Demartini2009expertspaces} take an entity's representation to be the sum of its documents:
\begin{equation}
\label{eq:expert-vsm}
\CandidateRepresentation{}_i = \sum_{\Document{}_j \in \CandidateDocuments[_i]{}} \VectorMap{}({\Document{}_j}),
\end{equation}
where $\CandidateRepresentation{}_i$ is the $\VectorDim{}$-dimensional vector representation of entity $\Candidate{}_i \in \Candidates{}$ and $\VectorMap{}$ is the function mapping a document to its vector space representation (e.g., \LSI{}). The dimensionality $\VectorDim{}$ depends on the underlying vector space. For simple bag-of-words representations, $\VectorDim{}$ is equal to the number of words in the vocabulary. For latent vector spaces (e.g., LSI), the $\VectorDim{}$-dimensional space encodes latent concepts and the choice of $\VectorDim{}$ is left to the user.

Vector space models for document retrieval are often constructed heuristically. E.g., Eq.~\ref{eq:expert-vsm} does not make optimal use of document-entity associations as document representations are added without taking into consideration the significance of words contained within them~\cite{Luhn1958significance}. And if many diverse documents are associated with an expert, then Eq.~\ref{eq:expert-vsm} is likely to succumb to the noise in these vectors and yield meaningless representations.

To address this problem, \citet{Le2014} introduced \DocToVec{} by adapting the \WordToVec{} models to incorporate the document a phrase occurs in. They optimize word and document embeddings jointly to predict a word given its context and the document the word occurs in. The key difference between \WordToVec{} and \DocToVec{} is that the latter considers an additional meta-token in the context that represents the document. Instead of performing dimensionality reduction on bag-of-words representations, \DocToVec{} learns representations from word phrases. Therefore, we use the \DocToVec{} model to learn expert embeddings by representing every expert $\Candidate{}_j \in \Candidates{}$ as a pseudo-document consisting of the concatenation of their associated documents $\CandidateDocuments[_j]{}$.

A different neural language model architecture than \DocToVec{} was proposed by \citet{VanGysel2016experts}, specifically for the expert finding task. For a given word $\Word{}_i$ and expert $\Candidate{}_j$:
\begin{equation}
\label{eq:expert-loglinear}
\ScoreFn{\Word{}_i}{\Candidate{}_j} = \exp{\left(\WordRepresentation{}_i^\intercal \cdot \CandidateRepresentation{}_j + \CandidateBias{}_j\right)},
\end{equation}
\newcommand{\ProbCandidateGivenWord}{\CondProb{\Candidates{} = \Candidate{}_j}{\Word{}_i}}%
where $\WordRepresentation{}_i$ ($\CandidateRepresentation{}_j$, resp.) are the latent $\VectorDim$-dimensional representations of word $\Word{}_i$ (and expert $\Candidate{}_j$, respectively) and $\CandidateBias{}_j$ is the bias scalar associated with expert $\Candidate{}_j$. Eq.~\ref{eq:expert-loglinear} can be interpreted as the unnormalized factor product of likelihood $\CondProb{\Word{}_i}{\Candidate{}_j}$ and prior $\Prob{\Candidate{}_j}$ in log-space. The score is then transformed to the conditional probability
\[
\ProbCandidateGivenWord{} = \frac{\ScoreFn{\Word{}_i}{\Candidate{}_j}}{\sum_{\Candidate{}_l \in \Candidates{}} \ScoreFn{\Word{}_i}{\Candidate{}_l}}.
\]
Unlike Eq.~\ref{eq:expert-vsm}, the conditional probability distribution $\ProbCandidateGivenWord{}$ will be skewed towards relevant experts if the word $\Word{}_i$ is significant as described by \citet{Luhn1958significance}. The parameters $\WordRepresentation{}_i$, $\CandidateRepresentation{}_j$ and $\CandidateBias{}_j$ are learned from the corpus using gradient descent. See \cite{VanGysel2016experts} for details.

Our focus lies on representations of entities $\CandidateRepresentation{}_j$ and how these correspond to structures within their domains (i.e., organizations for experts). These representations are estimated using a corpus only and can be interpreted as vectors in word embedding space that correspond to entities (i.e., people) instead of words. 


\section{Experimental set-up}

\subsection{Research questions}

\newcommand{\RQ}[2]{
	\begin{description}
	\item[\fontsize{10pt}{\baselineskip}\selectfont RQ#1] #2
	\end{description}
}

We investigate regularities within text-based entity vector spaces, using expert finding as our concrete test case, and ask how these representations correspond to structure in their respective domains. As indicated in the introduction of this \paper{}, we seek to answer the following research questions:

\RQ{1}{\ResearchQuestionOne{}}
Many organizations consist of smaller groups, committees or teams of experts who are appointed with a specific role. When we cluster expert representations, do the clusters correspond to these groups?

\medskip

\RQ{2}{\ResearchQuestionTwo{}}
The associations within expert domains encode a co-association graph structure. To what extent do the different expertise models encode this co-association between experts? In particular, if we rank experts according to their nearest neighbours, how does this ranking correspond to the academic co-author graph?

\subsection{Expert finding collections}
\label{sec:collections}

We use a subset of the expert finding collections of Chapter~\globalref{chapter:research-03}: the publicly-available expert finding collections provided by the World Wide Web Consortium (\BenchmarkWThreeC{}) and Tilburg University (\BenchmarkTU{}). We refer to Table~\globalref{research-03:tbl:benchmarks} (previous \paper{}) for an overview. Note that we do not use the CSIRO Enterprise Research Collection (CERC) in this \paper{} due to the lack of information about the structure of the organization.

\begin{description}
\item[\BenchmarkWThreeC{}] The \BenchmarkWThreeC{} collection was released as part of the 2005--2006 editions of the TREC Enterprise Track \cite{Craswell2005}. It contains a heterogeneous crawl of W3C's website (June 2004) and consists of mailing lists and discussion boards among others. In the 2005 edition, TREC released a list of working groups and their  members. Each working group is appointed to study and report on a particular aspect of the World Wide Web to enable the W3C to pursue its mission. We use the associations provided by \citet{VanGysel2016experts}, which they gathered by applying named entity recognition and linking these mentions to a list of organization members, as proposed by \citet{Balog2006experts}.
\item[\BenchmarkTU{}] The \BenchmarkTU{} collection consists of a crawl of a university's internal website and contains bi-lingual documents, such as academic publications, course descriptions and personal websites \citep{Berendsen2013profileassessment}. The document-candidate associations are part of the collection. For every member of the academic staff, their academic title is included as part of the collection.
\end{description}

\subsection{Implementations and parameters}

We follow a similar experimental set-up as previous work \cite{Balog2006experts,Demartini2009expertspaces,Mikolov2013compositionality,VanGysel2016experts}. For \LSI{}, \LDA{}, \WordToVec{} and \DocToVec{} we use the Gensim\footnote{\url{https://radimrehurek.com/gensim}} implementation, while for the log-linear model we use the Semantic Entity Retrieval Toolkit\footnote{\url{https://github.com/cvangysel/SERT}} (\SERT{}) \citep{VanGysel2017sert} that was released as part of Chapter~\globalref{chapter:research-03} and is more closely described in Appendix~\globalref{chapter:research-08}.

The corpora are normalized by lowercasing and removing punctuation and numbers. The vocabulary is pruned by removing stop words and  retaining the 60k most frequent words. We sweep exponentially over the vector space dimensionality ($\VectorDim{} = 32,\, 64,\, 128$ and $256$) of the methods under comparison. This allows us to evaluate the effect of differently-sized vector spaces and their modelling capabilities. 

For \WordToVec{}, a query/document is represented by its average word vector, which is effective for computing short text similarity \cite{Kenter2015shorttext}. We report both on the Continuous Bag-of-Words (CBOW) and Skip-gram (SG) variants of \WordToVec{}. 

For \LDA{}, we set $\alpha = \beta = 0.1$ and train the model for 100 iterations or until topic convergence is achieved. Default parameters are used in all other cases. Unlike \citet{VanGysel2016experts}, we do not initialize with pre-trained \WordToVec{} embeddings. 

For \LSI{}, \LDA{} and \WordToVec{}, expert representations are created from document representations according to Eq.~\ref{eq:expert-vsm}. 

In addition to text-based representations, we also include two baselines that do not consider textual data. For the first method (\GraphPCA{}), we construct a weighted, undirected co-association graph where the weight between two entities is given by the number of times they are co-associated. We then apply Principal Component Analysis to create a latent representation for every entity. Secondly, we include a baseline where experts are represented as a \Random{} vector sampled from a standard multivariate normal distribution.


\newcommand{\LatentModels}{\Random{}, \GraphPCA{}, \LSI{}, \LDA{}, \WordToVec{}, \DocToVec{} and \SERT{}}

\newcommand{\Significant}{^{*\phantom{**}}}
\newcommand{\MoreSignificant}{^{**\phantom{*}}}
\newcommand{\HighlySignificant}{^{***}}

\section{Regularities in entity vector spaces}

\newcommand{\RQRef}[1]{{RQ#1}}

We investigate regularities within latent text-based entity vector spaces. In particular, we first build latent representations for experts and ground these in the structure of the organizations where these experts are active. First, we cluster latent expert representations using different clustering techniques and compare the resulting clusters to committees in a standards organization of the World Wide Web (\RQRef{1}). We continue by investigating to what extent these representations encode entity relations (\RQRef{2}). We complement the answers to our research questions with an analysis of the prior (the scalar bias in Eq.~\ref{eq:expert-loglinear}) associated with every expert in one of the models we consider, \SERT{}, and compare this to their academic rank.

\subsection{Answers to research questions}
\newcommand{\RQAnswer}[3]{
	\begin{description}
	\item[\fontsize{10pt}{0}\selectfont RQ#1] {\fontsize{10pt}{0}\selectfont \textbf{#2}}%
	\end{description}%
	{ #3 }
}

\smallskip
\RQAnswer{1}{\ResearchQuestionOne{}}{%
\newcommand{\NumClusters}{K}
\begin{figure*}[!t]
	\centering
	\includegraphics[width=\textwidth]{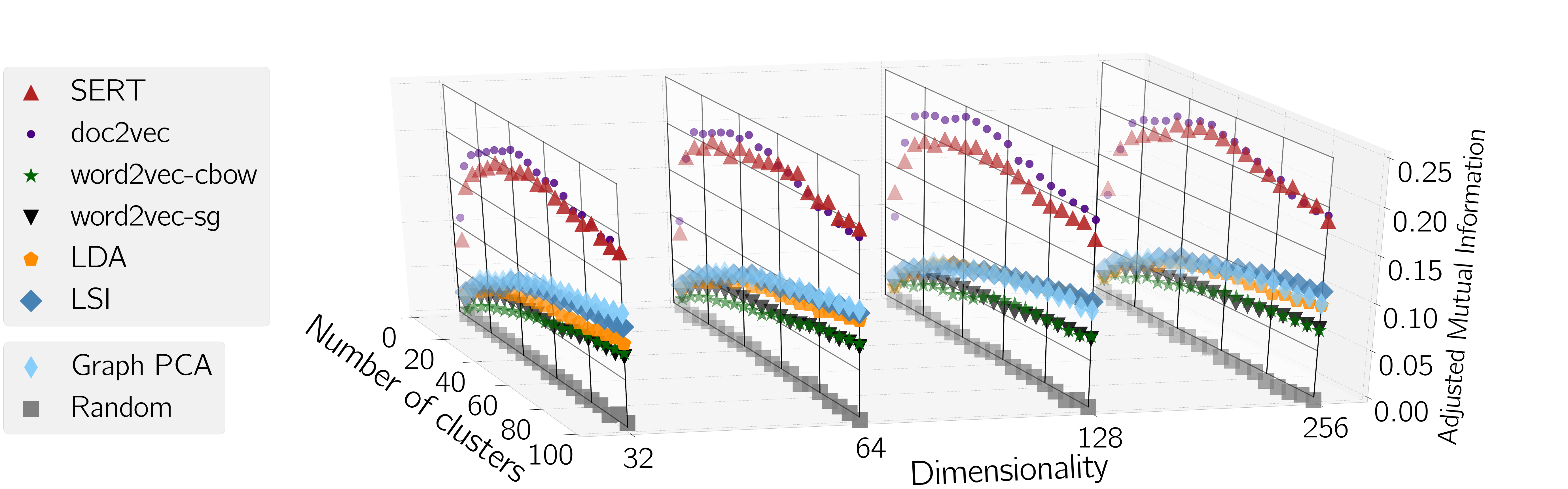}
	
	\smallskip
	\caption{Comparison of clustering capabilities of expert representations (\LatentModels{}) using $\NumClusters{}$-means for $10^0 \leq \NumClusters{} < 10^2$ (y-axis). The x-axis shows the dimensionality of the representations and the z-axis denotes the Adjusted Mutual Information.\label{fig:clustering}}
\end{figure*}
\noindent%
The World Wide Web Consortium (\BenchmarkWThreeC{}) consists of various working groups.\footnote{\url{http://www.w3.org/Consortium/activities}}
Each working group is responsible for a particular aspect of the WWW and consists of two or more experts. We use these working groups as ground truth for evaluating the ability of expert representations to encode similarity. The \BenchmarkWThreeC{} working groups are special committees that are established to produce a particular deliverable \citep[p. 492]{Robert2011rules} and are a way to gather experts from around the organization who share areas of expertise and who would otherwise not directly communicate. Working groups are non-hierarchical in nature and represent clusters of experts. Therefore, they can be used to evaluate to what extent entity representations can be used as feature vectors for clustering.

We cluster expert representations using $\NumClusters{}$-means \cite{MacQueen1967}. While $\NumClusters{}$-means imposes strong assumptions on cluster shapes (convexity and isotropism), it is still very popular today due to its linear time complexity, geometric interpretation and absence of hard to choose hyper-parameters (unlike spectral variants or DBSCAN). We cluster expert representations of increasing dimensionality $\VectorDim{}$ ($\VectorDim{} = 2^i$ for $5 \leq i < 9$) using a linear sweep over the number of clusters $\NumClusters{}$ ($10^0 \leq \NumClusters{} < 10^2$).

During evaluation we transform working group memberships to a hard clustering of experts by assigning every expert to the smallest working group to which they belong as we wish to find specialized clusters contrary to general clusters that contain many experts. We then use Adjusted Mutual Information, an adjusted-for-chance variant of Normalized Information Distance~\cite{Vinh2010}, to compare both clusterings. Adjusting for chance is important as non-adjusted measures (such as BCubed precision/recall\footnote{This can be verified empirically by computing BCubed measures for an increasing number of random partitions.} as presented by \citet{Amigo2009}) have the tendency to take on a higher value for a larger value of $\NumClusters{}$. Performing the adjustment allows us to compare clusterings for different values of $\NumClusters{}$. We repeat the $\NumClusters{}$-means clustering 10 times with different centroids initializations and report the average.

Fig.~\ref{fig:clustering} shows the clustering capabilities of the different representations for different values of $\NumClusters{}$ and vector space dimensionality. Ignoring the random baseline, representations built using \WordToVec{} perform worst. This is most likely due to the fact that document representations for \WordToVec{} are constructed by averaging individual word vectors. Next up, we observe a tie between \LSI{} and \LDA{}. Interestingly enough, the baseline that only considers entity-document associations and does not take into account textual content, \GraphPCA{}, outperforms all representations constructed from document-level vector space models (Eq.~\ref{eq:expert-vsm}). Furthermore, \DocToVec{} and \SERT{} perform best, regardless of vector space dimensionality, and consistently outperform the other representations. If we look at the vector space dimensionality, we see that the best clustering is created using 128-dimensional vector spaces. Considering the number of clusters, we see that \DocToVec{} and \SERT{} peak at about 40 to 60 clusters. This corresponds closely to the number of ground-truth clusters. The remaining representations (\WordToVec{}, \LSI{}, \LDA{}, \GraphPCA{}) only seem to plateau in terms of clustering performance at $\NumClusters{} = 100$, far below the clustering performance of the \DocToVec{} and \SERT{} representation methods. 

To answer our first research question, we conclude that expert representations can be used to discover structure within organizations. However, the quality of the clustering varies greatly and use of more advanced methods (i.e., \DocToVec{} or \SERT{}) is recommended.
}

\vspace*{\baselineskip}
\RQAnswer{2}{\ResearchQuestionTwo{}}{

\begin{table*}[!t]
\caption{Retrieval performance (\NDCG{} and \RPrec{}) when ranking experts for a query expert by the cosine similarity of expert representations (\LatentModels{}) for the \BenchmarkTU{} expert collection (\S\ref{sec:collections}) for an increasing representation dimensionality. The relevance labels are given by the number of times two experts were co-authors of academic papers. Significance of results is determined using a two-tailed paired Student t-test ($^{*} \, p < 0.10$, $^{**} \, p < 0.05$, $^{***} \, p < 0.01$) between the best performing model and second best performing method.
\label{tbl:graph}}
\smallskip

\centering
\resizebox{0.90\textwidth}{!}{%
\def\arraystretch{1.1125}
\begin{tabular}{l c c c c c c c c}%
\toprule%
$\text{Dimensionality } \VectorDim  = $&\multicolumn{2}{c}{32}&\multicolumn{2}{c}{64}&\multicolumn{2}{c}{128}&\multicolumn{2}{c}{256}\\%
&NDCG&R{-}Precision&NDCG&R{-}Precision&NDCG&R{-}Precision&NDCG&R{-}Precision\\%
\midrule%
Random&\nprounddigits{2}\npdecimalsign{.}\numprint{0.177236415094}$\phantom{\HighlySignificant}$&\nprounddigits{2}\npdecimalsign{.}\numprint{0.00996264150943}$\phantom{\HighlySignificant}$&\nprounddigits{2}\npdecimalsign{.}\numprint{0.177525660377}$\phantom{\HighlySignificant}$&\nprounddigits{2}\npdecimalsign{.}\numprint{0.00726377358491}$\phantom{\HighlySignificant}$&\nprounddigits{2}\npdecimalsign{.}\numprint{0.177086037736}$\phantom{\HighlySignificant}$&\nprounddigits{2}\npdecimalsign{.}\numprint{0.00762226415094}$\phantom{\HighlySignificant}$&\nprounddigits{2}\npdecimalsign{.}\numprint{0.179706603774}$\phantom{\HighlySignificant}$&\nprounddigits{2}\npdecimalsign{.}\numprint{0.0103903773585}$\phantom{\HighlySignificant}$\\%
Graph PCA&\nprounddigits{2}\npdecimalsign{.}\numprint{0.380876226415}$\phantom{\HighlySignificant}$&\nprounddigits{2}\npdecimalsign{.}\numprint{0.179528679245}$\phantom{\HighlySignificant}$&\nprounddigits{2}\npdecimalsign{.}\numprint{0.387263018868}$\phantom{\HighlySignificant}$&\nprounddigits{2}\npdecimalsign{.}\numprint{0.196906792453}$\phantom{\HighlySignificant}$&\nprounddigits{2}\npdecimalsign{.}\numprint{0.411703962264}$\phantom{\HighlySignificant}$&\nprounddigits{2}\npdecimalsign{.}\numprint{0.233318113208}$\phantom{\HighlySignificant}$&\nprounddigits{2}\npdecimalsign{.}\numprint{0.386152641509}$\phantom{\HighlySignificant}$&\nprounddigits{2}\npdecimalsign{.}\numprint{0.232906603774}$\phantom{\HighlySignificant}$\\%
\midrule
LSI&\nprounddigits{2}\npdecimalsign{.}\numprint{0.386723584906}$\phantom{\HighlySignificant}$&\nprounddigits{2}\npdecimalsign{.}\numprint{0.167349245283}$\phantom{\HighlySignificant}$&\nprounddigits{2}\npdecimalsign{.}\numprint{0.431202641509}$\phantom{\HighlySignificant}$&\nprounddigits{2}\npdecimalsign{.}\numprint{0.20766490566}$\phantom{\HighlySignificant}$&\nprounddigits{2}\npdecimalsign{.}\numprint{0.460599811321}$\phantom{\HighlySignificant}$&\nprounddigits{2}\npdecimalsign{.}\numprint{0.225984150943}$\phantom{\HighlySignificant}$&\nprounddigits{2}\npdecimalsign{.}\numprint{0.473885849057}$\phantom{\HighlySignificant}$&\nprounddigits{2}\npdecimalsign{.}\numprint{0.227925471698}$\phantom{\HighlySignificant}$\\%
LDA&\nprounddigits{2}\npdecimalsign{.}\numprint{0.437495849057}$\phantom{\HighlySignificant}$&\nprounddigits{2}\npdecimalsign{.}\numprint{0.186046603774}$\phantom{\HighlySignificant}$&\nprounddigits{2}\npdecimalsign{.}\numprint{0.446252075472}$\phantom{\HighlySignificant}$&\nprounddigits{2}\npdecimalsign{.}\numprint{0.199429245283}$\phantom{\HighlySignificant}$&\nprounddigits{2}\npdecimalsign{.}\numprint{0.464408490566}$\phantom{\HighlySignificant}$&\nprounddigits{2}\npdecimalsign{.}\numprint{0.219798867925}$\phantom{\HighlySignificant}$&\nprounddigits{2}\npdecimalsign{.}\numprint{0.523460754717}$\phantom{\HighlySignificant}$&\nprounddigits{2}\npdecimalsign{.}\numprint{0.277198301887}$\phantom{\HighlySignificant}$\\%
word2vec{-}sg&\nprounddigits{2}\npdecimalsign{.}\numprint{0.463946037736}$\phantom{\HighlySignificant}$&\nprounddigits{2}\npdecimalsign{.}\numprint{0.222067735849}$\phantom{\HighlySignificant}$&\nprounddigits{2}\npdecimalsign{.}\numprint{0.486528679245}$\phantom{\HighlySignificant}$&\nprounddigits{2}\npdecimalsign{.}\numprint{0.238368490566}$\phantom{\HighlySignificant}$&\nprounddigits{2}\npdecimalsign{.}\numprint{0.493312264151}$\phantom{\HighlySignificant}$&\nprounddigits{2}\npdecimalsign{.}\numprint{0.237186981132}$\phantom{\HighlySignificant}$&\nprounddigits{2}\npdecimalsign{.}\numprint{0.498984528302}$\phantom{\HighlySignificant}$&\nprounddigits{2}\npdecimalsign{.}\numprint{0.250447924528}$\phantom{\HighlySignificant}$\\%
word2vec{-}cbow&\nprounddigits{2}\npdecimalsign{.}\numprint{0.459394716981}$\phantom{\HighlySignificant}$&\nprounddigits{2}\npdecimalsign{.}\numprint{0.225746603774}$\phantom{\HighlySignificant}$&\nprounddigits{2}\npdecimalsign{.}\numprint{0.472065849057}$\phantom{\HighlySignificant}$&\nprounddigits{2}\npdecimalsign{.}\numprint{0.238029056604}$\phantom{\HighlySignificant}$&\nprounddigits{2}\npdecimalsign{.}\numprint{0.478136037736}$\phantom{\HighlySignificant}$&\nprounddigits{2}\npdecimalsign{.}\numprint{0.24816490566}$\phantom{\HighlySignificant}$&\nprounddigits{2}\npdecimalsign{.}\numprint{0.482222830189}$\phantom{\HighlySignificant}$&\nprounddigits{2}\npdecimalsign{.}\numprint{0.253563018868}$\phantom{\HighlySignificant}$\\%
doc2vec&\nprounddigits{2}\npdecimalsign{.}\numprint{0.346980754717}$\phantom{\HighlySignificant}$&\nprounddigits{2}\npdecimalsign{.}\numprint{0.138771320755}$\phantom{\HighlySignificant}$&\nprounddigits{2}\npdecimalsign{.}\numprint{0.357944716981}$\phantom{\HighlySignificant}$&\nprounddigits{2}\npdecimalsign{.}\numprint{0.15120490566}$\phantom{\HighlySignificant}$&\nprounddigits{2}\npdecimalsign{.}\numprint{0.362356415094}$\phantom{\HighlySignificant}$&\nprounddigits{2}\npdecimalsign{.}\numprint{0.160383773585}$\phantom{\HighlySignificant}$&\nprounddigits{2}\npdecimalsign{.}\numprint{0.352983584906}$\phantom{\HighlySignificant}$&\nprounddigits{2}\npdecimalsign{.}\numprint{0.145858679245}$\phantom{\HighlySignificant}$\\%
SERT&\nprounddigits{2}\npdecimalsign{.}\textbf{\numprint{0.526321698113}}$\HighlySignificant$&\nprounddigits{2}\npdecimalsign{.}\textbf{\numprint{0.288272830189}}$\HighlySignificant$&\nprounddigits{2}\npdecimalsign{.}\textbf{\numprint{0.544672075472}}$\HighlySignificant$&\nprounddigits{2}\npdecimalsign{.}\textbf{\numprint{0.311938679245}}$\HighlySignificant$&\nprounddigits{2}\npdecimalsign{.}\textbf{\numprint{0.534917735849}}$\HighlySignificant$&\nprounddigits{2}\npdecimalsign{.}\textbf{\numprint{0.303620566038}}$\HighlySignificant$&\nprounddigits{2}\npdecimalsign{.}\textbf{\numprint{0.532707169811}}$\phantom{\HighlySignificant}$&\nprounddigits{2}\npdecimalsign{.}\textbf{\numprint{0.306999245283}}$\Significant$\\%
\bottomrule%
\end{tabular}}
\smallskip
\end{table*}

\noindent%
The text-based entity representation problem is characterized by a bipartite graph of entities and documents where an edge denotes an entity-document association. This differs from entity finding settings where explicit entity-entity relations are available and fits into the scenario where representations have to be constructed from unstructured text only. If latent text-based entity representations encode co-associations, then we can use this insight for
\begin{inparaenum}[(1)]
	\item a better understanding of text-based entity representation models, and
	\item the usability of latent text-based entity representations as feature vectors in scenarios where relations between entities are important.
\end{inparaenum}

We evaluate the capacity of text-based expert representations to encode co-associations by casting the problem as a ranking task. Contrary to typical expert finding, where we rank experts according to their relevance to a textual query, for the purpose of answering \RQRef{2}, we rank experts according to their cosine similarity w.r.t.\ a query expert \citep{Balog2007similarexperts}. This task shares similarity with content-based recommendation based on unstructured data \citep{Pazzani2007recsys}.

In expert finding collections, document-expert associations can indicate many things. For example, in the \BenchmarkWThreeC{} collection, entity-document associations are mined from expert mentions \citep{Balog2006experts}. However, for the \BenchmarkTU{} collection, we know that a subset of associations corresponds to academic paper authorship. Therefore, we construct ranking ground-truth from paper co-authorship and take the relevance label of an expert to be the number of times the expert was a co-author with the query expert (excluding the query expert themselves). Our intention is to determine to what extent latent entity representations estimated from text can reconstruct the original co-author graph. Given that we estimate the latent entity representations using the complete \BenchmarkTU{} document collection, by design, our evaluation is contained within our training set for the purpose of this analysis.

Table~\ref{tbl:graph} shows \NDCG{} and \RPrec{} \cite[p.~158]{Manning2008book} for various representation models and dimensionality. \SERT{} performs significantly better than the other representations methods (except for the $256$-dimensional representations where significance was not achieved w.r.t.\ \LDA{}). \SERT{} is closely followed by \WordToVec{} (of which both variants score only slightly worse than \SERT{}), \LDA{} and \LSI{}. The count-based distributional methods (\LSI{}, \LDA{}) perform better as the dimensionality of the representations increases. This is contrary to \SERT{}, where retrieval performance is very stable across dimensionalities. Interestingly, \DocToVec{} performs very poorly at reconstructing the co-author graph and is even surpassed by the \GraphPCA{} baseline. This is likely due to the fact that \DocToVec{} is trained on expert profiles and is not explicitly presented with document-expert associations. The difference in performance between \DocToVec{} and \SERT{} for \RQRef{2} reflects a difference in architecture: while \SERT{} is directly optimized to discriminate between entities, \DocToVec{} models entities as context in addition to language. Hence, similarities and dissimilarities between entities are preserved much better by \SERT{}.

We answer our second research question as follows. Latent text-based entity representations do encode information about entity relations. However, there is a large difference in the performance of different methods. \SERT{} seems to encode the entity co-associations better than other methods, by achieving the highest performance independent of the vector space dimensionality.
}

\subsection{Analysis of the expert prior in the log-linear model}

\begin{figure*}[!th]
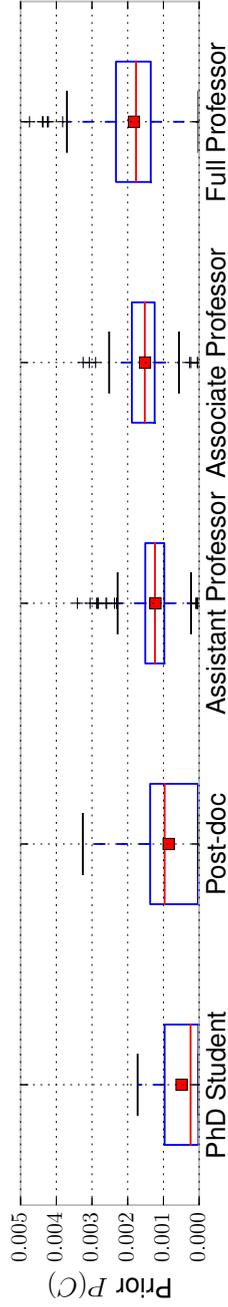

	\centering
	\myincludegraphics[width=0.90\textwidth]{resources/bias/sert_32.pdf}

	\smallskip
	\caption{Box plots of prior probabilities learned by \SERT{}, grouped by the experts' academic rank, for the \BenchmarkTU{} collection. We only show the prior learned for a \SERT{} model with $\VectorDim{} = 32$, as the distributions of models with a different representation dimensionality are qualitatively similar.\label{fig:bias}}
\end{figure*}

\noindent%
One of the semantic models that we consider, \SERT{}, learns a prior $\Prob{\Candidates{}}$ over entities. The remaining representation learning methods do not encode an explicit entity prior. It might be possible to extract a prior from generic entity vector spaces, e.g., by examining the deviation from the mean representation for every entity. However, developing such prior extraction methods are a topic of study by themselves and are out of scope for this \paper{}.

In the case of expert finding, this prior probability encodes a belief over experts without observing any evidence (i.e., query terms in \SERT{}). Which structural information does this prior capture? We now investigate the regularities encoded within this prior and link it back to the hierarchy among scholars in the Tilburg University collection. We estimate a \SERT{} model on the whole \BenchmarkTU{} collection and extract the prior probabilities:
\begin{equation}
\Prob{\Candidates{} = \Candidate{}_i} = \frac{\exp{(\CandidateBias{}_i)}}{\sum_l \exp{(\CandidateBias{}_l)}},
\end{equation}
where $\CandidateBias{}$ is the bias vector of the \SERT{} model in Eq.~\ref{eq:expert-loglinear}.

For \numprint{666} out of \numprint{977} experts in the \BenchmarkTU{} collection we have ground truth information regarding their academic rank \cite{Berendsen2013profileassessment}.\footnote{126 PhD Students, 49 Postdoctoral Researchers, 210 Assistant Professors, 89 Associate Professors and 190 Full Professors; we filtered out academic ranks that only occur once in the ground-truth, namely Scientific Programmer and Research Coordinator.} Fig.~\ref{fig:bias} shows box plots of the prior probabilities, learned automatically by the \SERT{} model from only text and associations, grouped by academic rank. Interestingly, the prior seems to encode the hierarchy amongst scholars at Tilburg University, e.g., Post-docs are ranked higher than PhD students. This is not surprising as it is quite likely that higher-ranked scholars have more associated documents.

The prior over experts in \SERT{} encodes rank within organizations. This is not surprising, as experts (i.e., academics in this experiment) of higher rank tend to occur more frequently in the expert collection. This observation unveils interesting insights about the expert finding task and consequently models targeted at solving it. Unlike unsupervised ad-hoc document retrieval where we assume a uniform prior and normalized document lengths, the prior over experts in the expert finding task is of much greater importance. In addition, we can use this insight to gain a better understanding of the formal language models for expertise retrieval \citep{Balog2006experts}. \citet{Balog2006experts} find that, for the expert finding task, the document-oriented language model performs better than an entity-oriented language model. However, the document-oriented model \citep{Balog2006experts} will rank experts with more associated documents higher than experts with few associated documents. On the contrary, the entity-oriented model of \citet{Balog2006experts}, imposes a uniform prior over experts. \SERT{} is an entity-oriented model and performs better than the formal document-oriented language model \citep{VanGysel2016experts}. This is likely due to the fact that \SERT{} learns an empirical prior over entities instead of making an assumption of uniformity, in addition to its entity-oriented perspective.

In the case of general entity finding, the importance of the number of associated documents might be of lesser importance. Other sources of prior information, such as link analysis \citep{Page1999}, recency \citep{Graus2016dynamic} and user interactions \citep{Schuth2016forum}, can be a better way of modelling entity importance than the length of entity descriptions.


\section{Summary}

In this \paper{} we have investigated the structural regularities contained within latent text-based entity representations. Entity representations were constructed from expert finding collections using methods from distributional semantics (\LSI{}), topic models (\LDA{}) and neural networks (\WordToVec{}, \DocToVec{} and \SERT{}). For \LSI{}, \LDA{} and \WordToVec{}, document-level representations were transformed to the entity scope according to the framework of \citet{Demartini2009expertspaces}. In the case of \DocToVec{} and \SERT{}, entity representations were learned directly. In addition to representations estimated only from text, we considered non-textual baselines, such as:
\begin{inparaenum}[(1)]
	\item random representations sampled from a Normal distribution, and
	\item the rows of the dimensionality-reduced adjacency matrix of the co-association graph.
\end{inparaenum}

We have found that text-based entity representations can be used to discover groups inherent to an organization. We have clustered entity representations using $K$-means and compared the obtained clusters with a ground-truth partitioning. No information about the organization is presented to the algorithms. Instead, these regularities are extracted by the documents associated with entities and published within the organization. Furthermore, we have evaluated the capacity of text-based expert representations to encode co-associations by casting the problem as a ranking task. We discover that text-based representations retain co-associations up to different extents. In particular, we find that \SERT{} entity representations encode the co-association graph better than the other representation learning methods. We conclude that this is due to the fact that \SERT{} representations are directly optimized to discriminate between entities. Lastly, we have shown that the prior probabilities learned by semantic models encode further structural information. That is, we find that the prior probability over experts (i.e., members of an academic institution), learned as part of a \SERT{} model, encodes academic rank. In addition, we discuss the similarities between \SERT{} and the document-oriented language model \citep{Balog2006experts} and find that the document association prior plays an important role in expert finding.

Our findings have shown insight into how different text-based entity representation methods behave in various applications. In particular, we find that the manner in which entity-document associations are encoded plays an important role. That is, representation learning methods that directly optimize the representation of the entity seem to perform best. When considering different neural representation learning models (\DocToVec{} and \SERT{}), we find that their difference in architecture allows them to encode different regularities. \DocToVec{} models an entity as context in addition to language, whereas \SERT{} learns to discriminate between entities given their language. Thus, \DocToVec{} can more adequately model the topical nature of entities, while \SERT{} more closely captures the similarities and dissimilarities between entities. In the case of expert finding, we find that the amount of textual data associated with an expert is a principal measure of expert importance.

While the focus of this \paper{} was the analysis of the structural regularities contained within latent entity representations, the focus of the following \paper{}s will be once again the modelling of latent vector spaces. In Chapter~\globalref{chapter:research-05}, we move from expert finding to a different entity retrieval problem: product search. Compared to expert finding, the product search scenario is characterized by a larger amount of retrievable entities. In addition, the amount of textual content per entity is much less than in the expert finding case.

}


\chapter{Learning Latent Vector Spaces for Product Search}
\label{chapter:research-05}

{
\ScopeLabels{research-05}
\ChapterRQ{3}

\def \paperImplementationUrl {\url{https://github.com/cvangysel/SERT}}

\def \ResearchQuestionOne {How do the parameters of \ModelName{} influence its efficacy?}
\def \ResearchQuestionTwo {How does \ModelName{} compare to latent vector models based on \LDA{}, \LSI{} and \WordToVec{}?}
\def \ResearchQuestionThree {How does \ModelName{} compare to a smoothed language model that applies lexical term matching?}
\def \ResearchQuestionFour {What is the benefit of incorporating \ModelName{} as a feature in a learning-to-rank setting?}

\def \ContributionOne {A latent vector model, \ModelName{}, that jointly learns the representations of words, entities and the relationship between the former, together with an open-source implementation.\footnote{\paperImplementationUrl}}
\def \ContributionTwo {A study of the influence of \ModelName{}'s parameters and how these influence its ability to discriminate between entities.}
\def \ContributionThree {An in-depth comparative analysis of the entity retrieval effectiveness of latent vector models.}
\def \ContributionFour {Insights in how \ModelName{} can improve retrieval performance in entity-oriented search engines.}
\def \ContributionFive {An analysis of the differences in performance between latent vector models by examining entity representations and mappings from queries to entity space.}

\def \FullModelName{Latent Semantic Entities}
\newcommand{\ModelName}{\LSE}

\newcommand{\QueryLikelihoodLM}{Query-likelihood Language Model}
\newcommand{\WordToVec}{\FullWordToVec}
\newcommand{\RankSVM}{Rank\ac{SVM}}

\newcommand{\assoc}{\text{candidates}(d)}

\newcolumntype{L}[1]{>{\raggedright\let\newline\\\arraybackslash\hspace{0pt}}m{#1}}
\newcolumntype{C}[1]{>{\centering\let\newline\\\arraybackslash\hspace{0pt}}m{#1}}
\newcolumntype{R}[1]{>{\raggedleft\let\newline\\\arraybackslash\hspace{0pt}}m{#1}}

\renewcommand{\log}{\text{log}}
\renewcommand{\exp}{\text{exp}}

\newcommand{\subfloat}[2][]{%
\begin{subfigure}[b]{0.185\paperheight}%
\centering%
{\setlength{\textwidth}{0.85\paperheight}#2}%
\caption{#1}%
\end{subfigure}%
}

\newcommand{\heatmap}[4]{%
	\def \inner {%
		\def \PlotPath {resources/heat_map/#1_#2_#3.pdf}%
		\IfFileExists{\PlotPath}{
			\includegraphics[width=0.495\paperwidth]{\PlotPath}}{
			\resizebox{0.225\columnwidth}{!}{\missingfigure{#3}}}%
	}%
	\ifstrequal{#4}{}{%
	\subfloat{\inner}%
	}{\begin{subfigure}[b]{0.45\textwidth}%
	\centering%
	\inner{}
	\caption{#4\label{fig:heatmap:#2:#1}}
	\end{subfigure}}
}

\newcommand{\vectorspace}[4]{
	\begin{subfigure}[b]{0.45\textwidth}%
	\def \PlotPath {resources/vector_space/#1_#2_#3.pdf}%
	\IfFileExists{\PlotPath}{
		\includegraphics[width=0.495\paperwidth]{\PlotPath}}{
		\resizebox{0.230\textwidth}{!}{\missingfigure{#4}}}
	\caption{#4\label{fig:vectorspace:#2:#1}}
	\end{subfigure}
}

\newcommand{\innerdeltaplot}[5]{
	\def \PlotPath {resources/delta/#1_differences.pdf}%
	\IfFileExists{\PlotPath}{
		\includegraphics[width=0.495\paperwidth]{\PlotPath}}{
		\resizebox{0.225\columnwidth}{!}{\missingfigure{#5}}}
}

\newcommand{\deltaplot}[5]{
	\begin{subfigure}[b]{0.45\textwidth}%
		\centering%
		\innerdeltaplot{#1}{#2}{#3}{#4}{ndcg}%
		\caption{#5\label{fig:delta:#2:#1}}%
	\end{subfigure}%
}

\newcommand{\centroids}[4]{%
	\def \inner {%
		\def \PlotPath {resources/centroids/#1_#2_#3.pdf}%
		\IfFileExists{\PlotPath}{
			\includegraphics[width=0.495\paperwidth]{\PlotPath}}{
			\resizebox{0.225\textwidth}{!}{\missingfigure{#4}}}
	}%
	\ifstrequal{#4}{}{%
	\begin{subfigure}[b]{0.45\textwidth}\inner\end{subfigure}%
	}{\begin{subfigure}[b]{0.45\textwidth}\inner\caption{#4\label{fig:centroid:#2:#1}}\end{subfigure}}%
}

\renewcommand{\LossFn}[1][]{\LossFnIdentifier{}\LossFnArguments[#1]{}}

\renewcommand{\newcommand}{\providecommand}

{
\renewcommand{\vfill}{}
\renewcommand{\pagebreak}{}

\section{Introduction}

Retail through online channels has become an integral part of consumers' lives \citep{PricewaterhouseCoopers}. In addition to using these online platforms that generate hundreds of billions of dollars in revenue \citep{Forrester}, consumers increasingly participate in multichannel shopping where they research items online before purchasing them in brick-and-mortar stores. Search engines are essential for consumers to be able to make sense of these large collections of products available online \citep{Jansen2006}. In the case of directed searching (in contrast to exploratory browsing), users formulate queries using characteristics of the product they are interested in (e.g., terms that describe the product's category) \citep{Rowley2000}. However, it is widely known that there exists a mismatch between queries and product representations where both use different terms to describe the same concepts \citep{Li2014}. Thus, there is an urgent need for better semantic matching methods.

Product search is a particular example of the more general entity finding task that is increasingly being studied. Other entity finding tasks considered recently include searching for people~\citep{Balog2012survey}, books~\citep{Gade2015} and groups~\citep{Liang2016}. Products are retrievable entities where every product is associated with a description and one or more user reviews. Therefore, we use the terms ``product'' and ``entity'' interchangeably in this \paper{}. However, there are two important differences between product search and the entity finding task as defined by \citet{DeVries2007inex}. First, in entity finding one retrieves entities of a particular type from large broad coverage multi-domain knowledge bases such as Wikipedia \citep{DeVries2007inex,Balog2012dbpedia}. In contrast, product search engines operate within a single domain which can greatly vary in size. Second, user queries in product search consist of free-form text \citep{Rowley2000}, as opposed to the semi-structured queries with additional type or relational constraints being used in entity finding~\citep{DeVries2007inex,Balog2011entitytrack}.

In this \paper{} we tackle the problem of discriminating between products based on the language (i.e., descriptions and reviews) they are associated with. Existing methods that are aimed at discriminating between entities based on textual data learn word representations using a language modelling objective or heuristically construct entity representations \citep{VanGysel2016experts,Demartini2009expertspaces}. Our approach directly learns two things: a unidirectional mapping between words and entities, as well as distributed representations of both words and entities. It does so in an unsupervised and automatic manner such that words that are strongly evidential for particular products are projected nearby those products.
While engineering of representations is important in information retrieval \citep{Balog2007expertprofiles,Demartini2009expertspaces,Bordes2011,Cai2015,Zhao2015,Graus2016dynamic}, unsupervised joint representation learning of words and entities has not received much attention.
We fill this gap.
Our focus on learning representations for an end-to-end task such as \emph{product search} is in contrast to the large volume of recent literature on word representation learning \citep{Turian2010} that has a strong focus on upstream components such as distributional semantics \citep{Mikolov2013word2vec,Pennington2014}, parsing \citep{Turian2010,Collobert2011scratch} and information extraction \citep{Turian2010,Collobert2011scratch}.
In addition, our focus on \emph{unsupervised} representation learning is in contrast to recent entity representation learning methods \citep{Bordes2011,Zhao2015} that heavily depend on precomputed entity relationships and cannot be applied in their absence.

In recent years, significant progress has been made concerning semantic representations of entities. We point out three key insights on which we build:
\begin{inparaenum}[(1)]
	\item Distributed representations \citep{Hinton1986} learned by discriminative neural networks reduce the curse of dimensionality and improve generalization. Latent features encapsulated by the model are shared by different concepts and, consequently, knowledge about one concept influences knowledge about others.
	\item Discriminative approaches outperform generative models if enough training data is available \citep{Ng2002,Baroni2014} as discriminative models solve the classification problem directly instead of solving a more general problem first \citep{Vapnik1998}.
	\item The unsupervised neural retrieval model we introduced in Chapter~\globalref{chapter:research-03} \citep{VanGysel2016experts} does not scale as they model a distribution over all retrievable entities; the approach is infeasible during training if the collection of retrievable entities is large.
\end{inparaenum}%

Building on these insights, we introduce \FullModelName{} (\ModelName{}), a method that learns separate representations of words and retrievable objects jointly for the case where mostly unstructured documents are associated with the objects (i.e., descriptions and user reviews for products) and without relying on predefined relationships between objects (e.g., knowledge graphs). \ModelName{} learns to discriminate between entities for a given word sequence by mapping the sequence into the entity representation space. Contrary to heuristically constructed entity representations \citep{Demartini2009expertspaces}, \ModelName{} learns the relationship between words and entities directly using gradient descent. Unlike the model we proposed in Chapter~\globalref{chapter:research-03} \citep{VanGysel2016experts}, we avoid computing the full probability distribution over entities; we do so by using noise-contrastive estimation.

\medskip
\noindent
The research questions we ask in this \paper{} towards answering \MainRQRef{3} are as follows:
\RQ{1}{\ResearchQuestionOne{}}
\RQ{2}{\ResearchQuestionTwo{}}
\RQ{3}{\ResearchQuestionThree{}}
\RQ{4}{\ResearchQuestionFour{}}

\ifthesis
\else
We contribute:
\begin{inparaenum}[(1)]
	\item \ContributionOne{}
	\item \ContributionTwo{}
	\item \ContributionThree{}
	\item \ContributionFour{}
	\item \ContributionFive{}
\end{inparaenum}
\fi

}

\section{Related work}
\label{section:related_work}

We refer to Section~\globalref{chapter:background:latent} of our background chapter (Chapter~\globalref{chapter:background}). On the topic of product search (\S\globalref{chapter:background:related:products}), \citet{Duan2015} study the problem of learning query intent representation for structured product entities. They emphasize that existing methods focus only on the query space and overlook critical information from the entity space and the connection in between. We agree that modelling the connection between query words and entities and propagating information from the entity representations back to words is essential. In contrast to their work, we consider the problem of learning representations for entities based on their associations with unstructured documents.

Learning the representations of entities is not new, and the topic is covered in Section~\globalref{chapter:background:related:entityreprs}. In contrast to existing methods for entity representation learning, we model representations of words and entities jointly in separate spaces, in addition to a mapping from word to entity representations, in an unsupervised manner.

In addition, latent semantic retrieval models (\S\globalref{chapter:background:related:latent}) are also relevant to this \paper{}. With Chapter~\globalref{chapter:research-03} we saw the introduction of an \acs{LSM} for entity retrieval, with an emphasis on expert finding; we noted that training the parameters of the log-linear model becomes infeasible when the number of entities increases. In this \paper{} we mitigate this problem by considering only a random sample of entities as negative examples during training. This allows us to efficiently estimate model parameters in large product retrieval collections, which is not possible using the approach we introduced in Chapter~\globalref{chapter:research-03} due to its requirement to compute a normalization constant over all entities.

In this \paper{}, we tackle the task of learning latent continuous vector representations for e-commerce products for the purpose of product search. The focus of this \paper{} lies in the language modelling and representation learning challenge. We learn distributed representations \citep{Hinton1986} of words and entities and a mapping between the two. At retrieval time, we rank entities according to the similarity of their latent representations to the projected representation of a query. Our model \ModelName{} is compared against existing entity-oriented latent vector representations that have been created using \LSI{}, \LDA{} and \WordToVec{}. We provide an analysis of model parameters and give insight in the quality of the joint representation space.

\renewcommand{\\}{ }

\vfill

\section{Latent vector spaces for\\entity retrieval}
\label{section:methodology}

\newcommand{\Prob}[1]{P(#1)}
\newcommand{\CondProb}[2]{\Prob{#1 \mid #2}}

\newcommand{\UnnormedProb}[1]{\tilde{P}(#1)}
\newcommand{\UnnormedCondProb}[2]{\UnnormedProb{#1 \mid #2}}

\newcommand{\SigmoidFn}[1]{\sigma{}(#1)}
\newcommand{\ExpFn}[1]{e^{#1}}

\newcommand{\DotProduct}[2]{#1 \cdot #2}
\newcommand{\Transpose}[1]{#1^\intercal}

\newcommand{\Length}[1]{|#1|}

\newcommand{\Entities}{X}
\newcommand{\Entity}{\MakeLowercase{\Entities{}}}

\newcommand{\NumEntities}{\Length{\Entities{}}}

\newcommand{\Vocabulary}{V}
\newcommand{\WordEmbedding}{\MakeLowercase{\Vocabulary}}

\newcommand{\VocabularySize}{\Length{\Vocabulary{}}}

\newcommand{\EntitySpace}{E}
\newcommand{\EntityEmbedding}{\MakeLowercase{\EntitySpace}}

\newcommand{\String}{s}
\newcommand{\Word}{w}

\newcommand{\SimilarityFn}{S_c}

\newcommand{\HashFn}{f}

\newcommand{\WordEmbeddingSize}{e_\Vocabulary{}}
\newcommand{\EntityEmbeddingSize}{e_\EntitySpace{}}

\newcommand{\VocabularyEmbeddingMatrix}{W_\WordEmbedding{}}
\newcommand{\MappingMatrix}{W}
\newcommand{\MappingBias}{b}
\newcommand{\EntityEmbeddingMatrix}{W_\EntityEmbedding{}}

\newcommand{\CategoricalTopic}{c}
\newcommand{\Query}{q}
\newcommand{\Term}{t}

\newcommand{\Beacon}{\tilde{\EntityEmbedding{}}}

\newcommand{\Rank}{\text{rank}}

\newcommand{\Documents}{D}
\newcommand{\Document}{\MakeLowercase{\Documents{}}}
\newcommand{\DocumentEmbedding}[1][]{\HashFn{}({\Document{}#1})}

\newcommand{\AssociatedDocuments}[1]{\Documents{}_{#1}}

\newcommand{\WindowSize}{n}

\newcommand{\NumNegativeExamples}{z}

\newcommand{\BatchSize}{m}

\newcommand{\LossFnIdentifier}{L}
\newcommand{\LossFnArguments}[1][]{(\VocabularyEmbeddingMatrix{}#1, \EntityEmbeddingMatrix{}#1, \MappingMatrix{}#1, \MappingBias{}#1)}
\newcommand{\LossFn}[1][]{\LossFnIdentifier{}\LossFnArguments[#1]{}}

\newcommand{\LearningRate}{\boldsymbol{\alpha}}

\tdplotsetmaincoords{0}{0}
\newcommand{\TwoDAxes}[1][]{
\begin{tikzpicture}
    [scale=3,
     tdplot_main_coords,
     axis/.style={->,black,thick}]

\coordinate (O) at (0,0);

\draw (0,0,0) -- (1,0,0);
\draw (0,0,0) -- (0,1,0);
\draw (1,0,0) -- (1,1,0);
\draw (0,1,0) -- (1,1,0);
#1
\end{tikzpicture}
}

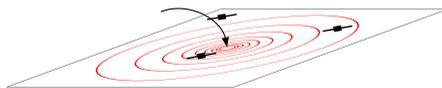
\begin{figure}
	\centering
	\begin{tikzpicture}[>=latex]%
		\node[cm={1,0,cos(20),sin(20),(0,0)}] (MetricSpace) {
			\TwoDAxes[
				\foreach \i in {1,...,9} {
					\draw[red] (0.5,0.5,0) circle (0.5*0.05cm^\i);
				}
				\foreach \i in {(0.48, 0.40), (0.75,0.75), (0.1,0.9)} {
					\draw \i node[cross=5pt,rotate=0,line width=0.75mm] {};
				}
			]{}
		};
		\draw[->] (-0.877582562,0.4794255386) to [bend left=50] (MetricSpace.center);
	\end{tikzpicture}

        \medskip
	\caption{Illustrative example of how entities are ranked in vector space models w.r.t. a projected query. Query $\Query{}$ is projected into entity space $\EntitySpace{}$ using mapping $\HashFn{}$ (black arrow) and entities (black crosses) are ranked according to their similarity in decreasing order.\label{fig:inference}}
\end{figure}

\begin{figure}
	\newcommand{\PrettyVector}[1]{
		\left[
	    \begin{matrix}
	     	#1
	    \end{matrix}
		\right]
	}
	\newcommand{\PrettyOneHotVector}{\PrettyVector{\\ 0 \\ \vdots{} \\ 1 \\ \vdots{} \\ 0 \\ \\}}
	\newcommand{\PrettyFilledVector}[1]{\PrettyVector{\\ \\ \\ #1 \\ \\ \\ \\}}
	\newcommand{\Sized}[2]{#1 \left\{ #2 \right.}
	\resizebox{\textwidth}{!}{
	\begin{tikzpicture}[>=latex]%
		\node[draw=white,line width=0] (Vocabulary) {$\Word \in \Vocabulary{}$};
		\node[draw=white,line width=0] (VocabularyEmbedding) [right=2cm of Vocabulary] {$\Sized{\WordEmbeddingSize{}}{\PrettyFilledVector{\WordEmbedding{}}}$};
		\node[draw=white,line width=0] (WordBeacon) [right=2.0cm of VocabularyEmbedding] {$\Sized{\EntityEmbeddingSize{}}{\PrettyFilledVector{\Beacon{}}}$};

		\node[cm={1,0,cos(20),sin(20),(0,0)}] (MetricSpace) [right=1.5cm of WordBeacon] {\TwoDAxes{}};

		\node[draw=white,line width=0] (EntityEmbedding) [right=1.5cm of MetricSpace] {$\Sized{\EntityEmbeddingSize{}}{\PrettyFilledVector{\EntityEmbedding{}}}$};
		\node[draw=white,line width=0] (Entity) [right=2cm of EntityEmbedding] {$\Entity \in \EntitySpace{}$};

		\draw[->] (Vocabulary) -> (VocabularyEmbedding) node[above,midway,text width=2cm,align=center] {Look-up embedding in\\$\VocabularyEmbeddingMatrix{}$};
		\draw[->] (VocabularyEmbedding) -> (WordBeacon) node[above,midway,text width=2cm, align=center] {Transform with\\\scriptsize{$\tanh{({\MappingMatrix{} \cdot \WordEmbedding{} + \MappingBias{})}}$}};
		\draw[->] (Entity) -> (EntityEmbedding) node[above,midway,text width=2cm,align=center] {Look-up embedding in\\$\EntityEmbeddingMatrix{}$};

		\draw[->] (EntityEmbedding) to [bend right=50] ($(MetricSpace.center)!0.50!(MetricSpace.north east)$);
		\draw[->] (WordBeacon) to [bend left=50] ($(MetricSpace.center)!0.50!(MetricSpace.south west)$);

		\begin{scope}[transparency group, opacity=0.75]
		\draw[<->,cyan] ($(MetricSpace.center)!0.50!(MetricSpace.south west)$) -- node[below=0,midway,black] {$\SimilarityFn{}(\Beacon{}, \EntityEmbedding{})$} ($(MetricSpace.center)!0.50!(MetricSpace.north east)$);
		\end{scope}

		\draw[decorate,decoration={brace,amplitude=10pt}]
([yshift=1.5cm]Vocabulary.west) -- ([yshift=1.5cm]WordBeacon.east) node [black,midway,yshift=0.75cm] {$\HashFn{}(\Word{})$};
	\end{tikzpicture}}

	\caption{Schematic representation of the \FullModelName{} model for a single word $\Word{}$. Word embeddings $\VocabularyEmbeddingMatrix{}$ ($\WordEmbeddingSize{}$-dim. for $\VocabularySize{}$ words), entity embeddings $\EntityEmbeddingMatrix{}$ ($\EntityEmbeddingSize{}$-dim. for $\NumEntities{}$ entities) and the mapping from words to entities ($\EntityEmbeddingSize{}$-by-$\WordEmbeddingSize{}$ matrix $\MappingMatrix{}$, $\EntityEmbeddingSize{}$-dim. vector $\MappingBias{}$) are learned using gradient descent.\label{fig:schema}}
\end{figure}

We first introduce a generalized formalism and notation for entity-oriented latent vector space models. After that, in \S\ref{section:methodology:model}, we introduce \FullModelName{}, a latent vector space model that jointly learns representations of words, entities and a mapping between the two directly, based on the idea that entities are characterized by the words they are associated with and vice versa. Product representations are constructed based on the n-grams the products are likely to generate based on their description and reviews, while word representations are based on the entities they are associated with and the context they appear in. We model the relation between word and product representations explicitly so that we can predict the product representation for a previously unseen word sequence.

\subsection{Background}
\label{section:methodology:background}

\newcommand{\ProjectedQuery}[1][]{\HashFn{}(\Query{}#1)}

We focus on a product retrieval setting in which a user wants to retrieve the most relevant products on an e-commerce platform. As in typical information retrieval scenarios, the user encodes their information need as a query $\Query{}$ and submits it to a search engine. Product search queries describe characteristics of the product the user is searching for, such as a set of terms that describe the product's category \citep{Rowley2000}.

Below, $\Entities{}$ denotes the set of entities that we consider. For every $\Entity{}_i \in \Entities{}$ we assume to have a set of associated documents $\AssociatedDocuments{\Entity{}_i}$. The exact relation between the entity and its documents depends on the problem setting. 
In this \paper{}, entities are products \citep{Nurmi2008,Duan2015} and documents associated with these products are descriptions and product reviews.

Latent vector space models rely on a function $\HashFn{}: \Vocabulary{}^+ \to \EntitySpace{}$ that maps a sequence of words (e.g., a query $\Query{}$ during retrieval) from a vocabulary $\Vocabulary{}$ to an $\EntityEmbeddingSize{}$-dimensional continuous entity vector space $\EntitySpace{} \subset \mathbb{R}^{\EntityEmbeddingSize{}}$. Every entity $\Entity{}_i \in \Entities{}$ has a corresponding vector representation $\EntityEmbedding{}_i \in \EntitySpace{}$. Let $\SimilarityFn{}: \EntitySpace{} \times \EntitySpace{} \to \mathbb{R}$ denote the cosine similarity between vectors in $\EntitySpace{}$. For a given query $\Query{}$, entities $\Entity{}_i$ are ranked in decreasing order of the cosine similarity between $\EntityEmbedding{}_i$ and the query projected into the space of entities, $\ProjectedQuery{}$. Fig.~\ref{fig:inference} illustrates how entities are ranked according to a projected query. For \LSI{}, $\HashFn{}$ is defined as the multiplication of the term-frequency vector representation of $\Query{}$ with the rank-reduced term-concept matrix and the inverse of the rank-reduced singular value matrix \citep{Deerwester1990lsi}. In the case of \LDA{}, $\HashFn{}$ becomes the distribution over topics conditioned on $\Query{}$ \citep{Blei2003}. This distribution is computed as the sum of the topic distributions conditioned on the individual words of $\Query{}$. In this \paper{}, the embedding $\HashFn{}$ is learned; see \S\ref{section:methodology:estimation} below.

Traditional vector space models operate on documents instead of entities. \citet{Demartini2009expertspaces} extend document-oriented vector spaces to entities by representing an entity as a weighted sum of the representations of their associated documents:
\begin{equation}
\label{eq:demartini}
\EntityEmbedding{}_i = \sum_{\Document{}_j \in \AssociatedDocuments{\Entity{}_i}} r_{i,j} \DocumentEmbedding[_j]{}
\end{equation}
where $\DocumentEmbedding[_j]{}$ is the vector representation of $\Document{}_j$ and $r_{i,j}$ denotes the relationship weight between document $\Document{}_j$ and entity $\Entity{}_i$. In this \paper{} we put $r_{i,j} = 1$ whenever $\Document{}_j \in \AssociatedDocuments{\Entity{}_i}$ for a particular $\Entity{}_i \in \Entities{}$ and $r_{i,j}=0$ otherwise, as determining the relationship weight between entities and documents is a task in itself.

\subsection{Latent semantic entities}
\label{section:methodology:model}

While Eq.~\ref{eq:demartini} adapts document-oriented vector space models to entities, in this \paper{} we define $\HashFn{}$ by explicitly learning (\S\ref{section:methodology:estimation}) the mapping between word and entity representations and the representations themselves:
\newcommand{\OneHotVector}{\delta{}_i}
\newcommand{\NormalizedBoWVector}{\frac{1}{\Length{\String{}}} \sum_{\Word{}_i \in \String{}} \OneHotVector{}}
\newcommand{\AvgWordEmbedding}{\VocabularyEmbeddingMatrix{} \cdot \NormalizedBoWVector{}}
\newcommand{\LinearCombination}{\MappingMatrix{} \cdot (\AvgWordEmbedding{}) + \MappingBias{}}
\begin{equation}
\label{eq:hash_fn}
\HashFn{}(\String{}) = \tanh\left(\LinearCombination{}\right)
\end{equation}
for a string $\String{}$ of constituent words $\Word{}_1, \ldots, \Word{}_{|\String{}|}$ (an n-gram extracted from a document or a user-issued query), where $\VocabularyEmbeddingMatrix{}$ is the $\WordEmbeddingSize{} \times \VocabularySize{}$ projection matrix that maps the averaged one-hot representations (i.e., a $|V|$-dimensional vector with element $i$ turned on and zero elsewhere) of word $\Word{}_i$, $\OneHotVector{}$, to its $\WordEmbeddingSize{}$-dimensional distributed representation. This is equivalent to taking the embeddings of the words in $\String{}$ and averaging them. In addition, $\MappingBias{}$ is a $\EntityEmbeddingSize{}$-dimensional bias vector, $\MappingMatrix{}$ is the $\EntityEmbeddingSize{} \times \WordEmbeddingSize{}$ matrix that maps averaged word embeddings to their corresponding position in entity space $\EntitySpace{}$ and $\tanh$ is the element-wise smooth hyperbolic tangent with range $(-1, 1)$. This transformation allows word embeddings and entity embeddings to be of a different dimensionality.

In other words, for a given string of words we take the representation of this string to be the average of the representations of the words it contains \citep{Mikolov2013word2vec,Le2014}. This averaged word representation is then transformed using a linear map ($\MappingMatrix{}$) and afterwards translated using $\MappingBias{}$. We then apply the hyperbolic tangent as non-linearity such that every component lies between $-1$ and $1$. First of all, this regularizes the domain of the space and avoids numerical instability issues that occur when the magnitude of the vector components becomes too large. Secondly, by making the function non-linear we are able to model non-linear class boundaries in the optimization objective that we introduce in the next section. We use $\EntityEmbeddingMatrix{}$ to denote the $|\Entities{}| \times \EntityEmbeddingSize{}$ matrix that holds the entity representations. Row $i$ of $\EntityEmbeddingMatrix{}$ corresponds to the vector representation, $\EntityEmbedding{}_i$, of entity $\Entity{}_i$. Fig.~\ref{fig:schema} depicts a schematic overview of the proposed model. The parameters $\VocabularyEmbeddingMatrix{}$, $\MappingMatrix{}$, $\MappingBias{}$ and $\EntityEmbeddingMatrix{}$ will be learned automatically using function approximation methods as explained below.

The model proposed in this section shares similarities with previous work on word embeddings and unsupervised neural retrieval models \citep{Mikolov2013word2vec,VanGysel2016experts}. However, its novelty lies in its ability to scale to large collections of entities and its underlying assumption that words and entities are embedded in spaces of different dimensionality:
\begin{inparaenum}[(1)]
	\item The model of \citep{Mikolov2013word2vec} has no notion of entity retrieval as it estimates a language model for the whole corpus.
	\item Similar to \citep{Mikolov2013word2vec}, Eq.~\ref{eq:hash_fn} aggregates words $\Word{}_i \in \String{}$ to create a single phrase representation of $\String{}$. However, in \citep{VanGysel2016experts}, a distribution $\CondProb{\Entities{}}{\Word{}_i}$ is computed for every $\Word{}_i$ independently and aggregation occurs using the factor product. This is infeasible during model training when the collection of retrievable objects becomes too large, as is the case for product search. In the next section (\S\ref{section:methodology:estimation}) we solve this problem by sampling.
	\item In both \citep{Mikolov2013word2vec,VanGysel2016experts} two sets of representations of the same dimensionality are learned for different types of objects with potentially different latent structures (e.g., words, word contexts and experts). As mentioned earlier, Eq.~\ref{eq:hash_fn} alleviates this problem by transforming one latent space to the other.
\end{inparaenum}

\subsection{Parameter estimation}
\label{section:methodology:estimation}

\newcommand{\NGram}[2][]{#1{\Word{}_{#21}}, \ldots, #1{\Word{}_{#2\WindowSize{}}}}
\newcommand{\NGramProjection}{\HashFn{}({\NGram{}})}

\newcommand{\IndexedNGram}{\NGram[]{j,}}
\newcommand{\IndexedNGramProjection}{\HashFn{}({\IndexedNGram{}})}

\newcommand{\TargetEntity}{\Entity{}_i}
\newcommand{\TargetEntityRepr}{\EntityEmbedding{}_i}

For a particular document $\Document \in \AssociatedDocuments{\Entity{}_i}$ associated with entity $\TargetEntity{}$, we generate n-grams $\IndexedNGram{}$ where $\WindowSize{}$ (window size) remains fixed during training. For every n-gram $\IndexedNGram{}$, we compute its projected representation $\IndexedNGramProjection{}$ in $\EntitySpace{}$ using $\HashFn{}$ (Eq.~\ref{eq:hash_fn}). The objective, then, is to directly maximize the similarity between the vector representation of the entity $\EntityEmbedding{}_i$ and the projected n-gram $f(w_{j,1}$, \ldots, $w_{j,n})$ with respect to $\SimilarityFn{}$ (\S\ref{section:methodology:background}), while minimizing the similarity between $\IndexedNGramProjection{}$ and the representations of non-associated entities. This allows the model to learn relations between neighbouring words in addition to the associated entity and every word.

However, considering the full set of entities for the purpose of discriminative training can be costly when the number of entities $\NumEntities{}$ is large. Therefore, we apply a variant of Noise-Contrastive Estimation (NCE) \citep{Gutmann2010,Mnih2012,Mnih2013,Mikolov2013compositionality} where we sample negative instances from a noise distribution with replacement. We use the uniform distribution over entities as noise distribution. Define
\newcommand{\SimilarityRandomVariable}{\mathcal{S}}
\newcommand{\ProbSimilarEntity}[1]{\CondProb{\SimilarityRandomVariable{}}{#1, \IndexedNGramProjection{}}}
\newcommand{\ProbMassEntity}[1]{\SigmoidFn{\DotProduct{#1}{\IndexedNGramProjection{}}}}
\begin{equation}
\ProbSimilarEntity{\TargetEntityRepr{}} = \ProbMassEntity{\TargetEntityRepr{}}
\end{equation}
as the similarity of two representations in latent entity space, where
\begin{equation*}
\SigmoidFn{t} = \frac{1}{1 + \ExpFn{-t}}
\end{equation*}
denotes the sigmoid function and $\SimilarityRandomVariable{}$ is an indicator binary random variable that says whether $\TargetEntity{}$ is similar to $\IndexedNGramProjection{}$.

We then approximate the probability of an entity $\TargetEntity{}$ given an n-gram by randomly sampling $\NumNegativeExamples{}$ contrastive examples:
\newcommand{\UnnormedProbTargetGivenContext}[2][i]{\UnnormedCondProb{\Entity{}_#1}{\NGram{#2}}}
\newcommand{\LogUnnormedProbTargetGivenContext}[2][i]{\log\UnnormedProbTargetGivenContext[#1]{#2}}
\begin{eqnarray}
\label{eq:instance_loss}
\lefteqn{\LogUnnormedProbTargetGivenContext{j,}} \\
& = & \log{\ProbSimilarEntity{\TargetEntityRepr{}}} \nonumber \\
&   & + \sum_{\substack{k=1, \\ \Entity{}_k \sim U(\Entities{})}}^{\NumNegativeExamples{}} \log{\left(1 - \ProbSimilarEntity{\EntityEmbedding{}_k}\right)} \nonumber
\end{eqnarray}
where $U(X)$ denotes the uniform distribution over entities $\Entities{}$, the noise distribution used in NCE \citep{Gutmann2010}. Eq.~\ref{eq:instance_loss} avoids iterating over all entities during parameter estimation as we stochastically sample $\NumNegativeExamples{}$ entities uniformly as negative training examples.\footnote{We exploit the special nature of our evaluation scenario where we know the unique association between documents and entities. The setup can easily be adapted to the more general case where a document is associated with multiple entities by extracting the same word sequences from the document for every associated entity.}

During model construction we maximize the log-probability \eqref{eq:instance_loss} using batched gradient descent. The loss function for a single batch of $\BatchSize{}$ instances $\left((\NGram[]{k,}), \Entity{}_k\right)$ consisting of n-grams sampled from documents $\AssociatedDocuments{\Entity{}_k}$ (see~\S\ref{sec:experimental:design}) and associated entity $\Entity{}_k$ is as follows:
\begin{eqnarray}
\label{eq:loss}
\lefteqn{\LossFn{}}
\nonumber \\
& = & - \frac{1}{\BatchSize{}} \sum_{k = 1}^\BatchSize{} \LogUnnormedProbTargetGivenContext[k]{k,} \nonumber \\
\nonumber \\
&   & + \frac{\lambda}{2 \BatchSize{}} \left( \sum_{i,j} \VocabularyEmbeddingMatrix{}_{i,j}^2 + \sum_{i,j} \EntityEmbeddingMatrix{}_{i,j}^2 + \sum_{i,j} \MappingMatrix{}_{i,j}^2 \right),
\end{eqnarray}
where $\lambda$ is a weight regularization parameter. Instances are shuffled before batches are created. The update rule for a particular parameter $\theta$ ($\VocabularyEmbeddingMatrix{}$, $\EntityEmbeddingMatrix{}$, $\MappingMatrix{}$ or $\MappingBias{}$) given a single batch of size $\BatchSize{}$ is:
\begin{equation}
\label{eq:update}
\theta^{(t+1)} = \theta^{(t)} - \LearningRate{}^{(t)} \odot \frac{\partial \LossFnIdentifier}{\partial \theta}\LossFnArguments[^{(t)}]{},
\end{equation}
where $\LearningRate{}^{(t)}$ and $\theta^{(t)}$ denote the per-parameter learning rate and parameter $\theta$ at time $t$, respectively. The learning rate $\LearningRate{}$ consists of the same number of elements as there are parameters; in the case of a global learning rate, all elements of $\LearningRate{}$ are equal to each other.
The derivatives of the loss function \eqref{eq:loss} are given in the Appendix.


\section{Experimental setup}
\label{section:setup}

\subsection{Research questions}
\label{section:setup:rq}

In this \paper{} we investigate the problem of constructing a latent vector model of words and entities by directly modelling the discriminative relation between entities and word context. As indicated in the introduction of this \paper{}, we seek to answer the following research questions:

\newcommand{\RQ}[2]{
	\begin{description}[topsep=0pt]
	\phantomsection\label{section:setup:rq#1}
	\item[RQ#1] #2
	\end{description}
}

\newcommand{\RQRef}[1]{\textbf{\hyperref[section:setup:rq#1]{RQ#1}}}

\RQ{1}{\ResearchQuestionOne{}}
In \S\ref{section:methodology} we introduced various hyper-parameters along with the definition of \FullModelName{}. We have the size of word representations $\WordEmbeddingSize{}$ and the dimensionality of the entity representations $\EntityEmbeddingSize{}$. During parameter estimation, the window size $\WindowSize{}$ influences the context width presented as evidence for a particular entity. What is the influence of these parameters on the effectiveness of \ModelName{} and can we identify relations among parameters?

\RQ{2}{\ResearchQuestionTwo{}}
Is there a single method that always performs best or does effectiveness differ per domain? Does an increase in the vector space dimensionality impact the effectiveness of these methods?
\RQ{3}{\ResearchQuestionThree{}}
How does \ModelName{} compare to language models on a per-topic basis? Are there particular topics that work especially well with either type of ranker?
\RQ{4}{\ResearchQuestionFour{}}
What if we combine popularity-based, exact matching and latent vector space features in a linear learning-to-rank setting? Do we observe an increase in effectiveness if we combine these features?

\subsection{Experimental design}
\label{sec:experimental:design}

\newcommand{\NoEmphHomeKitchen}{Home \& Kitchen}
\newcommand{\NoEmphClothing}{Clothing, Shoes \& Jewelry}
\newcommand{\NoEmphPetSupplies}{Pet Supplies}
\newcommand{\NoEmphSports}{Sports \& Outdoors}

\newcommand{\HomeKitchen}{\emph{\NoEmphHomeKitchen{}}}
\newcommand{\Clothing}{\emph{\NoEmphClothing{}}}
\newcommand{\PetSupplies}{\emph{\NoEmphPetSupplies{}}}
\newcommand{\Sports}{\emph{\NoEmphSports{}}}

To answer the research questions posed in \S\ref{section:setup:rq}, we evaluate \ModelName{} in an entity retrieval setting organized around Amazon products (see \S\ref{sec:benchmarks}). We choose to experiment with samples of Amazon product data \citep{McAuley2015-1,McAuley2015-2} for the following reasons:
\begin{inparaenum}[(1)]
	\item The collection contains heterogeneous types of evidential documents associated with every entity: descriptions as well as reviews.
	\item Every department (e.g., \HomeKitchen{}) constitutes a separate, self-contained domain.
	\item Within each department there is a hierarchical taxonomy that partitions the space of entities in a rich structure. We can use the labels associated with these partitions and the partitions themselves as ground truth during evaluation.
	\item Every department consists of a large number of products categorized over a large number of categories. Importantly, this allows us to construct benchmarks with an increasing number of entities.
	\item Every product has a variety of attributes that can be used as popularity-based features in a learning-to-rank setting.
\end{inparaenum}

\newcommand{\EntityFindingBenchmarksNoRef}{\HomeKitchen{}, \Clothing{}, \PetSupplies{} and \Sports{} product search benchmarks}
\newcommand{\EntityFindingBenchmarks}{\EntityFindingBenchmarksNoRef{} (\S\ref{sec:benchmarks})}

\begin{table*}[th]
\centering

\caption{Overview of the \EntityFindingBenchmarksNoRef{}. T and V denote the test and validation sets, respectively. Arithmetic mean and standard deviation are reported wherever applicable.\label{tbl:benchmarks}}

\IfFileExists{resources/statistics.tex}{
}{
\resizebox{\textwidth}{!}{\missingfigure{Statistics}}}

\end{table*}

To answer \RQRef{1} we investigate the relation between the dimensionality of the entity representations $\EntityEmbeddingSize{}$ and window size $\WindowSize{}$. The latter, the window size $\WindowSize{}$, controls the context width the model can learn from, while the former, the dimensionality of the entity representations $\EntityEmbeddingSize{}$, influences the number of parameters and expressive power of the model. We sweep exponentially over $\WindowSize{}$ ($2^i$ for $0 \leq i < 6$) and $\EntityEmbeddingSize{}$ ($2^i$ for $6 \leq i < 11$). \RQRef{2} is answered by comparing \ModelName{} with latent vector space model baselines (\S\ref{section:baselines}) for an increasing entity space dimensionality $\EntityEmbeddingSize{}$ ($2^i$ for $6 \leq i < 11$). For \RQRef{3}, we compare the per-topic paired differences between \ModelName{} and a lexical language model. In addition, we investigate the correlation between lexical matches in relevant entity documents and ranker preference. We address \RQRef{4} by evaluating \ModelName{} as a feature in a machine-learned ranking in addition to query-independent and lexical features.

\newcommand{\m}{\sqrt{\frac{6.0}{m + n}}}

The number of $\WindowSize{}$-grams sampled per entity $\Entity{} \in \Entities{}$ from associated documents $\AssociatedDocuments{\Entity{}}$ in every epoch (i.e., iteration of the training data) is equal to
$$%
\ceil*{\frac{1}{\Length{\Entities{}}} \sum_{\Document{} \in \Documents{}} \text{max}\left(\Length{\Document{}} - \WindowSize{} + 1, 0\right)},
$$%
where the $\Length{\cdot}$ operator is used interchangeably for the size of set $\Entities{}$ and the number of tokens in documents $\Document{} \in \Documents{}$. This implicitly imposes a uniform prior over entities (i.e., stratified sampling where every entity is of equal importance).
The word vocabulary $V$ is created for each benchmark by ignoring punctuation, stop words and case; numbers are replaced by a numerical placeholder token. We prune $V$ by only retaining the $2^{16}$ most-frequent words so that each word can be encoded by a 16-bit unsigned integer.
In terms of parameter initialization of the \FullModelName{} model, we sample the initial matrices $\VocabularyEmbeddingMatrix{}$, $\MappingMatrix{}$ (Eq.~\ref{eq:hash_fn}) and $\EntityEmbeddingMatrix{}$ uniformly in the range
$%
\left[ -\m, \m \; \right]
$%
for an $m \times n$ matrix, as this initialization scheme is known to improve model training convergence \citep{Glorot2010}, and take the bias vector $\MappingBias{}$ to be null. The number of word features is set to $\WordEmbeddingSize{}=300$, similar to \citep{Mikolov2013compositionality}.
We take the number of negative examples $\NumNegativeExamples{} = 10$ to be fixed. \citet{Mikolov2013compositionality} note that a value of $\NumNegativeExamples{}$ between 10 and 20 is sufficient for large data sets \citep{Mnih2013}.

We used Adam ($\alpha = 0.001, \beta_1 = 0.9, \beta_2 = 0.999$) \citep{Kingma2014adam} with batched gradient descent ($\BatchSize{}=4096$) and weight decay $\lambda=0.01$ during training on NVidia Titan X GPUs. Adam has been designed specifically for non-stationary, stochastic cost functions like the one we defined in Eq.~\ref{eq:instance_loss}. For every model, we iterate over the training data 15 times and choose the best epoch based on the validation sets (Table~\ref{tbl:benchmarks}).

\vfill
\subsection{Product search benchmarks}
\label{sec:benchmarks}

We evaluate on four samples from different product domains\footnote{A list of product identifiers, topics and relevance assessments can be found at \paperImplementationUrl{}.} (Amazon departments), each with of an increasing number of products: \HomeKitchen{} (\numprint{8192} products), \Clothing{} (\numprint{16384} products), \PetSupplies{} (\numprint{32768} products) and \Sports{} (\numprint{65536} products); see Table~\ref{tbl:benchmarks}. The documents associated with every product consist of the product description plus reviews provided by Amazon customers.

\citet[p.~24]{Rowley2000} describes directed product search as users searching for ``a producer's name, a brand or a set of terms which describe the category of the product.'' Following this observation, the test topics $\CategoricalTopic{}_i$ are extracted from the categories each product belongs to. Category hierarchies of less than two levels are ignored, as the first level in the category hierarchy is often non-descriptive for the product (e.g., in \Clothing{} this is the gender for which the clothes are designated). Products belonging to a particular category hierarchy are considered as relevant for its extracted topic. Products can be relevant for multiple topics. Textual representations $\Query{}_{\CategoricalTopic{}_i}$ of the topics based on the categories are extracted as follows. For a single hierarchy of categories, we tokenize the titles of its sub-categories and remove stopwords and duplicate words. For example, a digital camera lense found in the \emph{Electronics} department under the categorical topic \emph{Camera \& Photo} $\rightarrow$ \emph{Digital Camera Lenses} will be relevant for the textual query ``\emph{photo camera lenses digital}.'' Thus, we only have two levels of relevance. We do not index the categories of the products as otherwise the query would match the category and retrieval would be trivial.

\subsection{Evaluation measures and significance}
\label{section:evaluation}
\newcommand{\DCG}{DCG}
\newcommand{\IdealDCG}{perfect \DCG{}}
\newcommand{\IdealDCGCut}{\IdealDCG{}@100}
\newcommand{\NDCG}{N\DCG{}}
\newcommand{\NDCGCut}{\NDCG{}@100}
\newcommand{\MAPCut}{MAP@100}
\newcommand{\PrecisionCut}{Precision@k}

\newcommand{\Significant}{^{*}}
\newcommand{\MoreSignificant}{^{**}}
\newcommand{\HighlySignificant}{^{***}}

To measure retrieval effectiveness, we report Normalized Discounted Cumulative Gain (\NDCG{}). For \RQRef{4}, we additionally report \PrecisionCut{} ($k = 5, 10$).
Unless mentioned otherwise, significance of observed differences is determined using a two-tailed paired Student's t-test \citep{Smucker2007significance} ($\HighlySignificant{} \, p < 0.01$;  $\MoreSignificant{} \, p < 0.05$; $\Significant{} \, p < 0.1$).

\subsection{Methods used in comparisons}
\label{sec:other_methods}

We compare \FullModelName{} to state-of-the-art latent vector space models for entity retrieval that are known to perform semantic matching \citep{Li2014}. We also conduct a contrastive analysis between \ModelName{} and smoothed language models with exact matching capabilities.
\paragraph{Vector Space Models for entity finding}
\label{section:baselines}
\newcommand{\RelevantDocuments}[1]{\text{rel}_{#1}}
\newcommand{\Ideal}[1]{#1^*}
\citet{Demartini2009expertspaces} propose a formal model for finding entities using document vector space models (\S\ref{section:methodology:background}). We compare the retrieval effectiveness of \ModelName{} with baseline latent vector space models created using
\begin{inparaenum}[(1)]
	\item Latent Semantic Indexing (\LSI{}) \citep{Deerwester1990lsi} with TF-IDF term weighting,
	\item Latent Dirichlet Allocation (\LDA{}) \citep{Blei2003} with $\alpha = \beta = 0.1$, where a document is represented by its topic distribution, and
	\item \WordToVec{} \citep{Mikolov2013word2vec} with CBOW and negative sampling, where a query/document is represented by the average of its word embeddings (same for queries in \ModelName{}). Similar to \ModelName{}, we train \WordToVec{} for 15 iterations and select the best-performing model using the validation sets (Table~\ref{tbl:benchmarks}).
\end{inparaenum}
\paragraph{\QueryLikelihoodLM{}}
\label{section:baselines:lm}
For every entity a pro\-file-based statistical language model is constructed using maximum-likelihood estimation \citep{Vapnik1998,Liu2005,Balog2006experts}, which is then smoothed by the language model of the entire corpus. The retrieval score of entity $\Entity{}$ for query $\Query{}$ is defined as
\begin{equation}
\UnnormedCondProb{\Query{}}{\Entity{}} = \prod_{\Term{}_i \in \Query{}} P(\Term{}_i \mid \theta_{\Entity{}}),
\end{equation}
where $\CondProb{\Term{}}{\theta_{\Entity{}}}$ is the probability of term $\Term{}$ occurring in the smooth\-ed language model of $\Entity{}$ (Jelinek-Mercer smoothing \citep{Zhai2004smoothing}). Given a query $\Query{}$, entities are ranked according to $\UnnormedCondProb{\Query{}}{\Entity{}}$ in descending order.

\paragraph{Machine-learned ranking}
\label{sec:experimental:ltr}
RankSVM models~\citep{Joachims2002svm} in \S\ref{section:discussion:ltr} and~\ref{section:analysis:representations} are trained using stochastic gradient descent using the implementation of \citet{Sculley2009ranksvm}. We use default values for all parameters, unless stated otherwise. For the experiment investigating \ModelName{} as a feature in machine-learn\-ed ranking in \S\ref{section:discussion:ltr}, we construct training examples by using the relevant entities as positive examples. Negative instances are generated by sampling from the non-relevant entities with replacement until the class distribution is uniform.


\newcommand{\RQAnswer}[2]{
	\RQRef{#1}: #2%
}

\section{Results \& discussion}
\label{section:discussion}

We start by giving a high-level overview of our experimental results (\RQRef{1} and \RQRef{2}), followed by a comparison with lexical matching methods (\RQRef{3}) and the use of \ModelName{} as a ranking feature (\RQRef{4}) (see \S\ref{sec:experimental:design} for an overview of the experimental design).

\subsection{Overview of experimental results}

\begin{figure}[t]

\newcommand{\heatmap}[4]{%
	\def \inner {%
		\def \PlotPath {resources/heat_map/#1_#2_#3.pdf}%
		\IfFileExists{\PlotPath}{
			\includegraphics[width=0.225\textwidth]{\PlotPath}}{
			\resizebox{0.225\textwidth}{!}{\missingfigure{#3}}}%
	}%
	\ifstrequal{#4}{}{%
	\subfloat{\inner}%
	}{\subfloat[#4\label{fig:heatmap:#2:#1}]{\inner}}%
}

\heatmap{home_and_kitchen}{qrel_validation}{ndcg}{\NoEmphHomeKitchen{}\vspace*{\baselineskip}}
\hfill
\heatmap{clothing_shoes_and_jewelry}{qrel_validation}{ndcg}{\NoEmphClothing{}\vspace*{\baselineskip}}
\hfill
\scalebox{1.003125}{\heatmap{pet_supplies}{qrel_validation}{ndcg}{\NoEmphPetSupplies{}}}%
\hspace*{0.0925\textwidth}%
{\heatmap{sports_and_outdoors}{qrel_validation}{ndcg}{\NoEmphSports{}}}

\caption{Sensitivity analysis of \ModelName{} in terms of \NDCG{} for window size $\WindowSize{}$ and the size of entity representations $\EntityEmbeddingSize{}$ during parameter estimation (Eq.~\ref{eq:loss}) for models trained on \EntityFindingBenchmarks{} and evaluated on the \textbf{validation} sets.\label{fig:heatmap}}

\end{figure}

\RQAnswer{1}{Fig.~\ref{fig:heatmap} depicts a heat map for every combination of window size and entity space dimensionality evaluated on the validation sets (Table~\ref{tbl:benchmarks}).}
\noindent
Fig.~\ref{fig:heatmap} shows that neither extreme values for the dimensionality of the entity representations nor the context width alone achieve the highest performance on the validation sets.
Instead, a low-dimensional entity space (\numprint{128}- and \numprint{256}-dimen\-sional) combined with a medium-sized context window (\numprint{4}- and \numprint{8}-grams) achieve the highest NDCG. In the two largest benchmarks (Fig.~\ref{fig:heatmap:qrel_validation:pet_supplies},~\ref{fig:heatmap:qrel_validation:sports_and_outdoors}) we see that for \numprint{16}-grams, NDCG actually lowers as the dimensionality of the entity space increases. This is due to the model \emph{fitting} the optimization objective (Eq.~\ref{eq:loss}), which we use as an unsupervised surrogate of relevance, too well. That is, as the model is given more learning capacity (i.e., higher dimensional representations), it starts to learn more regularities of natural language which counteract retrieval performance.

\begin{figure}[t]

\newcommand{\vectorspace}[4]{
	\subfloat[#4\label{fig:vectorspace:#2:#1}]{%
		\def \PlotPath {resources/vector_space/#1_#2_#3.pdf}%
		\IfFileExists{\PlotPath}{
			\includegraphics[width=0.230\textwidth]{\PlotPath}}{
			\resizebox{0.230\textwidth}{!}{\missingfigure{#4}}}
	}
}

\vectorspace{home_and_kitchen}{qrel_test}{ndcg}{\NoEmphHomeKitchen{}\vspace*{\baselineskip}}
\hfill
\vectorspace{clothing_shoes_and_jewelry}{qrel_test}{ndcg}{\NoEmphClothing{}\vspace*{\baselineskip}}
\hfill
\hspace*{-0.0125\textwidth}%
\scalebox{1.085}{\vectorspace{pet_supplies}{qrel_test}{ndcg}{\NoEmphPetSupplies{}}}%
\hspace*{0.045\textwidth}%
\scalebox{1.05}{\vectorspace{sports_and_outdoors}{qrel_test}{ndcg}{\NoEmphSports{}}}

\caption{Comparison of \ModelName{} (with window size $\WindowSize{} = 4$) with latent vector space baselines (\LSI{}, \LDA{} and \WordToVec{}; \S\ref{section:baselines}) on \EntityFindingBenchmarks{} and evaluated on the \textbf{test} sets.
Significance (\S\ref{section:evaluation}) is computed between \ModelName{} and the baselines for each vector space size.\label{fig:vectorspace}}

\end{figure}

\RQAnswer{2}{Fig.~\ref{fig:vectorspace} presents a comparison between \ModelName{} (window size $\WindowSize{} = 4$) and vector space model baselines (\S\ref{section:baselines}) for increasing entity representation dimensionality ($2^i$ for $6 \leq i < 11$) on the test sets.}
\noindent
\ModelName{} significantly outperforms ($p < 0.01$) all baseline methods in most cases (except for Fig.~\ref{fig:vectorspace:qrel_test:home_and_kitchen} where $\EntityEmbeddingSize = 1024$). For the smaller benchmarks (Fig.~\ref{fig:vectorspace:qrel_test:home_and_kitchen},~\ref{fig:vectorspace:qrel_test:clothing_shoes_and_jewelry}), we see \LSI{} as the main competitor of \ModelName{}. However, as the training corpora become larger (in Fig.~\ref{fig:vectorspace:qrel_test:pet_supplies},~\ref{fig:vectorspace:qrel_test:sports_and_outdoors}), \WordToVec{} outperforms \LSI{} and becomes the main contester of \ModelName{}. On all benchmarks, \ModelName{} peaks when the entity representations are low-dimensional (\numprint{128}- or \numprint{256}-dimensional) and afterwards (for a higher dimensionality) performance decreases. On the other hand, \WordToVec{} stagnates in terms of \NDCG{} around representations of \numprint{512} dimensions and never achieves the same level as \ModelName{} did for one or two orders of magnitude (base 2) smaller representations. This is a beneficial trait of \ModelName{}, as high-dimensional vector spaces are undesirable due to their high computational cost during retrieval \citep{Weber1998}.

\subsection{A feature for machine-learned ranking}
\label{section:discussion:ltr}

We now investigate the use of \ModelName{} as a feature in a learning to rank setting \citep{Liu2011}. Latent vector space models are known to provide a means of semantic matching as opposed to a purely lexical matching~\citep{Li2014,VanGysel2016experts}. To determine to which degree this is indeed the case, we first perform a topic-wise comparison between \ModelName{} and a lexical language model, the \QueryLikelihoodLM{} (\QLM{}) \citep{Zhai2004smoothing}, as described in \S\ref{section:baselines:lm}. We optimize the parameters of \ModelName{} and \QLM{} on the validation sets for every benchmark (Table~\ref{tbl:benchmarks}). In the case of \ModelName{}, we select the model that performs best in Fig.~\ref{fig:heatmap}. For \QLM{}, we sweep over $\lambda$ linearly from $0.0$ to $1.0$ (inclusive) with increments of $0.05$.

\begin{figure}[bh!]

\newcommand{\innerdeltaplot}[5]{
	\def \PlotPath {resources/delta/#1_differences.pdf}%
		\IfFileExists{\PlotPath}{
			\includegraphics[width=0.225\textwidth]{\PlotPath}}{
			\resizebox{0.225\textwidth}{!}{\missingfigure{#5}}}
}

\newcommand{\deltaplot}[5]{
	\subfloat[#5\label{fig:delta:#2:#1}]{%
		\innerdeltaplot{#1}{#2}{#3}{#4}{ndcg}%
	}
}

\def \ModelConfiguration{4window.deep_pre_scalable_inproduct_tanh_512embedding_instance}
\newcommand{\BaselineModel}[1]{baseline_model_mle_lm.mle_lm_jm_#1_word_lm_uniform_prior}

\deltaplot{home_and_kitchen}{4window.deep_pre_scalable_inproduct_tanh_128embedding_instance}{\BaselineModel{0.15}}{qrel_test}{\NoEmphHomeKitchen{}}%
\hfill%
\deltaplot{clothing_shoes_and_jewelry}{4window.deep_pre_scalable_inproduct_tanh_1024embedding_instance}{\BaselineModel{0.15}}{qrel_test}{\NoEmphClothing{}}%
\hfill%
\deltaplot{pet_supplies}{4window.deep_pre_scalable_inproduct_tanh_256embedding_instance}{\BaselineModel{0.4}}{qrel_test}{\NoEmphPetSupplies{}}%
\hfill%
\deltaplot{sports_and_outdoors}{8window.deep_pre_scalable_inproduct_tanh_128embedding_instance}{\BaselineModel{0.1}}{qrel_test}{\NoEmphSports{}}%

\caption{Per-topic paired differences between \ModelName{} and \QueryLikelihoodLM{} for models trained on \EntityFindingBenchmarks{} and evaluated on the \textbf{test} sets. For every plot, the y-axis indicates $\Delta \text{\NDCG{}}$ between \ModelName{} and a \QueryLikelihoodLM{}. The x-axis lists the topics in the referenced benchmark in decreasing order of $\Delta \text{\NDCG{}}$ such that topics for which \ModelName{} performs better are on the left and vice versa for the \QueryLikelihoodLM{} on the right.\label{fig:delta}}

\vfill
\end{figure}

\begin{table}
	\centering

	\caption{Correlation coefficients between average IDF of lexically matched terms in documents associated with relevant entities and $\bigtriangleup{} \text{\NDCG{}}$. A negative correlation coefficient implies that queries consisting of more specific terms (i.e., low document freq.) that occur exactly in documents associated with relevant entities are more likely to benefit from \QLM{}, whereas other queries (with less specific terms or less exact matches) gain more from \ModelName{}. Significance is achieved for all benchmarks ($p < 0.01$) using a permutation test.\label{tbl:correlations}}

	\begin{tabular}{l c c}

	\toprule
	\textbf{Benchmark} & \textbf{Spearman $R$} & \textbf{Pearson $R$} \\
	\midrule

	\NoEmphHomeKitchen{} & \nprounddigits{2} \npdecimalsign{.} \numprint{-0.301271661356} & \nprounddigits{2} \npdecimalsign{.} \numprint{-0.348547810194} \\
	\NoEmphClothing{} & \nprounddigits{2} \npdecimalsign{.} \numprint{-0.396222193487} & \nprounddigits{2} \npdecimalsign{.} \numprint{-0.369019393753} \\
	\NoEmphPetSupplies{} & \nprounddigits{2} \npdecimalsign{.} \numprint{-0.165627764962} & \nprounddigits{2} \npdecimalsign{.} \numprint{-0.171408122876} \\
	\NoEmphSports{} & \nprounddigits{2} \npdecimalsign{.} \numprint{-0.338266676522} & \nprounddigits{2} \npdecimalsign{.} \numprint{-0.361384861078} \\

	\midrule

	\end{tabular}
\end{table}

\RQAnswer{3}{Fig.~\ref{fig:delta} shows the per-topic paired difference between \ModelName{} and \QLM{} in terms of \NDCG{}.}
\noindent
Topics that benefit more from \ModelName{} have a positive value on the y-axis, while those that prefer \QLM{} have a negative value. We can see that both methods perform similarly for many topics (where $\bigtriangleup = 0.0$). For certain topics one method performs substantially better than the other, suggesting that the two are complementary. To further quantify this, we investigate the relation between specific topic terms and their occurrence in documents relevant to these topics. That is, we measure the correlation between the per-topic $\bigtriangleup \text{\NDCG{}}$ (as described above) and the average inverse document frequency (IDF) of exact/lexically matched terms in the profile-based language model. In Table~\ref{tbl:correlations} we observe that queries that contain specific tokens (i.e., with high inverse document frequency) and occur exactly in documents associated with relevant products, benefit more from \QLM{} (lexical matches). Conversely, queries with less specific terms or without exact matches in the profiles of relevant products gain more from \ModelName{} (semantic matches).

This observation motivates the use of \ModelName{} as a ranking feature in addition to traditional language models.
Specifically, we now evaluate the use of \ModelName{} as a feature in a linear \RankSVM{} (\S\ref{sec:experimental:ltr}). Following \citet{Fang2010}, we consider query-independent (QI) popularity-based features in addition to features provided by \ModelName{} and \QLM{}. This allows us to consider the effect of the query-dependent features independent from their ability to model a popularity prior over entities. Table~\ref{fig:ltr-overview} lists the feature sets.

\begin{table}[t]
	\centering

	\caption{Overview of the feature sets used in the machine-learned ranking experiments.\label{fig:ltr-overview}}

	\small
	\begin{tabular}{c L{9cm}}

	\toprule
	\textbf{Features} & \textbf{Description} \\
	\midrule

	\textbf{QI} &
		Query-independent features:%
		\begin{inparaenum}[(1)]
			\item product price;
			\item product description length;
			\item reciprocal of the Amazon sales rank; and
			\item product PageRank scores based on four related product graphs (also bought, also viewed, bought together, buy after viewing).
		\end{inparaenum}
		\\
		\midrule
	\textbf{QLM} & \QueryLikelihoodLM{} using Jelinek-Mercer smoothing with $\lambda$ optimized on the validation set (Table~\ref{tbl:benchmarks}). Posterior $\CondProb{\Query{}}{\Entity{}}$ is used as a feature for entity $\Entity{}$ and query $\Query{}$. \\
	\midrule
	\textbf{\ModelName{}} & \FullModelName{} optimized on the validation set (Table~\ref{tbl:benchmarks}, Fig.~\ref{fig:heatmap}). Similarity $\SimilarityFn{}(\ProjectedQuery{}, \EntityEmbedding{})$ is used as a feature for entity $\Entity{}$, with vector representation $\EntityEmbedding{}$, and query $\Query{}$. \\
	\bottomrule
	\end{tabular}

\end{table}

\begin{table}[t]
	\centering
	\caption{Ranking performance results for query independent (QI) features, the \QueryLikelihoodLM{} (QLM) match feature, the \FullModelName{} (\ModelName{}) match feature and combinations thereof, weighted using \RankSVM{} (\S\ref{section:discussion:ltr}), evaluated on the test sets using 10-fold cross validation, for \EntityFindingBenchmarks{}. The hyperparameters of the individual query features (QLM and LSE) were optimized using the validation sets.
	Significance of the results (\S\ref{section:evaluation}) is computed between {QI + QLM + LSE} and {QI + QLM}.\label{tbl:ltr}}
	\IfFileExists{resources/results-ltr.tex}{
	\small
	\resizebox{0.65\textwidth}{!}{\begin{tabular}{c@{ }c@{ }c@{ }c}
\toprule
\multirow{2}{*}{} & \multicolumn{3}{c}{Home \& Kitchen} \\ 
& NDCG & P@5 & P@10 \\ 
\cmidrule(lr){2-4}
\multicolumn{1}{l}{QI} & $\phantom{}0.005$ & $\phantom{}0.002$ & $\phantom{}0.001$ \\ 
\multicolumn{1}{l}{QI + QLM} & $\phantom{}0.321$ & $\phantom{}0.180$ & $\phantom{}0.145$ \\ 
\multicolumn{1}{l}{QI + LSE} & $\phantom{}0.257$ & $\phantom{}0.121$ & $\phantom{}0.107$ \\ 
\multicolumn{1}{l}{QI + QLM + LSE} & $\phantom{\HighlySignificant{}}\textbf{0.352}\HighlySignificant{}$ & $\phantom{\MoreSignificant{}}\textbf{0.192}\MoreSignificant{}$ & $\phantom{\HighlySignificant{}}\textbf{0.157}\HighlySignificant{}$ \\ 
\cmidrule{1-4}
\multirow{2}{*}{} & \multicolumn{3}{c}{Clothing, Shoes \& Jewelry} \\ 
& NDCG & P@5 & P@10 \\ 
\cmidrule(lr){2-4}
\multicolumn{1}{l}{QI} & $\phantom{}0.002$ & $\phantom{}0.001$ & $\phantom{}0.001$ \\ 
\multicolumn{1}{l}{QI + QLM} & $\phantom{}0.177$ & $\phantom{}0.079$ & $\phantom{}0.068$ \\ 
\multicolumn{1}{l}{QI + LSE} & $\phantom{}0.144$ & $\phantom{}0.065$ & $\phantom{}0.057$ \\ 
\multicolumn{1}{l}{QI + QLM + LSE} & $\phantom{\HighlySignificant{}}\textbf{0.198}\HighlySignificant{}$ & $\phantom{\HighlySignificant{}}\textbf{0.094}\HighlySignificant{}$ & $\phantom{\HighlySignificant{}}\textbf{0.080}\HighlySignificant{}$ \\ 
\cmidrule{1-4}
\multirow{2}{*}{} & \multicolumn{3}{c}{Pet Supplies} \\ 
& NDCG & P@5 & P@10 \\ 
\cmidrule(lr){2-4}
\multicolumn{1}{l}{QI} & $\phantom{}0.003$ & $\phantom{}0.002$ & $\phantom{}0.002$ \\ 
\multicolumn{1}{l}{QI + QLM} & $\phantom{}0.250$ & $\phantom{}0.212$ & $\phantom{}0.199$ \\ 
\multicolumn{1}{l}{QI + LSE} & $\phantom{}0.268$ & $\phantom{}0.222$ & $\phantom{}0.214$ \\ 
\multicolumn{1}{l}{QI + QLM + LSE} & $\phantom{\HighlySignificant{}}\textbf{0.298}\HighlySignificant{}$ & $\phantom{\HighlySignificant{}}\textbf{0.255}\HighlySignificant{}$ & $\phantom{\HighlySignificant{}}\textbf{0.236}\HighlySignificant{}$ \\ 
\cmidrule{1-4}
\multirow{2}{*}{} & \multicolumn{3}{c}{Sports \& Outdoors} \\ 
& NDCG & P@5 & P@10 \\ 
\cmidrule(lr){2-4}
\multicolumn{1}{l}{QI} & $\phantom{}0.001$ & $\phantom{}0.001$ & $\phantom{}0.001$ \\ 
\multicolumn{1}{l}{QI + QLM} & $\phantom{}0.235$ & $\phantom{}0.183$ & $\phantom{}0.156$ \\ 
\multicolumn{1}{l}{QI + LSE} & $\phantom{}0.188$ & $\phantom{}0.132$ & $\phantom{}0.121$ \\ 
\multicolumn{1}{l}{QI + QLM + LSE} & $\phantom{\HighlySignificant{}}\textbf{0.264}\HighlySignificant{}$ & $\phantom{\HighlySignificant{}}\textbf{0.192}\HighlySignificant{}$ & $\phantom{\HighlySignificant{}}\textbf{0.172}\HighlySignificant{}$ \\ 
\bottomrule
\end{tabular}
}}{
	\resizebox{\textwidth}{!}{\missingfigure{LTR}}}
\end{table}

\RQAnswer{4}{Table~\ref{tbl:ltr} shows the results for different combinations of feature sets used in a machine-learned ranker, RankSVM.}
\noindent
The experiment was performed using 10-fold cross validation on the test sets (Table~\ref{tbl:benchmarks}). The combination using all features outperforms smaller subsets of features, on all metrics. We conclude that \FullModelName{} adds a signal that is complementary to traditional (lexical) language models, which makes it applicable in a wide range of entity-oriented search engines that use ranker fusion techniques.

\section{Analysis of representations}
\label{section:analysis:representations}

\newcommand{\IdealEntityEmbedding}{\EntityEmbedding{}^{*}_\CategoricalTopic{}}
\newcommand{\Approximate}[1]{\tilde{#1}}
\newcommand{\ApproximateIdealEntityEmbedding}{\tilde{\EntityEmbedding{}}^{*}_\CategoricalTopic{}}

Next, we analyse the entity representations $\EntityEmbedding{}_i$ of the vector space models independent of the textual representations by providing empirical lower-bounds on their maximal retrieval performance, followed by a comparison with their actual performance so as to measure the effectiveness of word-to-entity mapping $\HashFn{}$.

Fig.~\ref{fig:heatmap}~and~\ref{fig:vectorspace} show which levels of performance may be achieved by using the latent models to generate a ranking from textual queries (Eq.~\ref{eq:hash_fn}). But this is only one perspective. As entities are ranked according to their similarity with the projected query vector $\ProjectedQuery[_{\CategoricalTopic{}}]{}{}$, the performance for retrieving entities w.r.t.\ the textual representation of a topic $\CategoricalTopic{}$ depends on the structure of the entity space $\EntitySpace{}$, the ideal retrieval vector $\IdealEntityEmbedding{} \in \EntitySpace{}$ (i.e., the vector that optimizes retrieval performance), and the similarity between $\ProjectedQuery[_{\CategoricalTopic{}}]{}$ and $\IdealEntityEmbedding{}$.

How can we determine the ideal vector $\IdealEntityEmbedding{}$? First, we define it to be the vector for which the cosine similarity with each of the entity embeddings results in a ranking where relevant entities are ranked higher than non-relevant or unjudged entities. We approximate $\IdealEntityEmbedding{}$ by optimizing the pair-wise SVM objective \citep{Joachims2002svm,Sculley2009ranksvm}. That is, for every topic $\CategoricalTopic{}$ we construct a separate RankSVM model based on its ground-truth as follows. We only consider topics with at least two relevant entities, as topics with a single relevant entity have a trivial optimal retrieval vector (the entity representation of the single relevant entity). Using the notation of \citep{Joachims2002svm}, the normalized entity representations are used as features, and hence the feature mapping $\phi$ is defined as
\begin{equation*}
\phi(\CategoricalTopic{}, \Entity{}_i) = \frac{\EntityEmbedding{}_i}{\norm{\EntityEmbedding{}_i}_2} \text{ for all }\Entity{}_i \in \Entities{}.
\end{equation*}
The target ranking $r^*_\CategoricalTopic{}$ is given by the entities relevant to topic $\CategoricalTopic{}$. Thus, the features for every entity become the entity's normalized representation and its label is positive if it is relevant for the topic and negative otherwise. The pair-wise objective then finds a weight vector such that the ranking generated by ordering according to the vector scalar product between the weight vector and the normalized entity representations correlates with the target ranking $r^*_\CategoricalTopic{}$. Thus, our approximation of the \emph{ideal vector}, $\ApproximateIdealEntityEmbedding$, is given by the weight vector $w_\CategoricalTopic{}$ for every $\CategoricalTopic{}$.\footnote{Note that $\ApproximateIdealEntityEmbedding{}$ does not take into account the textual representations $\Query{}_\CategoricalTopic{}$ of topic $\CategoricalTopic{}$, but only the clustering of entities relevant to $\CategoricalTopic{}$ and their relation to other entities.}

\newcommand{\FullVersion}{false}

\begin{figure\if\FullVersion{}*\fi}[t]

\newcommand{\centroids}[4]{%
	\def \inner {%
		\def \PlotPath {resources/centroids/#1_#2_#3.pdf}%
		\IfFileExists{\PlotPath}{
			\includegraphics[width=0.225\textwidth]{\PlotPath}}{
			\resizebox{0.225\textwidth}{!}{\missingfigure{#4}}}
	}%
	\ifstrequal{#4}{}{%
	\subfloat{\inner}%
	}{\subfloat[#4\label{fig:heatmap:#2:#1}]{\inner}}%
}

\centroids{home_and_kitchen}{absolute}{ndcg}{\NoEmphHomeKitchen{}}
\if\FullVersion{}
\hfill
\centroids{clothing_shoes_and_jewelry}{absolute}{ndcg}{}
\fi
\hfill
\centroids{pet_supplies}{absolute}{ndcg}{\NoEmphPetSupplies{}}
\if\FullVersion{}
\hfill
\centroids{sports_and_outdoors}{absolute}{ndcg}{}
\fi

\caption{Comparison of the approximately ideal retrieval vector $\ApproximateIdealEntityEmbedding{}$ with the projected query retrieval vector $\ProjectedQuery{}$ for latent entity models built using \ModelName{}, \LSI{}, \LDA{} and \WordToVec{} (\S\ref{section:baselines}) on \if\FullVersion{}\EntityFindingBenchmarks{}\else{}\HomeKitchen{} and \PetSupplies{} product search benchmarks (\S\ref{sec:benchmarks})\fi{} and evaluated on the \textbf{test} sets\if\FullVersion{}\else{}. The plots for \Clothing{} and \Sports{} product search benchmarks are qualitatively similar to the ones shown\fi{}.
The figures show the absolute performance in terms of \NDCG{} of $\ApproximateIdealEntityEmbedding{}$ (dashed curves) and $\ProjectedQuery{}$ (solid curves); significance (\S\ref{section:evaluation}) for the results for the approximately ideal retrieval vectors $\ApproximateIdealEntityEmbedding{}$ is computed between \ModelName{} and the best-performing baseline for each vector space size and indicated along the x-axis.
\label{fig:idealcentroids}}

\end{figure\if\FullVersion{}*\fi}

What is the performance of this approximately ideal vector representation? And how far are our representations removed from it? Fig.~\ref{fig:idealcentroids} shows the absolute performance of $\ApproximateIdealEntityEmbedding{}$ (dashed curves) and $\ProjectedQuery{}$ (solid curves) in terms of \NDCG{}. Comparing the (absolute) difference between every pair of dashed and solid curves for a single latent model gives an intuition of how much performance in terms of \NDCG{} there is to gain by improving the projection function $\HashFn{}$ for that method. The approximately ideal vectors $\ApproximateIdealEntityEmbedding{}$ discovered for \ModelName{} outperform all baselines significantly. Interestingly, for representations created using \LDA{}, the optimal performance goes up while the actual performance stagnates. This indicates that a higher vector space dimensionality renders better representations using \LDA{}, however, the projection function $\HashFn{}$ is unable to keep up in the sense that projected query vectors are not similar to the representations of their relevant entities. The latent models with the best representations (\ModelName{} and \LSI{}) also have the biggest gap between $\ProjectedQuery{}$ and $\ApproximateIdealEntityEmbedding{}$ in terms of achieved \NDCG{}.

We interpret the outcomes of our analysis as follows. The entity space $\EntitySpace{}$ has more degrees of freedom to cluster entities more appropriately as the dimensionality of $\EntitySpace{}$ increases. Consequently, the query projection function $\HashFn{}$ is expected to learn a more complex function. In addition, as the dimensionality of $\EntitySpace{}$ increases, so does the modelling capacity of the projection function $\HashFn{}$ in the case of \ModelName{} and \LSI{} (i.e., the transformation matrices become larger) and therefore more parameters have to be learned. We conclude that our method can more effectively represent entities in a lower-dimensional space than \LSI{} by making better use of the vector space capacity. This is highly desirable, as the asymptotic runtime complexity of many algorithms operating on vector spaces increases at least linearly \citep{Weber1998} with the size of the vectors.


\vfill
\pagebreak

\vspace*{-3.5em}
\section{Summary}
\label{section:conclusions}

We have introduced \FullModelName{}, an unsupervised latent vector space model for product search. It jointly learns a unidirectional mapping between, and latent vector representations of, words and products. We have also defined a formalism for latent vector space models where latent models are decomposed into a map\-ping from word sequences to the product vector space, representations of products in that space, and a similarity function.
We have evaluated our model using Amazon product data, and compared it to state-of-the-art latent vector space models for product ranking (\LSI{}, \LDA{} and \WordToVec{}). \ModelName{} outperforms all baselines for lower-dimensional vector spaces.

In an analysis of the vector space models, we have compared the performance achieved with the ideal performance of the proposed product representations. We have shown that \ModelName{} constructs better product representations than any of the baselines. In addition, we have obtained important insights w.r.t.\ how much performance there is to gain by improving the individual components of latent vector space models. Future work can focus on improving the mapping from words to products by incorporating specialized features or increasing the mapping's complexity. In addition, semi-supervised learning may help specialize the vector space and mapping function for particular retrieval settings.

A comparison of \ModelName{} with a smoothed lexical language model unveils that the two methods make very different errors. Some directed product search queries require lexical matching, others benefit from the semantic matching capabilities of latent models. We have evaluated \ModelName{} as a feature in a machine-learned ranking setting and found that adding \ModelName{} to language models and popularity-based features significantly improves retrieval performance.

}


\chapter{Neural Vector Spaces for Unsupervised Information~Retrieval}
\label{chapter:research-06}

{
\ScopeLabels{research-06}
\ChapterRQ{4}

\newcommand{\BenchmarkFigureWidth}{0.200\textwidth}

\def \paperImplementationUrl {\url{https://github.com/cvangysel/cuNVSM}}

\newcommand{\DocToVec}{\ac{d2v}}
\newcommand{\WordToVec}{\ac{w2v}}
\newcommand{\CombineSelfInformation}{\texttt{\ac{si}}}
\newcommand{\CombineAdd}{\texttt{add}}
\newcommand{\WordToVecSgSi}{\WordToVec{}-\CombineSelfInformation{}}
\newcommand{\WordToVecSgAdd}{\WordToVec{}-\CombineAdd{}}
\newcommand{\ModelName}{\acs{NVSM}}
\newcommand{\FullModelName}{\acl{NVSM}}

\newcommand{\Dirichlet}{\texttt{\ac{d}}}
\newcommand{\JelinekMercer}{\texttt{\ac{jm}}}

\newcommand{\FullDirichlet}{Dirichlet}
\newcommand{\FullJelinekMercer}{Jelinek-Mercer}

\newcommand{\QLMDirichlet}{\QLM{} (\Dirichlet{})}
\newcommand{\QLMJelinekMercer}{\QLM{} (\JelinekMercer{})}

\newcommand{\TitlestatRelevant}{\texttt{titlestat\_rel}}
\newcommand{\MAPCut}{\ac{MAP}@1000}
\newcommand{\NDCGCut}{\ac{NDCG}@100}
\newcommand{\Precision}{\ac{P}@10}

\newcommand{\BenchmarkAP}{Associated Press 88-89}
\newcommand{\BenchmarkWSJ}{Wall Street Journal}
\newcommand{\BenchmarkLATimes}{LA Times}
\newcommand{\BenchmarkFT}{Financial Times}
\newcommand{\BenchmarkRobust}{Robust04}
\newcommand{\BenchmarkNYT}{New York Times}

\newcommand{\ShortBenchmarkAP}{AP88-89}
\newcommand{\ShortBenchmarkWSJ}{WSJ}
\newcommand{\ShortBenchmarkLATimes}{LA}
\newcommand{\ShortBenchmarkFT}{FT}
\newcommand{\ShortBenchmarkRobust}{\BenchmarkRobust{}}
\newcommand{\ShortBenchmarkNYT}{NY}

\newcommand{\Vocabulary}{V}

\newcommand{\Queries}{Q}
\newcommand{\Query}{\MakeLowercase{\Queries{}}}
\newcommand{\QueryTerm}[1]{t_{#1}}

\newcommand{\Documents}{D}
\newcommand{\Document}{\MakeLowercase{\Documents{}}}
\newcommand{\Word}{w}

\newcommand{\LatentDimension}[1]{{k_{#1}}}
\newcommand{\EmbeddingMatrix}[1]{{R_{#1}}}

\newcommand{\WordMatrix}{\EmbeddingMatrix{\Vocabulary{}}}
\newcommand{\WordEmbedding}[1]{\vec{\WordMatrix{}}^{(#1)}}
\newcommand{\LatentWordDim}{\LatentDimension{\Word{}}}

\newcommand{\DocumentMatrix}{\EmbeddingMatrix{\Documents{}}}
\newcommand{\DocumentEmbedding}[1]{\vec{\DocumentMatrix{}}^{(#1)}}
\newcommand{\LatentDocumentDim}{\LatentDimension{\Document{}}}

\newcommand{\TransformFn}{f}
\newcommand{\TransformMatrix}{W}

\newcommand{\NGramSize}{n}

\newcommand{\CompositionalFn}{g}

\newcommand{\HashFn}{h}

\newcommand{\ScoreFn}{\text{score}}

\newcommand{\ApproxProb}[1]{\tilde{P}\left(#1\right)}
\newcommand{\CondApproxProb}[2]{\ApproxProb{#1 \mid #2}}

\newcommand{\DotProduct}[2]{#1 \cdot #2}

\newcommand{\SampleExpectation}{\hat{\mathbb{E}}}
\newcommand{\SampleVariance}{\hat{\mathbb{V}}}

\newcommand{\ErrorBias}{\beta}
\newcommand{\NGramProjection}{\Apply{\Projection{}}{\BatchInstancePhrase{i}}}
\newcommand{\HardTanH}{\text{hard-tanh}}

\def \ResearchQuestionOne {How does \ModelName{} compare to other latent vector space models, such as \FullDocToVec{} \citep{Le2014}, \FullWordToVec{} \citep{Mikolov2013word2vec,Vulic2015monolingual}, \LSI{} \citep{Deerwester1990lsi}, \LDA{} \citep{Blei2003} and \LSE{} \citep{VanGysel2016products}, on the document retrieval task?}
\def \ResearchQuestionTwo {For what proportion of queries does \ModelName{} perform better than the other rankers?}
\def \ResearchQuestionThree {What gains does \ModelName{} bring when combined with a lexical \QLM{} and a competing state-of-the-art vector space model?}
\def \ResearchQuestionFour {Do \ModelName{}s exhibit regularities that we can link back to well-understood document collection statistics used in traditional retrieval models?}

\def \ContributionOne {A novel neural retrieval model, \ModelName{}, that is learned using gradient descent on a document collection.}
\def \ContributionTwo {Comparisons of lexical (\QLM{}) and semantic (\FullDocToVec{}, \FullWordToVec{}, \LSI{}, \LDA{}, \LSE{} and \ModelName{}) models on document retrieval test collections.}
\def \ContributionThree {An analysis of the internals of \ModelName{} to give insights in the workings of the model and the retrieval task.}
\def \ContributionFour {A highly-optimized open source C++/CUDA implementation of \ModelName{} that results in fast training and efficient memory usage.}
\def \ContributionFive {Advice on how to configure the hyperparameters of \ModelName{}.}

\renewcommand{\log}{\text{log}}
\renewcommand{\exp}{\text{exp}}

\newcommand{\subfloat}[2][]{%
\begin{subfigure}[b]{0.25\paperheight}%
\centering%
{\setlength{\textwidth}{1.25\paperheight}#2}%
\caption{#1}%
\end{subfigure}%
}

\newcommand{\Table}[4][]{
\begin{#2}[th!]
\centering
\caption{#1\label{tbl:#3_results}}
\scalebox{#4}{%
\renewcommand{\MAPCut}{MAP}%
\renewcommand{\NDCGCut}{NDCG}%
\renewcommand{\BenchmarkAP}{\ShortBenchmarkAP}%
\renewcommand{\BenchmarkWSJ}{\ShortBenchmarkWSJ}%
\renewcommand{\BenchmarkLATimes}{\ShortBenchmarkLATimes}%
\renewcommand{\BenchmarkFT}{\ShortBenchmarkFT}%
\renewcommand{\BenchmarkNYT}{\ShortBenchmarkNYT}%
\centering%
\begin{tabular}{c}%
\input{resources/#3.tex}%
\end{tabular}%
}%
\end{#2}%
}

\newcommand{\StretchedTable}[4][]{
\begin{#2}[th!]
\centering
\caption{#1\label{tbl:#3_results}}
\resizebox{#4}{!}{%
\renewcommand{\MAPCut}{MAP}%
\renewcommand{\NDCGCut}{NDCG}%
\renewcommand{\BenchmarkAP}{\ShortBenchmarkAP}%
\renewcommand{\BenchmarkWSJ}{\ShortBenchmarkWSJ}%
\renewcommand{\BenchmarkLATimes}{\ShortBenchmarkLATimes}%
\renewcommand{\BenchmarkFT}{\ShortBenchmarkFT}%
\renewcommand{\BenchmarkNYT}{\ShortBenchmarkNYT}%
\centering%
\begin{tabular}{c}%
\input{resources/#3.tex}%
\end{tabular}%
}%
\end{#2}%
}

\newcommand{\DoubleColumnTable}[2][]{

\begin{table*}[th!]
\centering
\caption{#1\label{tbl:#2_results}}
\resizebox{0.75\paperheight}{!}{%
\renewcommand{\MAPCut}{MAP}%
\renewcommand{\NDCGCut}{NDCG}%
\renewcommand{\BenchmarkAP}{\ShortBenchmarkAP}%
\renewcommand{\BenchmarkWSJ}{\ShortBenchmarkWSJ}%
\renewcommand{\BenchmarkLATimes}{\ShortBenchmarkLATimes}%
\renewcommand{\BenchmarkFT}{\ShortBenchmarkFT}%
\renewcommand{\BenchmarkNYT}{\ShortBenchmarkNYT}%
\centering%
\begin{tabular}{c}%
\input{resources/#2.tex}%
\end{tabular}%
}%
\end{table*}%

}

\newcommand{\SingleColumnTable}[2][]{

\begin{table}[th!]
\centering
\renewcommand{\arraystretch}{1.2}
\caption{#1\label{tbl:#2_results}}
\scalebox{0.90}{%
\renewcommand{\MAPCut}{MAP}%
\renewcommand{\NDCGCut}{NDCG}%
\renewcommand{\BenchmarkAP}{\ShortBenchmarkAP}%
\renewcommand{\BenchmarkWSJ}{\ShortBenchmarkWSJ}%
\renewcommand{\BenchmarkLATimes}{\ShortBenchmarkLATimes}%
\renewcommand{\BenchmarkFT}{\ShortBenchmarkFT}%
\renewcommand{\BenchmarkNYT}{\ShortBenchmarkNYT}%
\centering%
\renewcommand{\arraystretch}{0.8}%
\setlength{\tabcolsep}{3.5pt}%
\begin{tabular}{c}%
\input{resources/#2.tex}%
\end{tabular}%
}%
\end{table}%

}

\renewcommand{\newcommand}{\providecommand}


\section{Introduction}

The vocabulary mismatch between query and document poses a critical challenge in search \citep{Li2014}. The vocabulary gap occurs when documents and queries, represented as a bag-of-words, use different terms to describe the same concepts. While improved semantic matching methods are urgently needed, in order for these methods to be effective they need to be applicable at early stages of the retrieval pipeline. Otherwise, candidate documents most affected by the mismatch (i.e., relevant documents that do not contain any query terms) will simply remain undiscovered. \citet{Boytsov2016knn} show that (approximate) nearest neighbour algorithms \citep{Garcia2008fastknn,muja2014scalableknn} can be more efficient than classical term-based retrieval. This strongly motivates the design of semantic matching methods that represent queries and documents in finite-dimensional vector spaces.

Latent semantic models, such as \LSI{} \citep{Deerwester1990lsi}, fit in the finite-dimensional vector space paradigm needed for nearest neighbour retrieval. However, \LSI{} is known to retrieve non-relevant documents due to a lack of specificity \citep{Dumais95lsi}. The recent move towards learning word representations as part of neural language models \citep{Bengio2003} has shown impressive improvements in natural language processing (NLP) \citep{Collobert2011scratch,Mikolov2013word2vec,Graves2014speech}. Therefore, it is reasonable to explore these representation learning methods for information retrieval (IR) as well. Unfortunately, in the case of full text document retrieval, only few positive results have been obtained so far \citep{Craswell2016neuir}. We identify two causes for this shortfall.

First of all, IR tasks (e.g., document ranking) are fundamentally different from NLP tasks \citep{Craswell2016neuir}. NLP deals with natural language regularities (e.g., discovering long range dependencies), whereas IR involves satisfying a user's information need (e.g., matching a query to a document). Therefore, specialized solutions and architectures for IR are needed.

Secondly, in the bag-of-words paradigm, query/document matching is performed by counting term occurrences within queries and documents. Afterwards, the frequencies are adjusted using weighting schemes that favor term specificity \citep{Robertson1994bm25,Robertson2004idf}, used as parameter values in probabilistic frameworks \citep{Zhai2004smoothing} and/or reduced in dimensionality \citep{Blei2003,Deerwester1990lsi}. However, \citet{Baroni2014} noted that for NLP tasks, prediction-based models, learned from scratch using gradient descent, outperform count-based models. Similarly, the advances made by deep learning in computer vision \citep{Krizhevsky2012net} were not due to counting visual words \citep{Vidal2003sift} or by constructing complicated pipelines, but instead by optimizing a cost function using gradient descent and learning the model from scratch. However, for IR settings such as unsupervised news article retrieval, where one ranks documents in the absence of explicit or implicit relevance labels, prediction-based models have not received much attention. In order for deep learning to become feasible for unsupervised retrieval, we first need to construct an appropriate optimization objective.

In this \paper{} we introduce an optimization objective for learning latent representations of words and documents from scratch, in an unsupervised manner without relevance signals. Specifically, we introduce the \FullModelName{} (\ModelName{}) for document retrieval. The optimization objective of \ModelName{} mandates that word sequences extracted from a document should be predictive of that document. Learning a model of content in this way incorporates the following IR regularities:
\renewcommand{\paradescriptionlabel}[1]{\normalfont#1}%
\begin{inparadesc}%
	\item[semantic matching] (words occurring in each other's vicinity are learned to have similar representations), the
	\item[clustering hypothesis] (documents that contain similar language will have nearby representations in latent space), and
	\item[term specificity] (words associated with many documents will be neglected, as they have low predictive power).
\end{inparadesc}

One limitation of latent document vector spaces, including the \ModelName{} we introduce here, is that their asymptotic complexity is bounded by the number of documents (i.e., one vector for every document) \citep{Ai2016doc2veclm,VanGysel2016products,Chen2017corruption}. Consequently, the latent methods we consider in this \paper{} are only feasible to be constructed on document collections of medium scale. Therefore, we choose to evaluate our methods on article retrieval benchmarks (${\sim}200\text{k}$ to $500\text{k}$ documents each) from TREC \citep{Zhai2004smoothing,Harman1992tipster}.

\medskip
\noindent
The research questions we ask in this \paper{} towards answering \MainRQRef{4} are as follows:
\RQ{1}{\ResearchQuestionOne{}}
\RQ{2}{\ResearchQuestionTwo{}}
\RQ{3}{\ResearchQuestionThree{}}
\RQ{4}{\ResearchQuestionFour{}}


\section{Related work}

We refer to Section~\globalref{chapter:background:latent} of our background chapter (Chapter~\globalref{chapter:background}). Relevant to this \paper{} is the prior work on representation learning for document retrieval (\S\globalref{chapter:background:related:latent}) in addition to neural language modelling (\S\globalref{chapter:background:related:nlm}) for natural language processing and automatic speech recognition.

The contribution of this \paper{} over and above the related work discussed earlier is the following.
First, we substantially extend the \LSE{} model introduced in the previous chapter (Chapter~\globalref{chapter:research-05}) and evaluate it on document search collections from TREC \citep{TRECAdhoc}. Our improvements over the \LSE{} model are due to
\begin{inparaenum}[(1)]
	\item increased regularization, and
	\item accelerated training by reducing the internal covariate shift.
\end{inparaenum}

Second, we avoid \emph{irrational exuberance} known to plague the deep learning field \citep{Craswell2016neuir} by steering away from non-essential depth. That is, while we extend algorithms and techniques from deep learning, the model we present in this \paper{} is shallow.

Third, and contrary to previous work where information from pre-trained word embeddings is used to enrich the query/document representation of count-based models, our model, \ModelName{}, is learned directly from the document collection without explicit feature engineering. We show that \ModelName{} learns regularities known to be important for document retrieval from scratch. Our semantic vector space outperforms lexical retrieval models on some benchmarks. However, given that lexical and semantic models perform different types of matching, our approach is most useful as a supplementary signal to these lexical models.


\section{Learning semantic spaces}

In this section we provide the details of \ModelName{}. First, we give a birds-eye overview of the model and its parameters and explain how to rank documents for a given query. Secondly, we outline our training procedure and optimization objective. We explain the aspects of the objective that make it work in a retrieval setting. Finally, we go into the technical challenges that arise when implementing the model and how we solved them in our open-source release.

\subsection{The \FullModelName{}}

Our work revolves around unsupervised ad-hoc document retrieval where a user wishes to retrieve documents (e.g., articles) as to satisfy an information need encoded in query $\Query{}$. 

Below, a query $\Query{}$ consists of terms (i.e., words) $\QueryTerm{1}, \ldots, \QueryTerm{\Length{\Query{}}}$ originating from a vocabulary $\Vocabulary{}$, where $\Length{\cdot}$ denotes the length operator; $\Documents{}$ denotes the set of documents $\{\Document{}_1, \dots, \Document{}_{\Length{\Documents{}}}\}$. Every document $\Document{}_i \in \Documents{}$ consists of a sequence of words $\Word{}_{i,1}, \ldots, \Word{}_{i, \Length{\Document{}_i}}$ with $\Word{}_{i, j} \in \Vocabulary{}$, for all $1 \leq j \leq \Length{\Document{}_i}$.

We continue the work we started in Chapter~\globalref{chapter:research-05} \citep{VanGysel2016products} and learn low-dimensional representations of words, documents and the transformation between them from scratch. That is, instead of counting term frequencies for use in probabilistic frameworks \citep{Zhai2004smoothing} or applying dimensionality reduction to term frequency vectors \citep{Deerwester1990lsi,Wei2006lda}, we learn representations directly by gradient descent from sampled n-gram/document pairs extracted from the corpus. These representations are embedded parameters in matrices $\DocumentMatrix{} \in \mathbb{R}^{\Length{\Documents{}} \times \LatentDocumentDim{}}$ and $\WordMatrix{} \in \mathbb{R}^{\Length{\Vocabulary{}} \times \LatentWordDim{}}$ for documents $\Documents{}$ and vocabulary words $\Vocabulary{}$, respectively, such that $\vec{R}_\Vocabulary^{(i)}$ ($\vec{R}_\Documents^{(j)}$, respectively) denotes the $\LatentWordDim{}$-dimensional ($\LatentDocumentDim{}$-dim., respectively) vector representation of word $\Vocabulary{}_i$ (document $\Document{}_i$, respectively). 

As the word representations $\vec{R}_\Vocabulary^{(i)}$ and document representations $\vec{R}_\Documents^{(j)}$ are of different dimensionality, we require a transformation $\TransformFn{}: \mathbb{R}^\LatentWordDim{} \to \mathbb{R}^\LatentDocumentDim{}$ from the word feature space to the document feature space. In this \paper{} we take the transformation to be linear:
\begin{equation}
\label{eq:transform}
\TransformFn{}\left(\vec{x}\right) = \TransformMatrix{} \vec{x},
\end{equation}
where $\vec{x}$ is a $\LatentWordDim{}$-dimensional vector and $\TransformMatrix{}$ is a $\LatentDocumentDim{} \times \LatentWordDim{}$ parameter matrix that is learned using gradient descent (in addition to representation matrices $\WordMatrix{}$ and $\DocumentMatrix{}$).

\newcommand{\VerboseNGram}[1][\NGramSize{}]{\Word{}_1, \ldots, \Word{}_#1}

We compose a representation of a sequence of $\NGramSize{}$ words (i.e., an n-gram) $\VerboseNGram{}$ by averaging its constituent word representations:
\begin{equation}
\label{eq:averaging}
\CompositionalFn{}\left( \VerboseNGram{} \right) = \frac{1}{\NGramSize{}} \sum^{\NGramSize{}}_{i=1} 
\vec{R}_\Vocabulary^{(\Word{}_i)}.
\end{equation}
A query $\Query{}$ is projected to the document feature space by the composition of $\TransformFn{}$ and $\CompositionalFn{}$: $\Apply{\left(\TransformFn{} \circ \CompositionalFn{}\right)}{ \Query{}} = \Apply{\HashFn{}}{\Query{}}$. 

The matching score between a document $\Document{}$ and query $\Query{}$ is then given by the cosine similarity between their representations in document feature space:
\begin{equation}
\label{eq:score}
\Apply{\ScoreFn{}}{\Query{}, \Document{}} = \frac{\DotProduct{\Transpose{\Apply{\HashFn{}}{\Query{}}}}{{\vec{R}_\Documents^{(\Document)}}}}{ \EuclideanNorm{\Apply{\HashFn{}}{\Query{}}} \EuclideanNorm{\vec{R}_\Documents^{(\Document)}}} .
\end{equation}
We then proceed by ranking documents $\Document{} \in \Documents{}$ in decreasing order of $\Apply{\ScoreFn{}}{\Query{}, \Document{}}$ (Eq.~\ref{eq:score}) for a given query $\Query{}$. Note that the cosine similarity between two vectors is equivalent to their Euclidean distance if the vectors are normalized. Therefore, ranking according to Eq.~\ref{eq:score} can be formulated as an (approximate) nearest neighbour search problem in a metric space.

The model proposed here, \ModelName{}, is an extension of the \LSE{} model \citep{VanGysel2016products}. \ModelName{}/\LSE{} are different from existing unsupervised neural retrieval models learned from scratch due to their ability to scale to collections larger than expert finding collections \citep{VanGysel2016experts} (i.e., ${\sim}10\text{k}$ entities/documents) and the assumption that words and documents are embedded in different spaces \citep{Le2014}. Compared to word embeddings \citep{Mikolov2013word2vec,Pennington2014}, \ModelName{}/\LSE{} learn document-specific representations instead of collection-wide representations of language; see \S\globalref{research-05:section:methodology:model} for a more in-depth discussion. The algorithmic contribution of \ModelName{} over \LSE{} comes from improvements in the objective that we introduce in the next section.

\subsection{The objective and its optimization}
\label{sec:methodology:objective}

\newcommand{\Batch}{B}
\newcommand{\BatchSize}{m}
\newcommand{\BatchInstance}[1]{\Batch{}_{#1}}
\newcommand{\BatchInstancePhrase}[1]{\BatchInstance{#1}^{(p)}}
\newcommand{\BatchInstanceDocument}[1]{\BatchInstance{#1}^{(d)}}

\newcommand{\LossFn}{L}
\newcommand{\Parameters}{\Theta}

\newcommand{\Ideal}[1]{{#1}^*}

\newcommand{\NormFn}{\text{norm}}

We learn representations of words and documents using mini-batches of $\BatchSize{}$ n-gram/document pairs such that an n-gram representation---composed out of word representations---is projected nearby the document that contains the n-gram, similar to \LSE{} \citep{VanGysel2016products}. The word and n-gram representations, and the transformation between them are all learned simultaneously. This is in contrast to representing documents by a weighted sum of pre-trained representations of the words that it contains \citep{Vulic2015monolingual}. A mini-batch $\Batch{}$ is constructed as follows:
\begin{inparaenum}[(1)]
	\item Stochastically sample document $\Document{} \in \Documents{}$ according to $\Prob{\Documents{}}$. In this \paper{}, we assume $\Prob{\Documents{}}$ to be uniform, similar to \citep{Zhai2004smoothing}. Note that $\Prob{\Documents{}}$ can be used to incorporate importance-based information (e.g., document length).
	\item Sample a phrase of $\NGramSize{}$ contiguous words $\Word{}_1, \ldots{}, \Word{}_\NGramSize{}$ from document $\Document{}$.
	\item Add the phrase-document pair $\left(\Word{}_1, \ldots{}, \Word{}_\NGramSize{} ; \Document{} \right)$ to the batch.
	\item Repeat until the batch is full.
\end{inparaenum}

Given a batch $\Batch{}$, we proceed by constructing a differentiable, non-convex loss function $\LossFn{}$ of the parameters $\Parameters{}$ (e.g., $\WordMatrix{}, \DocumentMatrix{}$) and the parameter estimation problem is cast as an optimization problem that we approximate using stochastic gradient descent.

Denote $\BatchInstance{i}$ as the $i$-th n-gram/document pair of batch $\Batch{}$ and $\BatchInstancePhrase{i}$ ($\BatchInstanceDocument{i}$, respectively) as the n-gram (document, respectively) of pair $\BatchInstance{i}$. Further, we introduce an auxiliary function that L2-normalizes a vector of arbitrary dimensionality:
\begin{equation*}
\Apply{\NormFn{}}{\vec{x}} = \frac{\vec{x}}{\EuclideanNorm{\vec{x}}}.
\end{equation*}
\newcommand{\Projection}{T}%
\newcommand{\RawProjection}{\tilde{\Projection{}}}%
For an n-gram/document pair $\BatchInstance{i}$, the non-standardized projection of the n-gram into the $\LatentDocumentDim{}$-dimensional document feature space is as follows:
\begin{equation}
\label{eq:raw_projection}
\Apply{\RawProjection{}}{\BatchInstancePhrase{i}} = \Apply{\left(\TransformFn{} \circ \NormFn{} \circ \CompositionalFn{}\right)}{\BatchInstancePhrase{i}}.
\end{equation}
A few quick comments are in order. The function $\CompositionalFn{}$ (see Eq.~\ref{eq:averaging}) constructs an n-gram representation by averaging the word representations (embedded in the $\WordMatrix{}$ parameter matrix). This allows the model to learn semantic relations between words for the purpose of \textbf{semantic matching}. That is, the model does not learn from individual words, but instead it learns from an unordered sequence (order is not preserved as we sum in Eq.~\ref{eq:averaging}) of words that constitute meaning. As documents contain multiple n-grams and n-grams are made up of multiple words, semantic similarity between words and documents is learned. In addition, the composition function $\CompositionalFn{}$ in combination with L2-normalization $\Apply{\NormFn{}}{\cdot}$ causes words to compete in order to contribute to the resulting n-gram representation. Given that we will optimize the $\NGramSize{}$-gram representation to be close to the corresponding document (as we will explain below), words that are discriminative for the document in question will learn to contribute more to the n-gram representation (due to their discriminative power), and consequently, the L2-norm of the representations of discriminative words will be larger than the L2-norm of non-discriminative words. This incorporates a notion of \textbf{term specificity} into our model.

\newcommand{\BatchMean}{\ApplySquare{\SampleExpectation{}}{\Apply{\RawProjection{}}{\BatchInstancePhrase{i}}}}
\newcommand{\BatchVariance}{\ApplySquare{\SampleVariance{}}{\Apply{\RawProjection{}}{\BatchInstancePhrase{i}}}}

We then estimate the per-feature sample mean and variance
\[
\BatchMean{} 
\text{ and }
\BatchVariance{}
\]
over batch $\Batch{}$.
The standardized projection of n-gram/document pair $\BatchInstance{i}$ can then be obtained as follows:
\begin{equation}
\label{eq:projection}
\NGramProjection{} = \Apply{\HardTanH}{\frac{\Apply{\RawProjection{}}{\BatchInstancePhrase{i}} - \BatchMean{}}{\sqrt{\BatchVariance{}}} + \ErrorBias{}},
\end{equation}
where $\ErrorBias{}$ is a $\LatentDocumentDim{}$-dimensional bias vector parameter that captures doc\-ument-independent regularities corresponding to word frequency. While vector $\ErrorBias{}$ is learned during training, it is ignored during prediction (i.e., a nuisance parameter) similar to the position bias in click models \citep{Chapelle2009dbn} and score bias in learning-to-rank \citep{Joachims2002svm}. The standardization operation reduces the internal covariate shift \citep{Ioffe15bn}. That is, it avoids the complications introduced by changes in the distribution of document feature vectors during learning. In addition, as the document feature vectors are learned from scratch as well, the standardization forces the learned feature vectors to be centered around the null vector. Afterwards, we apply the hard-saturating nonlinearity $\HardTanH{}$ \citep{Gulcehre2016noisy} such that the feature activations are between $-1$ and $1$.

The objective is to maximize the similarity between
\begin{equation*}
\NGramProjection{}\text{ and }\vec{R}_\Documents^{(B_i^{(\Document)})}, 
\end{equation*}
while minimizing the similarity between $\NGramProjection{}$ and the representations of other documents. Therefore, a document is characterized by the concepts it contains, and consequently, documents describing similar concepts will cluster together, as postulated by the \textbf{clustering hypothesis} \citep{Rijsbergen1979}. \ModelName{} strongly relies on the clustering hypothesis as ranking is performed according to nearest neighbour retrieval (Eq.~\ref{eq:score}).

Considering the full set of documents $\Documents{}$ is often costly as $\Length{\Documents{}}$ can be large. Therefore, we apply an adjusted-for-bias variant of negative sampling \citep{VanGysel2016products,Mikolov2013compositionality,Gutmann2010}, where we uniformly sample negative examples from $\Documents{}$. Adopted from \citep{VanGysel2016products}, we define
\newcommand{\SimilarityRandomVariable}{\mathcal{S}}
\newcommand{\SigmoidFn}{\sigma}
\newcommand{\ExpFn}{\text{exp}}
\newcommand{\ProbSimilarEntity}[1]{\CondProb{\SimilarityRandomVariable{}}{#1, \BatchInstancePhrase{i}}}
\newcommand{\ProbMassEntity}[1]{\Apply{\SigmoidFn{}}{\DotProduct{#1}{\NGramProjection{}}}}
\newcommand{\NumNegativeExamples}{z}
\begin{equation}
\ProbSimilarEntity{\Document{}} = 
\sigma\left(\vec{R}_\Documents^{(\Document)}\cdot \NGramProjection{} \right)
\end{equation}
as the similarity of two representations in latent vector space, where
\begin{equation*}
\Apply{\SigmoidFn{}}{t} = \frac{1}{1 + \Apply{\ExpFn{}}{-t}}
\end{equation*}
denotes the sigmoid function and $\SimilarityRandomVariable{}$ is an indicator binary random variable that says whether the representation of document $\Document{}$ is similar to the projection of n-gram $\BatchInstancePhrase{i}$.

The probability of document $\BatchInstanceDocument{i}$ given phrase $\BatchInstancePhrase{i}$ is then approximated by uniformly sampling $\NumNegativeExamples{}$ contrastive examples:
\begin{eqnarray}
\label{eq:instance_loss}
\lefteqn{\log\CondApproxProb{\BatchInstanceDocument{i}}{\BatchInstancePhrase{i}} = \qquad\qquad} \\
& & \frac{\NumNegativeExamples{} + 1}{2 \NumNegativeExamples{}} \Bigg( \NumNegativeExamples{} \log\ProbSimilarEntity{\BatchInstanceDocument{i}} + \nonumber\\
& & \sum^{\NumNegativeExamples{}}_{\substack{k=1, \\ \Document{}_k \sim U(\Documents{})}} \log \left( 1.0 - \ProbSimilarEntity{\Document{}_k} \right) \Bigg), \nonumber
\end{eqnarray}
where $U(\Documents{})$ denotes the uniform distribution over documents $\Documents{}$, the distribution used for obtaining negative examples \citep{VanGysel2016products,Gutmann2010}. Then, the \textbf{loss function} we use to optimize our model is Eq.~\ref{eq:instance_loss} averaged over the instances in batch $\Batch{}$:
\begin{eqnarray}
\lefteqn{\Apply{\LossFn{}}{\WordMatrix{}, \DocumentMatrix{}, \TransformMatrix{}, \ErrorBias{} \mid \Batch{}}} \\
& = & \frac{1}{\BatchSize{}} \sum_{i=1}^{\BatchSize{}} \log\CondApproxProb{\BatchInstanceDocument{i}}{\BatchInstancePhrase{i}} \nonumber \\ \nonumber
& &\quad{}+ \frac{\lambda}{2 \BatchSize{}} \left( \sum_{i,j} \WordMatrix{}^2_{i,j} + \sum_{i,j} \DocumentMatrix{}^2_{i,j} + \sum_{i,j} \TransformMatrix{}^2_{i,j} \right),
\end{eqnarray}
where $\lambda$ is a weight regularization hyper-parameter. We optimize our parameters $\theta{}$ ($\WordMatrix{}$, $\DocumentMatrix{}$, $\TransformMatrix{}$ and $\ErrorBias{}$) using Adam~\citep{Kingma2014adam}, a first-order gradient-based optimization function for stochastic objective functions that is very similar to momentum. The update rule for parameter $\theta{}$ given a batch $\Batch{}$ at batch update time $t$ equals:
\newcommand{\FirstMoment}[1]{\hat{m}_{#1}^{(t)}}
\newcommand{\SecondMoment}[1]{\hat{v}_{#1}^{(t)}}
\begin{equation}
\label{eq:adam}
\theta{}^{(t + 1)} \leftarrow \theta{}^{(t)} - \alpha \frac{\FirstMoment{\theta{}}}{\sqrt{\SecondMoment{\theta{}}} + \epsilon},
\end{equation}
where $\FirstMoment{\theta{}}$ and $\SecondMoment{\theta{}}$, respectively, are the first and second moment estimates (over batch update times) \citep{Kingma2014adam} of the gradient of the loss $\Apply{\frac{\partial \LossFn}{\partial \theta{}}}{\WordMatrix{}, \DocumentMatrix{}, \TransformMatrix{}, \ErrorBias{} \mid \Batch{}}$ w.r.t.\ parameter $\theta{}$ at batch update time $t$ and $\epsilon = 10^{-8}$ is a constant to ensure numerical stability. The use of this optimization method causes every parameter to be updated with every batch, unlike regular stochastic gradient descent, where the only parameters that are updated are those for which there is a non-zero gradient of the loss. This is important in \ModelName{} due to the large number of word and document vectors.

The algorithmic contributions of \ModelName{} over the \LSE{} model that we introduced in the previous \paper{} are the components of the objective mentioned next. Eq.~\ref{eq:raw_projection} forces individual words to compete in order to contribute to the resulting n-gram representation. Consequently, non-discriminative words will have a small L2-norm. In Eq.~\ref{eq:projection} we perform standardization to reduce the internal covariate shift \citep{Ioffe15bn}. In addition, the standardization forces n-gram representations to distinguish themselves only in the dimensions that matter for matching. Frequent words are naturally prevalent in n-grams, however, they have low discriminative power as they are non-specific. The bias $\ErrorBias{}$ captures word frequency regularities that are non-discriminative for the semantic concepts within the respective n-gram/document pairs and allows the transformation in Eq.~\ref{eq:transform} to focus on concept matching. The re-weighting of the positive instance in Eq.~\ref{eq:instance_loss} removes a dependence on the number of negative examples $\NumNegativeExamples{}$ where a large $\NumNegativeExamples{}$ presented the model with a bias towards negative examples.

\subsection{Implementation}
\label{sec:impl}

The major cause of technical challenges of the \ModelName{} training procedure is not due to time complexity, but rather space restrictions. This is because we mitigate expensive computation by estimating vector space models using graphics processing units (GPUs). The main limitation of these massively-parallel computation devices is that they rely on their own memory units. Consequently, parameters and intermediate results of the training procedure need to persist in the GPU memory space. The asymptotic space complexity of the parameters equals:
\begin{equation*}
\Apply{O}{\underbrace{\Length{\Vocabulary{}} \times \LatentWordDim{}}_{\text{word representations } \WordMatrix{}} + \underbrace{\LatentDocumentDim{} \times \LatentWordDim{}}_{\text{transform } \TransformMatrix{}} + \underbrace{\Length{\Documents{}} \times \LatentDocumentDim{}}_{\text{document representations } \DocumentMatrix{}}}.
\end{equation*}
In addition, Eq.~\ref{eq:adam} requires us to keep the first and second moment of the gradient over time for every parameter in memory. Therefore, for every parameter, we retain three floating point values in memory at all times.
For example, if we have a collection of \numprint{1}M documents (256-dim.)\ with a vocabulary of \numprint{64}K terms (300-dim.), then the model has ${\sim}$\numprint{275}M parameters. Consequently, under the assumption of 32-bit floating point, the resident memory set has a best-case least upper bound of \numprint{3.30}GB memory. The scalability of our method---similar as with all latent vector space models---is determined by the number of documents within the collection. However, the current generation of GPUs---that typically boast around \numprint{12}GB of memory---can comfortably be used to train models of collections consisting of up to 2 million documents. In fact, next-generation GPUs have double the amount of memory---\numprint{24}GB---and this amount will likely increase with the introduction of future processing units \citep{Wiki2017nvidia}. This, and the development of distributed GPU technology \citep{Tensorflow2015whitepaper}, leads us to believe that the applicability of our approach to larger retrieval domains is simply a matter of time \citep{Moore1998cramming}.

In addition to the scarcity of memory on current generation GPUs, operations such as the averaging of word to n-gram representations (Eq.~\ref{eq:averaging}) can be performed in-place. However, these critical optimizations are not available in general-purpose machine learning toolkits. Therefore, we have implemented the \ModelName{} training procedure directly in C++/CUDA, such that we can make efficient use of sparseness and avoid unnecessary memory usage. In addition, models are stored in the open HDF5 format \citep{Folk2011hdf5} and the toolkit provides a Python module that can be used to query trained \ModelName{} models on the CPU. This way, a trained \ModelName{} can easily be integrated in existing applications. The toolkit is licensed under the permissive MIT open-source license.\footnote{\paperImplementationUrl{}\label{fn:ImplementationURL}}


\section{Experimental setup}
\label{sec:experimental}

\subsection{Research questions}

\newcommand{\RQ}[2]{
    \begin{description}[topsep=0pt,leftmargin=0.8cm]
    \phantomsection\label{section:setup:rq#1}
    \item[RQ#1] #2
    \end{description}
}

\newcommand{\RQRef}[1]{\textbf{\hyperref[section:setup:rq#1]{RQ#1}}}

In this \paper{} we investigate the viability of neural representation learning methods for semantic matching in document search. As indicated in the introduction of this \paper{}, we seek to answer the following research questions:

\RQ{1}{\ResearchQuestionOne{}}

\noindent%
In particular, how does it compare to other methods that represent queries/documents as low-dimensional vectors? What is the difference in performance with purely lexical models that perform exact term matching and represent queries/documents as bag-of-words vectors?

\RQ{2}{\ResearchQuestionTwo{}}

\noindent%
Does \ModelName{} improve over other retrieval models only on a handful of queries? Instead of computing averages over queries, what if we look at the pairwise differences between rankers?

\RQ{3}{\ResearchQuestionThree{}}

\noindent%
Can we use the differences that we observe in \RQRef{2} between \ModelName{}, \QLM{} and other latent vector space models to our advantage to improve retrieval performance? Can we pinpoint where the improvements of \ModelName{} come from?

\RQ{4}{\ResearchQuestionFour{}}

\noindent%
Traditional retrieval models such as generative language models are known to incorporate corpus statistics regarding term specificity \citep{Sparck1972specificity,Robertson2004idf} and document length \citep{Zhai2004smoothing}. Are similar corpus statistics present in \ModelName{} and what does this tell us about the ranking task?

\subsection{Benchmark datasets \& experiments}
\label{sec:benchmarks}

In this \paper{} we are interested in query/document matching. Therefore, we evaluate \ModelName{} on newswire article collections from TREC. Other retrieval domains, such as web search or social media, deal with various aspects, such as document freshness/importance and social/hyperlink/click graphs, that may obfuscate the impact of matching queries to documents.
Therefore, we follow the experimental setup of \citet{Zhai2004smoothing} and use four article retrieval sub-collections from the TIPSTER corpus \citep{Harman1992tipster}: \BenchmarkAP{} (\ShortBenchmarkAP{}), \BenchmarkFT{} (\ShortBenchmarkFT{}), \BenchmarkLATimes{} (\ShortBenchmarkLATimes{}) and \BenchmarkWSJ{} (\ShortBenchmarkWSJ{}) \citep{Harman1993document}. In addition, we consider the \BenchmarkRobust{} collection that constitutes of Disk 4/5 of the TIPSTER corpus without the Congressional Record and the \BenchmarkNYT{} collection that consists of articles written and published by the New York Times between 1987 and 2007. For evaluation, we take topics \numprint{50}--\numprint{200} from TREC 1--3\footnote{We only consider judgments corresponding to each of the sub-collections.\label{footnote:subcollections}} (\ShortBenchmarkAP{}, \ShortBenchmarkWSJ{}), topics \numprint{301}--\numprint{450} from TREC 6--8\footnotemark[\thefootnote]{} (\ShortBenchmarkFT{}, \ShortBenchmarkLATimes{}) \cite{TRECAdhoc}, topics \numprint{301}--\numprint{450}, \numprint{601}--\numprint{700} from \BenchmarkRobust{} \citep{Voorhees2005robust} and the 50 topics assessed by NIST (a subset of the \BenchmarkRobust{} topics judged for the \ShortBenchmarkNYT{} collection) during the TREC 2017 Common Core Track \citep{Allan2017treccommon}. From each topic we take its title as corresponding query. Topics without relevant documents are filtered out. We randomly create a split\footnote{The validation/test splits can be found at \paperImplementationUrl{}.} of validation (20\%) and test (80\%) queries (with the exception of the \ShortBenchmarkNYT{} collection). For the \ShortBenchmarkNYT{} collection, we select the hyperparameter configuration that optimizes the \BenchmarkRobust{} validation set on the \BenchmarkRobust{} collection and we take the 50 queries assessed by NIST and their judgments---specifically created for the \ShortBenchmarkNYT{} collection---as our test set. This way, method hyperparameters are optimized on the validation set (as described in Section~\ref{sec:retrievalmodels}) and the retrieval performance is reported on the test set; see Table~\ref{tbl:adhoc_stats}.

The inclusion of early TREC collections (\ShortBenchmarkAP{}, \ShortBenchmarkFT{}, \ShortBenchmarkLATimes{} and \ShortBenchmarkWSJ{}) is motivated by the fact that during the first few years of TREC, there was a big emphasis on submissions where the query was constructed manually from each topic and interactive feedback was used \citep{Harman1993document}. That is, domain experts repeatedly formulated manual queries using the full topic (title, description, narrative), observed the obtained rankings and then reformulated their query in order to obtain better rankings. Consequently, these test collections are very useful when evaluating semantic matches as relevant documents do not necessarily contain topic title terms. From TREC-5 and onwards, less emphasis was put on rankings generated using interactive feedback and shifted towards automated systems only \citep[footnote 1]{Harman1996trec5}. In fact, within the \BenchmarkRobust{} track, only automated systems were submitted \citep[\S2]{Voorhees2005robust} due to the large number of (a) documents ($\sim$500K) and (b) topics (250). In the 2017 Common Core Track \citep{Allan2017treccommon}, interactive rankings were once again submitted by participants, in addition to rankings obtained by the latent vector space model presented in this \paper{}.

\newcommand{\Domain}[3]{
\subsubsection{#2}
\label{sec:#1}
#3}

\newcommand{\DomainStatsTable}[1]{
\begin{table*}[th]
\centering
\renewcommand{\arraystretch}{1.2}

\caption{Overview of the retrieval benchmarks. T and V denote the test and validation sets, respectively. Arithmetic mean and standard deviation are reported wherever applicable.\label{tbl:#1_stats}}

\scalebox{0.85}{%
\centering%
\begin{tabular}{c}%
\input{resources/#1_stats.tex}%
\end{tabular}}
\end{table*}%
}

\newcommand{\CF}[1]{\text{CF}_{#1}}
\newcommand{\TF}[1]{\text{TF}_{#1}}
\newcommand{\IDF}[1]{\text{IDF}_{#1}}
{
\renewcommand{\BenchmarkAP}{\ShortBenchmarkAP}
\renewcommand{\BenchmarkWSJ}{\ShortBenchmarkWSJ}
\renewcommand{\BenchmarkLATimes}{\ShortBenchmarkLATimes}
\renewcommand{\BenchmarkFT}{\ShortBenchmarkFT}
\renewcommand{\BenchmarkNYT}{\ShortBenchmarkNYT}

\begin{table*}[th]
\centering
\renewcommand{\arraystretch}{1.2}

\caption{Overview of the retrieval benchmarks. T and V denote the test and validation sets, respectively. Arithmetic mean and standard deviation are reported wherever applicable.\label{tbl:adhoc_stats}}

\scalebox{0.85}{%
\centering%
\begin{tabular}{c}%
\begin{tabular}{lccc}%
\toprule%
& \BenchmarkAP{} & \BenchmarkFT{} & \BenchmarkLATimes{} \\%
\midrule%
\textbf{Collection} (training) \\%
Documents & \phantom{}\numprint{164597}  & \phantom{}\numprint{210158}  & \phantom{}\numprint{131896} \\%
Document length & {\renewcommand{\arraystretch}{0.8}\setlength{\tabcolsep}{8pt}\begin{tabular}[t]{ll}&\phantom{}\nprounddigits{2}\npdecimalsign{.}\numprint{461.6340698797704}$\,\pm\,$\phantom{}\nprounddigits{2}\npdecimalsign{.}\numprint{243.61286335206674}  \\ \end{tabular}} & {\renewcommand{\arraystretch}{0.8}\setlength{\tabcolsep}{8pt}\begin{tabular}[t]{ll}&\phantom{}\nprounddigits{2}\npdecimalsign{.}\numprint{399.6802120309481}$\,\pm\,$\phantom{}\nprounddigits{2}\npdecimalsign{.}\numprint{366.4147351756061}  \\ \end{tabular}} & {\renewcommand{\arraystretch}{0.8}\setlength{\tabcolsep}{8pt}\begin{tabular}[t]{ll}&\phantom{}\nprounddigits{2}\npdecimalsign{.}\numprint{502.1807333050321}$\,\pm\,$\phantom{}\nprounddigits{2}\npdecimalsign{.}\numprint{519.5806529063212}  \\ \end{tabular}}\\%
Unique terms & \nprounddigits{2}\npdecimalsign{.}\numprint{\xintFloat [3]{267318}}  & \nprounddigits{2}\npdecimalsign{.}\numprint{\xintFloat [3]{305310}}  & \nprounddigits{2}\npdecimalsign{.}\numprint{\xintFloat [3]{267156}} \\%
\textbf{Queries} (testing) & {\renewcommand{\arraystretch}{1.0}\begin{tabular}[t]{ll} &(T) \phantom{}\numprint{119}  \\  &(V)\phantom{0}\numprint{30}  \\ \end{tabular}} & {\renewcommand{\arraystretch}{1.0}\begin{tabular}[t]{ll} &(T) \phantom{}\numprint{116}  \\  &(V)\phantom{0}\numprint{28}  \\ \end{tabular}} & {\renewcommand{\arraystretch}{1.0}\begin{tabular}[t]{ll} &(T) \phantom{}\numprint{113}  \\  &(V)\phantom{0}\numprint{30}  \\ \end{tabular}}\\%
Query terms & {\renewcommand{\arraystretch}{0.8}\setlength{\tabcolsep}{8pt}\begin{tabular}[t]{ll}\phantom{(T)}&\phantom{00}\nprounddigits{2}\npdecimalsign{.}\numprint{5.06}$\,\pm\,$\phantom{00}\nprounddigits{2}\npdecimalsign{.}\numprint{3.14}  \\ \end{tabular}} & {\renewcommand{\arraystretch}{0.8}\setlength{\tabcolsep}{8pt}\begin{tabular}[t]{ll}\phantom{(T)}&\phantom{00}\nprounddigits{2}\npdecimalsign{.}\numprint{2.50}$\,\pm\,$\phantom{00}\nprounddigits{2}\npdecimalsign{.}\numprint{0.69}  \\ \end{tabular}} & {\renewcommand{\arraystretch}{0.8}\setlength{\tabcolsep}{8pt}\begin{tabular}[t]{ll}\phantom{(T)}&\phantom{00}\nprounddigits{2}\npdecimalsign{.}\numprint{2.48}$\,\pm\,$\phantom{00}\nprounddigits{2}\npdecimalsign{.}\numprint{0.69}  \\ \end{tabular}}\\%
Relevant documents & {\renewcommand{\arraystretch}{0.8}\setlength{\tabcolsep}{8pt}\begin{tabular}[t]{ll}(T)&\phantom{}\nprounddigits{2}\npdecimalsign{.}\numprint{111.45}$\,\pm\,$\phantom{}\nprounddigits{2}\npdecimalsign{.}\numprint{136.01}  \\ (V)&\phantom{0}\nprounddigits{2}\npdecimalsign{.}\numprint{86.43}$\,\pm\,$\phantom{0}\nprounddigits{2}\npdecimalsign{.}\numprint{72.63}  \\ \end{tabular}} & {\renewcommand{\arraystretch}{0.8}\setlength{\tabcolsep}{8pt}\begin{tabular}[t]{ll}(T)&\phantom{0}\nprounddigits{2}\npdecimalsign{.}\numprint{34.91}$\,\pm\,$\phantom{0}\nprounddigits{2}\npdecimalsign{.}\numprint{42.94}  \\ (V)&\phantom{0}\nprounddigits{2}\npdecimalsign{.}\numprint{30.46}$\,\pm\,$\phantom{0}\nprounddigits{2}\npdecimalsign{.}\numprint{26.97}  \\ \end{tabular}} & {\renewcommand{\arraystretch}{0.8}\setlength{\tabcolsep}{8pt}\begin{tabular}[t]{ll}(T)&\phantom{0}\nprounddigits{2}\npdecimalsign{.}\numprint{24.83}$\,\pm\,$\phantom{0}\nprounddigits{2}\npdecimalsign{.}\numprint{34.31}  \\ (V)&\phantom{0}\nprounddigits{2}\npdecimalsign{.}\numprint{24.30}$\,\pm\,$\phantom{0}\nprounddigits{2}\npdecimalsign{.}\numprint{21.19}  \\ \end{tabular}}\\%
\\%
\midrule%
\end{tabular}%
\\%
\begin{tabular}{lccc}%
& \BenchmarkNYT{} & \BenchmarkRobust{} & \BenchmarkWSJ{} \\%
\midrule%
\textbf{Collection} (training) \\%
Documents & \phantom{}\numprint{1855658}  & \phantom{}\numprint{528155}  & \phantom{}\numprint{173252} \\%
Document length & {\renewcommand{\arraystretch}{0.8}\setlength{\tabcolsep}{8pt}\begin{tabular}[t]{ll}&\phantom{}\nprounddigits{2}\npdecimalsign{.}\numprint{572.1752801432273}$\,\pm\,$\phantom{}\nprounddigits{2}\npdecimalsign{.}\numprint{605.8240047996469}  \\ \end{tabular}} & {\renewcommand{\arraystretch}{0.8}\setlength{\tabcolsep}{8pt}\begin{tabular}[t]{ll}&\phantom{}\nprounddigits{2}\npdecimalsign{.}\numprint{479.7217653908449}$\,\pm\,$\phantom{}\nprounddigits{2}\npdecimalsign{.}\numprint{869.2695174032617}  \\ \end{tabular}} & {\renewcommand{\arraystretch}{0.8}\setlength{\tabcolsep}{8pt}\begin{tabular}[t]{ll}&\phantom{}\nprounddigits{2}\npdecimalsign{.}\numprint{447.5146491815376}$\,\pm\,$\phantom{}\nprounddigits{2}\npdecimalsign{.}\numprint{454.7468232217638}  \\ \end{tabular}}\\%
Unique terms & \nprounddigits{2}\npdecimalsign{.}\numprint{\xintFloat [3]{1351413}}  & \nprounddigits{2}\npdecimalsign{.}\numprint{\xintFloat [3]{782799}}  & \nprounddigits{2}\npdecimalsign{.}\numprint{\xintFloat [3]{250164}} \\%
\textbf{Queries} (testing) & {\renewcommand{\arraystretch}{1.0}\begin{tabular}[t]{ll} &(T) \phantom{0}\numprint{50}  \\  & \\ \end{tabular}} & {\renewcommand{\arraystretch}{1.0}\begin{tabular}[t]{ll} &(T) \phantom{}\numprint{200}  \\  &(V)\phantom{0}\numprint{49}  \\ \end{tabular}} & {\renewcommand{\arraystretch}{1.0}\begin{tabular}[t]{ll} &(T) \phantom{}\numprint{120}  \\  &(V)\phantom{0}\numprint{30}  \\ \end{tabular}}\\%
Query terms & {\renewcommand{\arraystretch}{0.8}\setlength{\tabcolsep}{8pt}\begin{tabular}[t]{ll}\phantom{(T)}&\phantom{00}\nprounddigits{2}\npdecimalsign{.}\numprint{6.58}$\,\pm\,$\phantom{00}\nprounddigits{2}\npdecimalsign{.}\numprint{0.70}  \\ \end{tabular}} & {\renewcommand{\arraystretch}{0.8}\setlength{\tabcolsep}{8pt}\begin{tabular}[t]{ll}\phantom{(T)}&\phantom{00}\nprounddigits{2}\npdecimalsign{.}\numprint{5.28}$\,\pm\,$\phantom{00}\nprounddigits{2}\npdecimalsign{.}\numprint{0.74}  \\ \end{tabular}} & {\renewcommand{\arraystretch}{0.8}\setlength{\tabcolsep}{8pt}\begin{tabular}[t]{ll}\phantom{(T)}&\phantom{00}\nprounddigits{2}\npdecimalsign{.}\numprint{5.05}$\,\pm\,$\phantom{00}\nprounddigits{2}\npdecimalsign{.}\numprint{3.14}  \\ \end{tabular}}\\%
Relevant documents & {\renewcommand{\arraystretch}{0.8}\setlength{\tabcolsep}{8pt}\begin{tabular}[t]{ll}(T)&\phantom{}\nprounddigits{2}\npdecimalsign{.}\numprint{180.04}$\,\pm\,$\phantom{}\nprounddigits{2}\npdecimalsign{.}\numprint{132.74}  \\  \\ \end{tabular}} & {\renewcommand{\arraystretch}{0.8}\setlength{\tabcolsep}{8pt}\begin{tabular}[t]{ll}(T)&\phantom{0}\nprounddigits{2}\npdecimalsign{.}\numprint{70.33}$\,\pm\,$\phantom{0}\nprounddigits{2}\npdecimalsign{.}\numprint{73.68}  \\ (V)&\phantom{0}\nprounddigits{2}\npdecimalsign{.}\numprint{68.27}$\,\pm\,$\phantom{0}\nprounddigits{2}\npdecimalsign{.}\numprint{77.41}  \\ \end{tabular}} & {\renewcommand{\arraystretch}{0.8}\setlength{\tabcolsep}{8pt}\begin{tabular}[t]{ll}(T)&\phantom{0}\nprounddigits{2}\npdecimalsign{.}\numprint{96.99}$\,\pm\,$\phantom{0}\nprounddigits{2}\npdecimalsign{.}\numprint{93.30}  \\ (V)&\phantom{}\nprounddigits{2}\npdecimalsign{.}\numprint{101.93}$\,\pm\,$\phantom{}\nprounddigits{2}\npdecimalsign{.}\numprint{117.65}  \\ \end{tabular}}\\%
\\%
\bottomrule%
\end{tabular}%
\\%
\end{tabular}}
\end{table*}%

}

We address \RQRef{1} by comparing \ModelName{} to latent retrieval models (detailed in \S\ref{sec:retrievalmodels}). In addition, we perform a per-query pairwise comparison of methods where we look at what method performs best for each query in terms of \MAPCut{} (\RQRef{2}). A method performs better or worse than another if the absolute difference in \MAPCut{} exceeds $\delta = 0.01$; otherwise, the two methods perform similar. To address \RQRef{3}, we consider the combinations (\S\ref{sec:ltr}) of \QLM{} with \ModelName{} and the strongest latent vector space baseline of \RQRef{1}. That is, \FullWordToVec{} where the summands are weighted using self-information. In addition, we look at the correlation between per-query \TitlestatRelevant{} (see \S\ref{sec:evaluationmeasures}) and the pairwise differences in \MAPCut{} between \ModelName{} and all the other retrieval models. A positive correlation indicates that \ModelName{} is better at lexical matching than the other method, and vice versa for a negative correlation. For \RQRef{4}, we examine the relation between the collection frequency $\CF{\Word{}}$ and the L2-norm of their word embeddings $\Apply{\CompositionalFn{}}{\Word{}}$ for all terms $\Word{} \in \Vocabulary{}$.

\subsection{Retrieval models considered for comparison}
\label{sec:retrievalmodels}

The document collection is first indexed by Indri\footnote{Stopwords are removed using the standard stopword list of Indri.} \citep{Strohman2005indri}. Retrieval models not implemented by Indri access the underlying tokenized document collection using \texttt{pyndri} \citep{VanGysel2017pyndri}. This way, all methods compared in this \paper{} parse the text collection consistently. 

\subsubsection{Models compared}

The key focus of this \paper{} is the alleviation of the vocabulary gap in information retrieval and consequently, in theory, we score all documents in each collection for every query. In practice, however, we rely on nearest neighbor search algorithms to retrieve the top-k documents \citep{Boytsov2016knn}. Note that this is in contrast to many other semantic matching methods \citep{Zuccon2015nntm,Ai2016doc2veclm,Nalisnick2016desm} that have only been shown to perform well in document re-ranking scenarios where an initial pool of candidate documents is retrieved using a lexical matching method. However, candidate documents most affected by the vocabulary gap (i.e., relevant documents that do not contain any query terms) will simply remain undiscovered in a re-ranking scenario and consequently we compare \ModelName{} only to latent vector space models that can be queried using a nearest neighbor search.

The following latent vector space models are compared:
\begin{enumerate}
	\item \FullDocToVec{} (\DocToVec{}) \citep{Le2014} with the distributed memory architecture. The pre-processing of document texts to learn latent document representations is a topic of study by itself and its effects are outside the scope of this work. Consequently, we disable vocabulary filtering and frequent word subsampling in order to keep the input to all representation learning algorithms consistent. We sweep the one-sided window size and the embedding size respectively in partitions $\{x / 2 \mid x = 4, 6, 8, 10, 12, 16, 24, 32\}$ and $\{64, 128, 256\}$ on the validation set. Models are trained for \numprint{15} iterations on the validation set and we select the model iteration that performs best on the validation set. Documents are ranked in decreasing order of the cosine similarity between the document representation and the average of the word embeddings in the query.
	\item \FullWordToVec{} (\WordToVec{}) \citep{Mikolov2013word2vec,Vulic2015monolingual} with the Skip-Gram architecture. We follow the method introduced by \citet{Vulic2015monolingual} where query/document representations are constructed by composing the representations of the words contained within them. We consider both the unweighted sum (\CombineAdd{}) and the sum of vectors weighted by the term's self-information (\CombineSelfInformation{}). Self-information is a term specificity measure similar to Inverse Document Frequency (IDF) \citep{Cover2012informationtheory}. The hyperparameters of \FullWordToVec{} are swept in the same manner as \FullDocToVec{}.
	\item Latent Semantic Indexing (\LSI{}) \citep{Deerwester1990lsi} with TF-IDF weighting and the number of topics $K \in \{64, 128, 256\}$ optimized on the validation set.
	\item Latent Dirichlet Allocation (\LDA{}) \citep{Blei2003} with $\alpha = \beta = 0.1$ and the number of topics $K \in \{64, 128, 256\}$ optimized on the validation set. We train the model for 100 iterations or until topic convergence is achieved. Documents are ranked in decreasing order of the cosine similarity between the query topic distribution and the document topic distribution.
	\item Representation learning methods \LSE{} \citep{VanGysel2016products} and \ModelName{} (this \paper{}). For hyperparameters, we largely follow the findings of \citep{VanGysel2016products}: word representation dim. $\LatentWordDim{} = 300$, number of negative examples $\NumNegativeExamples{} = 10$, learning rate $\alpha = 0.001$, regularization lambda $\lambda = 0.01$. For \LSE{}, batch size $\BatchSize{} = 4096$ (as in \citep{VanGysel2016products}), while for \ModelName{} the batch size $\BatchSize{} = 51200$ (empirically determined on a holdout document collection that we did not include in this \paper{}). The dimensionality of the document representations $\LatentDocumentDim{} \in \{64, 128, 256\}$ and the n-gram size $\NGramSize{} \in \{4, 6, 8, 10, 12, 16, 24, 32\}$ are optimized on the validation set. Similar to \DocToVec{}, models are trained for \numprint{15} iterations on the training set and we select the model iteration that performs best on the validation set; a single iteration consists of $\ceil{\frac{1}{\BatchSize{}} \sum_{\Document{} \in \Documents{}} \left(\Length{\Document{}} - \NGramSize{} + 1 \right)}$ batches.
\end{enumerate}

In addition, we consider lexical language models (\QLM{}) \citep{Zhai2004smoothing} using the Indri engine with both Dirichlet (\Dirichlet{}) and Jelinek-Mercer (\JelinekMercer{}) smoothing; smoothing hyperparameters $\mu \in \splitatcommas{\{125, 250, 500, 750, 1000, 2000, 3000, 4000, 5000\}}$ and $\lambda \in \{ x \mid k \in \mathbb{N}_{>0}, k \leq 20, x = {k}/{20}\}$, respectively, are optimized on the validation set. The retrieval effectiveness of \QLM{} is provided as a point of reference in \RQRef{1}. For \RQRef{2}, the \QLM{} is used as a lexical retrieval model that is fused with latent vector space models to provide a mixture of lexical and semantic matching.

For the latent vector spaces (\DocToVec{}, \LSI{}, \LDA{}, \LSE{} and \ModelName{}), the vocabulary size is limited to the top-60k most frequent words as per \citep{VanGysel2016products}, given that latent methods rely on word co-occurrence in order to learn latent relations. For \FullDocToVec{}, \FullWordToVec{}, \LSI{} and \LDA{} we use the Gensim\footnote{\url{https://github.com/RaRe-Technologies/gensim}} implementation; the neural representation learning methods use our open source CUDA implementation described in \S\ref{sec:impl}; see footnote~\ref{fn:ImplementationURL}.

\subsubsection{Combinations of \QLM{} and latent features}
\label{sec:ltr}

We combine individual rankers by performing a grid search on the weights of a linear combination using 20-fold cross validation on the test sets (Table~\ref{tbl:adhoc_stats}). For \QLM{}, feature values are the log-probabilities of the query given the document, while for the latent features (\ModelName{} and \WordToVecSgSi{}), we use the cosine similarity between query/document representations. For every feature weight, we sweep between $0.0$ and $1.0$ with increments of $0.0125$ on the fold training set. Individual features are normalized per query such that their values lie between $0$ and $1$. We select the weight configuration that achieves highest Mean Average Precision on the training set and use that configuration to score the test set. During scoring of the fold test set, we take the pool of the top-1k documents ranked by the individual features as candidate set.

\subsection{Evaluation measures and statistical significance}
\label{sec:evaluationmeasures}

\newcommand{\CorrelationSignificant}{^{\dagger}}

\newcommand{\Significant}{^{*}}
\newcommand{\MoreSignificant}{^{**}}
\newcommand{\HighlySignificant}{^{***}}
\newcommand{\UnknownSignificant}{\phantom{\HighlySignificant{}}}

To address \RQRef{1}, \RQRef{2} and \RQRef{3}, we report Mean Average Precision at rank 1000 (\MAPCut{}), Normalized Discounted Cumulative Gain at rank 100 (\NDCGCut{}) and Precision at rank 10 (\Precision{}) to measure retrieval effectiveness. For \RQRef{3}, we also look at the per-query \TitlestatRelevant{} \citep{Buckley2007bias}, the expected normalized term overlap between query and document. All evaluation measures are computed using TREC's official evaluation tool, {\tt trec\_eval}.\footnote{\url{https://github.com/usnistgov/trec_eval}} Wherever reported, significance of observed differences is determined using a two-tailed paired Student's t-test \citep{Smucker2007significance} ($\HighlySignificant{} \, p < 0.01$;  $\MoreSignificant{} \, p < 0.05$; $\Significant{} \, p < 0.1$). For correlation coefficients, significance is determined using a permutation test ($\CorrelationSignificant{} \, p < 0.01$). For \RQRef{4}, we use Welch's t-test to determine whether the mean L2-norm of mid-frequency (middle-50\%) words is significantly different from the mean L2-norm of low- (bottom 25\%) and high-frequency (top 25\%) words.


\newcommand{\RQAnswer}[3]{
\begin{description}[topsep=0pt,parsep=0pt,leftmargin=0.8cm]
	\item[\RQRef{#1}] #2
\end{description}
\noindent #3
}

\newcommand{\Table}[4][]{
\begin{#2}[th!]
\centering
\renewcommand{\arraystretch}{1.2}
\caption{#1\label{tbl:#3_results}}
\scalebox{#4}{%
\renewcommand{\MAPCut}{MAP}%
\renewcommand{\NDCGCut}{NDCG}%
\renewcommand{\BenchmarkAP}{\ShortBenchmarkAP}%
\renewcommand{\BenchmarkWSJ}{\ShortBenchmarkWSJ}%
\renewcommand{\BenchmarkLATimes}{\ShortBenchmarkLATimes}%
\renewcommand{\BenchmarkFT}{\ShortBenchmarkFT}%
\renewcommand{\BenchmarkNYT}{\ShortBenchmarkNYT}%
\centering%
\renewcommand{\arraystretch}{0.8}%
\setlength{\tabcolsep}{3.5pt}%
\begin{tabular}{c}%
\input{resources/#3.tex}%
\end{tabular}%
}%
\end{#2}%
}

\newcommand{\DoubleColumnTable}[2][]{

\begin{table*}[th!]
\centering
\renewcommand{\arraystretch}{1.2}
\caption{#1\label{tbl:#2_results}}
\scalebox{0.835}{%
\renewcommand{\MAPCut}{MAP}%
\renewcommand{\NDCGCut}{NDCG}%
\renewcommand{\BenchmarkAP}{\ShortBenchmarkAP}%
\renewcommand{\BenchmarkWSJ}{\ShortBenchmarkWSJ}%
\renewcommand{\BenchmarkLATimes}{\ShortBenchmarkLATimes}%
\renewcommand{\BenchmarkFT}{\ShortBenchmarkFT}%
\renewcommand{\BenchmarkNYT}{\ShortBenchmarkNYT}%
\centering%
\renewcommand{\arraystretch}{0.8}%
\setlength{\tabcolsep}{3.5pt}%
\begin{tabular}{c}%
\input{resources/#2.tex}%
\end{tabular}%
}%
\end{table*}%

}

\newcommand{\SingleColumnTable}[2][]{

\begin{table}[th!]
\centering
\renewcommand{\arraystretch}{1.2}
\caption{#1\label{tbl:#2_results}}
\scalebox{0.90}{%
\renewcommand{\MAPCut}{MAP}%
\renewcommand{\NDCGCut}{NDCG}%
\renewcommand{\BenchmarkAP}{\ShortBenchmarkAP}%
\renewcommand{\BenchmarkWSJ}{\ShortBenchmarkWSJ}%
\renewcommand{\BenchmarkLATimes}{\ShortBenchmarkLATimes}%
\renewcommand{\BenchmarkFT}{\ShortBenchmarkFT}%
\renewcommand{\BenchmarkNYT}{\ShortBenchmarkNYT}%
\centering%
\renewcommand{\arraystretch}{0.8}%
\setlength{\tabcolsep}{3.5pt}%
\begin{tabular}{c}%
\input{resources/#2.tex}%
\end{tabular}%
}%
\end{table}%

}

\section{Results}

First, we present a comparison between methods (\RQRef{1}) on ad-hoc document retrieval, followed by a per-query comparison between methods (\RQRef{2}) and a combination experiment where we combine latent features with the lexical \QLM{} (\RQRef{3}). We then relate regularities learned by the model to traditional retrieval statistics (\RQRef{4}).

\subsection{Performance of \ModelName{}}
\DoubleColumnTable[Comparison of \ModelName{} with lexical (\QLM{} with \FullDirichlet{} and \FullJelinekMercer{} smoothing) and latent (\DocToVec{}, \WordToVec{}, \LSI{}, \LDA{} and \LSE{}) retrieval models (\S\ref{sec:retrievalmodels}) on article search benchmarks (\S\ref{sec:benchmarks}). Significance (\S\ref{sec:evaluationmeasures}) is computed between \WordToVecSgSi{} and \ModelName{}. Bold values indicate the highest measure value for latent features.]{adhoc}

\RQAnswer{1}{%
Table~\ref{tbl:adhoc_results} shows the retrieval results for ad-hoc document retrieval on the newswire article collections (\S\ref{sec:benchmarks}).
}{%
We see that \ModelName{} outperforms all other latent rankers on all benchmarks. In particular, \ModelName{} significantly outperforms (\MAPCut{}) the \FullWordToVec{}-based method that weighs word vectors according to self-information (significance is not achieved on \ShortBenchmarkNYT{}). This is an interesting observation as \ModelName{} is trained from scratch without the use of hand-engineered features (i.e., self-information). Compared to the lexical \QLM{}, \ModelName{} performs better on the \ShortBenchmarkAP{} and \ShortBenchmarkWSJ{} benchmarks. However, it is known that no single ranker performs best on all test sets~\citep{Shaw1994,Liu2011}. In addition, \ModelName{} is a latent model that performs a different type of matching than lexical models. Therefore, we first examine the per-query differences between rankers (\RQRef{2}) and later we will examine the complementary nature of the two types of matching by evaluating combinations of different ranking features (\RQRef{3}).
}

\begin{figure*}[t]
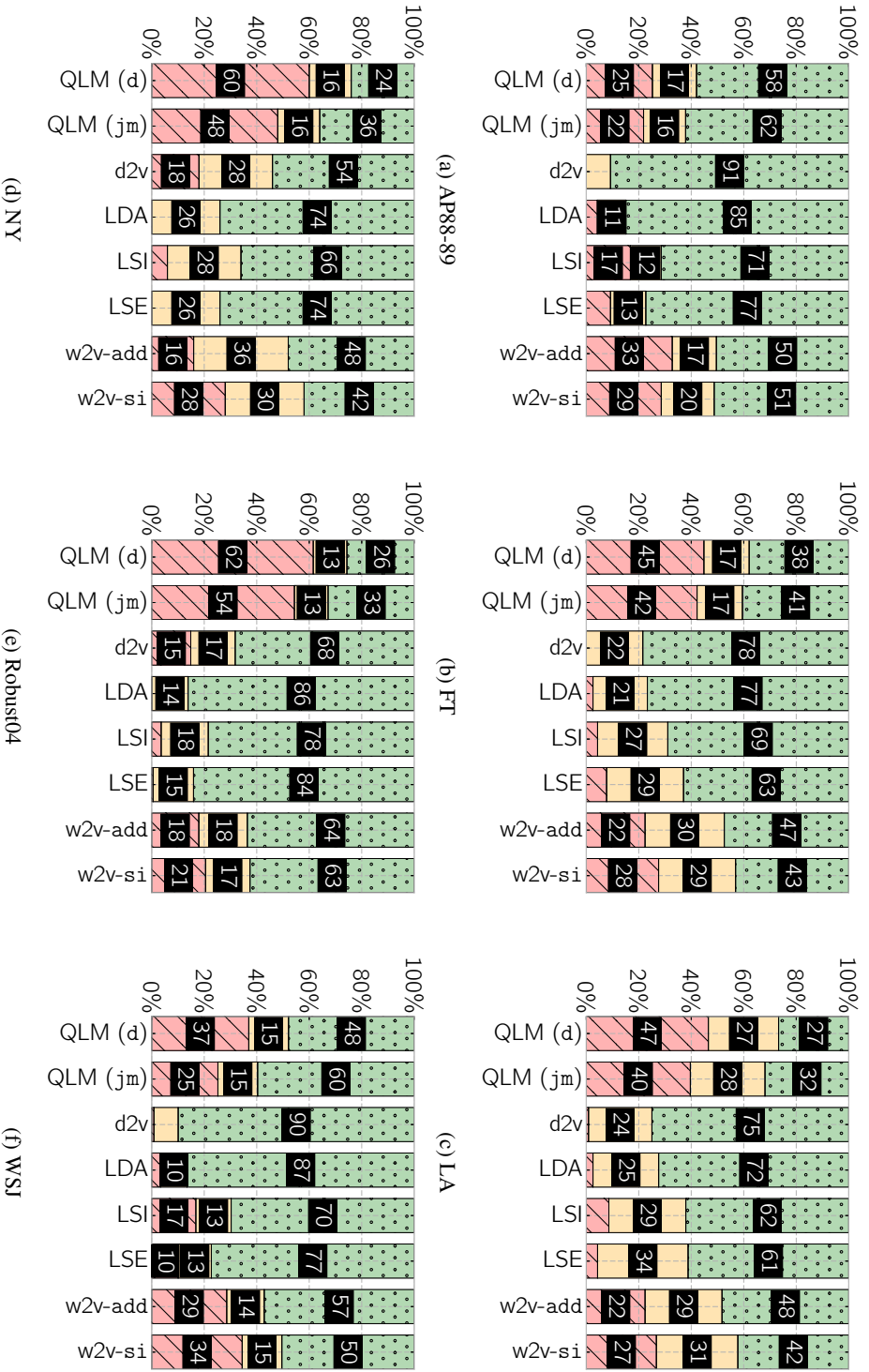


\newcommand{\differences}[2]{%
	\def \inner {%
		\def \PlotPath {resources/analysis/differences/#1/winners.pdf}%
		\IfFileExists{\PlotPath}{%
			\includegraphics[width=\BenchmarkFigureWidth{}]{\PlotPath}}{%
			\resizebox{\BenchmarkFigureWidth{}}{!}{\missingfigure{#1}}}%
	}%
	\ifstrequal{#2}{}{%
	\subfloat{\inner}%
	}{\subfloat[#2\label{fig:differences:#1}]{\inner}}%
}

\centering
\differences{ap_88_89}{\ShortBenchmarkAP{}}%
\hfill%
\differences{ft}{\ShortBenchmarkFT{}}%
\hfill%
\differences{latimes}{\ShortBenchmarkLATimes{}}

\centering
\differences{nyt}{\ShortBenchmarkNYT{}}%
\hfill%
\differences{disk4_disk5_no-cr}{\ShortBenchmarkRobust{}}%
\hfill%
\differences{wsj}{\ShortBenchmarkWSJ{}}%

\caption{Per-query pairwise ranker comparison between \ModelName{} and the \QLMDirichlet{}, \QLMJelinekMercer{}, \DocToVec{}, \WordToVec{}, \LSI{}, \LDA{} and \LSE{} rankers. For every bar, the dotted/green area, solid/orange and red/slashed areas respectively depict the portion of queries for which \ModelName{} outperforms, ties or loses against the other ranker. One ranker outperforms the other if the absolute difference in \MAPCut{} between both rankers exceeds $\delta$.\label{fig:differences}}

\end{figure*}

\begin{oldtable}[th!]
\centering
\caption{Evaluation of latent features (\DocToVec{}, \WordToVec{}, \LSI{}, \LDA{}, \LSE{} and \ModelName{}) as a complementary signal to \QLMDirichlet{} (\S\ref{sec:ltr}). Significance (\S\ref{sec:evaluationmeasures}) is computed between \QLMDirichlet{} + \WordToVecSgSi{} and the best performing combination.\label{tbl:adhoc_combinations_results}}
\resizebox{0.90\textwidth}{!}{%
\renewcommand{\MAPCut}{MAP}%
\renewcommand{\NDCGCut}{NDCG}%
\renewcommand{\BenchmarkAP}{\ShortBenchmarkAP}%
\renewcommand{\BenchmarkWSJ}{\ShortBenchmarkWSJ}%
\renewcommand{\BenchmarkLATimes}{\ShortBenchmarkLATimes}%
\renewcommand{\BenchmarkFT}{\ShortBenchmarkFT}%
\renewcommand{\BenchmarkNYT}{\ShortBenchmarkNYT}%
\centering%
\begin{tabular}{c}%
\begin{tabular}{l c c c}%
\toprule%
&\multicolumn{3}{c}{\BenchmarkAP{}}\\%
&\MAPCut&\NDCGCut&\Precision\\%
\cmidrule(lr){2-4}%
\QLM{} (\Dirichlet{})&\nprounddigits{3}\npdecimalsign{.}\npthousandsep{.}\numprint{0.2155805085}$\phantom{\HighlySignificant}$\phantom{\begin{footnotesize}%
$\phantom{1} (+0\%)$%
\end{footnotesize}}&\nprounddigits{3}\npdecimalsign{.}\npthousandsep{.}\numprint{0.3695194915}$\phantom{\HighlySignificant}$\phantom{\begin{footnotesize}%
$\phantom{1} (+0\%)$%
\end{footnotesize}}&\nprounddigits{3}\npdecimalsign{.}\npthousandsep{.}\numprint{0.3915254237}$\phantom{\HighlySignificant}$\phantom{\begin{footnotesize}%
$\phantom{1} (+0\%)$%
\end{footnotesize}}\\%
\QLMDirichlet{} + \WordToVecSgSi{}&\nprounddigits{3}\npdecimalsign{.}\npthousandsep{.}\numprint{0.2786381356}$\phantom{\HighlySignificant}$\begin{footnotesize}%
$(+29\%)$%
\end{footnotesize}&\nprounddigits{3}\npdecimalsign{.}\npthousandsep{.}\numprint{0.4371398305}$\phantom{\HighlySignificant}$\begin{footnotesize}%
$(+18\%)$%
\end{footnotesize}&\nprounddigits{3}\npdecimalsign{.}\npthousandsep{.}\numprint{0.4500000000}$\phantom{\HighlySignificant}$\begin{footnotesize}%
$(+14\%)$%
\end{footnotesize}\\%
\QLMDirichlet{} + \ModelName{}&\nprounddigits{3}\npdecimalsign{.}\npthousandsep{.}\numprint{0.2885135593}$\phantom{\HighlySignificant}$\begin{footnotesize}%
$(+33\%)$%
\end{footnotesize}&\nprounddigits{3}\npdecimalsign{.}\npthousandsep{.}\numprint{0.4443661017}$\phantom{\HighlySignificant}$\begin{footnotesize}%
$(+20\%)$%
\end{footnotesize}&\nprounddigits{3}\npdecimalsign{.}\npthousandsep{.}\numprint{0.4728813559}$\phantom{\HighlySignificant}$\begin{footnotesize}%
$(+20\%)$%
\end{footnotesize}\\%
\midrule%
\QLMDirichlet{} + \WordToVecSgSi{} + \ModelName{}&\nprounddigits{3}\npdecimalsign{.}\npthousandsep{.}\textbf{\numprint{0.3069923729}}$\HighlySignificant$\begin{footnotesize}%
$(+42\%)$%
\end{footnotesize}&\nprounddigits{3}\npdecimalsign{.}\npthousandsep{.}\textbf{\numprint{0.4662991525}}$\HighlySignificant$\begin{footnotesize}%
$(+26\%)$%
\end{footnotesize}&\nprounddigits{3}\npdecimalsign{.}\npthousandsep{.}\textbf{\numprint{0.4983050847}}$\HighlySignificant$\begin{footnotesize}%
$(+27\%)$%
\end{footnotesize}\\%
\midrule%
\end{tabular}%
\\%
\begin{tabular}{l c c c}%
&\multicolumn{3}{c}{\BenchmarkFT{}}\\%
&\MAPCut&\NDCGCut&\Precision\\%
\cmidrule(lr){2-4}%
\QLM{} (\Dirichlet{})&\nprounddigits{3}\npdecimalsign{.}\npthousandsep{.}\numprint{0.2395116071}$\phantom{\UnknownSignificant}$\phantom{\begin{footnotesize}%
$\phantom{1} (+0\%)$%
\end{footnotesize}}&\nprounddigits{3}\npdecimalsign{.}\npthousandsep{.}\numprint{0.3813714286}$\phantom{\UnknownSignificant}$\phantom{\begin{footnotesize}%
$\phantom{1} (+0\%)$%
\end{footnotesize}}&\nprounddigits{3}\npdecimalsign{.}\npthousandsep{.}\numprint{0.2955357143}$\phantom{\UnknownSignificant}$\phantom{\begin{footnotesize}%
$\phantom{1} (+0\%)$%
\end{footnotesize}}\\%
\QLMDirichlet{} + \WordToVecSgSi{}&\nprounddigits{3}\npdecimalsign{.}\npthousandsep{.}\numprint{0.2506375000}$\phantom{\UnknownSignificant}$\begin{footnotesize}%
$\phantom{1} (+4\%)$%
\end{footnotesize}&\nprounddigits{3}\npdecimalsign{.}\npthousandsep{.}\numprint{0.3932848214}$\phantom{\UnknownSignificant}$\begin{footnotesize}%
$\phantom{1} (+3\%)$%
\end{footnotesize}&\nprounddigits{3}\npdecimalsign{.}\npthousandsep{.}\numprint{0.3133928571}$\phantom{\UnknownSignificant}$\begin{footnotesize}%
$\phantom{1} (+6\%)$%
\end{footnotesize}\\%
\QLMDirichlet{} + \ModelName{}&\nprounddigits{3}\npdecimalsign{.}\npthousandsep{.}\numprint{0.2508553571}$\phantom{\UnknownSignificant}$\begin{footnotesize}%
$\phantom{1} (+4\%)$%
\end{footnotesize}&\nprounddigits{3}\npdecimalsign{.}\npthousandsep{.}\numprint{0.4007991071}$\phantom{\UnknownSignificant}$\begin{footnotesize}%
$\phantom{1} (+5\%)$%
\end{footnotesize}&\nprounddigits{3}\npdecimalsign{.}\npthousandsep{.}\numprint{0.3223214286}$\phantom{\UnknownSignificant}$\begin{footnotesize}%
$\phantom{1} (+9\%)$%
\end{footnotesize}\\%
\midrule%
\QLMDirichlet{} + \WordToVecSgSi{} + \ModelName{}&\nprounddigits{3}\npdecimalsign{.}\npthousandsep{.}\textbf{\numprint{0.2583098214}}$\UnknownSignificant$\begin{footnotesize}%
$\phantom{1} (+7\%)$%
\end{footnotesize}&\nprounddigits{3}\npdecimalsign{.}\npthousandsep{.}\textbf{\numprint{0.4057178571}}$\UnknownSignificant$\begin{footnotesize}%
$\phantom{1} (+6\%)$%
\end{footnotesize}&\nprounddigits{3}\npdecimalsign{.}\npthousandsep{.}\textbf{\numprint{0.3223214286}}$\UnknownSignificant$\begin{footnotesize}%
$\phantom{1} (+9\%)$%
\end{footnotesize}\\%
\midrule%
\end{tabular}%
\\%
\begin{tabular}{l c c c}%
&\multicolumn{3}{c}{\BenchmarkLATimes{}}\\%
&\MAPCut&\NDCGCut&\Precision\\%
\cmidrule(lr){2-4}%
\QLM{} (\Dirichlet{})&\nprounddigits{3}\npdecimalsign{.}\npthousandsep{.}\numprint{0.1976504587}$\phantom{\HighlySignificant}$\phantom{\begin{footnotesize}%
$\phantom{1} (+0\%)$%
\end{footnotesize}}&\nprounddigits{3}\npdecimalsign{.}\npthousandsep{.}\numprint{0.3483926606}$\phantom{\HighlySignificant}$\phantom{\begin{footnotesize}%
$\phantom{1} (+0\%)$%
\end{footnotesize}}&\nprounddigits{3}\npdecimalsign{.}\npthousandsep{.}\numprint{0.2394495413}$\phantom{\MoreSignificant}$\phantom{\begin{footnotesize}%
$\phantom{1} (+0\%)$%
\end{footnotesize}}\\%
\QLMDirichlet{} + \WordToVecSgSi{}&\nprounddigits{3}\npdecimalsign{.}\npthousandsep{.}\numprint{0.2117724771}$\phantom{\HighlySignificant}$\begin{footnotesize}%
$\phantom{1} (+7\%)$%
\end{footnotesize}&\nprounddigits{3}\npdecimalsign{.}\npthousandsep{.}\numprint{0.3604954128}$\phantom{\HighlySignificant}$\begin{footnotesize}%
$\phantom{1} (+3\%)$%
\end{footnotesize}&\nprounddigits{3}\npdecimalsign{.}\npthousandsep{.}\numprint{0.2357798165}$\phantom{\MoreSignificant}$\begin{footnotesize}%
$\phantom{1} ({-}1\%)$%
\end{footnotesize}\\%
\QLMDirichlet{} + \ModelName{}&\nprounddigits{3}\npdecimalsign{.}\npthousandsep{.}\numprint{0.2201183486}$\phantom{\HighlySignificant}$\begin{footnotesize}%
$(+11\%)$%
\end{footnotesize}&\nprounddigits{3}\npdecimalsign{.}\npthousandsep{.}\numprint{0.3759082569}$\phantom{\HighlySignificant}$\begin{footnotesize}%
$\phantom{1} (+7\%)$%
\end{footnotesize}&\nprounddigits{3}\npdecimalsign{.}\npthousandsep{.}\numprint{0.2440366972}$\phantom{\MoreSignificant}$\begin{footnotesize}%
$\phantom{1} (+1\%)$%
\end{footnotesize}\\%
\midrule%
\QLMDirichlet{} + \WordToVecSgSi{} + \ModelName{}&\nprounddigits{3}\npdecimalsign{.}\npthousandsep{.}\textbf{\numprint{0.2255330275}}$\HighlySignificant$\begin{footnotesize}%
$(+14\%)$%
\end{footnotesize}&\nprounddigits{3}\npdecimalsign{.}\npthousandsep{.}\textbf{\numprint{0.3778522936}}$\HighlySignificant$\begin{footnotesize}%
$\phantom{1} (+8\%)$%
\end{footnotesize}&\nprounddigits{3}\npdecimalsign{.}\npthousandsep{.}\textbf{\numprint{0.2504587156}}$\MoreSignificant$\begin{footnotesize}%
$\phantom{1} (+4\%)$%
\end{footnotesize}\\%
\midrule%
\end{tabular}%
\\%
\begin{tabular}{l c c c}%
&\multicolumn{3}{c}{\BenchmarkNYT{}}\\%
&\MAPCut&\NDCGCut&\Precision\\%
\cmidrule(lr){2-4}%
\QLM{} (\Dirichlet{})&\nprounddigits{3}\npdecimalsign{.}\npthousandsep{.}\numprint{0.1878600000}$\phantom{\HighlySignificant}$\phantom{\begin{footnotesize}%
$\phantom{1} (+0\%)$%
\end{footnotesize}}&\nprounddigits{3}\npdecimalsign{.}\npthousandsep{.}\numprint{0.3180400000}$\phantom{\MoreSignificant}$\phantom{\begin{footnotesize}%
$\phantom{1} (+0\%)$%
\end{footnotesize}}&\nprounddigits{3}\npdecimalsign{.}\npthousandsep{.}\numprint{0.4860000000}$\phantom{\MoreSignificant}$\phantom{\begin{footnotesize}%
$\phantom{1} (+0\%)$%
\end{footnotesize}}\\%
\QLMDirichlet{} + \WordToVecSgSi{}&\nprounddigits{3}\npdecimalsign{.}\npthousandsep{.}\numprint{0.2061560000}$\phantom{\HighlySignificant}$\begin{footnotesize}%
$\phantom{1} (+9\%)$%
\end{footnotesize}&\nprounddigits{3}\npdecimalsign{.}\npthousandsep{.}\numprint{0.3332000000}$\phantom{\MoreSignificant}$\begin{footnotesize}%
$\phantom{1} (+4\%)$%
\end{footnotesize}&\nprounddigits{3}\npdecimalsign{.}\npthousandsep{.}\numprint{0.4940000000}$\phantom{\MoreSignificant}$\begin{footnotesize}%
$\phantom{1} (+1\%)$%
\end{footnotesize}\\%
\QLMDirichlet{} + \ModelName{}&\nprounddigits{3}\npdecimalsign{.}\npthousandsep{.}\numprint{0.2217960000}$\phantom{\HighlySignificant}$\begin{footnotesize}%
$(+18\%)$%
\end{footnotesize}&\nprounddigits{3}\npdecimalsign{.}\npthousandsep{.}\textbf{\numprint{0.3552700000}}$\MoreSignificant$\begin{footnotesize}%
$(+11\%)$%
\end{footnotesize}&\nprounddigits{3}\npdecimalsign{.}\npthousandsep{.}\numprint{0.5200000000}$\phantom{\MoreSignificant}$\begin{footnotesize}%
$\phantom{1} (+6\%)$%
\end{footnotesize}\\%
\midrule%
\QLMDirichlet{} + \WordToVecSgSi{} + \ModelName{}&\nprounddigits{3}\npdecimalsign{.}\npthousandsep{.}\textbf{\numprint{0.2224680000}}$\HighlySignificant$\begin{footnotesize}%
$(+18\%)$%
\end{footnotesize}&\nprounddigits{3}\npdecimalsign{.}\npthousandsep{.}\numprint{0.3526440000}$\phantom{\MoreSignificant}$\begin{footnotesize}%
$(+10\%)$%
\end{footnotesize}&\nprounddigits{3}\npdecimalsign{.}\npthousandsep{.}\textbf{\numprint{0.5260000000}}$\MoreSignificant$\begin{footnotesize}%
$\phantom{1} (+8\%)$%
\end{footnotesize}\\%
\midrule%
\end{tabular}%
\\%
\begin{tabular}{l c c c}%
&\multicolumn{3}{c}{\BenchmarkRobust{}}\\%
&\MAPCut&\NDCGCut&\Precision\\%
\cmidrule(lr){2-4}%
\QLM{} (\Dirichlet{})&\nprounddigits{3}\npdecimalsign{.}\npthousandsep{.}\numprint{0.2235169231}$\phantom{\HighlySignificant}$\phantom{\begin{footnotesize}%
$\phantom{1} (+0\%)$%
\end{footnotesize}}&\nprounddigits{3}\npdecimalsign{.}\npthousandsep{.}\numprint{0.3876805128}$\phantom{\HighlySignificant}$\phantom{\begin{footnotesize}%
$\phantom{1} (+0\%)$%
\end{footnotesize}}&\nprounddigits{3}\npdecimalsign{.}\npthousandsep{.}\numprint{0.4153846154}$\phantom{\HighlySignificant}$\phantom{\begin{footnotesize}%
$\phantom{1} (+0\%)$%
\end{footnotesize}}\\%
\QLMDirichlet{} + \WordToVecSgSi{}&\nprounddigits{3}\npdecimalsign{.}\npthousandsep{.}\numprint{0.2318030769}$\phantom{\HighlySignificant}$\begin{footnotesize}%
$\phantom{1} (+3\%)$%
\end{footnotesize}&\nprounddigits{3}\npdecimalsign{.}\npthousandsep{.}\numprint{0.3987071795}$\phantom{\HighlySignificant}$\begin{footnotesize}%
$\phantom{1} (+2\%)$%
\end{footnotesize}&\nprounddigits{3}\npdecimalsign{.}\npthousandsep{.}\numprint{0.4276923077}$\phantom{\HighlySignificant}$\begin{footnotesize}%
$\phantom{1} (+2\%)$%
\end{footnotesize}\\%
\QLMDirichlet{} + \ModelName{}&\nprounddigits{3}\npdecimalsign{.}\npthousandsep{.}\numprint{0.2465482051}$\phantom{\HighlySignificant}$\begin{footnotesize}%
$(+10\%)$%
\end{footnotesize}&\nprounddigits{3}\npdecimalsign{.}\npthousandsep{.}\numprint{0.4113035897}$\phantom{\HighlySignificant}$\begin{footnotesize}%
$\phantom{1} (+6\%)$%
\end{footnotesize}&\nprounddigits{3}\npdecimalsign{.}\npthousandsep{.}\textbf{\numprint{0.4476923077}}$\HighlySignificant$\begin{footnotesize}%
$\phantom{1} (+7\%)$%
\end{footnotesize}\\%
\midrule%
\QLMDirichlet{} + \WordToVecSgSi{} + \ModelName{}&\nprounddigits{3}\npdecimalsign{.}\npthousandsep{.}\textbf{\numprint{0.2472841026}}$\HighlySignificant$\begin{footnotesize}%
$(+10\%)$%
\end{footnotesize}&\nprounddigits{3}\npdecimalsign{.}\npthousandsep{.}\textbf{\numprint{0.4121948718}}$\HighlySignificant$\begin{footnotesize}%
$\phantom{1} (+6\%)$%
\end{footnotesize}&\nprounddigits{3}\npdecimalsign{.}\npthousandsep{.}\numprint{0.4461538462}$\phantom{\HighlySignificant}$\begin{footnotesize}%
$\phantom{1} (+7\%)$%
\end{footnotesize}\\%
\midrule%
\end{tabular}%
\\%
\begin{tabular}{l c c c}%
&\multicolumn{3}{c}{\BenchmarkWSJ{}}\\%
&\MAPCut&\NDCGCut&\Precision\\%
\cmidrule(lr){2-4}%
\QLM{} (\Dirichlet{})&\nprounddigits{3}\npdecimalsign{.}\npthousandsep{.}\numprint{0.2040453782}$\phantom{\HighlySignificant}$\phantom{\begin{footnotesize}%
$\phantom{1} (+0\%)$%
\end{footnotesize}}&\nprounddigits{3}\npdecimalsign{.}\npthousandsep{.}\numprint{0.3510411765}$\phantom{\HighlySignificant}$\phantom{\begin{footnotesize}%
$\phantom{1} (+0\%)$%
\end{footnotesize}}&\nprounddigits{3}\npdecimalsign{.}\npthousandsep{.}\numprint{0.3983193277}$\phantom{\UnknownSignificant}$\phantom{\begin{footnotesize}%
$\phantom{1} (+0\%)$%
\end{footnotesize}}\\%
\QLMDirichlet{} + \WordToVecSgSi{}&\nprounddigits{3}\npdecimalsign{.}\npthousandsep{.}\numprint{0.2544008403}$\phantom{\HighlySignificant}$\begin{footnotesize}%
$(+24\%)$%
\end{footnotesize}&\nprounddigits{3}\npdecimalsign{.}\npthousandsep{.}\numprint{0.4101957983}$\phantom{\HighlySignificant}$\begin{footnotesize}%
$(+16\%)$%
\end{footnotesize}&\nprounddigits{3}\npdecimalsign{.}\npthousandsep{.}\numprint{0.4537815126}$\phantom{\UnknownSignificant}$\begin{footnotesize}%
$(+13\%)$%
\end{footnotesize}\\%
\QLMDirichlet{} + \ModelName{}&\nprounddigits{3}\npdecimalsign{.}\npthousandsep{.}\numprint{0.2482420168}$\phantom{\HighlySignificant}$\begin{footnotesize}%
$(+21\%)$%
\end{footnotesize}&\nprounddigits{3}\npdecimalsign{.}\npthousandsep{.}\numprint{0.3960613445}$\phantom{\HighlySignificant}$\begin{footnotesize}%
$(+12\%)$%
\end{footnotesize}&\nprounddigits{3}\npdecimalsign{.}\npthousandsep{.}\numprint{0.4252100840}$\phantom{\UnknownSignificant}$\begin{footnotesize}%
$\phantom{1} (+6\%)$%
\end{footnotesize}\\%
\midrule%
\QLMDirichlet{} + \WordToVecSgSi{} + \ModelName{}&\nprounddigits{3}\npdecimalsign{.}\npthousandsep{.}\textbf{\numprint{0.2710823529}}$\HighlySignificant$\begin{footnotesize}%
$(+32\%)$%
\end{footnotesize}&\nprounddigits{3}\npdecimalsign{.}\npthousandsep{.}\textbf{\numprint{0.4257823529}}$\HighlySignificant$\begin{footnotesize}%
$(+21\%)$%
\end{footnotesize}&\nprounddigits{3}\npdecimalsign{.}\npthousandsep{.}\textbf{\numprint{0.4563025210}}$\UnknownSignificant$\begin{footnotesize}%
$(+14\%)$%
\end{footnotesize}\\%
\bottomrule%
\end{tabular}%
\end{tabular}%
}%
\end{oldtable}%

\subsection{Query-level analysis}
\label{sec:results:querylevel}
\RQAnswer{2}{%
Fig.~\ref{fig:differences} shows the distribution of queries where one individual ranker performs better than the other ($\abs{\Delta \text{\MAPCut{}}} > \delta$).
}{%
We observe similar trends across all benchmarks where \ModelName{} performs best compared to all latent rankers. One competing vector space model, \WordToVecSgSi{}, stands out as it is the strongest baseline and  performs better than \ModelName{} on $20$ to $35\%$ of queries, however, \ModelName{} still beats \WordToVecSgSi{} overall and specifically on $40$ to $55\%$ of queries. Moreover, Fig.~\ref{fig:differences} shows us that \QLM{} and \ModelName{} make very different errors. This implies that the combination of \QLM{}, \WordToVecSgSi{} and \ModelName{} might improve performance even further.

While answering \RQRef{1}, we saw that for some benchmarks, latent methods (i.e., \ModelName{}) perform better than lexical methods. While the amount of semantic matching needed depends on various factors, such as the query intent being informational or navigational/transactional \citep{Broder2002taxonomy}, we do see in Fig.~\ref{fig:differences} that \ModelName{} performs considerably better amongst latent methods in cases where latent methods perform poorly (e.g., \BenchmarkRobust{} in Fig.~\ref{fig:differences:disk4_disk5_no-cr}). Can we shine more light on the difference between \ModelName{} and existing latent methods? We answer this question in the second half of the next section.
}

\newcommand{\DeltaMAPCut}{\Delta{}_{\text{\MAPCut{}}}}
\SingleColumnTable[Correlation coefficients between \TitlestatRelevant{} and $\DeltaMAPCut{}$ between \ModelName{} and the other methods. A positive correlation indicates that \ModelName{} is better at lexical matching, while a negative correlation indicates that \ModelName{} is worse at lexical matching than the alternative. Significance (\S\ref{sec:evaluationmeasures}) is computed using a permutation test.]{titlestat_correlations}

\subsection{Semantic vs.\ lexical matching}
\RQAnswer{3}{%
Beyond individual rankers, we now also consider combinations of the two best-performing vector space models with \QLM{} \citep{Shaw1994} (see \S\ref{sec:ltr} for details) in Table~\ref{tbl:adhoc_combinations_results}.
}{%
If we consider the \QLM{} paired with either \WordToVecSgSi{} or \ModelName{}, we see that the combination involving \ModelName{} outperforms the combination with \WordToVecSgSi{} on four out of six benchmarks (\ShortBenchmarkAP{}, \ShortBenchmarkLATimes{}, \ShortBenchmarkNYT{}, \ShortBenchmarkRobust{}). However, Figure~\ref{fig:differences} shows that \ModelName{} and \WordToVecSgSi{} outperform each other on different queries as well. Can we use this difference to our advantage?

The addition of \ModelName{} to the \QLM{} + \WordToVecSgSi{} combination yields an improvement in terms of \MAPCut{} on all benchmarks. Significance is achieved in five out of six benchmarks. In the case of \ShortBenchmarkNYT{} and \BenchmarkRobust{}, the combination of all three rankers (\QLM{} + \WordToVecSgSi{} + \ModelName{}) performs at about the same level as the combination of \QLM{} + \ModelName{}. However, the addition of \ModelName{} to the \QLM{} + \WordToVecSgSi{} combination still creates a significant improvement over just the combination involving \QLM{} and \WordToVec{} only. For \ShortBenchmarkFT{}, the only benchmark where no significance is achieved, we do see that the relative increase in performance nearly doubles from the addition of \ModelName{}. Consequently, we can conclude that the \ModelName{} adds an additional matching signal.

Let us return to the question raised at the end of the previous section (\S\ref{sec:results:querylevel}): what exactly does the \ModelName{} add in terms of content matching? We investigate this question by determining the amount of semantic matching needed. For each query, we compute \TitlestatRelevant{} (\S\ref{sec:evaluationmeasures}), the expected normalized overlap between query terms and the terms of relevant document. If \TitlestatRelevant{} is close to \numprint{1.0} for a particular query, then the query requires mostly lexical matching; on the other hand, if \TitlestatRelevant{} is near \numprint{0.0} for a query, then none of the query's relevant document contain the query terms and semantic matching is needed. We continue by examining the per-query pairwise difference ($\DeltaMAPCut{}$) between \ModelName{} and the remaining lexical (\QLM{}) and latent (\DocToVec{}, \LDA{}, \LSI{}, \LSE{}) features. If $\DeltaMAPCut{} > 0$, then \ModelName{} performs better than the other method and vice versa if $\DeltaMAPCut{} < 0$. Table~\ref{tbl:titlestat_correlations_results} shows the Pearson correlation between \TitlestatRelevant{} and $\DeltaMAPCut{}$. A positive correlation, as is the case for \DocToVec{}, \WordToVec{}, \LDA{}, \LSI{} and \LSE{}, indicates that \ModelName{} performs better on queries that require lexical matching. Conversely, a negative correlation, such as observed for both variants of \QLM{}, indicates that \QLM{} performs better on queries that require lexical matching than \ModelName{}. Combining this observation with the conclusion to \RQRef{2} (i.e., \ModelName{} generally improves upon latent methods), we conclude that, in addition to semantic matching, \ModelName{} also performs well in cases where lexical matching is needed and thus contributes a hybrid matching signal.}

\subsection{\ModelName{} and Luhn significance}

{
\newcommand{\NormFigureWidth}{0.450\textwidth}

\newcommand{\inner}[1]{}
\newcommand{\regularities}[3]{%
	\renewcommand{\inner}[1]{%
		\def \PlotPath {resources/analysis/model-regularities/#1/##1/#2.png}%
		\IfFileExists{\PlotPath}{%
			\resizebox{\NormFigureWidth{}}{!}{\includegraphics[width=\NormFigureWidth{}]{\PlotPath}}}{%
			\resizebox{\NormFigureWidth{}}{!}{\missingfigure{#1}}}%
	}%
	\begin{subfigure}[b]{\textwidth}%
	\centering
	\inner{lse}~~~\inner{nvsm}%
	\ifstrequal{#3}{}{}{\caption{#3\label{fig:regularities:#1}}}%
	\end{subfigure}%
}

\newcommand{\Header}{%
\begin{subfigure}[b]{\textwidth}%
\centering
\parbox{\NormFigureWidth{}}{\hspace{1.75em}\centering\textbf\LSE{}}~~~\parbox{\NormFigureWidth{}}{\hspace{1.75em}\centering\textbf\ModelName{}}
\end{subfigure}%
}

\newcommand{\CaptionText}{Scatter plots of term frequency in the document collections and the L2-norm of the \LSE{} (left) and \ModelName{} (right) representations of these terms. In the case of \LSE{} (left scatter plots for every benchmark), we observe that the L2-norm of term representations grows linearly with the term collection frequency, and consequently, high-frequency terms are of greater importance within \LSE{} representations. For \ModelName{} (right scatter plot for every benchmark), we observe that terms of mid-frequency (middle 50\%) have a statistically significant ($p < 0.01$) higher L2-norm (\S\ref{sec:evaluationmeasures}) and consequently are of greater importance for retrieval.}

\begin{figure}[t!]
\centering

\Header{}
\setcounter{subfigure}{0}

\regularities{ap_88_89}{words.collection_frequency_embedding_norm}{\ShortBenchmarkAP{}}%

\regularities{ft}{words.collection_frequency_embedding_norm}{\ShortBenchmarkFT{}}%

\regularities{latimes}{words.collection_frequency_embedding_norm}{\ShortBenchmarkLATimes{}}%

\caption{\CaptionText{}\label{fig:regularities}}

\end{figure}%
\begin{figure}[t!]
\ContinuedFloat
\centering

\Header{}
\setcounter{subfigure}{3}

\regularities{nyt}{words.collection_frequency_embedding_norm}{\ShortBenchmarkNYT{}}%

\regularities{disk4_disk5_no-cr}{words.collection_frequency_embedding_norm}{\ShortBenchmarkRobust{}}%

\regularities{wsj}{words.collection_frequency_embedding_norm}{\ShortBenchmarkWSJ{}}%

\makeatletter
\renewcommand\fnum@figure{\figurename~\thefigure~(cont'd)}
\makeatother

\caption{\CaptionText{}}

\end{figure}
}

If \ModelName{} performs better at lexical matching than other latent vector space models, does it then also contain regularities associated with term specificity?

\RQAnswer{4}{%
Fig.~\ref{fig:regularities} shows the L2-norm of individual term representations for \LSE{} (left scatter plots) and \ModelName{} (right scatter plots).
}{%
\citet{Luhn1958significance} measures the significance of words based on their frequency. They specify a lower and upper frequency cutoff to exclude frequent and infrequent words. For \ModelName{} (scatter plot on the right for every benchmark), we find that infrequent and frequent terms have a statistically significant ($p < 0.01$) smaller L2-norm than terms of medium frequency (Section~\ref{sec:evaluationmeasures}). This observation is further motivated by the shape of the relation between collection frequency in the collection and the L2-norm of term representations in Fig.~\ref{fig:regularities}. The key observation that---within \ModelName{} representations---terms of medium frequency are of greater importance (i.e., higher L2-norm) than low- or high-frequency terms closely corresponds to the theory of \citeauthor{Luhn1958significance} significance. Particularly noteworthy is the fact that the \ModelName{} learned this relationship from an unsupervised objective directly, without any notion of relevance. The scatter plots on the left for every benchmark in Fig.~\ref{fig:regularities} shows the same analysis for \LSE{} term representations. Unlike with \ModelName{}, we observe that the L2-norm of term representations grows linearly with the term collection frequency, and consequently, high-frequency terms are of greater importance within \LSE{} representations. Therefore, the key difference between \ModelName{} and \LSE{} is that \ModelName{} learns to better encode term specificity.
}

\section{Discussion and analysis}

In this section, we investigate the impact of the judgement bias (\S\ref{sec:analysis:judgment}). We then proceed by giving guidelines on how to deploy \ModelName{} in the absence of a validation set (\S\ref{sec:analysis:deployment}).

\subsection{An investigation of judgement bias}
\label{sec:analysis:judgment}

In what capacity does the judgement bias inherent to the construction of TREC test collections affect the evaluation of novel retrieval models such as \ModelName{}? The relevance assessments of TREC test collections are created using a pooling strategy of the rankings produced by participants of the TREC ad-hoc track \citep{TRECAdhoc}. This strategy is known to introduce a pooling bias \citep{Hawking2000overview,Buckley2007bias,Lipani2016bias}: rankings that have fewer documents judged may be scored lower just because of this fact. Does judgement bias influence the evaluation of \ModelName{} compared to lexical language models?

\begin{figure*}[t]
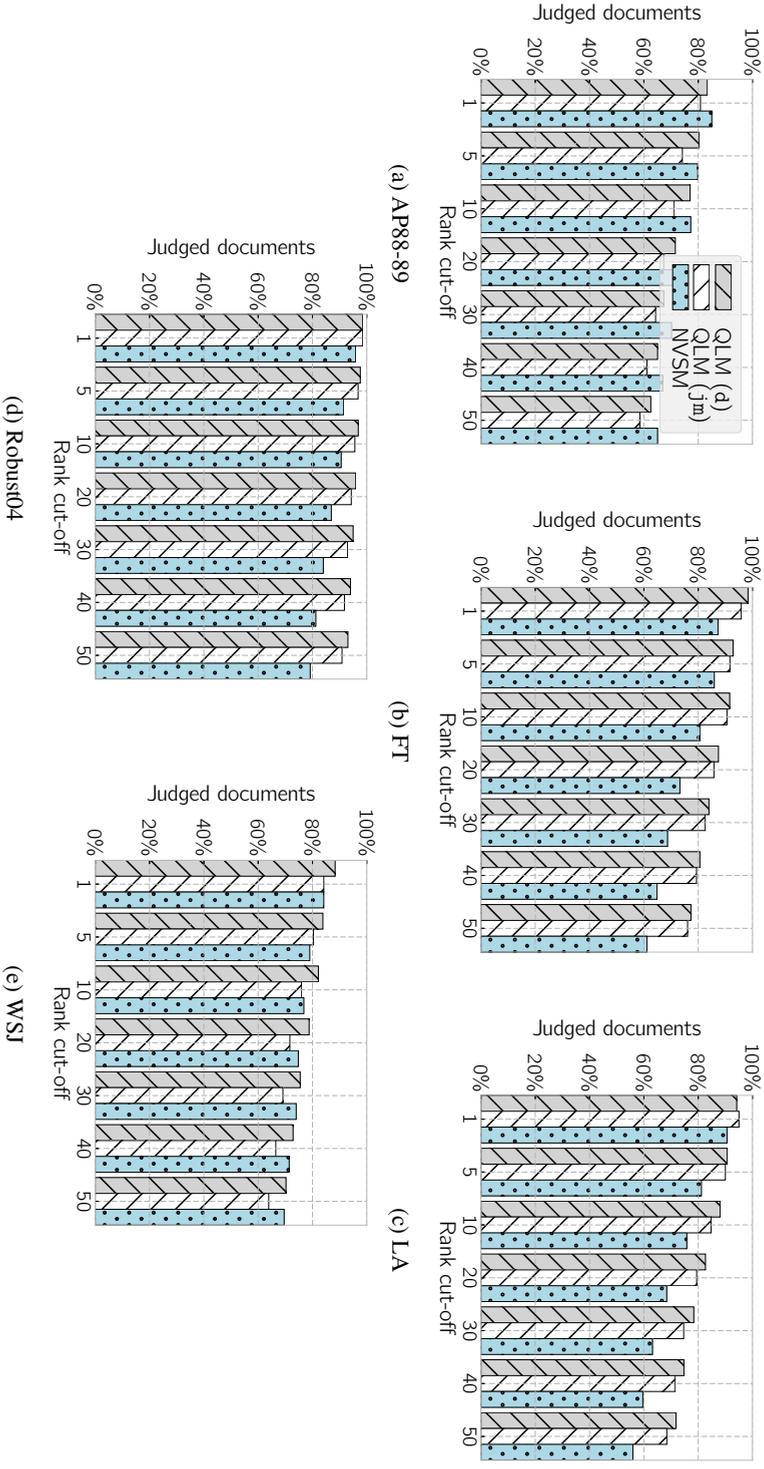


\newcommand{\poolbias}[2]{%
    \def \inner {%
        \def \PlotPath {resources/analysis/pool-bias/#1/unjudged-qrel_test.pdf}%
        \IfFileExists{\PlotPath}{%
            \myincludegraphics[width=\BenchmarkFigureWidth{}]{\PlotPath}}{%
            \resizebox{\BenchmarkFigureWidth{}}{!}{\missingfigure{#1}}}%
    }%
    \ifstrequal{#2}{}{%
    \subfloat{\inner}%
    }{\subfloat[#2\label{fig:poolbias:#1}]{\inner}}%
}

\centering
\poolbias{ap_88_89}{\ShortBenchmarkAP{}}%
\hfill%
\poolbias{ft}{\ShortBenchmarkFT{}}%
\hfill%
\poolbias{latimes}{\ShortBenchmarkLATimes{}}%
\hfill%
\poolbias{disk4_disk5_no-cr}{\ShortBenchmarkRobust{}}%
\hspace*{0.05\paperheight}%
\poolbias{wsj}{\ShortBenchmarkWSJ{}}%

\caption{Investigation of the average number of judged document (relevant and non-relevant) at different ranks for bag-of-words methods (\QLM{}) and \ModelName{} on article search benchmarks (\S\ref{sec:benchmarks}).\label{fig:poolbias}}

\end{figure*}

Fig.~\ref{fig:poolbias} shows the average number of judged documents at different ranks for bag-of-words methods (\QLM{}) and the \ModelName{} on the article search benchmarks (\S\ref{sec:benchmarks}). Here we omit the \ShortBenchmarkNYT{} benchmark, as \ModelName{}-based rankings were used during the construction of the \ShortBenchmarkNYT{} test collection. We can see that for \ShortBenchmarkAP{} and \ShortBenchmarkWSJ{}, all methods have approximately the same number of judged documents over all rank cut-offs. However, on \ShortBenchmarkWSJ{}, \QLM{} with Dirichlet smoothing has more judged documents for lower rank cut-offs (cut-off 1, 5, 10 and 20) and therefore we can conclude that there is a bias in the evaluation towards this method.

Interestingly, on the \ShortBenchmarkAP{} and \ShortBenchmarkWSJ{} benchmarks, as the rank cut-off increases from 1--50, we see that the \ModelName{} retrieves more, or the same number of, judged documents as the lexical methods in Fig.~\ref{fig:poolbias:ap_88_89}--\ref{fig:poolbias:wsj}. This indicates that the gains obtained by \ModelName{} are likely to originate from the torso of the ranking instead of the top. The judgement bias is even more prevalent on the \ShortBenchmarkFT{} and \ShortBenchmarkLATimes{} benchmarks. There, we observe a difference of about 10\% fewer judged documents, consistent across all rank cut-offs, for \ModelName{}. 

We conclude this analysis as follows. There is an indication of a judgement bias against \ModelName{} and more judgements are required in order to determine the true ranking amongst methods. In addition, ignoring unjudged documents does not solve the problem, as newer methods (i.e., \ModelName{}) are more likely to retrieve relevant unjudged documents than the methods used to construct the original pools (e.g., \QLM{}).

\subsection{Unsupervised deployment}
\label{sec:analysis:deployment}

In our experiments, we use a validation set for model selection (training iterations and hyperparameters). However, in many cases relevance labels are unavailable. Fortunately, \ModelName{} learns representations of the document collection directly and does not require query-document relevance information. How can we choose values for the hyperparameters of \ModelName{} in the absence of a validation set?
\begin{figure*}[t]
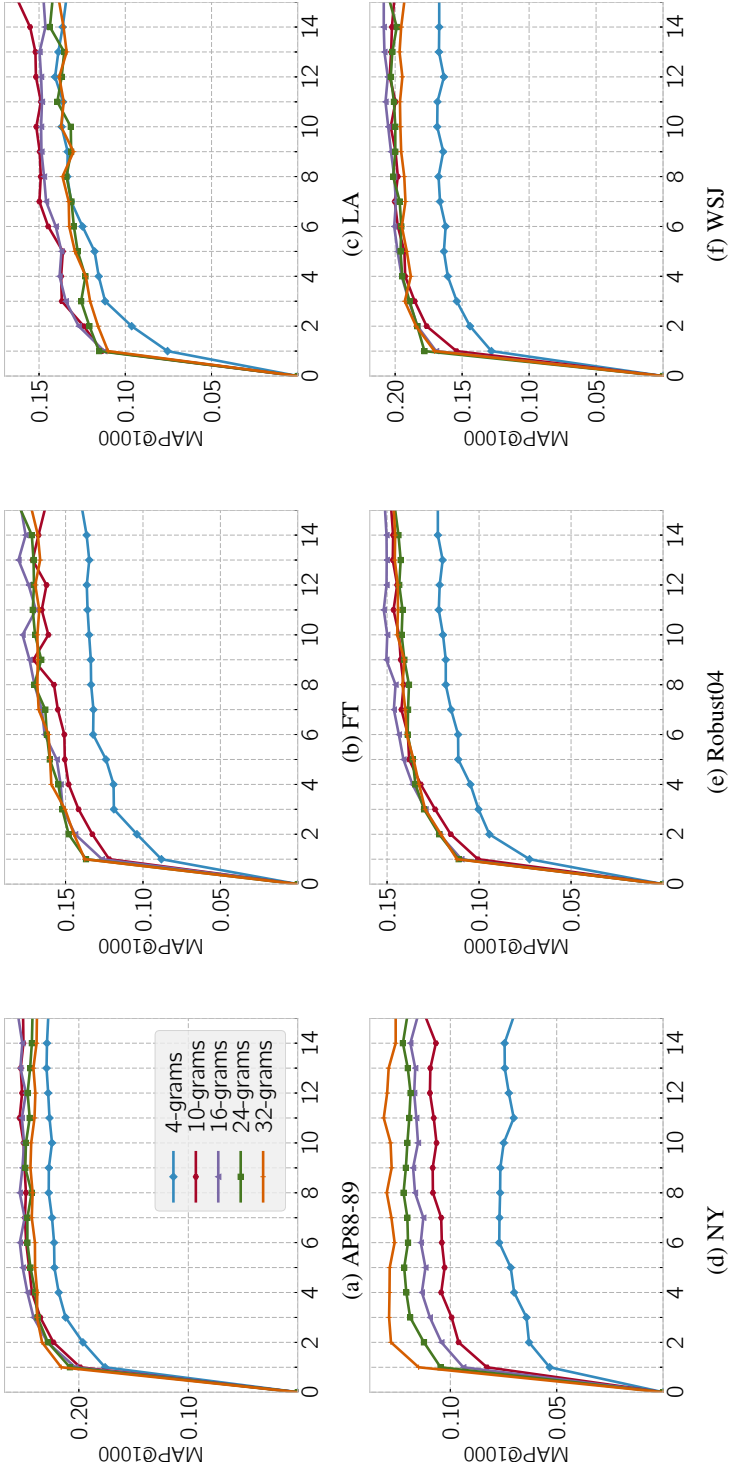


\newcommand{\convergence}[2]{%
    \def \inner {%
        \def \PlotPath {resources/analysis/convergence/#1/#1-qrel_test.pdf}%
        \IfFileExists{\PlotPath}{%
            \includegraphics[width=0.195\textwidth]{\PlotPath}}{%
            \resizebox{\BenchmarkFigureWidth{}}{!}{\missingfigure{#1}}}%
    }%
    \ifstrequal{#2}{}{%
    \subfloat{\inner}%
    }{\subfloat[#2\label{fig:convergence:#1}]{\inner}}%
}

\centering
\convergence{ap_88_89}{\ShortBenchmarkAP{}}%
\hfill%
\convergence{ft}{\ShortBenchmarkFT{}}%
\hfill%
\convergence{latimes}{\ShortBenchmarkLATimes{}}%

\centering
\convergence{nyt}{\ShortBenchmarkNYT{}}%
\hfill%
\convergence{disk4_disk5_no-cr}{\ShortBenchmarkRobust{}}%
\hfill%
\convergence{wsj}{\ShortBenchmarkWSJ{}}%

\caption{Test set \MAPCut{} as training progresses on article search benchmarks with document space dimensionality $\LatentDocumentDim{} = 256$. We see that \MAPCut{} converges to a fixed performance level with differently-sized $\NGramSize{}$-grams (here we show $\NGramSize{} = 4, 10, 16, 24, 32$; the curves for the remaining values $\NGramSize{} = 6, 8, 12$ are qualitatively similar and omitted to avoid clutter).\label{fig:convergence}}

\vspace*{.25\baselineskip}
\end{figure*}
For the majority of hyperparameters (\S\ref{sec:retrievalmodels}) we follow the setup of previous work \citep{VanGysel2016experts,VanGysel2016products}. We are, however, still tasked with the problem of choosing
\begin{inparaenum}[(a)]
    \item the number of training iterations,
    \item the dimensionality of the document representations $\LatentDocumentDim{}$, and
    \item the size of the $\NGramSize{}$-grams used for training.
\end{inparaenum}
We choose the dimensionality of the document representations $\LatentDocumentDim{} = 256$ as the value was reported to work well for \LSE{} \citep{VanGysel2016products}. Fig.~\ref{fig:convergence} shows that \MAPCut{} converges as the number of training iterations increases for different $\NGramSize{}$-gram widths. Therefore, we train \ModelName{} for \numprint{15} iterations and select the last iteration model.
\newcommand{\EnsembleScore}[1]{\Apply{\ScoreFn{}_\text{ensemble}}{\Query{}, \Document{}}}
\newcommand{\Gram}[1]{#1\text{-grams}}
\newcommand{\NGramScoreFn}[1]{\ScoreFn{}_{\Gram{#1}}}
\newcommand{\KGramScoreRandom}[1]{\Apply{\NGramScoreFn{#1}}{\Query{}, \Documents{}}}
\newcommand{\KGramScore}[1]{\Apply{\NGramScoreFn{#1}}{\Query{}, \Document{}}}
\newcommand{\NGramSizes}{\mathcal{N}}
{
    \let\ModelNameSingle\ModelName
    \newcommand{\ModelNameEnsemble}{8 \ModelNameSingle{}s (ensemble)}

    \renewcommand{\ModelName}{1 \ModelNameSingle{} (cross-validated)}

\begin{table*}[th!]
\centering
\renewcommand{\arraystretch}{1.2}
\caption{Comparison with single cross-validated \ModelNameSingle{} and ensemble of \ModelNameSingle{} through the unsupervised combination of models trained on differently-sized $\NGramSize{}$-grams ($\NGramSizes{} = \{ 2, 4, 8, 10, 12, 16, 24, 32 \}$). Significance (\S\ref{sec:evaluationmeasures}) is computed between \ModelNameSingle{} and the ensemble of \ModelNameSingle{}.\label{tbl:adhoc_ensemble_results}}
\scalebox{0.85}{%
\renewcommand{\MAPCut}{MAP}%
\renewcommand{\NDCGCut}{NDCG}%
\renewcommand{\BenchmarkAP}{\ShortBenchmarkAP}%
\renewcommand{\BenchmarkWSJ}{\ShortBenchmarkWSJ}%
\renewcommand{\BenchmarkLATimes}{\ShortBenchmarkLATimes}%
\renewcommand{\BenchmarkFT}{\ShortBenchmarkFT}%
\renewcommand{\BenchmarkNYT}{\ShortBenchmarkNYT}%
\centering%
\renewcommand{\arraystretch}{0.8}%
\setlength{\tabcolsep}{3.5pt}%
\begin{tabular}{c}%
\begin{tabular}{l c c c c c c}%
\toprule%
&\multicolumn{3}{c}{\BenchmarkAP{}}&\multicolumn{3}{c}{\BenchmarkFT{}}\\%
&\MAPCut&\NDCGCut&\Precision&\MAPCut&\NDCGCut&\Precision\\%
\cmidrule(lr){2-4}%
\cmidrule(lr){5-7}%
\ModelName{}&\nprounddigits{3}\npdecimalsign{.}\npthousandsep{.}\numprint{0.2567949580}$\phantom{\HighlySignificant}$&\nprounddigits{3}\npdecimalsign{.}\npthousandsep{.}\numprint{0.4163218487}$\phantom{\HighlySignificant}$&\nprounddigits{3}\npdecimalsign{.}\npthousandsep{.}\numprint{0.4235294118}$\phantom{\HighlySignificant}$&\nprounddigits{3}\npdecimalsign{.}\npthousandsep{.}\numprint{0.1697578947}$\phantom{\HighlySignificant}$&\nprounddigits{3}\npdecimalsign{.}\npthousandsep{.}\numprint{0.2985500000}$\phantom{\HighlySignificant}$&\nprounddigits{3}\npdecimalsign{.}\npthousandsep{.}\numprint{0.2359649123}$\phantom{\HighlySignificant}$\\%
\ModelNameEnsemble{}&\nprounddigits{3}\npdecimalsign{.}\npthousandsep{.}\textbf{\numprint{0.2820042017}}$\HighlySignificant$&\nprounddigits{3}\npdecimalsign{.}\npthousandsep{.}\textbf{\numprint{0.4533504202}}$\HighlySignificant$&\nprounddigits{3}\npdecimalsign{.}\npthousandsep{.}\textbf{\numprint{0.4663865546}}$\HighlySignificant$&\nprounddigits{3}\npdecimalsign{.}\npthousandsep{.}\textbf{\numprint{0.2120315789}}$\HighlySignificant$&\nprounddigits{3}\npdecimalsign{.}\npthousandsep{.}\textbf{\numprint{0.3518596491}}$\HighlySignificant$&\nprounddigits{3}\npdecimalsign{.}\npthousandsep{.}\textbf{\numprint{0.2815789474}}$\HighlySignificant$\\%
\midrule%
\end{tabular}%
\\%
\begin{tabular}{l c c c c c c}%
&\multicolumn{3}{c}{\BenchmarkLATimes{}}&\multicolumn{3}{c}{\BenchmarkNYT{}}\\%
&\MAPCut&\NDCGCut&\Precision&\MAPCut&\NDCGCut&\Precision\\%
\cmidrule(lr){2-4}%
\cmidrule(lr){5-7}%
\ModelName{}&\nprounddigits{3}\npdecimalsign{.}\npthousandsep{.}\numprint{0.1676036364}$\phantom{\HighlySignificant}$&\nprounddigits{3}\npdecimalsign{.}\npthousandsep{.}\numprint{0.3015336364}$\phantom{\HighlySignificant}$&\nprounddigits{3}\npdecimalsign{.}\npthousandsep{.}\numprint{0.2109090909}$\phantom{\HighlySignificant}$&\nprounddigits{3}\npdecimalsign{.}\npthousandsep{.}\numprint{0.1170420000}$\phantom{\UnknownSignificant}$&\nprounddigits{3}\npdecimalsign{.}\npthousandsep{.}\numprint{0.2078180000}$\phantom{\HighlySignificant}$&\nprounddigits{3}\npdecimalsign{.}\npthousandsep{.}\numprint{0.2960000000}$\phantom{\Significant}$\\%
\ModelNameEnsemble{}&\nprounddigits{3}\npdecimalsign{.}\npthousandsep{.}\textbf{\numprint{0.1880645455}}$\HighlySignificant$&\nprounddigits{3}\npdecimalsign{.}\npthousandsep{.}\textbf{\numprint{0.3313945455}}$\HighlySignificant$&\nprounddigits{3}\npdecimalsign{.}\npthousandsep{.}\textbf{\numprint{0.2300000000}}$\HighlySignificant$&\nprounddigits{3}\npdecimalsign{.}\npthousandsep{.}\textbf{\numprint{0.1206640000}}$\UnknownSignificant$&\nprounddigits{3}\npdecimalsign{.}\npthousandsep{.}\textbf{\numprint{0.2375800000}}$\HighlySignificant$&\nprounddigits{3}\npdecimalsign{.}\npthousandsep{.}\textbf{\numprint{0.3240000000}}$\Significant$\\%
\midrule%
\end{tabular}%
\\%
\begin{tabular}{l c c c c c c}%
&\multicolumn{3}{c}{\BenchmarkRobust{}}&\multicolumn{3}{c}{\BenchmarkWSJ{}}\\%
&\MAPCut&\NDCGCut&\Precision&\MAPCut&\NDCGCut&\Precision\\%
\cmidrule(lr){2-4}%
\cmidrule(lr){5-7}%
\ModelName{}&\nprounddigits{3}\npdecimalsign{.}\npthousandsep{.}\numprint{0.1502525510}$\phantom{\HighlySignificant}$&\nprounddigits{3}\npdecimalsign{.}\npthousandsep{.}\numprint{0.2872454082}$\phantom{\HighlySignificant}$&\nprounddigits{3}\npdecimalsign{.}\npthousandsep{.}\numprint{0.2974489796}$\phantom{\HighlySignificant}$&\nprounddigits{3}\npdecimalsign{.}\npthousandsep{.}\numprint{0.2079558333}$\phantom{\HighlySignificant}$&\nprounddigits{3}\npdecimalsign{.}\npthousandsep{.}\numprint{0.3509033333}$\phantom{\HighlySignificant}$&\nprounddigits{3}\npdecimalsign{.}\npthousandsep{.}\numprint{0.3716666667}$\phantom{\HighlySignificant}$\\%
\ModelNameEnsemble{}&\nprounddigits{3}\npdecimalsign{.}\npthousandsep{.}\textbf{\numprint{0.1706806122}}$\HighlySignificant$&\nprounddigits{3}\npdecimalsign{.}\npthousandsep{.}\textbf{\numprint{0.3231954082}}$\HighlySignificant$&\nprounddigits{3}\npdecimalsign{.}\npthousandsep{.}\textbf{\numprint{0.3306122449}}$\HighlySignificant$&\nprounddigits{3}\npdecimalsign{.}\npthousandsep{.}\textbf{\numprint{0.2251033333}}$\HighlySignificant$&\nprounddigits{3}\npdecimalsign{.}\npthousandsep{.}\textbf{\numprint{0.3852450000}}$\HighlySignificant$&\nprounddigits{3}\npdecimalsign{.}\npthousandsep{.}\textbf{\numprint{0.4233333333}}$\HighlySignificant$\\%
\bottomrule%
\end{tabular}%
\end{tabular}%
}%
\end{table*}%

}
The final remaining question is the choice of $\NGramSize{}$-gram size used during training. This parameter has a big influence on model performance as it determines the amount of context from which semantic relationships are learned. Therefore, we propose to combine different vector spaces trained using different $\NGramSize{}$-gram widths as follows. We write $\NGramSizes{}$ for the set of all $k$ for which we construct an \ModelName{} using $k$-grams. For a given query $\Query{}$, we rank documents $\Document{} \in \Documents{}$ in descending order of:
\begin{equation}
\label{eq:ensemble}
\EnsembleScore{} = \sum_{k \in \NGramSizes{}} \frac{\KGramScore{k} - \mu_{\Gram{k}, \Query{}}}{\sigma_{\Gram{k}, \Query{}}},
\end{equation}
where $\KGramScore{k}$ is Eq.~\ref{eq:score} for \ModelName{} of $\Gram{k}$ and
\begin{equation}
\begin{split}
\mu_{\Gram{k}, \Query{}} & = \ApplySquare{\SampleExpectation}{\KGramScoreRandom{k}} \\
\sigma_{\Gram{k}, \Query{}} & = \sqrt{\ApplySquare{\SampleVariance}{\KGramScoreRandom{k}}},
\end{split}
\end{equation}
denote the sample expectation and sample variance over documents $\Documents{}$ that are estimated on the top-$1000$ documents returned by the individual models, respectively. That is, we rank documents according to the sum of the standardized scores of vector space models trained with different $\NGramSize{}$-gram widths. The score aggregation in Eq.~\ref{eq:ensemble} is performed without any a priori knowledge about the $\NGramSize{}$-gram sizes. Table~\ref{tbl:adhoc_ensemble_results} lists the performance of the unsupervised ensemble, where every model was trained for 15 iterations, against a single cross-validated model. We see that the unsupervised ensemble always outperforms (significantly in terms of \MAPCut{} for all benchmarks except \ShortBenchmarkNYT{}) the singleton model. Hence, we can easily deploy \ModelName{} without any supervision and, surprisingly, it will perform better than individual models optimized on a validation set.


\section{Summary}

We proposed the \FullModelName{} (\ModelName{}) that learns representations of a document collection in an unsupervised manner. 

We showed that \ModelName{} performs better than existing latent vector space/bag-of-words approaches. \ModelName{} performs lexical and semantic matching in a latent space. \ModelName{} provides a complementary signal to lexical language models. In addition, we showed that \ModelName{} automatically learns a notion of term specificity. Finally, we gave advice on how to select values for the hyperparameters of \ModelName{}. Interestingly, an unsupervised ensemble of multiple models trained with different hyperparameters performs better than a single cross-validated model.

The evidence that \ModelName{} provides a notion of lexical matching tells us that latent vector space models are not limited to only semantic matching. While the framework presented in this \paper{} focuses on a single unsupervised objective, additional objectives (i.e., document/document or query/document similarity) can be incorporated to improve retrieval performance.

The \LSE{} model \citep{VanGysel2016products}---introduced in Chapter~\globalref{chapter:research-05}---improved the learning time complexity of earlier entity retrieval models (Chapter~\globalref{chapter:research-03}) \citep{VanGysel2016experts} such that they scale to ${\sim}100\text{k}$ retrievable items (i.e., entities). However, as shown in Table~\ref{tbl:adhoc_results}, \LSE{} performs poorly on article retrieval benchmarks. In this \paper{}, we extend \LSE{} and learn vector spaces of ${\sim}500\text{k}$ documents that perform better than existing latent vector spaces. As mentioned in the introduction, the main challenge for latent vector spaces is their limited scalability to large document collections due to space complexity. The observation that retrieval is not only impacted by the vector space representation of the relevant document, but also of the documents surrounding it, raises non-trivial questions regarding the distribution of document vectors over multiple machines. While there have been efforts towards distributed training of neural models, the application of distributed learning algorithms is left for future work. The unsupervised objective that learns from word sequences is limited by its inability to deal with very short documents. While this makes the unsupervised objective less applicable in domains such as web search, unsupervised bag-of-words approaches have the opposite problem of degrading performance when used to search over long documents. With respect to incremental indexing, there is currently no theoretically sound way to obtain representations for new documents that were added to the collection after the initial estimation of a \ModelName{}. In the case of \LDA{} or \LSI{}, representations for new documents can be obtained by transforming bag-of-words vectors to the latent space. However, as the \LDA{}/\LSI{} transformation to the latent space is not updated after estimating the \LDA{}/\LSI{} model using the initial set of documents, this procedure can be catastrophic when topic drift occurs. For \FullDocToVec{}, one way to obtain a representation for a previously-unseen document is to keep all parameters fixed and train the representation of the new document using the standard training algorithm \citep{Rehurek2010gensim}. This approach can also be used in the case of \LSE{} or \ModelName{}. However, there are no guarantees that the obtained representation will be of desirable quality. In addition, the same problem remains as with the bag-of-words methods. That is, the previously-mentioned incremental updating mechanism is likely to fail when topic drift occurs.
}

}

\bookmarksetup{startatroot}
\addtocontents{toc}{\bigskip}


\chapter{Conclusions}
\label{chapter:conclusions}

In this dissertation, we have devoted six research chapters to address two limitations of the inverted index that contribute to the vocabulary gap between query and document. This vocabulary gap causes decreased retrieval effectiveness that occurs due to:
\begin{inparaenum}[(1)]
	\item the use of complex textual structures as an unfiltered query with many---possibly misleading---terms and
	\item queries and their relevant documents that use different words to describe the same concepts. 
\end{inparaenum}
Specifically, Part~\ref{part:formulation} considers the task of formulating an effective query from a complex textual structure (i.e., search sessions, email threads) such that the formulated query is more focused and better satisfies the information need. Part~\ref{part:latent} is dedicated to latent vector spaces that allow us to bridge the semantic vocabulary gap.

In this final chapter, we revisit the research questions we answered and summarize our findings in Section~\ref{sec:findings}. In Section~\ref{sec:future}, we discuss limitations of our work and directions for future work.

\section{Main findings}
\label{sec:findings}

\newcommand{\MainRQAnswer}[3]{

\medskip

\begin{description}\item[RQ#1] #2\end{description}
#3}

We now revisit our research questions introduced in Chapter~\ref{chapter:introduction} and summarize our findings.

\MainRQAnswer{1}{\MainFirstRQ{}}{%
To answer our first question, we performed an analysis of TREC search session logs in Chapter~\ref{chapter:research-01} and introduced a frequency-based query term weighting method that summarizes the user's information need. We found that the semantic vocabulary gap, the subject of the next question, is prevalent in session search and that methods restricted to the user-specified query terms face a very strict performance ceiling.
The focus of Chapter~\ref{chapter:research-02} was to formulate a query from an email thread. In particular, we focused on the case where an incoming email contains a request for content and a query needs to be formulated to retrieve the relevant attachment from a repository. The information need in the email scenario is less clear than in session search, as email messages have multiple aspects and the request for content can be implicit. We introduced a methodology for constructing a pseudo test/training collection from email collections, including the construction of silver-standard training queries, and proposed a neural network architecture that learns to select query terms from the incoming request message.

The principal shortcoming of the query formulation methods explored in \MainRQRef{1} is that the extracted query terms are limited to the terms present in the complex textual structures. In session search, query reformulations performed by the user can succinctly describe the information need as the user directly interacts with the search engine. However, for the case of email attachment recommendation where the interaction between requester and search engine is indirect and possibly even unknown to the requester, the conditions for a serious discrepancy between the content request and the document repository have been met. Consequently, the semantic mismatch between query/document was addressed in the following questions.}

\MainRQAnswer{2}{\MainSecondRQ{}}{%
We introduced a latent semantic model for the expert finding task in Chapter~\ref{chapter:research-03}. We showed that our latent model, that consists of word/expert representations and a prior over experts, outperforms state-of-the-art retrieval models and contributes a complementary signal to lexical models. In Chapter~\ref{chapter:research-04}, we investigated the structural regularities that are present in entity vector spaces. We showed that entity (i.e., expert individuals) representations estimated from associated texts alone can be used as feature vectors for clustering and recommendation. In addition, we showed that the prior over entities of the model introduced in Chapter~\ref{chapter:research-03} encodes entity salience.

However, the expert finding task is characterized by a few thousand entities that are each represented by a sizeable collection of documents that cover their expertise. This setting is ideal for latent semantic models like ours that learn from context due to the abundance of textual content. In addition, queries used to search in expert finding are informational \citep{Broder2002taxonomy}, cover a broad topic and consequently have a need for semantic matching. Note that the case when the user knows the name of the expert (i.e., known-item search) is different from the expert finding setting. This brings us to our following question: can we adapt our latent model to larger entity domains where textual content is scarce? We answered this question as part of \MainRQRef{3}.}

\MainRQAnswer{3}{\MainThirdRQ{}}{%
Chapter~\ref{chapter:research-05} introduced \acl{LSE} (\acs{LSE}), a modification of our latent model for expert finding where queries and entities are represented in a latent metric space. This implies that we no longer learn an importance prior over entities. In addition, the scalability of the learning mechanism was improved by sampling. We evaluated \acs{LSE} in a product search setting and showed that it improves retrieval effectiveness as a complementary signal next to product saliency and lexical matching in a product-oriented search engine.

The solution to \MainRQRef{3} was a training mechanism that is based on sampling rather than considering the full set of entities. Nearest neighbour algorithms can be used to rank entities with a time complexity sub-linear w.r.t. the number of entities. However, entity ranking is only a small part of information retrieval. How do our latent vector spaces operate in a more traditional setting, such as news article retrieval?}

\MainRQAnswer{4}{\MainFourthRQ{}}{%
With Chapter~\ref{chapter:research-06}, we saw the introduction of an extension to \acs{LSE}: the \acl{NVSM} (\acs{NVSM}). We evaluated \acs{NVSM} on article retrieval benchmarks from TREC and showed that it outperforms all of the existing state-of-the-art latent vector spaces. In addition, we showed that \acs{NVSM} significantly improves retrieval effectiveness when added as a complimentary feature in addition to a lexical language model and another latent vector space model. Consequently, \acs{NVSM} contributes an additional signal. A comparative analysis showed that \acs{NVSM} performs better on queries than other latent vector spaces when the queries require a greater extent of lexical matching. Further investigation resulted in the observation that \acs{NVSM} learns a notion of Luhn significance, a quantity known to be important for retrieval. In particular, we found that the L2-norm of mid-frequency words is significantly larger (and thus, the words are of greater importance) than low- and high-frequency words.}

\medskip

To conclude this section, we reflect on the first chapter of this dissertation and repeat the two major drawbacks of the inverted index that motivated our research:
\begin{inparaenum}[(1)]
	\item Queries consisting of many terms induce high computational costs, while there often exists a shorter query that is more effective in fulfilling an information need. We addressed this drawback in Part~\ref{part:formulation} of this dissertation by answering \MainRQRef{1}. In particular, we focused on the case where we wish to formulate a query from a complex textual structure. However, we also found that term-based matching by itself is not sufficient to fulfil information needs as there exists a vocabulary gap between the user query and the relevant document.
	\item Part~\ref{part:latent} of this dissertation was dedicated to bridging the vocabulary gap. The use of term-based matching to build an initial candidate set of documents may incorrectly classify relevant documents that do not contain query terms as irrelevant. We addressed this issue in \MainRQRef{2}, \MainRQRef{3} and \MainRQRef{4} by the development of latent vector spaces where queries and documents are matched according to their semantics rather than exact term occurrences. Consequently, term-based retrieval may be complemented by semantic vector spaces that can be queried through a nearest neighbour search in a low-dimensional vector space.
\end{inparaenum}

\section{Future work}
\label{sec:future}

This dissertation resulted in insights and algorithms for bridging the vocabulary gap in IR. However, the research performed as part of this dissertation raised more questions than it answered. In this section, we summarize the limitations of our work and conclude this dissertation with directions for future work.

The limitations of Part~\ref{part:formulation} of this dissertation are as follows.
\begin{inparaenum}[(a)]
\item In Part~\ref{part:formulation} we only considered terms occurring within the complex textual structures as candidates. While this prevents the term candidate set from becoming too large, it does limit the ability for methods to formulate expressive queries in the case where textual data is scarce.
\item The retrieval model used in Part~\ref{part:formulation}, a language model with Dirichlet smoothing, is ubiquitous in retrieval systems. However, smoothing allows the search engine to deal with verbose queries \citep{Zhai2004smoothing} that contain terms absent from the messages. Consequently, our findings may change when considering other retrieval model classes, such as boolean models or semantic matching models.
\end{inparaenum}

Considering Part~\ref{part:latent} of this dissertation, its limitations are as follows:
\begin{inparaenum}[(a)]
\item The largest retrieval collection used in Part~\ref{part:latent} of this dissertation consists of half a million documents. While we obtained promising results on larger collections of up to two million documents (not included in this dissertation), the question remains of whether latent vector spaces are applicable in large retrieval scenarios. The question of applicability applies to two separate aspects. The first aspect pertains to the training of latent vector space models. More specifically, the training of latent vector space models is limited by their space complexity that grows linearly with the number of document terms. Secondly, the question remains whether (approximate) nearest neighbour search algorithms are actually efficient enough to perform retrieval in real-time. This brings us to the second limitation of Part~\ref{part:latent}.
\item In this dissertation, we assumed that the modelling of latent vector spaces is separated from the development of nearest neighbour algorithms that are used to query them. Particularly, we did not evaluate the effect of using approximate nearest neighbour algorithms---which are likely required for real-time querying---on retrieval effectiveness. However, in this dissertation we performed our retrieval evaluation using offline test collections. In addition, the focus of this dissertation lies on the modelling side of things. Consequently, the use of exact nearest neighbour algorithms is justified in this dissertation.
\end{inparaenum}

\bigskip

\noindent
To address these limitations, we identify the following directions for future work:
\vspace*{-1\baselineskip}
\paragraph{Query formulation} We explored the task of formulating queries from complex textual structures, with applications to session search and email. We first discuss directions for the applications and then focus on the general task. In the case of session search (Chapter~\ref{chapter:research-01}), there is still much room for improvement by re-weighting query terms. Future work should focus on better lexical query models for session search, in addition to semantic matching and tracking the dynamics of contextualized semantics in search.

For email attachment recommendation (Chapter~\ref{chapter:research-02}), future work includes the incorporation of social connections in the email domain where a social graph can be constructed from email interactions and entity mentions. In addition, structured queries with operators searching different fields (e.g., recipients, subject) can improve performance. Finally, we assumed a single model for all mailboxes. However, per-mailbox specialized models are likely to generate better queries. Overall, the query/document mismatch is prevalent when formulating queries from complex textual structures and methods restricted to lexical query modelling face a very strict performance ceiling. Consequently, future work should focus on formulating queries using the full set of terms, instead of only those occurring in the structures. This is non-trivial as retrieval systems deal with very large dictionaries, and therefore, the effect of including a single term is hard to estimate.

\paragraph{Latent representations as feature vectors} In Chapter~\ref{chapter:research-04} we explored using latent entity representations as feature vectors in other applications, such as clustering, recommendation and determining entity salience. Future work includes the use of text-based entity representations in end-to-end applications. For example, in social networks these methods can be applied to cluster users or to induce graphs based on thread participation or hashtag usage. In addition, text-based entity representations can be used as item feature vectors in recommendation systems. Beyond text-only entity collections, there is also a plenitude of applications where entity relations are available. While there has been some work on learning latent representations from entity relations \citep{Bordes2011,Zhao2015}, little attention has so far been given to combining textual evidence and entity relations. Therefore, we identify two additional directions for future work. First, an analysis showing in what capacity entity representations estimated from text alone encode entity-entity relations (beyond the co-associations considered in this work). Secondly, the incorporation of entity-entity similarity in the construction of latent entity representations.

\paragraph{Latent vector spaces for information retrieval}
Part~\ref{part:latent} of this dissertation is centered around the construction of latent vector spaces for the retrieval task. While significant progress was made, more questions and directions for future work arise:
\begin{inparaenum}[(1)]
\item How can we further improve the retrieval effectiveness of latent vector spaces? Future work includes adding additional model expressiveness through depth or width. In addition, multiple nearest neighbour searches for every query term could be performed in parallel and consequently, the influence of individual query terms can be combined in a more expressive way.
\item Additional signals of relevance (e.g., query/document pairs) or similarity, such as entity/entity similarity \citep{Reinanda2017phd}, can be incorporated during training. Modelling objects beyond entities and query terms, such as users within a personalization context \citep{Ai2017personalizedproducts} is also a promising direction.
\item Can we scale up the latent vector spaces to hundreds of millions of documents? This direction introduces both engineering and modelling challenges. One way to scale up the existing vector spaces presented in this dissertation is to train multiple models in parallel on sub-samples of the full document collection.
\item How do our latent vector spaces perform in online settings? In particular, what is the effect of approximate nearest neighbour algorithms on retrieval performance? This direction comes with non-trivial engineering challenges and is best performed when one has access to a platform with actual users.
\end{inparaenum}

\part*{Appendices}
\appendix
{
\newcommand{\paper}{appendix}

\newminted{python}{mathescape, numbersep=5pt, autogobble, framesep=2mm, fontsize=\footnotesize}

\newfloat{algorithm}{t}{lop}
\floatname{algorithm}{Code snippet}


\chapter{Pyndri: A Python Interface to the Indri Search Engine}
\label{chapter:research-07}

{
\ScopeLabels{research-07}

\section{Introduction}

Research in Artificial Intelligence progresses at a rate proportional to the time it takes to implement an idea. Therefore, it is natural for researchers to prefer scripting languages (e.g., Python) over conventional programming languages (e.g., C++) as programs implemented using the latter are often up to three factors longer (in lines of code) and require twice as much time to implement \citep{Prechelt2000}. Python, an interactive scripting language that emphasizes readability, has risen in popularity due to its wide range of scientific libraries (e.g., NumPy), built-in data structures and holistic language design \citep{Koepke2010}.

There is still, however, a lack of an integrated Python library dedicated to information retrieval research. Researchers often implement their own procedures to parse common file formats, perform tokenization, token normalization that encompass the overall task of corpus indexing. \citet{Uysal2014} show that text classification algorithms can perform significantly differently, depending on the level of preprocessing performed. Existing frameworks, such as NLTK \citep{Loper2002NLTK}, are primarily targeted at processing natural language as opposed to retrieving information and do not scale well. At the algorithm level, small implementation differences can have significant differences in retrieval performance due to floating point errors \citep{Goldberg1991fp}. While this is unavoidable due to the fast-paced nature of research, at least for seminal algorithms and models, standardized implementations are needed.
\section{Introducing Pyndri}

Fortunately, the IR community has developed a series of indexing frameworks (e.g., Galago, Lucene, Terrier) that correctly implement a wide range of retrieval models. The Indri search engine \citep{Strohman2005indri} supports complex queries involving \emph{evidence combination} and the ability to specify a wide variety of constraints involving \emph{proximity}, \emph{syntax}, \emph{extracted entities} and \emph{document structure}. Furthermore, the framework has been efficiently implemented using C++ and was designed from the ground up to support \emph{very large databases}, \emph{optimized query execution} and \emph{fast and concurrent indexing}. A large subset of the retrieval models \citep{Zhai2001smoothing,Balog2006experts,Bendersky2010sdm,Guan2013qcm,VanGysel2016sessions} introduced over the course of history can be succinctly formulated as an Indri query. However, to do so in an automated manner, up until now researchers were required to resort to C++, Java or shell scripting. C++ and Java, while excellent for production-style systems, are slow and inflexible for the fast prototyping paradigm used in research. Shell scripting fits better in the research paradigm, but offers poor string processing functionality and can be error-prone. Besides, shell scripting is unsuited if one wants to evaluate a large number of complex queries or wishes to extract documents from the repository as this incurs overhead, causing avoidable slow execution. Existing Python libraries for indexing and searching, such as PyLucene, Whoosh or ElasticSearch, do not support the rich Indri language and functionality required for rapid prototyping.

\begin{algorithm}
\begin{minted}[frame=lines]{python}
index = pyndri.Index('/opt/local/clueweb09')

for int_doc_id in range(index.document_base(),
                        index.maximum_document()):
    ext_doc_id, doc_tokens = index.document(int_doc_id)
\end{minted}
\caption{Tokenized documents in the index can be iterated over. The \texttt{ext\_doc\_id} variable in the inner loop will equal the document identifier (e.g.,  \texttt{clueweb09-en0039-05-00000}), while the \texttt{doc\_tokens} points to a tuple of integers that correspond to the document term identifiers.\label{example:iteration}}
\end{algorithm}

We fill this gap by introducing pyndri, a lightweight interface to the Indri search engine. Pyndri offers read-only access at two levels in a given Indri index.

\subsection{Low-level access to document repository}
First of all, pyndri allows the retrieval of tokenized documents stored in the index repository. This allows researchers to avoid implementing their own format parsing as Indri supports all major formats used in IR, such as the trectext, trecweb, XML documents and Web ARChive (WARC) formats. Furthermore, standardized tokenization and normalization of texts is performed by Indri and is no longer a burden to the researcher. Code snippet~\ref{example:iteration} shows how a researcher can easily access documents in the index. Lookup of internal document identifiers given their external name is provided by the \texttt{Index.document\_ids} function.

\begin{algorithm}
\begin{minted}[frame=lines]{python}
index = pyndri.Index('/opt/local/clueweb09')
dictionary = pyndri.extract_dictionary(index)

_, int_doc_id = index.document_ids(
    ['clueweb09-en0039-05-00000'])
print([dictionary[token_id]
       for token_id in index.document(int_doc_id)[1]])
\end{minted}
\caption{A specific document is retrieved by its external document identifier. The index dictionary can be queried as well. In the above example, a list of token strings corresponding to the document's contents will be printed to \texttt{stdout}.\label{example:retrieval}}
\end{algorithm}

The \texttt{dictionary} of the index (Code snippet~\ref{example:retrieval}) can be accessed from Python as well. Beyond bi-directional token-to-identifier translation, the dictionary contains corpus statistics such as term and document frequencies as well. The combination of index iteration and dictionary interfacing integrates conveniently with the Gensim\footnote{\url{https://radimrehurek.com/gensim}} package, a collection of topic and latent semantic models such as LSI \citep{Deerwester1990lsi} and word2vec \citep{Mikolov2013word2vec}. In particular for word2vec, this allows for the training of word embeddings on a corpus while avoiding the tokenization mismatch between the index and word2vec. In addition to tokenized documents, pyndri also supports retrieving various corpus statistics such as document length and corpus term frequency.

\begin{algorithm}
\begin{minted}[frame=lines]{python}
index = pyndri.Index('/opt/local/clueweb09')

for int_doc, score in index.query('obama family tree'):
    # Do stuff with the document.
\end{minted}
\caption{Simple queries can be fired using a simple interface. Here we query the index for topic \texttt{wt09-1} from the TREC 2009 Web Track using the Indri defaults (Query Language Model (QLM) with Dirichlet smoothing, $\mu = 2500$).\label{example:simple_query}}
\end{algorithm}

\subsection{Querying Indri from Python}
Secondly, pyndri allows the execution of Indri queries using the index. Code snippet~\ref{example:simple_query} shows how one would query an index using a topic from the TREC 2009 Web Track using the Indri default retrieval model.
\begin{algorithm}
\begin{minted}[frame=lines]{python}
index = pyndri.Index('/opt/local/clueweb09')
query_env = pyndri.QueryEnvironment(
    index, rules=('method:dirichlet,mu:5000',))

results = query_env.query(
    '#weight( 0.70 obama 0.20 family 0.10 tree )',
    document_set=map(
        operator.itemgetter(1),
        index.document_ids([
            'clueweb09-en0003-55-31884',
            'clueweb09-en0006-21-20387',
            'clueweb09-enwp01-75-20596',
            'clueweb09-enwp00-64-03709',
            'clueweb09-en0005-76-03988'
        ])),
    results_requested=3,
    include_snippets=True)

for int_doc_id, score, snippet in results:
    # Do stuff with the document and snippet.
\end{minted}
\caption{Advanced querying of topic \texttt{wt09-1} with custom smoothing rules, using a weighted-QLM. Only a subset of documents is searched and we impose a limit on the size of the returned list. In addition to the document identifiers and their retrieval score, the function now returns snippets of the documents where the query terms match.\label{example:advanced_query}}
\end{algorithm}
Beyond simple terms, the \texttt{query()} function fully supports the Indri Query Language.\footnote{\url{http://lemurproject.org/lemur/IndriQueryLanguage.php}}

In addition, we can specify a subset of documents to query, the number of requested results and whether or not snippets should be returned. In Code snippet~\ref{example:advanced_query} we create a \texttt{QueryEnvironment}, with a set of custom smoothing rules. This allows the user to apply fine-grained smoothing settings (i.e., per-field granularity).
\section{Summary}

In this \paper{} we introduced pyndri, a Python interface to the Indri search engine. Pyndri allows researchers to access tokenized documents from Indri using a convenient Python interface. By relying on Indri for tokenization and normalization, IR researchers are no longer burdened by this task. In addition, complex retrieval models can easily be implemented by constructing them in the Indri Query Language in Python and querying the index. This will make it easier for researchers to release their code, as Python is designed to be readable and cross-platform. We hope that with the release of pyndri, we will stimulate \textbf{reproducible}, \textbf{open} and \textbf{fast-paced} IR research. More information regarding the available API and installation instructions can be found on Github.\footnote{\url{https://github.com/cvangysel/pyndri}}
}

\chapter{Semantic Entity Retrieval Toolkit}
\label{chapter:research-08}

{
\ScopeLabels{research-08}

\newcommand{\paperImplementationUrl}{https://github.com/cvangysel/SERT}

\section{Introduction}

The unsupervised learning of low-dimensional, semantic representations of words and entities has recently gained attention for the entity-oriented tasks of expert finding \citep{VanGysel2016experts} and product search \citep{VanGysel2016products}. Representations are learned from a document collection and domain-specific associations between documents and entities. Expert finding is the task of finding the right person with the appropriate skills or knowledge \citep{Balog2012survey} and an association indicates document authorship (e.g., academic papers) or involvement in a project (e.g., annual progress reports). In the case of product search, an associated document is a product description or review \citep{VanGysel2016products}.

In this \paper{} we describe the Semantic Entity Retrieval Toolkit (\SERT{}) that provides implementations of our previously published entity representation models \citep{VanGysel2016experts,VanGysel2016products}. Beyond a unified interface that combines different models, the toolkit allows for fine-grained parsing configuration and GPU-based training through integration with Theano \citep{Lasagne2015,Theano2016}. Users can easily extend existing models or implement their own models within the unified framework. After model training, \SERT{} can compute matching scores between an entity and a piece of text (e.g., a query). This matching score can then be used for ranking entities, or as a feature in a downstream machine learning system, such as the learning to rank component of a search engine. In addition, the learned representations can be extracted and used as feature vectors in entity clustering or recommendation tasks \citep{VanGysel2017expertregularities}. The toolkit is licensed under the permissive MIT open-source license.\footnote{The toolkit is licensed under the permissive MIT open-source license and can be found at \url{\paperImplementationUrl}.}
\section{The toolkit}

\begin{figure}
\centering
\begin{tikzpicture}
\node [draw, text width=2.5cm, align=center] (prepare) at (-0.85, 0) {\textbf{Text processing}\\(prepare; \S\ref{sec:prepare})};
\node [draw, text width=2.3cm, align=center] (train) at (2.25, 0) {\textbf{Repr. learning}\\(train; \S\ref{sec:train})};
\node [draw, text width=1.8cm, align=center] (query) at (5, 0) {\textbf{Inference}\\(query; \S\ref{sec:query})};

\node [above = 0.10cm of train] {\textbf{\acl{SERT}}};

\draw [->, line width=0.25mm] (prepare) -- (train);
\draw [->, line width=0.25mm] (train) -- (query);
\end{tikzpicture}

\caption{Schematic overview of the different pipeline components of SERT. The collection is parsed, processed and packaged in a numerical format using the \emph{prepare} (\S\ref{sec:prepare}) utility. Afterwards, the \emph{training} (\S\ref{sec:train}) utility learns representations of entities and words and the \emph{query} (\S\ref{sec:query}) utility is used to compute matching scores between entities and queries.\label{fig:overview}}

\end{figure}
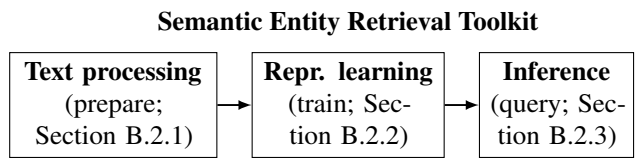

\SERT{} is organized as a pipeline of utilities as depicted in Fig.~\ref{fig:overview}. First, a collection of documents and entity associations is processed and packaged using a numerical format (\S\ref{sec:prepare}). Low-dimensional representations of words and entities are then learned (\S\ref{sec:train}) and afterwards the representations can be used to make inferences (\S\ref{sec:query}).

\subsection{Collection parsing and preparation}
\label{sec:prepare}

To begin, \SERT{} constructs a vocabulary that will be used to tokenize the document collection. Non-significant words that are too frequent (e.g., stopwords), noisy (e.g., single characters) and rare words are filtered out. Words that do not occur in the dictionary are ignored. Afterwards, word sequences are extracted from the documents and stored together with the associated entities in the numerical format provided by NumPy \citep{VanDerWalt2011numpy}. Word sequences can be extracted consecutively or a stride can be specified to extract non-consecutive windows. In addition, a hierarchy of word sequence extractors can be applied to extract skip-grams, i.e., word sequences where a number of tokens are skipped after selecting a token \citep{Guthrie2006skipgram}. To support short documents, a special-purpose padding token can be used to fill up word sequences that are longer than a particular document.

After word sequence extraction, a weight can be assigned to each word sequence/entity pair that can be used to re-weight the training objective. For example, in the case of expert finding \citep{VanGysel2016experts}, this weight is the reciprocal of the document length of the document where the sequence was extracted from. This avoids a bias in the objective towards long documents. An alternative option that exists within the toolkit is to resample word sequence/entity pairs such that every entity is associated with the same number of word sequences, as used for product search \citep{VanGysel2016products}.

\subsection{Representation learning}
\label{sec:train}

\begin{algorithm}
\begin{minted}[frame=lines]{python}
class ExampleModel(VectorSpaceLanguageModelBase):

    def __init__(self, *args, **kwargs):
        super(ExampleModel, self).__init__(
            *args, **kwargs)

        # Define model architecture.
        input_layer = InputLayer(
            shape=(self.batch_size, self.window_size))

        ...

        def loss_fn(pred, actual, _):
            # Compute symbolic loss between
            # predicted/actual entities.

        # The framework deals with underlying boilerplate.
        self._finalize(loss_fn, ....)

    def get_representations(self):
        # Returns the representations and parameters
        # to be extracted.
\end{minted}
\caption{Illustrative example of the \SERT{} model interface. The full interface supports more functionality omitted here for brevity. Users can define a symbolic graph of computation using the Theano library \citep{Theano2016} in combination with Lasagne \citep{Lasagne2015}.\label{snippet:model}}
\vspace*{-\baselineskip}
\end{algorithm}

After the collection has been processed and packaged in a machine-friendly format, representations of words and entities can be learned. The toolkit includes implementations of state-of-the-art representation learning models that were applied to expert finding \citep{VanGysel2016experts} and product search \citep{VanGysel2016products}.
Users of the toolkit can use these implementations to learn representations out-of-the-box or adapt the algorithms to their needs. In addition, users can implement their own models by extending an interface provided by the framework. Code snippet~\ref{snippet:model} shows an example of a model implemented in the \SERT{} toolkit where users can define a symbolic cost function that will be optimized using Theano \citep{Theano2016}. Due to the component-wise organization of the toolkit (Fig.~\ref{fig:overview}), modelling and text processing are separated from each other. Consequently, researchers can focus on modelling and representation learning only. In addition, any improvements to the collection processing (\S\ref{sec:prepare}) collectively benefits all models implemented in \SERT{}.

\subsection{Entity ranking \& other uses of the representations}
\label{sec:query}

Once a model has been trained, \SERT{} can be used to rank entities w.r.t. a textual query. The concrete implementation used to rank entities depends on the model that was trained. In the most generic case, a matching score is computed for every entity and entities are ranked in decreasing order of his score. However, in the special case when the model is interpreted as a metric vector space \citep{VanGysel2016products,Boytsov2016knn}, \SERT{} casts entity ranking as a $k$-nearest neighbour problem and uses specialized data structures for retrieval \citep{Kibriya2007knn}. After ranking, \SERT{} outputs the entity rankings as a TREC-compatible file that can be used as input to the \texttt{trec\_eval}\footnote{\url{https://github.com/usnistgov/trec_eval}} evaluation utility.

Apart from entity ranking, the learned representations and model-specific parameters can be extracted conveniently from the models through the interface\footnote{See \texttt{get\_representations} in Snippet~\ref{snippet:model}.} and used for down-stream tasks such as clustering, recommendation and determining entity importance as shown in \citep{VanGysel2017expertregularities}.

\section{Summary}

In this \paper{} we described the \acl{SERT}, a toolkit that learns latent representations of words and entities. The toolkit contains implementations of state-of-the-art entity representations algorithms \citep{VanGysel2016products,VanGysel2016experts} and consists of three components: text processing, representation learning and inference. Users of the toolkit can easily make changes to existing model implementations or contribute their own models by extending an interface provided by the \SERT{} framework.

Future work includes integration with Pyndri \citep{VanGysel2017pyndri} such that document collections indexed with Indri can transparently be used to train entity representations. In addition, integration with machine learning frameworks besides Theano, such as TensorFlow and PyTorch, will make it easier to integrate existing models into \SERT{}.

}
}

}

\backmatter

\renewcommand{\bibsection}{\chapter{Bibliography}}
\renewcommand{\bibname}{Bibliography}
\markboth{Bibliography}{Bibliography}
\renewcommand{\bibfont}{\footnotesize}
\setlength{\bibsep}{0pt}

\bibliographystyle{abbrvnat}
\bibliography{thesis}

\makeatletter\@openrightfalse
\chapter{Samenvatting}

Zoekmachines zijn enorm afhankelijk van methodes gebaseerd op overlappende lexicale termen: methodes die zoekopdrachten en documenten voorstellen als een verzameling van woorden en die de gelijkenis tussen zoekopdracht en een document berekenen aan de hand van zoektermen die exact voorkomen in het document. Tekst---een document of een zoekopdracht---wordt voorgesteld als een zak van de woorden die voorkomen in de tekst, waarbij grammatica of woordvolgorde genegeerd wordt, maar de woordfrequentie behouden blijft. Wanneer de gebruiker een zoekopdracht opgeeft, sorteert de zoekmachine documenten aan de hand van een relevantie score, die onder meer bepaald wordt door de mate van overeenkomst tussen termen die voorkomen in de zoekopdracht en het document. Hoewel methodes gebaseerd op lexicale termen intu\"{i}tief en effectief zijn in de praktijk, steunen ze erg hard op de hypothese dat documenten waarin de zoekopdracht exact voorkomt relevant zijn voor de zoekopdracht, ongeacht de betekenis van de zoekopdracht en de bedoelingen van de gebruiker. Omgekeerd, methodes gebaseerd op lexicale termen veronderstellen dat documenten die geen enkele zoekopdracht term bevatten irrelevant zijn tot de zoekopdracht van de gebruiker. Maar het is bekend dat een hoge graad van overeenkomst op het term-niveau niet noodzakelijk relevantie impliceert, en omgekeerd, dat documenten die geen zoekopdracht termen bevatten wel relevant kunnen zijn. Bijgevolg bestaat er een \emph{woordenschat kloof} tussen zoekopdrachten en documenten die voorkomt als beide verschillende termen gebruiken om dezelfde concepten te beschrijven. Het is het bestrijden van het effect dat voortgebracht wordt door deze woordenschat kloof dat het onderwerp is van deze dissertatie.

In het eerste deel van deze dissertatie formuleren we zoekopdrachten---voor het ophalen van documenten met methodes gebaseerd op lexicale term overeenkomst---van complexe en heterogene tekstuele structuren (zoeksessies en email conversaties) om zo goed mogelijk de informatiebehoefte van de gebruiker te vervullen. In dit scenario is het gebruiken van de volledige tekstuele structuur als zoekopdracht
\begin{inparaenum}[(a)]
	\item computationeel kostbaar gezien de lange lengte van de zoekopdracht en
	\item gevoelig tot het onjuist classificeren van document als irrelevant door de aanwezigheid van vervuilende termen in de tekstuele structuur die niet voorkomen in relevant documenten.
\end{inparaenum}
Zoeksessies bestaan uit een opeenvolging van gebruikersinteracties, resultaten getoond door de zoekmachine en zijn gewoonlijk het teken van een complexe informatiebehoefte van de gebruiker. Het is dan de bedoeling om een tekstuele zoekopdracht te formuleren die de informatiebehoefte kan vervullen en de sessie voldoende beschrijft. In het geval van email conversaties is de informatiebehoefte van de gebruiker minder duidelijk. We richten ons op het specifieke geval waar een binnenkomende email een vraag voor inhoud, bijvoorbeeld een document of een hyperlink, bevat. We willen vervolgens een zoekopdracht formuleren die het correcte item ophaalt uit een verzameling documenten. Het doel van het onderzoek van het eerste deel van deze dissertatie is het formuleren van zoekopdrachten die
\begin{inparaenum}[(a)]
	\item korter zijn en
	\item betere resultaten ophalen dan als we de gehele tekstuele structuur gebruiken als zoekopdracht.
\end{inparaenum}

In het tweede deel van deze dissertatie, vermijden we methodes die gebaseerd zijn op de overeenkomst van lexicale termen tussen zoekopdrachten en documenten. Dit staat ons toe om de woordenschat kloof te vermijden door te vergelijken op basis van semantische concepten in plaats van lexicale termen. Documenten en zoekopdrachten worden voorgesteld als vectoren in een ruimte van lage dimensionaliteit. Een zoekopdracht, die door de gebruiker uitgedrukt wordt in termen, wordt geprojecteerd naar een latente vectorruimte van zoekopdrachten. Vervolgens wordt de latente representatie van de zoekopdracht geprojecteerd naar de latente ruimte van documenten. Documenten worden vervolgens gesorteerd in afnemende volgorde van de gelijkenis van hun latente representatie met de geprojecteerde representatie van de zoekopdracht. We focussen op zoekdomeinen waarvan bekend is dat semantisch vergelijken belangrijk is: entiteiten en nieuwsartikelen.

\chapter{Summary}

Search engines rely heavily on term-based approaches that represent queries and documents as bags of words. Text---a document or a query---is represented by a bag of its words that ignores grammar and word order, but retains word frequency counts. When presented with a search query, the engine then ranks documents according to their relevance scores by computing, among other things, the matching degrees between query and document terms. While term-based approaches are intuitive and effective in practice, they are based on the hypothesis that documents that exactly contain the query terms are highly relevant regardless of query semantics. Inversely, term-based approaches assume documents that do not contain query terms as irrelevant. However, it is known that a high matching degree at the term level does not necessarily mean high relevance and, vice versa, documents that match null query terms may still be relevant. Consequently, there exists a \emph{vocabulary gap} between queries and documents that occurs when both use different words to describe the same concepts. It is the alleviation of the effect brought forward by this vocabulary gap that is the topic of this dissertation.

In the first part of this dissertation, we formulate queries---for retrieval using term-based approaches---from complex and heterogeneous textual structures (search sessions and email threads) so as to fulfill the user's information need. In this scenario, issuing the full textual structure as a query is
\begin{inparaenum}[(a)]
	\item computationally expensive due to the long query length and
	\item prone to classifying documents as false negatives for retrieval due to noisy terms part of the textual structure that do not occur in relevant documents.
\end{inparaenum}
Search sessions consist of a sequence of user interactions, search engine responses and are indicative of a complex information need. The task is then to formulate a textual query to satisfy the overall information need that characterizes the session. In the case of email threads, the explicit information need is less straightforward. We focus on a particular case where an incoming email message is a request for content. The task is then to formulate a query that correctly retrieves the appropriate item from a document repository. Overall, the common goal of the research performed in first part of this dissertation is to formulate queries that are
\begin{inparaenum}[(a)]
	\item shorter and
	\item exhibit more effective retrieval than taking the full textual structure as a query.
\end{inparaenum}

In the second part of this dissertation, we steer away from term-based approaches. This allows us to avoid the vocabulary gap altogether by matching based on semantic concepts rather than lexical terms. Documents and queries are represented by low-dimensional representations in a latent vector space. Term-based queries---entered by the user---are first projected into the low-dimensional query space. Subsequently, the latent representation of the user query is projected to the latent space of documents. Documents are then ranked in decreasing order of similarity between the document's representation and the user query representation. We focus on retrieval domains where semantic matching is known to be important: entities and news articles.

\@openrighttrue\makeatother

\ifdefined\isprint
\else
\includepdf[fitpaper=false, noautoscale, offset=9.178cm -0.293cm]{cover/cover}
\fi

\end{document}